# Higher Order Accurate Space-Time Schemes for Computational Astrophysics – Part I – Finite Volume Methods

By
Dinshaw S. Balsara (dbalsara@nd.edu) Physics and ACMS Departments, University of Notre Dame

**Abstract**

As computational astrophysics comes under pressure to become a precision science, there is an increasing need to move to high accuracy schemes for computational astrophysics. The algorithmic needs of computational astrophysics are indeed very special. The meshes used in computational astrophysics may not be very complicated, however, the computational methods have to face up to other challenges. The methods need to be robust and preserve the positivity of density and pressure. Relativistic flows should remain sub-luminal. Oftentimes, magnetohydrodynamics (MHD) or its more extended variants are needed, with the result that the method should preserve the divergence-free constraint on the magnetic field. The methods should take well to adaptive mesh refinement. All these requirements place additional pressures on a computational astrophysics code, which are usually not felt by a traditional fluid dynamics code. Hence the need for a specialized review on higher order schemes for computational astrophysics.

The focus here is on weighted essentially non-oscillatory (WENO) schemes, discontinuous Galerkin (DG) schemes and PNPM schemes. WENO schemes are higher order extensions of traditional second order finite volume schemes which are already familiar to most computational astrophysicists. They start with the mean value in a zone and build up all the higher order moments of the solution within a zone. At third order, they are most similar to piecewise parabolic method (PPM) schemes, which are also included in this review in order to provide an important point of comparison and contrast with WENO schemes. DG schemes, on the other hand, evolve all the moments of the solution, with the result that they are more accurate than WENO schemes. But this accuracy comes at the price of accommodating to a timestep that decreases with increasing order of accuracy. PNPM schemes occupy a compromise position between WENO and PNPM schemes. They evolve an $N^{th}$ order spatial polynomial, while reconstructing higher order terms up to $M^{th}$ order. As a result, the timestep can be larger.

Time-dependent astrophysical codes need to be accurate in space and time with the result that the high order spatial accuracy has to be utilized in conjunction with high order temporal accuracy. This is realized with the help of SSP-RK (strong stability preserving Runge-Kutta) schemes and ADER (Arbitrary DERivative in space and time) schemes. The most popular approaches to SSP-RK and ADER schemes are also described.

MHD comes with its own unique challenges. For some very good reasons, astrophysicists prefer constrained-transport (CT) schemes. The reconstruction has to be modified to include the divergence-free constraint on the magnetic field. Moreover, multidimensional Riemann solvers have to be designed to provide a truly multidimensional notion of upwinding. The need for constraint-preserving schemes also changes the nature of the DG and PNPM update. Part II of this review will focus on these issues.

The style of this review is to assume that readers have a basic understanding of hyperbolic systems and one-dimensional Riemann solvers. Such an understanding can be acquired from a sequence of prepackaged lectures available from http://www.nd.edu/~dbalsara/Numerical-PDE-Course. We then build on this understanding to give the reader a practical introduction to the schemes described here. The emphasis is on computer-implementable ideas, not necessarily on the underlying theory, because it was felt that this would be most interesting to most computational astrophysicists.

**I) Introduction**

The first overarching goal of this review is to document several higher order methods that can now be applied to simulations in computational astrophysics. In that sense, the review seeks to bring the computational astrophysics community and the higher order numerical methods community closer together. Even this is a daunting task because computational astrophysics has its own inner requirements. For example, for some very good reasons, computational astrophysicists prefer to have mimetic schemes for non-relativistic magnetohydrodynamics (MHD) and relativistic MHD (RMHD). Likewise, astrophysical computations usually involve stiff source terms and non-ideal effects. For that reason, this review has been split into two parts. Part I, which is this review, introduces higher order finite volume methods to the greater computational astrophysics community. Part II, which will be a subsequent review, with present many nuances in constraint preserving schemes along with treatment of stiff source terms to the computational astrophysics community.

The second overarching goal is to show the astrophysics community that astrophysics codes are easy to understand if they are studied from the inside out. In other words, all these computational astrophysical fluid dynamics codes are based on a common core of algorithms. Usually, young computational astrophysicists are taught about a code from the outside in. I.e. they learn what the inputs are and what the outputs ought to be for a specific code; but the inner workings of the code remain a mystery. By understanding the common algorithmic core, the computational astrophysical fluid dynamics codes can be demystified.

The methods presented in this review have been developed in the literature over the last several years. However, this review differs from other reviews because astrophysicists like to minimize mathematical notation and they also like to make the numerical method operationally explicit. This review minimizes the mathematical notation and displays all formulae explicitly, as much as possible. In some instances, making the numerical methods more transparent for astrophysicists has also yielded important innovations and simplifications that are catalogued here. Each useful method is followed by a box that explicitly catalogues the major steps that go into implementing the method. A sequence of pedagogically designed lectures on this topic is also available on the author's website (http://www.nd.edu/~dbalsara/Numerical-PDE-Course).

Because of the scope of this review, we divide this introduction into four parts. The first part focuses on the partial differential equations (PDE) systems of interest in astrophysics, cosmology and relativity. The second part focuses on achieving spatially high order of accuracy for hyperbolic PDE systems. The third part focuses on achieving high order of temporal accuracy. The fourth part gives us some useful preliminaries on hyperbolic systems.

**I.1) Focus on the PDE systems of Interest to Computational Astrophysics**

From its start in the 1970s, computational astrophysics has blossomed into a vibrant field that has been applied to many sub-disciplines of astrophysics, cosmology and numerical relativity. While it would be impossible to make a comprehensive list of all these sub-disciplines, these sub-disciplines include most types of origins questions. Thus, computational astrophysicists simulate the origins of the cosmos through cosmological simulations, the origin of stars and planetary systems around stars, the turbulent environments in molecular clouds and the interstellar medium, accretion processes around stars, compact objects and black holes, convection in stars, nova and supernova explosions and the interaction of neutron stars and black holes to produce gravitational radiation. In all these fields, simulating for the origin and evolution of an astrophysical system entails accurately evolving given initial conditions forward in time with spatial and temporal accuracy. The availability of PetaScale computers and the intended availability of ExaScale supercomputers in the next five years ensures that ever more detailed computations will be undertaken. Furthermore, the presence of ground-based and space-based observational facilities that can measure astrophysical processes with precision puts some pressure on computational astrophysics to move towards becoming a precision science. Astrophysicists have realized that turbulence regulates various astrophysical processes, like star formation, stellar convection and the physics of galactic interstellar medium. Accurately simulating turbulence also requires the use of highly accurate numerical methods.

There is an emerging interest amongst astrophysicists to carry out precise, high-accuracy simulations to support observational projects. Powerful, massively-parallel computers and GPU co-processors also make it possible to invest in computational methods that might be a little more computationally costly but provide a much more precise answer. The differential gain in accuracy per unit increase in computational cost is such as to favor the implementation of high accuracy schemes for computational astrophysics on modern computational architectures. Most astrophysical codes simulate a hyperbolic system with perhaps some additional contributions from parabolic terms or stiff source terms. For that reason too, the focus here is on hyperbolic systems. Most questions about origins in computational astrophysics entail simulating the time-evolution of astrophysical objects. For that reason, we are interested in time-dependent higher order methods for the simulation of hyperbolic systems.

The hyperbolic systems of interest include, but are not restricted to, the Euler equations, the non-relativistic magnetohydrodynamic (MHD) equations, relativistic hydrodynamics (RHD) and relativistic MHD (RMHD). Initial interest focused on the Euler equations (Godunov 1959, van Leer 1974, 1977, 1979, Norman, Wilson and Barton 1980, Roe 1981, Harten 1983, Woodward and Colella 1984, Colella and Woodward 1984, Sweby 1984, Osher and Chakravarthy 1984, Tadmor 1989, Colella 1990, Berger and Colella 1989, Stone and Norman 1992a, Colella and Sekora 2008, McCorquodale and Colella 2011). However, it soon became apparent that the Euler equations were just one specific instance of a hyperbolic system. Appendix A gives useful information about the Euler equations viewed as a hyperbolic system.

Non-relativistic MHD next saw an initial spurt of interest where it was treated as a hyperbolic system (Brio and Wu 1988, Stone and Norman 1992b, Dai and Woodward 1994, Ryu and Jones 1995, Roe & Balsara 1996, Cargo and Gallice 1997, Balsara 1998a,b, Falle, Komissarov and Joarder 1998, Gurski 2004, Li 2005, Crockett *et al.* 2005, Miyoshi and Kusano 2005, Fuchs *et al.* 2011, Chandrashekar and Klingenberg 2016, Winters and Gassner 2016, Winters *et al.* 2017, Dergis *et al.* 2017). The realization that the magnetic field should be divergence-free (Brackbill and Barnes 1980, Brackbill 1985, Brecht *et al.* 1981, Evans and Hawley 1989, DeVore 1991) has prompted a lot of subsequent work in the field of constrained transport (CT) schemes for MHD (Ryu *et al.* 1998, Dai and Woodward 1998, Balsara and Spicer 1999a,b, Balsara 2001a, 2004, 2009, Londrillo and DelZanna 2004, Gardiner and Stone 2005, 2008, Balsara et al. 2009, 2013, Dumbser et al. 2013 Balsara and Dumbser 2015, Xu *et al.* 2016). Recently, the need to achieve multidimensional upwinding has led to the development of multidimensional Riemann solvers that are efficient and easy to implement (Balsara 2010, 2012a, 2014, 2015, Balsara, Dumbser and Abgrall 2014, Vides *et al.* 2015, Balsara *et al.* 2016, Balsara and Nkonga 2017). Appendix B gives useful information about the MHD equations viewed as a hyperbolic system.

Soon after the onset of interest in MHD, there was also a burst of interest in developing higher order Godunov schemes for relativistic hydrodynamics (Eulderink 1993, Balsara 1994, Font *et al.* 1994, Martí and Müller 1994, Eulderink and Mellema 1995, Falle and Komissarov 1996, Aloy *et al.* 1999, Pons *et al.* 2000, Rezzolla and Zanotti 2001, Font 2003, Martí and Müller 2003 Ryu *et al*. 2006). That interest in relativistic hydrodynamics transitioned into a burgeoning interest in numerical relativistic MHD which continues to this day (Anile 1989, Komissarov 1999, Balsara 2001b, Koide *et al*. 2001, Gammie *et al*. 2003, Giacomazzo and Rezzolla 2006, 2007, DelZanna *et al*. 2003, 2007, Noble *et al.* 2006, Komissarov 2006, Mignone and Bodo 2006, Tchekhovskoy et al. 2007, Mignone et al. 2009, Dumbser and Zanotti 2009, Anton et al. 2010, Beckwith and Stone 2011, McKinney *et al.* 2014, Kim and Balsara 2014, Radice, Rezzolla and Galeazzi 2014, Zanotti and Dumbser 2016, White, Stone and Gammie 2016, Balsara and Kim 2016). Higher order schemes and multidimensional Riemann solvers have also been developed for relativistic MHD (RMHD). Appendix C gives some useful pointers for the RHD and RMHD equations.

**I.2) Numerical Methods for Higher Order Spatial Accuracy**

The previous paragraphs have paid due attention to the most important PDE systems of interest in astrophysics. To be sure, there are many further systems of PDEs that will become interesting to astrophysicists in the future. Let us now turn our attention to the solution methodologies. Astrophysicists have been amongst the earliest developers of numerical methods for fluid dynamics (LeBlanc and Wilson 1970, Norman, Wilson and Barton 1980, Hawley, Smarr and Wilson 1984). However, the distinction of being the most prescient developer of fluid dynamics methods falls to Bram van Leer, who started his intellectual life as an astronomer and subsequently left the field! In an intellectual tour de force, van Leer (1974, 1977, 1979) developed a second order accurate extension to a first order accurate method by Godunov (1959). This launched the field of higher order Godunov schemes which have gone on to become the most successful class of methods for numerically treating all manner of hyperbolic systems of partial differential equations (PDEs). van Leer's 1979 paper has been cited over 5000 times at the time of this writing! Higher order Godunov methods offer robust performance over a broad range of physical conditions and for a large number of hyperbolic PDE systems. They do have their pitfalls, but their pitfalls have been well-documented in the literature and suitable fixes that overcome those pitfalls have been devised. For that reason, this review focuses on higher order Godunov schemes. Progress in this field came rapidly on the heels of van Leer's seminal papers. Since Godunov methods rely on Riemann solvers to provide upwinding and entropy-enforcement at discontinuities, a large number of very efficient Riemann solvers have been devised (Rusanov 1961, van Leer 1979, Roe 1981, Harten 1983, Osher and Solomon 1982, Harten, Lax and van Leer 1983, Colella 1985, Einfeldt, 1988, Einfeldt *et al.* 1991, Toro, Spruce and Speares 1994a,b, Batten *et al.* 1997, Liou et al. 1990, Liou and Steffen 1993, Liou 1996, 1998, 2006, Zha and Bilgen 1993, Ismail and Roe 2009, Toro and Vázquez-Cendón 2012, Chandrashekhar 2013, Dumbser and Balsara 2016). It was also soon realized that higher order Godunov methods achieve their stability because they restrict the reconstructed profiles within each zone so as to avoid producing spurious extrema. This gave rise to the emergence of total variation diminishing (TVD) schemes (Harten 1983, Sweby 1984, Tadmor 1989) which used piecewise linear reconstructed profiles within each zone. Inclusion of parabolic reconstruction profiles, instead of linear ones, gave rise to the piecewise parabolic method (PPM) (Woodward and Colella 1984, Colella and Woodward 1984, Colella and Sekora 2008, McCorquodale and Colella 2011). PPM has proved to be very popular with astronomers because it gives reasonably good quality solutions at a modest computational cost.

The PPM method introduced the idea of "reconstruction by primitive" which subsequently formed an integral part of essentially non-oscillatory (ENO) schemes (Harten et al. 1986, Shu and Osher 1988, 1989). ENO schemes provided a pathway to increasingly high orders of accuracy. However, the early ENO schemes had their own deficiencies owing to the sudden shifts in the reconstruction stencil (Rogerson and Meiburg 1990). With the advent of weighted essentially non-oscillatory (WENO) schemes a natural path was found for designing schemes of increasingly order accuracy (Liu, Osher and Chan 1994, Jiang and Shu 1996, Friedrichs 1998, Balsara and Shu 2000, Levy, Puppo and Russo 2000, Deng and Zhang 2005, Käser and Iske 2005, Henrick, Aslam and Powers 2006, Dumbser and Käser 2007, Borges *et al.* 2008, Shu 2009, Gerolymos, Sénéchal & Vallet 2009, Castro *et al.* 2011, Liu and Zhang 2014, Zhu and Qiu 2016, Balsara, Garain and Shu 2016, Cravero and Semplice 2016). WENO schemes will form an important fraction of this review, partly because of their intrinsic interest and partly because of their role as limiters for the DG schemes that we will introduce very shortly. A WENO scheme spatially reconstructs all the moments (except the $0^{th}$ moment) of an $M^{th}$ order polynomial so as to provide $(M+1)^{th}$ order of spatial accuracy.

The observant reader may well ask whether all of these moments can be reasonably reconstructed? The quick answer is that indeed they can be reconstructed. However, one can still ask whether there is a way of evolving all these moments consistent with the dynamics? This is where discontinuous Galerkin (DG) schemes step in because they give us a logical way of evolving all the higher moments in a way that is consistent with the dynamics. Let us consider a simple example that enables us to compare and contrast WENO schemes with DG schemes. A fourth order CWENO (centered WENO) scheme would reconstruct the linear, parabolic and cubic moments at each timestep while evolving only the zone averaged value of the flow variable (i.e. the zeroth moment) at each timestep. On the other hand, a fourth order DG scheme would use a Galerkin projection procedure to develop evolutionary equations for the evolution of not just the zeroth moment, but also the first, second and third moments (i.e., the linear, parabolic and cubic terms in one dimension). These additional evolutionary equations can be designed consistent with the governing PDE (i.e. the underlying dynamics). The reader can now appreciate why a third order DG scheme would be more accurate than its WENO counterpart. This advantage persists at all orders.

Having obtained an intuitive background on DG schemes, let us now document the history of these schemes. DG methods occupy an intermediate place between full Galerkin/Spectral methods which assume that the solution is described by a basis set that extends over the whole domain (think of a Fourier method) and a finite volume TVD/PPM/WENO method which assumes that the solution is specified by slabs of fluid within each zone at the beginning of each timestep. In a DG method, the solution is specified as having a certain number of moments within each zone at the beginning of each timestep. DG schemes were initially invented for solving neutron transport problems (Reid and Hill 1973). Understanding how to incorporate many of the nicer features of finite volume methods in DG schemes took over a decade of development. Cockburn and Shu (1989) made the first breakthrough for scalar hyperbolic conservation laws with the following three advances. First, by endowing each element (zone) with an $M^{th}$ order polynomial, they were able to show that $(M+1)^{th}$ order of spatial accuracy could be achieved. Second, to match the spatial accuracy, they proposed the use of an $(M+1)^{th}$ order Runge-Kutta timestepping for the time evolution. Third, they generalized a slope limiter method to yield TVB (total variation bounded) limiting. Extension to systems and to multiple dimensions came in Cockburn, Lin and Shu (1989) and Cockburn, Hou and Shu (1990) and Cockburn and Shu (1998) where it was realized that fluxes from Riemann solvers could be used at zone boundaries in order to provide upwinding and to stabilize the scheme.

The DG methods have the following four significant advantages which make them attractive for computational astrophysics:- (i) DG methods of arbitrarily high order can be formulated. (ii) DG methods are highly parallelizable. (iii) DG methods can handle complicated geometries. (iv) DG methods take very well to adaptive mesh refinement (AMR). Furthermore, the degree of the approximating polynomial can be easily changed from one element to the other. The former spatial refinement is often referred to as h-adaptivity, where "h" stands for the size of a cell. The adaptivity in the approximating polynomial is referred to as p-adaptivity, where "p" stands for order of the approximating polynomial. As a result, while simpler finite volume methods can undergo h-adaptivity on an AMR mesh, a DG scheme has the potential to undergo hp-adaptivity (Biswas, Devine and Flaherty 1994). For other reviews of DG schemes, see (Cockburn, Shu and Karniadakis 2000, Hesthaven and Warburton 2008). As of this writing, DG schemes have begun to make inroads in computational astrophysics, cosmology and general relativity (Mocz *et al*. 2015, Schaal et al. 2015, Zanotti *et al*. 2015, Teukolsky *et al*. 2016, Kidder *et al*. 2017, Balsara and Käppeli 2017).

We will examine DG schemes in the course of this review. Like all numerical schemes for treating non-linear hyperbolic systems, DG schemes need some form of non-linear limiting. Indeed, the quality of a DG scheme depends strongly on the limiter that is being used. If the limiter is invoked too frequently, it damages the quality of the solution. If the limiter is invoked less than it needs to be invoked, the code develops spurious oscillations that have a negative effect on the solution. Several limiters have been presented over the years (Biswas, Devine and Flaherty 1994, Burbeau, Sagaut, Bruneau 2001, Qiu and Shu 2004, 2005, Balsara *et al*. 2007, Krivodonova 2007, Zhu *et al*. 2008, Xu, Liu & Shu 2009a,b,c, Xu & Lin 2009, Xu *et al.* 2011, Zhu and Qiu 2011, Zhong and Shu 2013, Zhu *et al*. 2013, Dumbser *et al*. 2014). The problem is that there has been no coalescence of consensus around any one particular limiter. For that reason, we will present two viable strategies for limiting DG schemes. The first strategy is based on WENO limiting; it is simple to retrofit into any pre-existing DG code and seems to work well (Zhong and Shu 2013, Zhu *et al*. 2013). This WENO limiter acts in an *a priori* fashion in the sense that the limiter is applied to troubled zones that need limiting *before* (at the beginning of) taking a DG timestep. Since limiters are applied at the beginning of taking a timestep in all other schemes for solving hyperbolic PDEs, this is the traditional style of using limiters. That makes the WENO limiter for DG schemes easy to retrofit into pre-existing codes. The other approach consists of the MOOD (Multi-dimensional Optimal Order Detection) method (Clain et al. 2011, Diot *et al*. 2012, 2013, Dumbser *et al*. 2014). The MOOD limiter is an *a posteriori* limiter in the sense that one initially takes a timestep without invoking any limiter. As a result, some of the zones that should have been limited, will indeed be corrupted by the end of a timestep. *After* the timestep has been taken one identifies the corrupted zones, i.e. the zones where a limiter should have been invoked (but wasn't). Then one tries to backtrack and redo the timestep in those zones that got corrupted. This process of backtracking and redoing can indeed take place more than once. Needless to say, MOOD limiting results in a DG code that is recursive and difficult to implement. The one virtue of MOOD limiting for DG is, however, that one only invokes the limiter in those zones where it is absolutely needed. Unlike the WENO limiter, which may apply more limiting than the absolute minimum that is needed, MOOD limiting will usually apply just the minimum amount of limiting. Since the MOOD limiter is based on heuristics, one cannot however claim that it always applies the minimum amount of limiting. On idealized problems, MOOD limiting for DG schemes has produced charming results.

As the order of accuracy of a DG scheme is increased, the permissible CFL decreases (Zhang and Shu 2005, Liu et al. 2008). The previous two citations showed this for zone-centered DG methods that apply to conservation laws. An analogous reduction in permissible timestep occurs for face-centered DG schemes for constrained transport (CT) of the magnetic field (Yang and Li 2015, Balsara and Käppeli 2017). To give but one example, the permissible CFL number for a DG scheme that is fourth order accurate in space and time can be as small as 0.14! For this reason, we invented PNPM methods (Dumbser *et al*. 2008). The PNPM scheme evolves an $N^{th}$ order spatial polynomial, while reconstructing higher order terms up to $M^{th}$ order. Let us consider fourth order methods as an example. A P0P3 method is fourth order accurate and is effectively a fourth order CWENO scheme with a maximum CFL number of 1.0 in one-dimension. A P1P3 method evolves the zone averaged value as well as the first moment, while reconstructing the second and third moments. It has a maximum CFL number that is comparable to a second order in space and time DG method of 0.33. A P2P3 scheme evolves the zone averaged value as well as the first and second moments, while reconstructing the third moments. It has a maximum CFL number that is comparable to a third order in space and time DG method of 0.17. A P3P3 method is basically a fourth order DG method with a CFL of 0.10 when spatial and temporal accuracies are matched. We see, therefore, that it might be beneficial to use PNPM schemes with N<M. Experience has shown that P1PM or P2PM schemes often give most of the sought-after accuracy of a PMPM scheme. This has been borne out via numerical experiments in Dumbser *et al*. (2008) for conservation laws and in Balsara and Käppeli (2017) for DG schemes for constrained transport of magnetic fields. For this reason, PNPM schemes will also form part of our study.

At least for now, the mesh structures used in computational astrophysics are simple, though there is also an emerging interest in methods that use Voronoi tessellations and Delaunay triangulations in astrophysics (Springel 2010, Vogelsberger *et al*. 2012, Florinski *et al*. 2013, Balsara and Dumbser 2015, Mocz *et al*. 2015, Xu *et al*. 2016). For that reason, we will focus this version of the living review on structured meshes.

**I.3) Numerical Methods for Higher Order Temporal Accuracy**

Unlike the plethora of numerical methods for achieving higher order spatial accuracy, the methods for achieving high order of temporal accuracy are somewhat fewer. The most popular methods these days split into two dominant styles. There are the Runge-Kutta methods and the ADER (Arbitrary DERivative in space and time) methods. We briefly introduce them in the two succeeding paragraphs and we will describe them in detail later on in this review.

Runge-Kutta (RK) methods rely on discretizing the PDE in time in a fashion that is quite similar to the temporal discretization of an ordinary differential equation (ODE). The Runge-Kutta discretization of a time-dependent ODE splits the time evolution into a sequence of stages, each of which is only first order in time. The entire sequence of stages does indeed retain the desired order of temporal accuracy. In a similar fashion, the Runge-Kutta discretization of a time-dependent PDE also splits the time evolution into a sequence of stages. Each individual stage is high order accurate in space, but only first order accurate in time. As before, the entire sequence of stages does indeed retain the designed temporal accuracy. One almost always wants each stage to be non-oscillatory or even TVD. The strong-stability preserving (SSP) variant of RK methods guarantee that if each stage is TVD then the entire scheme will be TVD. As a result, these methods are known as RK-SSP methods. Such methods are available for treating hyperbolic systems without stiff source terms (Shu and Osher 1988, 1989, Shu 1988, Gottlieb *et al*. 2001, Spiteri and Ruuth 2002, 2003, Gottlieb 2005, Gottlieb, Ketcheson and Shu 2011) and also hyperbolic systems with stiff source terms (Pareschi and Russo 2005, Hunsdorfer and Ruuth 2007, Kupka et al. 2011). These methods tend to be popular because each stage is practically identical to the previous stage, resulting in a simple implementation. For that reason, we will describe some of the most popular SSP-RK methods in this review.

While simplicity is the strong suit of RK-SSP methods, many of the steps in a multi-stage RK method are unnecessary. Consider the example of a three stage RK scheme, it requires the reconstruction to be done thrice and also the Riemann solvers to be invoked thrice. ADER schemes present a better alternative where the reconstruction is only done once and the Riemann solvers are invoked a fewer number of times. As a result, ADER schemes are computationally less expensive. Modern ADER schemes derive from two alternative antecedents. On the one hand, there is the generalized Riemann problem (GRP) (van Leer 1979, Ben-Artzi 1989, Ben-Artzi and Birman 1990, Ben-Artzi and Falcovitz 1984, 2003, Qian *et al*. 2014) which seeks to understand the evolution of the Riemann problem when the flow variables on either side of it have linear or quadratic variation in space. One strain of ADER schemes derive from the development of the GRP (Titarev & Toro 2002, 2005, Toro and Titarev 2002, Montecinos *et al*. 2012, Montecinos and Toro 2014). Another strain of ADER schemes derive from the second order Lax-Wendroff procedure (Lax and Wendroff 1960, Colella 1985) and its higher order extensions (Harten *et al*. 1996). Modern ADER schemes that stem from the Lax-Wendroff procedure rely on a very efficient Galerkin projection to iteratively solve the Cauchy problem within each zone (Dumbser *et al*. 2008, Balsara *et al*. 2009, Balsara *et al*. 2013, Dumbser *et al*. 2013, Balsara and Kim 2016). In other words, given all the spatial moments of the reconstruction within a zone up to some level of spatial accuracy, the ADER predictor step tells us how the solution within that zone will evolve forward in time with a comparable accuracy in space and time. Modern ADER schemes of the latter type are easy to implement and converge very fast. Indeed, the methods are provably convergent with or without stiff source terms (Jackson 2017). This makes them much more efficient in comparison to SSP-RK methods (Balsara et al. 2013). For that reason, we will focus on ADER methods in this review.

**I.4) Brief Background on Hyperbolic Systems**

In this review we will be principally interested in the numerical solution of hyperbolic conservation laws of interest to computational astrophysics. We will instantiate our solution methodologies explicitly in two dimensions, because three dimensional extensions follow trivially. Thus consider the *M*-component conservation law

$$U_t + F(U)_x + G(U)_y = 0 \tag{1}$$

Here U is the vector of "*M*" conserved variables and F(U) and G(U) are the corresponding fluxes in the *x* and *y*-directions. The conservation law is hyperbolic for *x*-directional variations if we can write

$$A \equiv \frac{\partial F(U)}{\partial U} = R \, \Lambda \, L \tag{2}$$

Where A is an *M*×*M* characteristic matrix, $\Lambda$ is a diagonal matrix with an ordered set of real eigenvalues and *R* and *L* are a complete set of right and left eigenvectors. For multidimensional problems, we want a similar set of real eigenvalues to exist regardless of the direction in which we analyze the hyperbolic nature of the conservation law. In practical terms, it implies that a similar characteristic decomposition can be made for the matrix $B \equiv \partial G(U)/\partial U$. Eqn. (1) can be discretized in a finite volume fashion on a mesh. Let the mesh be uniform with zones of size $\Delta x$ and $\Delta y$ in the two directions. Let (*i,j*) denote the zone centers of the mesh and (*i+1/2,j*) and (*i,j+1/2*) denote the centers of the x and y-faces of the mesh as shown in Fig. 1. Numerically evolving eqn. (1) entails taking a time step of size $\Delta t$ which takes us from a time $t^n$ to a time $t^{n+1} = t^n + \Delta t$ as

$$\overline{U}_{i,j}^{n+1} = \overline{U}_{i,j}^n - \frac{\Delta t}{\Delta x}\left(\overline{F}_{i+1/2,j}^{n+1/2} - \overline{F}_{i-1/2,j}^{n+1/2}\right) - \frac{\Delta t}{\Delta y}\left(\overline{G}_{i,j+1/2}^{n+1/2} - \overline{G}_{i,j-1/2}^{n+1/2}\right) \tag{3}$$

In eqn. (3) we define the conserved variable $\overline{U}_{i,j}^n$ as a volumetric average over a rectangular zone and the numerical fluxes $\overline{F}_{i+1/2,j}^{n+1/2}$ and $\overline{G}_{i,j+1/2}^{n+1/2}$ as the space-time averages over the faces of the mesh as follows

$$\overline{U}_{i,j}^n \equiv \frac{1}{\Delta x \, \Delta y} \int_{y=-\Delta y/2}^{y=\Delta y/2} \int_{x=-\Delta x/2}^{x=\Delta x/2} U(x,y,t^n) \, dx \, dy \; ;$$

$$\overline{F}_{i+1/2,j}^{n+1/2} \equiv \frac{1}{\Delta t \, \Delta y} \int_{t=t^n}^{t=t^{n+1}} \int_{y=-\Delta y/2}^{y=\Delta y/2} F(\Delta x/2, y, t) \, dy \, dt \; ; \quad \overline{F}_{i-1/2,j}^{n+1/2} \equiv \frac{1}{\Delta t \, \Delta y} \int_{t=t^n}^{t=t^{n+1}} \int_{y=-\Delta y/2}^{y=\Delta y/2} F(-\Delta x/2, y, t) \, dy \, dt \; ;$$

$$\overline{G}_{i,j+1/2}^{n+1/2} \equiv \frac{1}{\Delta t \, \Delta x} \int_{t=t^n}^{t=t^{n+1}} \int_{x=-\Delta x/2}^{x=\Delta x/2} G(x, \Delta y/2, t) \, dx \, dt \; ; \quad \overline{G}_{i,j-1/2}^{n+1/2} \equiv \frac{1}{\Delta t \, \Delta x} \int_{t=t^n}^{t=t^{n+1}} \int_{x=-\Delta x/2}^{x=\Delta x/2} G(x, -\Delta y/2, t) \, dx \, dt$$

$$\tag{4}$$

Recall that the Lax-Wendroff theorem tells us that consistent and stable schemes that are written in conservation form will indeed propagate shocks at the correct physical speed.

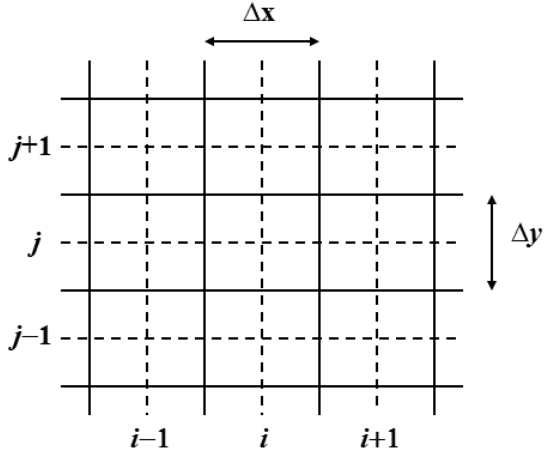

*Fig. 1 showing the zone-centered mesh on which a hyperbolic conservation law is discretized. Conserved variables are collocated at zone-centers, i.e. the intersections of dashed lines. Fluxes are collocated at face-centers, i.e. the intersections of dashed lines with solid lines.*

The prior Introduction has introduced us to the next two important ingredients. We were introduced to the importance of monotonicity preserving reconstruction. Extensive information on monotonicity preserving reconstruction is also given in Chapters 2 and 3 of the author's website. The non-linear hybridization provided by TVD reconstruction is a very good way of getting past the limitations of Godunov's theorem. (Godunov's theorem says that the only linear schemes that can be constructed for monotone advection are indeed first order ones.) The same concept is important for linear hyperbolic systems where the system can be decomposed into characteristic variables. When viewed in characteristic variables, the time-evolution of an $M$-component linear hyperbolic system in one dimension is equivalent to the scalar advection of $M$ characteristic variables as follows

$$w_t^m + \lambda^m w_x^m = 0 \quad \text{with} \quad w^m(x) = l^m \, \mathrm{U}(x) \quad \forall \quad m = 1,..,M \tag{5}$$

Here $\lambda^m$ is the $m^{\text{th}}$ eigenvalue, $l^m$ is the $m^{\text{th}}$ eigenvector and $w^m$ is the $m^{\text{th}}$ characteristic variable. For non-linear hyperbolic systems, we are not quite so lucky. Eqn. (5) is not globally true, even in one dimension. However, we will see that the characteristic decomposition that is available within each zone can be used to make a local version of eqn. (5) which holds true for one time step within a zone. We will see that local characteristic decompositions can be used with good effect in numerical schemes. The monotonicity preserving reconstruction produces jumps at zone boundaries. A physically consistent way of resolving the jumps is through the Riemann problem. The Riemann problem simultaneously gives us an upwinded solution that also satisfies an entropy principle. The dissipation provided by the Riemann problem was seen to be essential for treating discontinuities in conservation laws. Extensive references to the Riemann problem were given in the Introduction, and more information is available in Chapters 4, 5 and 6 of the author's website. Monotonicity preserving reconstruction as well as the Riemann problem will be used as building blocks when constructing successful schemes for the numerical solution of hyperbolic conservation laws.

The present review follows a certain line of development for the solution of hyperbolic conservation laws. The schemes catalogued here are called *higher order*

*Godunov schemes* and are by far the most popular and well-developed solution methodology for this class of problem. Such schemes are robust and can handle shocks of almost any strength. They are relatively fast and work well in multi-dimensions, making them the workhorse of choice. In their essentials, they do not rely on any adjustable parameters, though various means for improving the solution quality are well known. Because these methods have seen extensive development and use, the instances where they have deficiencies are well-known (Quirk 1994) and good workarounds have been developed. There are, however, interesting alternatives that each have their selling points. *Flux corrected transport* schemes (Boris and Book 1976, Zalesak 1981, Oran and Boris 1987) are an interesting forerunner of higher order Godunov schemes that have been used with success for reactive flow. *Central schemes* (Swanson and Turkel 1992, Levy, Puppo and Russo 2000, Kurganov and Tadmor 2000, Kurganov, Noelle and Petrova 2001) use ideas on upwinding from Godunov schemes but bypass the use of the Riemann solver. While their use of a dual mesh increases the programming complexity, bypassing the Riemann problem may be desirable when the Riemann problem is computationally expensive. *Spectral schemes* (Canuto *et al.* 2007, Gottlieb and Orzag 1977) offer high accuracies for problems with simple geometries and boundary conditions. *Compact schemes* (Lele 1994) offer low dispersion error and have proved useful for turbulence research. *Wavelet-based schemes* for solving PDEs rely on the fast wavelet transform (Daubechies 1992). They have also reached a level of maturity where they can adaptively solve certain CFD problems to a desired level of accuracy (Rastigejev & Paolucci 2006, Zikoski 2011). Any such list of worthy numerical methods will always be incomplete, so we beg the reader's indulgence for any omissions.

Many of the popular astrophysics codes have focused on second order of accuracy, though we have often alluded to the advantages of schemes with higher order accuracy. Explaining and understanding a second order scheme is pedagogically simple. As a result, we will briefly open some of the sections in this review with a second order variant of a Godunov scheme. However, robust higher order Godunov schemes that go well beyond second order accuracy are now commonplace. For that reason, we also present methods that go beyond second order. Conceptually, the design and implementation of any scheme that goes beyond second order requires one to pay careful attention to the same set of issues. For this reason, we will instantiate the schemes at third order. A student who understands the issues at third order will find it easy to go beyond third order if needed. The relevant literature base for schemes that go beyond second order is also cited in the text.

It is assumed that the reader is familiar with the eigenstructure of the hyperbolic systems being considered. However, Appendix A gives a thorough discussion of the eigenstructure for the Euler equations. Appendix B gives a similarly thorough discussion of the eigenstructure for the MHD equations and points out some of the nuances in understanding the eigenstructure of this much larger and more complicated hyperbolic system. Appendix C briefly mentions the RHD and RMHD equations and gives pointers to the literature. Usually, an exposure to the eigenstructure for one or two hyperbolic systems is sufficient to give the reader the gist of the idea; and the Euler and MHD equations are the two equations we discuss in Appendices A and B. It is also assumed that the reader has some working familiarity with Riemann solvers. However, Appendix D gives a quick introduction to the HLL Riemann solver. Appendix E gives a practical, implementation-oriented sketch of the HLLI Riemann solver (Dumbser and Balsara 2016), which can indeed be applied to any hyperbolic system with exceedingly good results. Because of its extreme simplicity and generality, as well as its ability to give superb results at a very low computational cost, it is hoped that the HLLI Riemann solver will become a workhorse in computational astrophysics.

This review can be read in different ways depending on the reader's learning goals. If the learning goal is to become familiar with second order schemes, which tend to be simpler, then one can get by with the following Sub-sections: II.1 on TVD reconstruction, IV.1 and IV.2 on second order Runge-Kutta timestepping, V.1 on second order predictor-corrector schemes and VIII for numerical examples. Of course, one should also read the introductory parts of the sections that lead into the above-mentioned sub-sections. The reader who wants to make a quick, first pass through this review may well want to take in just the previously mentioned sub-sections. The rest of Section II as well as all of Section III make a thorough study of PPM and WENO reconstruction strategies. The rest of Sections IV and V give details on making efficient implementations of higher order Runge-Kutta and ADER timestepping respectively. Discontinuous Galerkin schemes also see extensive use in several computational areas and are, therefore, discussed in Section VI. As the emphasis shifts to simulations with greater fidelity, the issues of positivity discussed in Section VII assume greater importance and should be incorporated into codes. Sections VIII aned IX provide accuracy analysis and the results of several stringent test problems that use the methods described here. Section X draws some conclusions.

**II) Reconstructing the Solution for Conservation Laws – Part I, TVD and PPM Reconstruction**

At the beginning of a time step, most higher order Godunov schemes start with a mesh function that is made up of the zone-averaged conserved variables as prescribed on a mesh. The conserved variables are evolved for a timestep using eqn. (3). Taking several timesteps, each of which is bounded by the CFL number, enables us to evolve the conservation law in time. Some higher order Godunov schemes retain and evolve higher order moments of the mesh function within each zone (van Leer 1979, Cockburn & Shu 1989, 1998, Lowrie, Roe and van Leer 1995, Cockburn, Karniadakis and Shu 2000, Qiu and Shu 2004, 2005, Schwartzkopff, Dumbser & Munz 2004, Balsara *et al.* 2007, Dumbser *et al*. 2008, Xu, Liu & Shu 2009a,b). For such schemes, known as *discontinuous Galerkin* schemes, the conserved variables, as well as all their higher moments, are evolved in time. However, in the interest of reducing the memory footprint, most schemes simply idealize the solution as a sequence of slabs of fluid within each zone. The process of endowing these slabs with a meaningful sub-structure is known as the *reconstruction* problem. By reconstructing the solution, we hope to resolve the often contradictory requirements of increasing the order of accuracy of the solution that is represented within each zone while simultaneously preventing the solution from developing spurious oscillations in the vicinity of strong discontinuities. Schemes that rely on reconstruction to endow the mesh function with sub-structure have been studied very extensively in the literature. The happy consequence is that they can be served up as a general-purpose building block for numerical treatment of hyperbolic conservation laws.

In this section we focus on schemes which reconstruct the solution based on the TVD principles; for details, please see Chapter 3 of the author's website. In the next section we will focus on schemes that refrain from truncating local extrema when it is justified. The PPM scheme discussed in this section straddles these two design philosophies since the modern versions of PPM indeed do not truncate local extrema.

## II.1) TVD Reconstruction in Conserved, Primitive or Characteristic Variables

*Piecewise linear (TVD) reconstruction* in the context of linear hyperbolic systems has been explained in detail in Chapter 3 of this author's website. On a two dimensional mesh, like the one shown in Fig. 1, we want the solution vector in each zone (*i,j*) to have a piecewise linear variation in each direction. Consequently, at some time $t^n$ in the zone (*i,j*) we start with a zone-averaged solution vector $\bar{U}_{i,j}^n$, and such solution vectors are specified in all zones. Obtaining a piecewise linear reconstruction in each zone means that we want the mesh function $\{\bar{U}_{i,j}^n\}$ to have linear variation as follows

$$U_{i,j}^n(\tilde{x},\tilde{y}) = \bar{U}_{i,j}^n + \Delta_x \bar{U}_{i,j}\, \tilde{x} + \Delta_y \bar{U}_{i,j}\, \tilde{y} \quad \text{where } \tilde{x} \equiv (x - x_i)/\Delta x \;;\; \tilde{y} \equiv (y - y_j)/\Delta y \quad (6)$$

Here $(x_i, y_j)$ is the centroid of zone (*i,j*) and $(\tilde{x}, \tilde{y}) \in [-1/2, 1/2] \times [-1/2, 1/2]$ are local coordinates that we define in the same zone. The vectors $\Delta_x \bar{U}_{i,j}$ and $\Delta_y \bar{U}_{i,j}$ hold the piecewise linear variation of the mesh function within the zone (*i,j*). The three ways to carry out this piecewise linear reconstruction that are explored in this section are, reconstruction in the conserved variables, reconstruction in the primitive variables and reconstruction in the characteristic variables. Each has its strengths and uses and we catalogue them below.

Reconstruction can be easily enforced componentwise on the conserved variables. For reasons of simplicity, let $\bar{u}_{i,j}^m$ denote the $m^{\text{th}}$ component of the vector $\bar{U}_{i,j}^n$. (The superscript "$n$" from $\bar{U}_{i,j}^n$ is being dropped in $\bar{u}_{i,j}^m$, because the components are only being considered at a given time.) Then piecewise linear *reconstruction of the conserved variables* simply consists of specifying $\Delta_x \bar{u}_{i,j}^m$ and $\Delta_y \bar{u}_{i,j}^m$ in the ensuing formula

$$u_{i,j}^m(\tilde{x}, \tilde{y}) = \bar{u}_{i,j}^m + \Delta_x \bar{u}_{i,j}^m\, \tilde{x} + \Delta_y \bar{u}_{i,j}^m\, \tilde{y} \quad (7)$$

When such a specification is provided for all of the components of $\bar{U}_{i,j}^n$, we say that the solution has been reconstructed. Let "*Limiter* (a,b)" denote any slope limiter, where "a" and "b" are the left and right-biased slopes. (The box at the end of this sub-section provides a smorgasbord of limiters!) The easiest way to achieve our goal is to limit the variation in each of the components of $\bar{U}_{i,j}^n$ as follows

$$\Delta_x \bar{u}_{i,j}^m = \text{Limiter} \left( \bar{u}_{i+1,j}^m - \bar{u}_{i,j}^m, \bar{u}_{i,j}^m - \bar{u}_{i-1,j}^m \right) \quad ; \quad \Delta_y \bar{u}_{i,j}^m = \text{Limiter} \left( \bar{u}_{i,j+1}^m - \bar{u}_{i,j}^m, \bar{u}_{i,j}^m - \bar{u}_{i,j-1}^m \right)$$
(8)

This gives us a piecewise linear reconstruction strategy where the limiter has been applied to the conserved variables. This is the fastest form of limiting.

In some problems, like fluid dynamics, a premium is placed on retaining positive densities and pressures in the reconstruction. In such situations, it helps to reconstruct the profile within a zone using the primitive variables. Let $V_{i,j}^n$ denote the vector of primitive variables that is obtained from the vector of conserved variables $\bar{U}_{i,j}^n$. Let $v_{i,j}^m$ denote the $m^{\text{th}}$ component of the vector $V_{i,j}^n$. *Reconstruction of the primitive variables* is then trivially obtained by setting $\bar{u} \to v$ in eqns. (7) and (8).

For some problems it is very beneficial to resort to piecewise linear *reconstruction of the characteristic variables*. To see this, notice from eqn. (5) that the system decomposes into a set of scalar advection problems only when the problem is decomposed in characteristic variables. Thus limiting on the characteristic variables is conceptually well justified. The other two forms of limiting, i.e. componentwise limiting on the conserved or primitive variables, are not as well justified. Furthermore, different wave families may have different properties; some may be linearly degenerate (e.g. contact discontinuity in Euler flow) while others may be genuinely non-linear (shocks in Euler flow). In order to devise a good solution strategy, different families of waves may have to be limited slightly differently. For example, the profile of a discontinuity in a linearly degenerate wave family may need to be sharpened. This can be accomplished by using a compressive limiter. Because of their tendency to self-steepen, genuinely non-linear wave families do not need any such improvement; consequently, a less compressive limiter might be appropriate for such wave families. However, it is worth recalling that if the hyperbolic system is non-convex, as is the case for MHD and RMHD, the non-linear wave families might give rise to their own further pathologies.

Reconstructing the characteristic variables is a little more intricate. Notice from eqn. (5) that for a linear problem, the left and right eigenvectors as well as the eigenvalues are constant. As a result, the characteristic equation, $w_t^m + \lambda^m w_x^m = 0$, is valid at all points along the *x*-axis. For a nonlinear problem, the eigenvalues as well as the eigenvectors depend on the solution $\bar{U}_{i,j}^n$ within a zone, and they change as the solution changes in time. However, we can still make a local linearization around a given state, and for zone (*i,j*) that state is $\bar{U}_{i,j}^n$. Thus the $m^{\text{th}}$ eigenvalue can be written as $\lambda^m \left( \bar{U}_{i,j}^n \right)$ and the $m^{\text{th}}$ right and left eigenvectors are written as $l^m \left( \bar{U}_{i,j}^n \right)$ and $r^m \left( \bar{U}_{i,j}^n \right)$ respectively. The dependence of the eigenvectors on the solution $\bar{U}_{i,j}^n$, around which we linearize the problem, has been made explicit. Any solution vector, even the ones from the zones that are to the right or left of the zone *(i,j)*, can now be projected into the eigenspace that has been formed by the eigenvectors that are defined at the zone of interest. To make it explicit, please realize that the set of left eigenvectors in zone $(i+1, j)$, given by $\left\{ l^m \left( \overline{U}_{i+1,j}^n \right) : m = 1, ..., M \right\}$, will not be orthonormal with the set of right eigenvectors in zone $(i, j)$, given by $\left\{ r^m \left( \overline{U}_{i,j}^n \right) : m = 1, ..., M \right\}$. Consequently, because of the solution-dependence in the eigenvectors, we realize that each zone defines its own local eigenspace. We want to project the characteristic variables from the neighboring zones in the local eigenspace of the zone that we are considering.

Let us detail the *x*-variation; the *y*-variation can be obtained in an analogous fashion. We describe the process of making a characteristic reconstruction in three easy steps. First, for a TVD reconstruction we only need the two neighboring characteristic variables in addition to the central one. So we can use the left eigenvector $l^m \left( \overline{U}_{i,j}^n \right)$ from the zone *(i,j)* to locally project the characteristic variables in the $m^{\text{th}}$ characteristic field as

$$w_{i,j;L}^m = l^m \left( \overline{U}_{i,j}^n \right) \cdot \overline{U}_{i-1,j}^n \; ; \; w_{i,j;C}^m = l^m \left( \overline{U}_{i,j}^n \right) \cdot \overline{U}_{i,j}^n \; ; \; w_{i,j;R}^m = l^m \left( \overline{U}_{i,j}^n \right) \cdot \overline{U}_{i+1,j}^n \quad \forall \quad m = 1, ..., M \tag{9}$$

The subscripts "*L*", "*C*" and "*R*" refer to the zone that is left of the central zone, the central zone itself and the zone that is right of the central zone. This has to be done for all the characteristic fields in zone $(i, j)$. Second, the local *x*-variation in the $m^{\text{th}}$ characteristic field can now be written as

$$\Delta w_{i,j}^m = Limiter \left( w_{i,j;R}^m - w_{i,j;C}^m, w_{i,j;C}^m - w_{i,j;L}^m \right) \quad \forall \quad m = 1, ..., M \tag{10}$$

This should be done for all the characteristic fields in zone $(i, j)$. Third, the *x*-variation in the mesh function can now be obtained by projecting the variation in the characteristic fields into the local space of right eigenvectors $r^m \left( \overline{U}_{i,j}^n \right)$ in the zone *(i,j)* as follows

$$\Delta_x \overline{U}_{i,j} = \sum_{m=1}^{M} \Delta w_{i,j}^m \; r^m \left( \overline{U}_{i,j}^n \right) \tag{11}$$

Our use of the word "local" in describing eqns. (9) to (11) is intentional. Notice that despite its conceptual elegance, the characteristic limiting described in eqns. (9) to (11) involves matrix-vector multiplies in the first and third steps. If the hyperbolic system is large, these matrix operations can add to the computational complexity. In its defense, however, it is worth pointing out that characteristic limiting usually gives better *entropy enforcement* than componentwise limiting on the conserved or primitive variables. In other words, when the initial conditions have arbitrary discontinuities, those discontinuities will be most rapidly resolved into their entropy-satisfying simple wave solutions if characteristic limiting is used. This completes our description of characteristic limiting for TVD schemes.

It is also useful to point out that the PPM and WENO limiting that follow in the next two sub-sections require larger stencils. In that case, eqn. (9) can be extended to a stencil that includes more than just the immediately neighboring zones. For example, if we have a five zone stencil centered around zone $(i,j)$, we would include the characteristic variables $l^m(\overline{U}_{i,j}^n) \cdot \overline{U}_{i-2,j}^n$ and $l^m(\overline{U}_{i,j}^n) \cdot \overline{U}_{i+2,j}^n$ in eqn. (9).

It is interesting to ask what sort of results we get with the reconstruction schemes catalogued in this sub-section. It is easiest to demonstrate the effect of reconstruction on scalar advection because advection is indeed free of the effects of non-linear terms. To that end, Jiang and Shu (1996) constructed a very useful test problem. It consists of solving the advection equation, $u_t + u_x = 0$, on the interval $[-1,1]$ in periodic geometry. The advected profile is described by

$$\begin{aligned}
u(x, t=0) &= \frac{1}{6}\left[ G(x, \beta, z-\delta) + G(x, \beta, z+\delta) + 4\,G(x, \beta, z) \right] & -0.8 \le x \le -0.6 \\
&= 1 & -0.4 \le x \le -0.2 \\
&= 1 - |10(x - 0.1)| & 0.0 \le x \le 0.2 \\
&= \frac{1}{6}\left[ F(x, \alpha, a-\delta) + F(x, \alpha, a+\delta) + 4\,F(x, \alpha, a) \right] & 0.4 \le x \le 0.6 \\
&= 0 & \text{otherwise}
\end{aligned}$$

Here the functions "F" and "G" are given by

$$F(x, \alpha, a) = \sqrt{\max\left(1 - \alpha^2 (x-a)^2, 0\right)} \quad ; \quad G(x, \beta, z) = e^{-\beta(x-z)^2}$$

The constants in the above equations are given by

$$a = 0.5 \; ; \; z = -0.7 \; ; \; d = 0.005 \; ; \; a = 10 \; ; \; b = \frac{\log 2}{36\,d^2}$$

The problem has several shapes that are difficult to advect with fidelity. From left to right the shapes consist of : 1) a combination of Gaussians, 2) a square wave, 3) a sharply peaked triangle and 4) a half ellipse. It is a stringent test problem because it has a combination of functions that are not smooth and functions that are smooth but sharply peaked. The Gaussians differ from the triangle in that the Gaussians' profile actually has an inflection in the second derivative. A good numerical method that can advect information with a high level of fidelity must be able to preserve the specific features of this problem.

The problem was initialized on a mesh of 400 zones and was run for a simulation time of 10 which corresponds to five traversals around the mesh. In doing so, the features catalogued in the above equations were advected over 2000 mesh points. The problem was run with a CFL number of 0.6. (We will introduce third and fourth order accurate Runge-Kutta time stepping in Section IV.) In all instances, we used a Runge-Kutta time stepping scheme with temporal accuracy that matched the spatial accuracy of the reconstruction strategy.

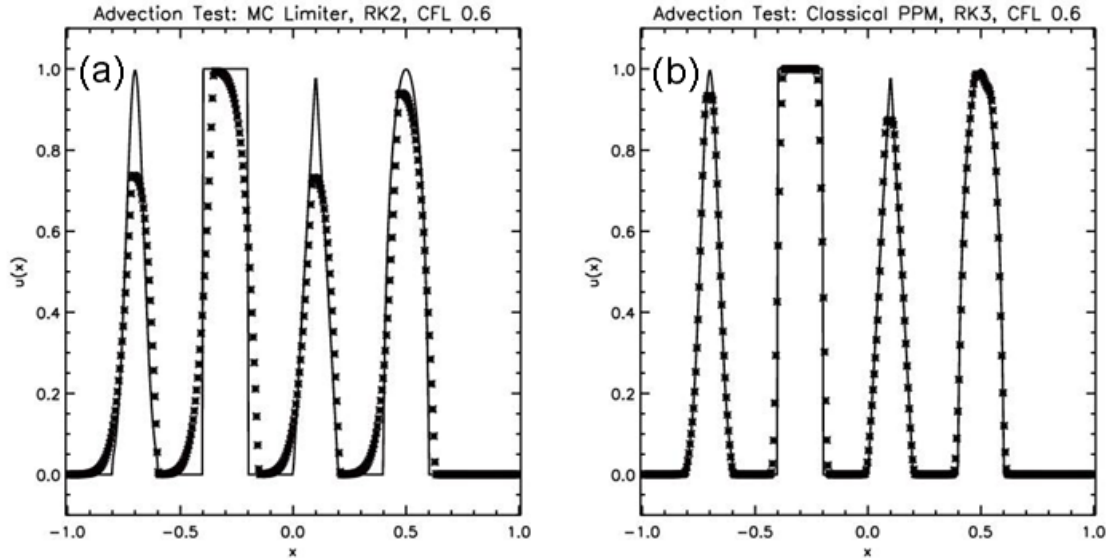

Fig. 2a shows the advection test catalogued in the text when the MC limiter was used with a second order Runge-Kutta scheme. Fig. 2b shows the same when the classical PPM reconstruction was used with a third order Runge-Kutta scheme. The solid line shows the analytic solution, the crosses show the computed solution.

Fig. 2a shows the result for the MC limiter, which yields second order accurate spatial reconstruction, along with a temporally second order accurate Runge-Kutta scheme. The solid line shows the analytic solution, the overlaid crosses show the computed result. Despite the MC limiter being one of the better limiters, we see that the resulting profile shows substantial degradation. None of the profiles has been preserved in such a way that their original shape can be distinguished by the end of the simulation. We also see a strong loss of symmetry in the resulting profiles, which we can understand because the scheme that was used was an upwind-biased scheme. The MC limiter is amongst the best general-purpose TVD limiters, yet we see that the quality of the solution is rather poor. This gives us added motivation to study the better reconstruction strategies in the next few Sections.

**More on Limiters**

It helps to catalogue many of the popularly used limiters here along with their attribution. Thus with *a* and *b* specifying the left and right slopes respectively the slope limiters can be written as

Minmod (Roe 1986):
$$\text{minmod}(a,b) = \frac{1}{2}\left(\text{sgn}(a) + \text{sgn}(b)\right) \min\left(|a|, |b|\right)$$

van Leer (van Leer 1974):
$$\text{vanleer}(a,b) = \left(\text{sgn}(a) + \text{sgn}(b)\right) \frac{a\,b}{|a| + |b|}$$

Monotonized Central (MC)(van Leer 1977):
$$\text{MC}(a,b) = \frac{1}{2}\left(\text{sgn}(a) + \text{sgn}(b)\right) \min\left(\frac{1}{2}|a+b|,\ 2|a|,\ 2|b|\right)$$

$\text{MC}_\beta$:
$$\text{MC}_\beta(a,b) = \frac{1}{2}\left(\text{sgn}(a) + \text{sgn}(b)\right) \min\left(\frac{1}{2}|a+b|,\ \beta|a|,\ \beta|b|\right) \quad 1 \leq \beta \leq 2$$

Superbee (Roe 1986):
$$\text{Superbee}(a,b) = \frac{1}{2}\left(\text{sgn}(a) + \text{sgn}(b)\right) \max\left(\min(2|a|,|b|),\ \min(|a|,2|b|)\right)$$

Sweby (Sweby 1984):
$$\text{Superbee}_\beta(a,b) = \frac{1}{2}\left(\text{sgn}(a) + \text{sgn}(b)\right) \max\left(\min(\beta|a|,|b|),\ \min(|a|,\beta|b|)\right) \quad 1 \leq \beta \leq 2$$

The limiters are given here in a form that is most efficient when implementing them on modern computers with modern languages. Notice that the MC class of limiters have the advantage that they can retrieve the centered slope $(a+b)/2$ when the left and right slopes do not constrain the slope limiting process. The centered slope is the most stable slope that one can provide for smooth variations in the flow. Compared to the left and right slopes, it is also the most accurate slope. The MC class of limiters provide a special advantage over the other limiters in the vicinity of smooth flow because they permit us to retrieve a centered slope. The minmod limiter is the most stable of these limiters in the presence of strong discontinuities, with the *vanLeer* and MC limiters also performing ably on large classes of problems. While the *superbee* limiter by Roe (1986) can produce charming results for certain types of linear advection problems, it can also be a temperamental performer on problems with strong shocks.

Notice that the $\text{MC}_\beta$ limiter reduces to the minmod limiter when $\beta = 1$ and reverts to the MC limiter when $\beta = 2$. One may, therefore, ask what is being controlled by the parameter $\beta$. The ensuing two figures show us the difference between the minmod and MC limiters graphically. The dashed line in both figures shows the mesh function. The solid lines show the reconstructed profiles for the minmod and MC limiters in the figures to the left and right respectively. In this example, the slope produced by the MC limiter is twice as large as the slope produced by the minmod limiter. Without introducing any new extrema, the MC limiter has produced the steeper, i.e. more compressed, profile with smaller jumps at zone boundaries. Consequently, the MC limiter produces sharper profiles.

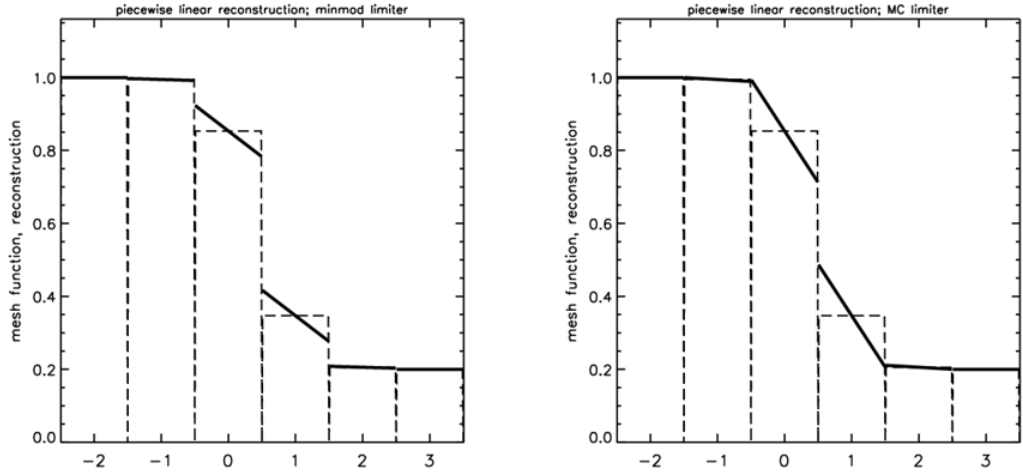

*The dashed lines show the mesh function and the solid lines show the piecewise linear reconstructed profile. The figure to the left was produced with the minmod limiter, the figure to the right was produced with the MC limiter.*

There are several flow features, such as an entropy wave in hydrodynamics or an Alfven wave in ideal MHD or RMHD, where the flow feature should ideally propagate unchanged over very long distances. By using the eigenvectors it is possible to detect where such features occur in the flow. Compressive limiters can be very useful in designing schemes that allow such features to propagate over long distances on a computational mesh without much change. Schemes that pay more attention to the reconstruction problem, like the PPM, WENO or DG schemes offer an even more elegant solution to the problem of accurate advection.

**II.2) Going Beyond Piecewise Linear Reconstruction: Piecewise Parabolic (PPM) Reconstruction**

The desire to improve on piecewise linear reconstruction drove the development of the *piecewise parabolic method* (PPM) (Colella and Woodward 1984, Colella and Sekora 2009, McCorquodale & Colella 2011). An excellent review of PPM has been provided by Woodward (1986) and several stringent test problems for compressible fluid flow have been documented in Woodward and Colella (1984). In this sub-section we document the classical formulation of PPM from Colella and Woodward (1984), while leaving recent extensions (McCorquodale & Colella 2011) for the reader's self-study. It is also interesting to point out that PPM is a forerunner of a class of schemes (Leonard, Lock and MacVean 1995, Suresh and Huynh 1997) that attempt to produce a higher order reconstructed profile within a zone and then use neighboring zones to endow the profile with monotonicity preserving properties.

The PPM method is best illustrated by showing how the reconstructed profile evolves in a set of zones as the steps in the PPM reconstruction procedure are applied to an initial mesh function. To that end, the dotted line in Fig. 3a shows the function $u(x) = 1.2 + \tanh\left((0.65 - x)/0.3\right)$ which mimics a shock profile over the domain $x \in [-2.5, 3.5]$. The domain is spanned by six zones of unit size and the hyperbolic tangent function is shown with the dotted line in Fig. 3a. Let $\bar{u}_{i-2}$, $\bar{u}_{i-1}$, $\bar{u}_i$, $\bar{u}_{i+1}$, $\bar{u}_{i+2}$ and $\bar{u}_{i+3}$ denote the values of the mesh function for the zones that are centered at $x = -2, -1, 0, 1, 2$ and $3$ respectively. We label these zones from "$i-2$" to "$i+3$", and our goal is to demonstrate the steps in the PPM reconstruction especially as they are applied to zone "$i$" which spans $x \in [-0.5, 0.5]$. The mesh function is shown with dashed lines in Fig. 3a. A third order, i.e. parabolic, reconstruction in the $i^{th}$ zone, centered at $x = 0$, is most easily enforced by using Legendre polynomials as follows

$$u_i(x) = \bar{u}_i + \hat{u}_x\, x + \hat{u}_{xx}\left(x^2 - \frac{1}{12}\right) \tag{12}$$

The linear and quadratic Legendre polynomials in the above formula provide the two-fold advantages of orthogonality and a zero average value. As a result, the zone average of $u_i(x)$ over the $i^{th}$ zone is given by $\bar{u}_i$. In PPM, one focuses on the values of the interpolated function at the zone boundaries. Thus for the zone being considered, we have the right and left extrapolated edge values of the parabolic profile defined by $u_{i;R}$ and $u_{i;L}$, see Fig. 3a. Along with the mean value $\bar{u}_i$, these three values uniquely specify the parabolic profile in eqn. (12) so that for each zone we have

$$\hat{u}_x = u_{i;R} - u_{i;L} \quad ; \quad \hat{u}_{xx} = 3\, u_{i;R} - 6\, \bar{u}_i + 3\, u_{i;L} \tag{13}$$

The finite difference-like form of eqn. (13) is readily apparent. One has still to specify $u_{i;R}$ and $u_{i;L}$ at the zone edges with third or better accuracy in order for the reconstruction in eqn. (12) to be third order accurate. We do that next.

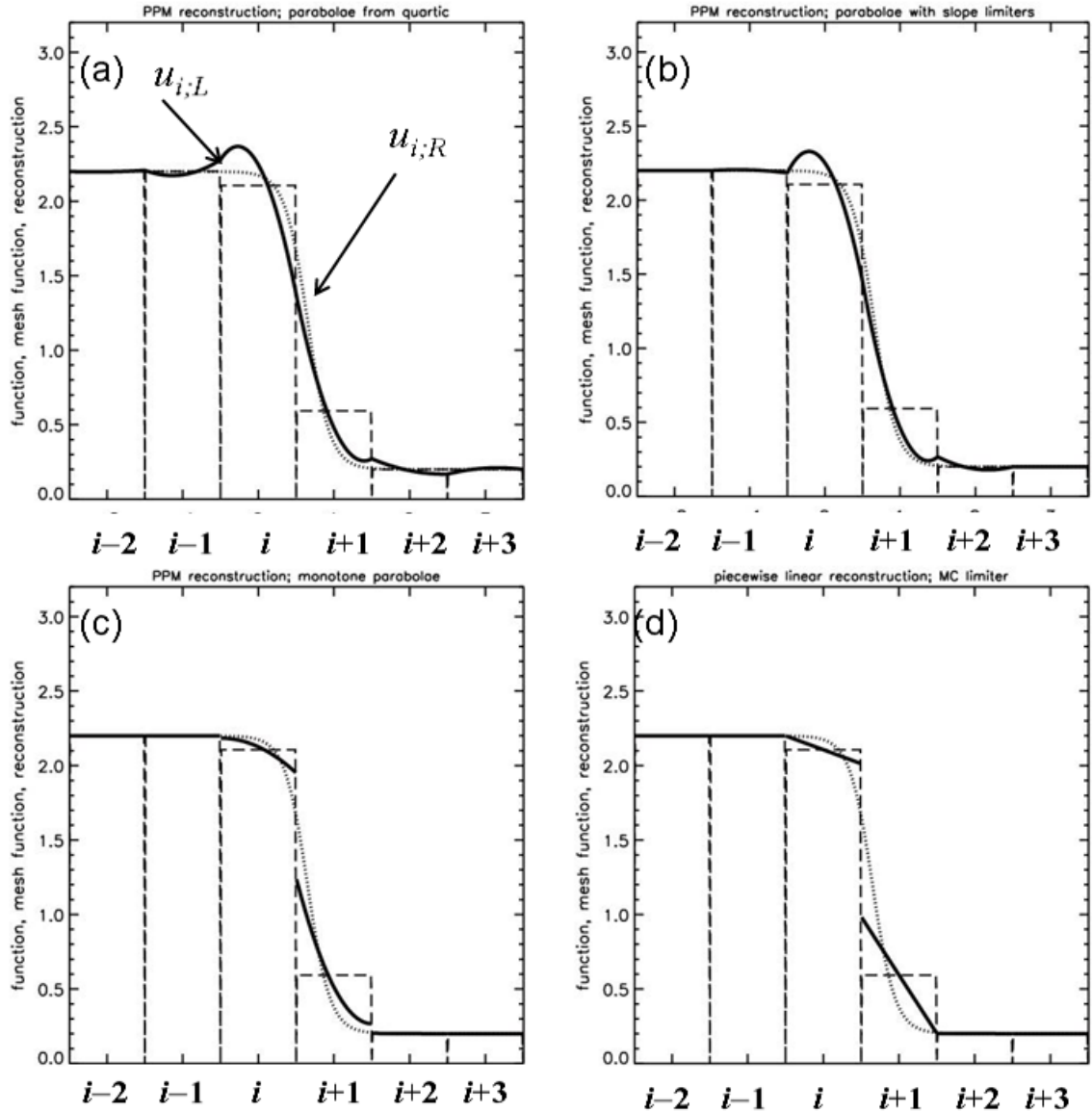

Figs. 3a, 3b and 3c illustrate the steps in the classical PPM reconstruction. The dotted curve shows the original function, the dashed lines show the mesh function and the solid curves show the reconstructed function. Fig. 3a shows the parabolae within each zone that are derived from the original quartic without limiting. Fig. 3b shows the parabolae that use the limited slopes. Fig. 3c shows the final PPM reconstruction with monotonicity preserving parabolae. Fig. 3d shows the piecewise linear reconstruction with MC limiter.

Let us focus on the process of obtaining $u_{i;R}$. In classical PPM we begin by specifying this value with fourth order accuracy. Thus one defines a cubic polynomial $q(x) = q_0 + q_1 x + q_2 x^2 + q_3 x^3$ that spans the domain $x \in [-1.5, 2.5]$, i.e. the zones from "$i-1$" to "$i+2$". The coefficients of the cubic are easily obtained by enforcing the following four consistency conditions

$$\int_{-1.5}^{-0.5} q(x)dx = \bar{u}_{i-1} \;\; ; \;\; \int_{-0.5}^{0.5} q(x)dx = \bar{u}_i \;\; ; \;\; \int_{0.5}^{1.5} q(x)dx = \bar{u}_{i+1} \;\; ; \;\; \int_{1.5}^{2.5} q(x)dx = \bar{u}_{i+2} \quad (14)$$

The above approach is known as *reconstruction via primitive function*. It is the standard method for obtaining higher order reconstructions. The resulting value of $u_{i;R}$ can be easily obtained from $q(x=0.5)$ and we write it in two illustrative formats below.

$$u_{i;R} = \frac{7}{12}(\bar{u}_i + \bar{u}_{i+1}) - \frac{1}{12}(\bar{u}_{i-1} + \bar{u}_{i+2})$$

$$u_{i;R} = \bar{u}_i + \frac{1}{2}(\bar{u}_{i+1} - \bar{u}_i) - \frac{1}{6}(\Delta u_{i+1} - \Delta u_i) \text{ with } \Delta u_{i+1} = \frac{1}{2}(\bar{u}_{i+2} - \bar{u}_i) \quad (15)$$

$$\text{and } \Delta u_i = \frac{1}{2}(\bar{u}_{i+1} - \bar{u}_{i-1})$$

We see that $\Delta u_{i+1}$ and $\Delta u_i$ in eqn. (15) are simply the undivided differences. Formulae similar to the above one can be used to obtain $u_{i-1;R}$, $u_{i+1;R}$ and so on in the adjoining zones. By setting $u_{i;L} = u_{i-1;R}$ and so on, we can specify all the parabolae in all the zones. In other words, we assert that the left extrapolated edge value in one zone is equal to the right extrapolated edge value in the zone to the left of it. The right and left extrapolated edge values, $u_{i;R}$ and $u_{i;L}$, will then be fourth order accurate. The resulting parabolae are shown by the solid curve in Fig. 3a. Fig. 3a shows the parabolae within each zone that have been obtained from the original quartic in eqn. (15) without limiting. These parabolae are only being shown by way of illustration and are never used in classical PPM. We clearly see that the parabolic profiles introduce several new extrema in the reconstructed function, making them an unsuitable starting point for a monotonicity preserving reconstruction. As shown in Fig. 3a, they also do not produce any jumps at the zone boundaries despite the fact that Fig. 3a represents a discontinuous profile. Consequently, a Riemann solver would not generate entropy and help stabilize the reconstructed piecewise parabolic profiles shown by the solid curves in Fig. 3a.

The reconstruction in Fig. 3a introduces too many extrema in several zones, which is unacceptable. The second formula in eqn. (15) suggests a way out. Since $\Delta u_{i+1}$ and $\Delta u_i$ are simply undivided differences, we replace them with the slopes coming from an MC limiter. Thus we get

$$\Delta u_{i+1} = MC(\bar{u}_{i+2} - \bar{u}_{i+1}, \bar{u}_{i+1} - \bar{u}_i) \;\; ; \;\; \Delta u_i = MC(\bar{u}_{i+1} - \bar{u}_i, \bar{u}_i - \bar{u}_{i-1}) \quad (16)$$

Notice that the MC limiter has the property that when the mesh function is smooth, $\Delta u_{i+1}$ and $\Delta u_i$ from eqn. (16) exactly reduce to their centered equivalents in eqn. (15). Consequently, for smooth mesh functions, eqn. (15) will stay fourth order accurate. The slopes from eqn. (16) are used in the second formula in eqn. (15) to yield $u_{i;R}$. Analogous formulae give all the extrapolated right edge values. The extrapolated right edge values can then be used to obtain the extrapolated left edge values by enforcing $u_{i;L} = u_{i-1;R}$ and so on at all the zones. The resulting parabolae are shown by the solid curve in Fig. 3b and we can easily see that they have substantially fewer extrema within the zones compared to Fig. 3a. These parabola, with slopes that have been limited, are used as a starting point for the reconstruction. We see from Fig. 3b that the profiles within each zone do have some extrema. Furthermore, their values do match up at the zone boundaries. These reconstructed profiles would still be unsuitable for use within a higher order Godunov scheme because the Riemann solver relies on the existence of jumps at zone boundaries to introduce the extra dissipation that is needed at shocks. We clearly see from Fig. 3b that monotonicity should be enforced within each zone and, in doing that, we will also obtain the jumps at the zone boundaries that represent discontinuities.

The last step in PPM, therefore, consists of enforcing monotonicity within each zone. For our example profile in Fig. 3b we see that zones "$i$", "$i+1$" and "$i+2$" introduce new extrema in the reconstructed profile. The first, and most natural, enforcement of a monotonicity condition indeed consists of requiring that the zone average $\bar{u}_i$ must stay within $\min(u_{i;L}, u_{i;R})$ and $\max(u_{i;L}, u_{i;R})$. I.e., we require that the parabolic profile should not introduce new extrema. When such a condition is applied to the zone "$i+2$", we see from Fig. 3b that the reconstructed profile would be immediately flattened. This is borne out in Fig. 3c. Thus the first condition for enforcing monotonicity that we apply to all the zones is given by

$$u_{i;L} \to \bar{u}_i \quad \text{and} \quad u_{i;R} \to \bar{u}_i \quad \text{if} \quad (u_{i;R} - \bar{u}_i)(\bar{u}_i - u_{i;L}) \leq 0 \tag{17a}$$

While the above choice is suggested by Colella and Woodward (1984), this author's own preference for the above equation would be

$$u_{i;L} \to \bar{u}_i - \Delta u_i/2 \quad \text{and} \quad u_{i;R} \to \bar{u}_i + \Delta u_i/2 \quad \text{if} \quad (u_{i;R} - \bar{u}_i)(\bar{u}_i - u_{i;L}) \leq 0 \tag{17b}$$

We see, however, that the two zones labeled by "$i$" and "$i+1$" in Fig. 3b would be unaffected by the above condition. These two zones do have new extrema within them that were not present in the original profile. To diagnose the extrema that are introduced in those two zones, we have to realize that eqn. (12) has its extremum at $x_e = -\hat{u}_x/(2\hat{u}_{xx})$. Thus the reason we see a new extremum in the zone "$i$" which is centered at $x = 0$ stems from the fact that $-0.5 < -\hat{u}_x/(2\hat{u}_{xx}) < 0.5$ for that zone. In other words, if one detects the existence of a new extremum within a zone then one should be willing to reduce the curvature, $|\hat{u}_{xx}|$, of the parabola within that zone. I.e., if $x_e$ is negative then reducing $|\hat{u}_{xx}|$ without changing the sign of $\hat{u}_{xx}$ will eventually shift the extremum past $x = -0.5$; if $x_e$ is positive then reducing $|\hat{u}_{xx}|$ without changing the sign of $\hat{u}_{xx}$ will eventually shift the extremum past $x = 0.5$. For the zone "$i$" under consideration, reducing $|\hat{u}_{xx}|$ will immediately cause the maximum or the minimum of the parabola to lie outside (or at the boundary of) the domain [–0.5,0.5]. Colella and Woodward (1984) provide closed-form expressions that detect when the curvature needs to be reduced for a parabolic profile. When such a reduction in the curvature is deemed necessary, they also provide explicit formulae for reducing the curvature by modifying one or the other edge extrapolated states. We repeat those formulae here. Consequently, the second condition for enforcing monotonicity, which is also applied to each of the zones, is given by

$$u_{i;L} \to 3\,\bar{u}_i - 2\,u_{i;R} \quad \text{if } \left(u_{i;R} - u_{i;L}\right)\left(\bar{u}_i - \frac{1}{2}\left(u_{i;R} + u_{i;L}\right)\right) > \frac{\left(u_{i;R} - u_{i;L}\right)^2}{6}$$

$$u_{i;R} \to 3\,\bar{u}_i - 2\,u_{i;L} \quad \text{if } -\frac{\left(u_{i;R} - u_{i;L}\right)^2}{6} > \left(u_{i;R} - u_{i;L}\right)\left(\bar{u}_i - \frac{1}{2}\left(u_{i;R} + u_{i;L}\right)\right)$$

(18)

Once the above two conditions are applied at each of the zones, we see from the solid curve in Fig. 3c that all the zones have a monotone, piecewise parabolic profile. By comparing Figs. 3b and 3c, one can even observe that the maximum of the $i^{\text{th}}$ zone has been shifted to its left boundary while the minimum of the $(i+1)^{\text{th}}$ zone is shifted to its right boundary. For each zone, we can use the extrapolated right and left edge values along with the zone average in eqn. (13) to obtain the final, reconstructed parabolic profile, i.e. eqn. (12).

    Notice that the final piecewise parabolic reconstruction in Fig. 3c has jumps at the zone boundaries that represent a discontinuity. If the discontinuity represents a jump in a linearly degenerate wave field then it is desirable to minimize the jumps, and therefore the dissipation, at zone boundaries. Fig. 3d shows the piecewise linear profile that one obtains by applying an MC limiter to the same mesh function. We see that the jumps at zone boundaries in Fig. 3d are much larger than those in Fig. 3c. As a result, PPM represents contact discontinuities in fluid flow much better than its piecewise linear cousins. If the discontinuity is a shock then the Riemann solver will be able to introduce additional dissipation to stabilize the shock. By virtue of its being a monotonicity preserving scheme, PPM does indeed introduce the requisite jumps in flow variables at zone boundaries. However, for a strong shock, the jumps at zone boundaries may be less than the amount that is needed to fully stabilize the shock. As a result, proper treatment of a strong shock in PPM requires a flattener algorithm (Colella and Woodward 1984). By locally detecting the existence of a shock and flattening the flow profiles at the shock, one increases the jumps at zone boundaries and, therefore, the local dissipation. We will learn more about this in the next section. This completes our description of PPM reconstruction.

    Fig. 2b shows the result from our advection test when classical PPM reconstruction was used. Since PPM nominally produces a third order accurate reconstruction for smooth flow, it was used along with a temporally third order accurate

Runge-Kutta time stepping scheme. We clearly see a substantial improvement in Fig. 2b relative to Fig 2a, which shows that an investment in good reconstruction strategies pays rich dividends. The Gaussian, triangle and elliptical profiles can be clearly distinguished from each other. The top of the ellipse does show some upwind biasing. The square wave is crisply represented with few zones across its boundaries.

> **Implementing PPM Reconstruction**:
> The steps for carrying out PPM reconstruction are as follows:
> <u>Step 1</u>: Construct the limited slopes using eqn. (16) in each zone. Use them in the second equation in eqn. (15) to obtain $u_{i;R}$. Set $u_{i+1;L} = u_{i;R}$, i.e. in this step the reconstruction does not introduce discontinuities at the zone boundaries.
> <u>Step 2</u>: Reset $u_{i;R}$ and $u_{i;L}$ in each zone by applying eqns. (17) and (18) in that sequence. This does introduce discontinuities at zone boundaries.
> <u>Step 3</u>: Use $\bar{u}_i$, $u_{i;R}$ and $u_{i;L}$ within each zone to obtain the coefficients $\hat{u}_x$ and $\hat{u}_{xx}$ from eqn. (13). Eqn. (12) then gives the final, reconstructed piecewise parabolic profile within each zone.

## III) Reconstructing the Solution for Conservation Laws – Part II, WENO Reconstruction

The previous section has shown us that reconstructing the solution from a given mesh function is an intricate problem and can have a great deal of bearing on the quality of our numerical solution. In his early paper, van Leer (1979) had anticipated that it might be possible to reconstruct the solution with better than second order accuracy leading to schemes that go beyond second order. Indeed, the PPM scheme of Colella & Woodward (1984) was a step in that direction. The original PPM scheme was restricted to second order accuracy by the use of a monotonicity preserving limiter (Woodward 1986) and subsequent variants of PPM, see McCorquodale & Colella (2011), represent an effort to go beyond second order accuracy. We, therefore, see that the limiters that provide stability at discontinuities by enforcing the TVD property also restrict the accuracy of the numerical method. The limiter simply clips local extrema and, when such a limiter is applied at every time step in a long-running simulation, it degrades the accuracy of the method.

*Essentially non-oscillatory* (ENO) schemes represent an effort to go beyond second order by totally circumventing the harsher effects of TVD limiting. They are based on the realization that in order to avoid clipping extrema and thus degrading the accuracy, one has to accept a reconstruction strategy that may introduce local extrema within a zone as long as no new oscillations are introduced and as long as the solution remains numerically stable. The original ENO schemes were formulated as finite volume methods in Harten et al. (1987) and efficient finite difference versions of the same were provided in Shu & Osher (1988, 1989). The finite difference formulations have the advantage of speed when applied to uniform (or smooth), structured meshes. The finite volume schemes, while somewhat slower, are more versatile and can take well to a wide variety of structured or unstructured meshes, including adaptive meshes that change to accommodate a changing solution. Unlike TVD and PPM schemes, all of which were formulated in a small number of papers, there have been a few generations or ENO-type schemes where each generation improved on the deficiencies of the previous generation. The *weighted essentially non-oscillatory* (WENO) schemes that see modern use stem from the work of Liu, Osher & Chan (1994) and Jiang & Shu (1996). WENO schemes are especially suitable for problems that simultaneously contain strong discontinuities along with complex, smooth solution features. Finite difference WENO schemes have been formulated that go up to eleventh order in Balsara & Shu (2000). Efficient finite volume formulations of WENO reconstruction are now available for structured meshes (Balsara *et al.* 2009, 2013) and unstructured meshes (Friedrichs 1998, Hu & Shu (1999), Dumbser & Käser 2007, Zhang & Shu 2009). For a superb review of WENO schemes, see Shu (2009). As with PPM, WENO reconstruction methods work well in strong shock situations if coupled with a good flattener algorithm (Colella & Woodward 1984, Balsara *2012b*). We will introduce flatteners in Section VII. *Compact WENO* schemes which minimize the dispersion error (Lele 1992) have also been formulated for simulating high Mach number turbulence (Zhang, Jiang & Shu 2008). Shu (2009) has catalogued a plethora of science and engineering problems where WENO schemes have been used with great success.

**III.1) Weighted Essentially Non-Oscillatory (WENO) Reconstruction in One Dimension**

We have seen that the minmod slope limiter selects the limited slope either by looking to the left of a zone or by looking to the right of a zone. In other words, we may think of a zone and its neighbor to the left as providing a left-biased stencil and the same zone along with its neighbor to the right as providing a right-biased stencil. Either of the two stencils can, in principle, provide a second order accurate reconstruction in the central zone and the minmod limiter chooses the stencil with the smaller slope. WENO reconstruction takes this concept a lot further by carrying out a very sophisticated analysis of the solution that is available on all the possible stencils. We have also seen that the minmod slope limiter achieves its stability via non-linear hybridization, i.e. the final slope is a strongly non-linear function of the right and left-biased slopes. WENO schemes also achieve their stability via non-linear hybridization, the only difference being that a more refined process is used for achieving the non-linear hybridization. So, to summarize, WENO schemes carry out a much more sophisticated stencil analysis along with a more refined non-linear hybridization.

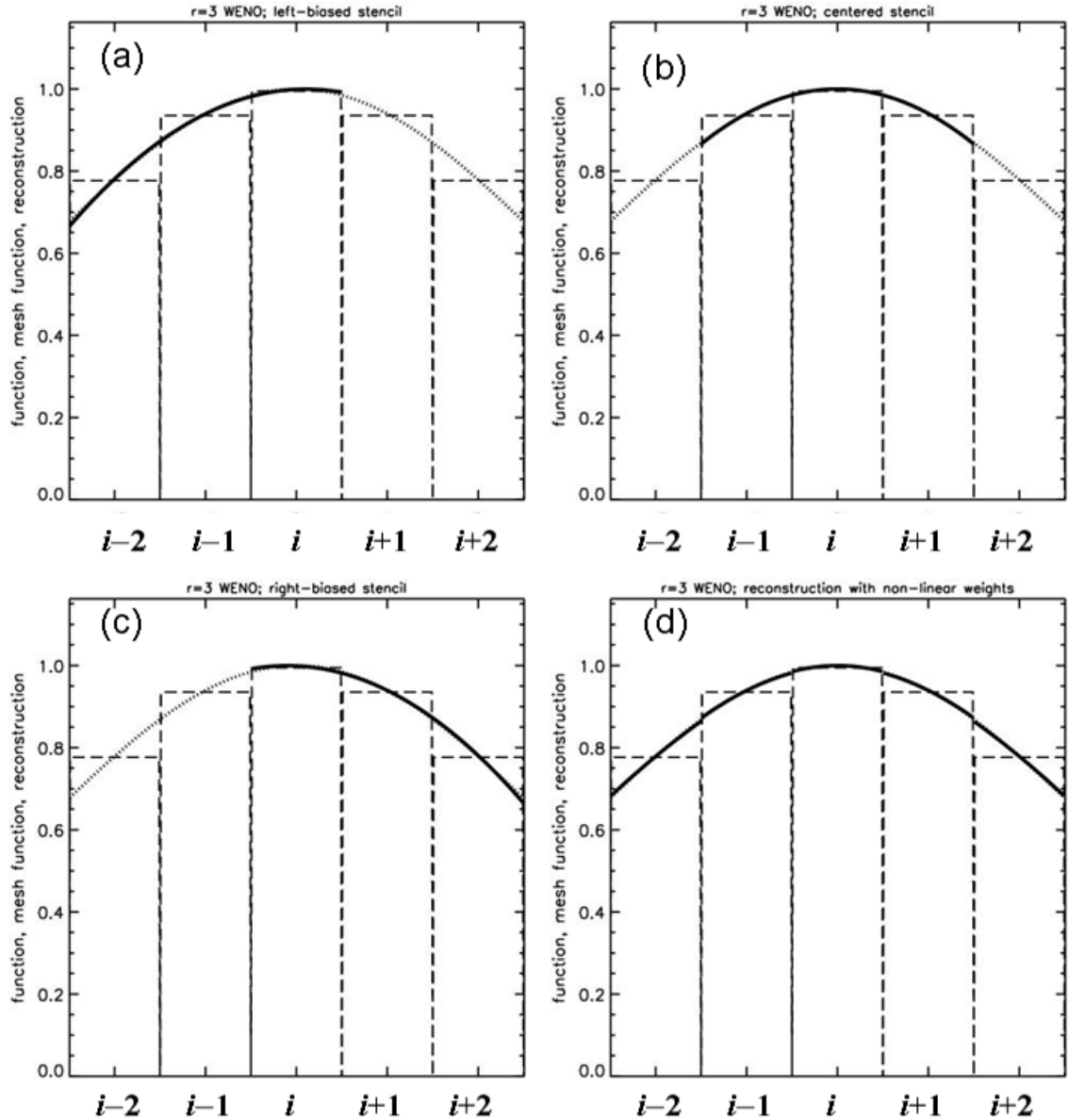

*Figs. 4a, 4b and 4c show the WENO reconstruction from the left-biased, centrally-biased and right-biased stencils for the central zone of the Gaussian profile. Fig. 4d shows the WENO reconstruction in all the zones with non-linear weights. The dotted curve shows the original function, the dashed lines show the mesh function and the solid curves show the reconstructed function.*

The easiest way to introduce WENO reconstruction is by relying on a couple of visually motivated examples in one dimension. Thus Fig. 4 introduces the process of reconstructing the Gaussian function $u(x) = e^{-(x/4)^2}$ while Fig. 5 does the same for the hyperbolic tangent function that was used in the previous section. The dotted lines in Figs. 4 and 5 show the original function. We consider a five zone mesh spanning the domain $x \in [-2.5, 2.5]$, where all the zones have unit extent. We label these zones from "$i-2$" to "$i+2$", and our goal is to demonstrate the steps in the WENO reconstruction as they are applied to zone "$i$". We are interested in the third order accurate WENO reconstruction within the central zone which spans $x \in [-0.5, 0.5]$. The mesh functions for each of the two profiles being considered are shown in Figs. 4 and 5 with dashed lines. We can see that the Gaussian is represented by a smoothly-varying function on the mesh while the shock is represented as a discontinuity. Let $\bar{u}_{i-2}$, $\bar{u}_{i-1}$, $\bar{u}_i$, $\bar{u}_{i+1}$ and $\bar{u}_{i+2}$ denote the values of the mesh function for the zones that are centered at $x = -2, -1, 0, 1$ and $2$ respectively. A third order, i.e. quadratic, reconstruction in the central zone is most easily enforced by using Legendre polynomials as follows

$$u_i(x) = \bar{u}_i + \hat{u}_x x + \hat{u}_{xx}\left(x^2 - \frac{1}{12}\right) \tag{19}$$

The central zone is the zone "$i$" and it is taken to be centered at $x = 0$. As with PPM, the linear and quadratic Legendre polynomials in the above formula provide the dual advantages of orthogonality and a zero average value. Higher order extensions as well as multidimensional extensions of eqn. (19), with the same nice orthogonality property, are given in Balsara et al. (2009, 2013) and Balsara, Garain and Shu (2016). The problem of reconstructing the solution consists of arriving at a properly limited specification of $\hat{u}_x$ and $\hat{u}_{xx}$, i.e. the first and second moments of eqn (19).

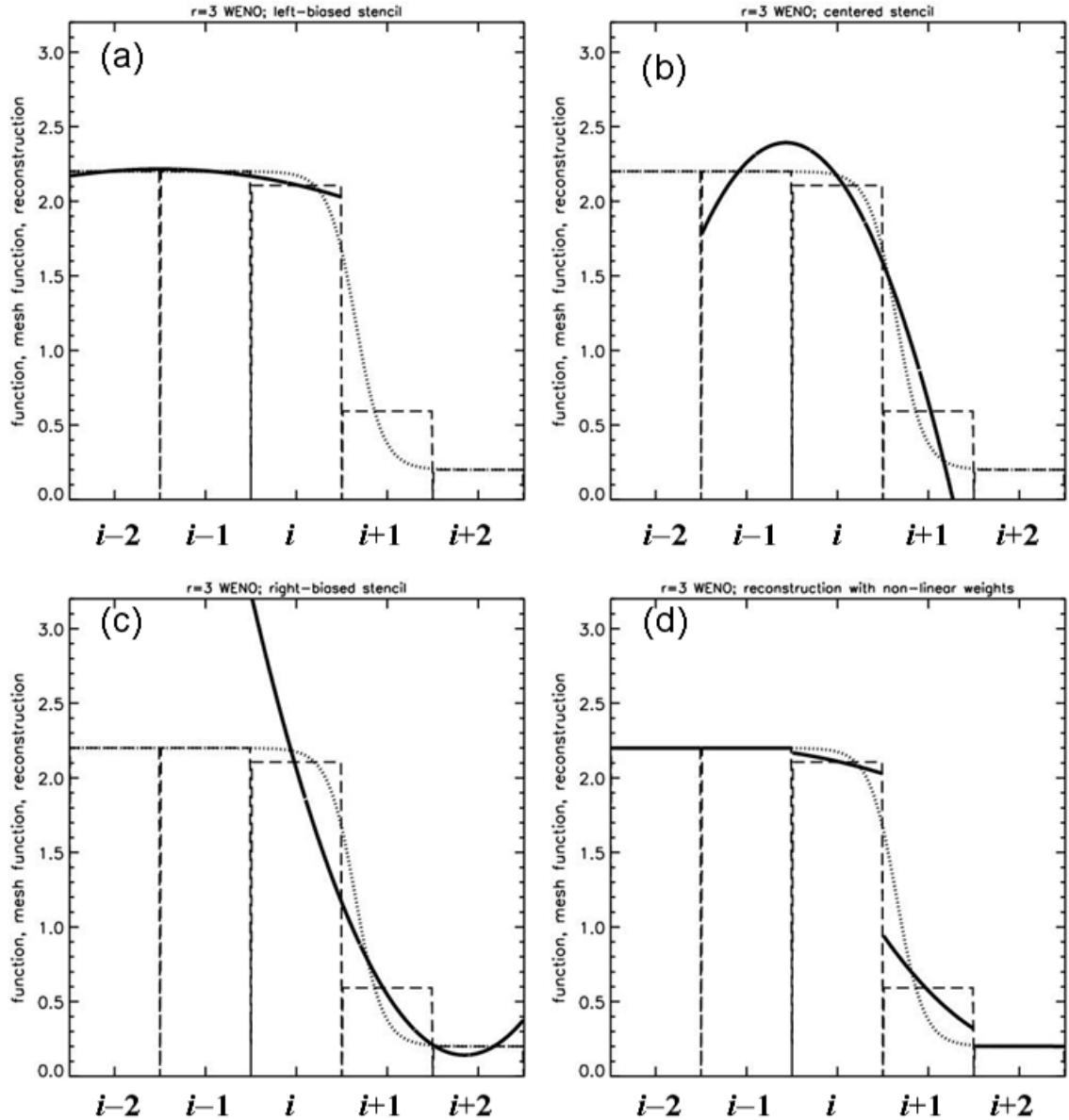

*Figs. 5a, 5b and 5c show the WENO reconstruction from the left-biased, centrally-biased and right-biased stencils for the central zone of the shock profile. Fig. 5d shows the WENO reconstruction in all the zones with non-linear weights. The dotted curve shows the original function, the dashed lines show the mesh function and the solid curves show the reconstructed function.*

Just as piecewise linear TVD reconstruction relied on examining two stencils, each with a width of two zones, piecewise quadratic reconstruction consists of looking at three possible stencils, each of which has a width of three zones. Since we focus on the reconstruction in the central zone of Figs. 4 and 5, we only choose stencils that completely cover the zone of interest. Thus we have a left-biased stencil which spans the interval $x \in [-2.5, 0.5]$ and depends on the zones $\{i-2, i-1, i\}$. The left-biased reconstruction is specified by the quadratic polynomial

$$u_{i;L}(x) = \bar{u}_i + \hat{u}_{L;x}\, x + \hat{u}_{L;xx}\left(x^2 - \frac{1}{12}\right) \tag{20}$$

The left-biased reconstruction is obtained by enforcing the following consistency conditions (i.e. a reconstruction via primitive function):

$$\int_{-2.5}^{-1.5} u_{i;L}(x)\,dx = \bar{u}_{i-2} \;;\; \int_{-1.5}^{-0.5} u_{i;L}(x)\,dx = \bar{u}_{i-1} \tag{21}$$

$$\Rightarrow \hat{u}_{L;x} = -2\,\bar{u}_{i-1} + 0.5\,\bar{u}_{i-2} + 1.5\,\bar{u}_i \;;\; \hat{u}_{L;xx} = 0.5\,\bar{u}_{i-2} - \bar{u}_{i-1} + 0.5\,\bar{u}_i$$

In other words, we require that the reconstructed polynomial correctly represents each of the three zone-averaged values in the left-biased stencil. We see that the conditions in eqn. (21) fully determine the coefficients in the left-biased reconstruction in eqn. (20). The solid curve in Fig. 4a shows the left-biased reconstruction for the Gaussian profile. Since the Gaussian is very smooth, we see that the left-biased reconstruction approximates it quite well. Fig. 5a shows the same for the shock profile. In this case, the left-biased reconstruction is also non-oscillatory within the zone of interest. Notice too, that some structure is still retained within the central zone despite there being a discontinuity at that zone. We realize, therefore, that if the final reconstruction approximates the reconstructed profile from the left-biased stencil most closely in Fig. 5, we will get a properly upwinded reconstruction that is also non-oscillatory.

The centrally-biased stencil spans the interval $x \in [-1.5, 1.5]$ and depends on the zones $\{i-1, i, i+1\}$. The centrally-biased reconstruction is specified by

$$u_{i;C}(x) = \bar{u}_i + \hat{u}_{C;x}\, x + \hat{u}_{C;xx}\left(x^2 - \frac{1}{12}\right) \tag{22}$$

The centrally-biased reconstruction is obtained by enforcing the following consistency conditions

$$\int_{-1.5}^{-0.5} u_{i;C}(x)\,dx = \bar{u}_{i-1} \;;\; \int_{0.5}^{1.5} u_{i;C}(x)\,dx = \bar{u}_{i+1} \tag{23}$$

$$\Rightarrow \hat{u}_{C;x} = 0.5\,(\bar{u}_{i+1} - \bar{u}_{i-1}) \;;\; \hat{u}_{C;xx} = 0.5\,\bar{u}_{i-1} - \bar{u}_i + 0.5\,\bar{u}_{i+1}$$

The solid curves in Figs. 4b and 5b show the centrally-biased reconstruction for the Gaussian and shock profiles. As before, we see that the Gaussian is approximated very well by the central stencil. In fact, the central stencil is also the one which endows maximal stability and accuracy for smooth flow. As a result, the Gaussian example has shown us that our reconstruction should have the property that it gravitates to the central stencil when the mesh function is smooth. We see, however, that the centrally-biased stencil does a very poor job of reconstructing the shock's profile. Indeed it introduces a spurious extremum, with the result that its influence on the final reconstruction should be strongly suppressed.

The right-biased stencil spans the interval $x \in [-0.5, 2.5]$ and depends on the zones $\{i, i+1, i+2\}$. The right-biased reconstruction is specified by

$$u_{i;R}(x) = \bar{u}_i + \hat{u}_{R;x} \, x + \hat{u}_{R;xx} \left( x^2 - \frac{1}{12} \right) \tag{24}$$

The right-biased reconstruction is obtained by enforcing the following consistency conditions

$$\int_{0.5}^{1.5} u_{i;R}(x) \, dx = \bar{u}_{i+1} \; ; \; \int_{1.5}^{2.5} u_{i;R}(x) \, dx = \bar{u}_{i+2} \; ; $$
$$\Rightarrow \hat{u}_{R;x} = -1.5 \, \bar{u}_i + 2 \, \bar{u}_{i+1} - 0.5 \, \bar{u}_{i+2} \; ; \; \hat{u}_{R;xx} = 0.5 \, \bar{u}_i - \bar{u}_{i+1} + 0.5 \, \bar{u}_{i+2} \tag{25}$$

The solid curves in Figs. 4c and 5c show the right-biased reconstruction for the Gaussian and shock profiles. We see that the Gaussian is approximated quite well by the right-biased stencil. Given our comments on stability and accuracy, we realize that it is best to gravitate to the central stencil despite the fact that all the three stencils produce an almost equally good reconstruction for the Gaussian profile. As expected, the shock profile is approximated very poorly by the right-biased stencil. Consequently, for the shock profile, best safety lies in relying predominantly on the left-biased stencil.

The previous three paragraphs have brought us to the realization that the choice of stable stencil depends on analyzing the smoothness properties of the reconstructed polynomial in the zone of interest. In other words, the stencil should be chosen in a solution-dependent fashion. Just as the minmod slope limiter chooses the stencil with the smallest slope, our estimation of the smoothness of each our three stencils should depend on the moments of the three reconstructed polynomials in eqns. (20), (22) and (24). Since the quadratic reconstruction can have a non-zero second derivative, the first and second derivatives should both participate equally in constructing a measure of the smoothness of a reconstruction. This prompted Jiang and Shu (1996) to build *smoothness indicators* for the reconstruction. (For a fourth order accurate WENO scheme, the smoothness indicators would include the third derivatives, and so on.) To take the example of the left-biased stencil, we define its smoothness indicator as

$$IS_L = \int_{-0.5}^{0.5} \left[ \left( \frac{d \, u_{i;L}(x)}{dx} \right)^2 + \left( \frac{d^2 \, u_{i;L}(x)}{dx^2} \right)^2 \right] dx \; \Rightarrow \; IS_L = \hat{u}_{L;x}^2 + \frac{13}{3} \hat{u}_{L;xx}^2 \tag{26}$$

Similar definitions for the other two stencils yield

$$IS_C = \hat{u}_{C;x}^2 + \frac{13}{3} \hat{u}_{C;xx}^2 \qquad \text{and} \qquad IS_R = \hat{u}_{R;x}^2 + \frac{13}{3} \hat{u}_{R;xx}^2 \qquad (27)$$

We see from eqn. (26) that "smoothness indicator" might be something of a misnomer since a higher value for the smoothness indicator implies that the stencil under consideration actually produces larger first and second derivatives, i.e. it is less smooth. However, the nomenclature is well-established in the literature and we accept it as it is.

Scanning Fig. 4, we see that all three stencils should have similar smoothness indicators for the Gaussian problem. In such a situation, it is not advisable to pick the single stencil that has the lowest value of the smoothness indicator because even the tiniest changes in the smoothness indicator can cause stencils to discretely switch back and forth from one time step to the next, thereby producing numerically generated oscillations (Rogerson and Meiburg 1990). A better strategy would be to blend (i.e. make a convex combination of) all the available stencils while giving the central stencil a much higher weight when all the smoothness indicators are roughly equal. Fig. 5, for the shock problem, shows that the left-biased stencil has much smaller first and second derivatives compared to the centrally-biased and right-biased stencils. Consequently, it should have a much smaller smoothness indicator than the other two stencils. In order to pick out the left-biased stencil for the shock problem, we need to weight the stencils in inverse proportion to their smoothness indicators. Economical strategies that accomplish all this do exist. The *non-linear weights*, $\bar{w}_L$, $\bar{w}_C$ and $\bar{w}_R$ are given by

$$w_L = \frac{\gamma_L}{(IS_L + \varepsilon)^p} \; ; \; w_C = \frac{\gamma_C}{(IS_C + \varepsilon)^p} \; ; \; w_R = \frac{\gamma_R}{(IS_R + \varepsilon)^p} \; ;$$

$$\bar{w}_L = \frac{w_L}{w_L + w_C + w_R} \; ; \; \bar{w}_C = \frac{w_C}{w_L + w_C + w_R} \; ; \; \bar{w}_R = \frac{w_R}{w_L + w_C + w_R} \qquad (28)$$

Here $\varepsilon$ is a small number, which may be solution-dependent, and is usually set to $10^{-12}$. The coefficients $\gamma_L$, $\gamma_C$ and $\gamma_R$ are referred to as the *linear weights*. Once the non-linear weights are obtained from eqn. (28), the final reconstructed profile in eqn. (19) is given by

$$\hat{u}_x = \bar{w}_L \hat{u}_{L;x} + \bar{w}_C \hat{u}_{C;x} + \bar{w}_R \hat{u}_{R;x} \quad ; \quad \hat{u}_{xx} = \bar{w}_L \hat{u}_{L;xx} + \bar{w}_C \hat{u}_{C;xx} + \bar{w}_R \hat{u}_{R;xx} \qquad (29)$$

There is some flexibility in the specification of the linear weights and they are usually specified based on the goals that one wants to accomplish. In the next two paragraphs we catalogue some of the popular choices for the linear weights.

For finite volume WENO schemes, it is best to aim for greater stability. One approach (Friedrichs 1998, Levy, Puppo & Russo 2000, Dumbser & Käser 2007) consists of emphasizing the role of the central stencil by taking $\gamma_L = \gamma_R = 1$ and setting $\gamma_C$ in the range of 50 to 400 with $p = 4$. Such a scheme is often referred to as a *central WENO* (CWENO) scheme. Increasing $\gamma_C$ increases the central biasing in the scheme. I.e., for most forms of smooth flow all three stencils will have comparable smoothness indicators and we will mostly rely on the central stencil with its greater stability. The difference between $\gamma_C$ and $\gamma_L, \gamma_R$ is modest. As a result, when discontinuities are present in the flow, the smoothness indicators will be vastly smaller for the stencil with the smoothest solution. In that situation eqn. (28) will select that stencil. Yet another approach by Martin *et al.* (2006) uses the linear weights to minimize the dispersion error in turbulence calculations. It has also been suggested that *p* should increase with increasing order. It is worth noting that the choices catalogued in this paragraph are most relevant to finite volume WENO schemes (the schemes of interest here), where the resulting reconstruction will only be third order accurate.

The reconstructed profile in a finite volume scheme should represent the solution at all points within the zone. In a finite difference scheme, however, we only need to evaluate the solution and its fluxes at given points on the mesh. For finite difference schemes, this opens the door to optimizing the linear weights differently so that accuracy is improved. The choice in Jiang and Shu (1996) and Balsara and Shu (2000) consists of realizing that when the flow is smooth, one can make a convex combination of the three smaller stencils to obtain a larger stencil spanning the zones $\{i-2, i-1, i, i+1, i+2\}$. For smooth flow, and with the right convex combination, the larger stencil can provide fifth order accuracy! Optimal, i.e. fifth, order of accuracy is obtained for finite difference formulations by setting $\gamma_L = 0.1$, $\gamma_C = 0.6$ and $\gamma_R = 0.3$ with a choice of $p = 2$. This can be very important for improving the accuracy of rightward propagating waves to fifth order. Mechanistically, when the flow is smooth, we have $IS_L \cong IS_C \cong IS_R$ in eqn. (28) so that the non-linear weights $\bar{w}_L$, $\bar{w}_C$ and $\bar{w}_R$ equal the optimal linear weights $\gamma_L$, $\gamma_C$ and $\gamma_R$ respectively. When the flow is not smooth, the accuracy improvement is relinquished. Henrick, Aslam and Powers (2006) showed that a mapping function needs to be applied to the non-linear weights in eqn. (28) in order to circumvent a loss of accuracy at critical points, i.e. points where the first or higher derivatives can become zero. (Setting $\gamma_L = 0.3$, $\gamma_C = 0.6$ and $\gamma_R = 0.1$ maximizes the accuracy of the reconstruction at the left zone boundary. This permits leftward propagating waves to do so with fifth order accuracy, when those waves are smooth.) It is important to point out that this accuracy improvement is only most effective when considering finite difference schemes, which are not the direct point of focus here.

Notice that the final $\hat{u}_x$ and $\hat{u}_{xx}$ that we obtain from eqn. (29) and use in eqn. (19) have a strongly non-linear dependence on the solution. This is how WENO schemes achieve their non-linear hybridization. The solid lines in Figs. 4d and 5d show the reconstructed profiles for the Gaussian and shock profiles. We see that the reconstructed polynomial for the Gaussian follows the original Gaussian function extremely well without clipping the maximum, as well it should for a smooth profile. From Fig. 5d for the shock profile we see that the reconstructed polynomial for the $i^{th}$ zone that is centered at $x = 0$ is non-oscillatory, retains a small amount of curvature and is obtained, for the most part, from the left-biased stencil which is the only stable stencil in this problem. Comparing Fig. 5d with the analogous Fig. 3c for PPM, we see that the third order WENO reconstruction indeed produces larger jumps at shocks. Thus the WENO reconstruction indeed does provide good stabilization at shocks while leaving other extrema intact.

While we have catalogued the formulation in physical space, WENO schemes can also be formulated for the characteristic variables. All the early formulations of WENO schemes in Jiang and Shu (1996) and Balsara and Shu (2000) were in characteristic variables. Qiu & Shu (2002) have shown that there might be some advantage formulating the reconstruction problem in characteristic variables. Eqns. (9), (10) and (11) have shown us how the reconstruction problem can be cast in characteristic variables. For structured meshes, WENO reconstruction is most easily formulated in modal space and Balsara *et al.* (2009, 2013) and Balsara, Garain and Shu (2016) have provided easily implementable closed form expressions for WENO reconstruction up to very high orders. The choice of non-linear hybridization described in eqn. (28) is not the only one there is. Jiang & Shu (1996), Balsara & Shu (2000), Henrick, Aslam and Powers (2006), Borges *et al.* (2008), Gerolymos, Sénéchal and Vallet (2009) and Hu, Wang & Adams (2010), Castro *et al.* (2011), Fan *et al.* (2014a,b) have shown that different strategies for evaluating the non-linear weights may be used in one dimension with a resultant increase in the order of accuracy of the scheme. I.e., if one wishes to have a finite difference scheme then reconstruction with $r^{th}$ order accurate polynomials can be made to yield a scheme with an overall accuracy of $2r-1$ for smooth flow. Such schemes appear in the literature under a variety of variant names like WENO-M, WENO-Z and WENO-η. A similar increase in accuracy can be achieved for WENO schemes on unstructured meshes (Zhang & Shu 2009). Shi, Hu and Shu (2002) and Mignone (2014) also discuss the case where the mesh is non-uniform. Divergence-free WENO reconstruction of vector fields is discussed in Balsara (2004, 2009) and Balsara *et al.* (2013). This completes our description of WENO reconstruction in one dimension.

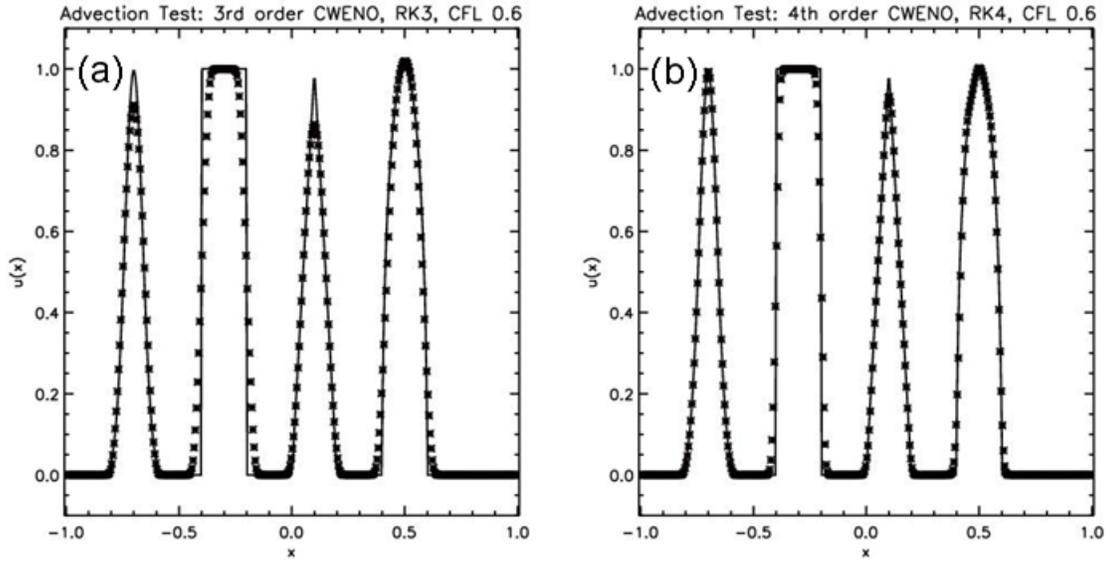

*Fig. 6a shows the advection test catalogued in the text when the third order CWENO reconstruction was used with a third order Runge-Kutta scheme. Fig. 6b shows the same when the fourth order CWENO reconstruction was used with a fourth order Runge-Kutta scheme. The solid line shows the analytic solution, the crosses show the computed solution.*

Figs. 6a and 6b show the results of our one-dimensional advection test when third and fourth order accurate CWENO reconstruction was used along with a Runge-Kutta time stepping of matching accuracy. We see that both schemes reproduce the correct solution very well without any spurious overshoots and undershoots. In Figs. 6a and 6b we can clearly distinguish the shape of each profile from the other. The Gaussian and triangular profiles show crisp extrema which have not been clipped. The ellipse does not show any flattening at the top of its profile, nor does it show any upwind bias. The profile of the square wave has been preserved very crisply by the fourth order scheme and slightly less so by the third order scheme. The PPM scheme in Fig. 2b represents the square wave profile as sharply as the fourth order CWENO scheme because both schemes start the reconstruction with a fourth order accurate representation of the boundary values. The fourth order CWENO scheme does, however, do a superlative job of preserving the extrema in the Gaussian and triangular profiles. One can use the optimal weights described in Jiang and Shu (1996) and Balsara and Shu (2000) to improve the formal order of accuracy, which also improves the performance of WENO schemes on the present test problem.

**Implementing One-Dimensional WENO Reconstruction**:
Step 1: For each zone "$i$", use eqn. (21) to obtain $\hat{u}_{L;x}$ and $\hat{u}_{L;xx}$. Likewise, for each zone, use eqn. (23) to obtain $\hat{u}_{C;x}$ and $\hat{u}_{C;xx}$. Similarly, for each zone, use eqn. (25) to obtain $\hat{u}_{R;x}$ and $\hat{u}_{R;xx}$.

Step 2: For each zone "$i$", use eqns. (26) and (27) to obtain $IS_L$, $IS_C$ and $IS_R$.

Step 3: For each zone "$i$", use eqn. (28) to obtain $\bar{w}_L$, $\bar{w}_C$ and $\bar{w}_R$. When specifying the linear weights in eqn. (28), use one of the choices catalogued in the paragraphs before eqn. (29).

Step 4: Eqn. (29) can then be used within each zone to obtain the moments $\hat{u}_x$ and $\hat{u}_{xx}$ for that zone. When those moments are used in eqn. (19), the one-dimensional WENO reconstruction is complete.

**Steepening the Profiles of Linearly Degenerate Discontinuities**

The examples in Figs. 2 and 6 have shown that even higher order schemes tend to spread out the boundaries of the square wave profile. This is a generic feature of all higher order Godunov methods when they are called on to crisply represent discontinuities in linearly degenerate wave fields. Contact discontinuities in fluid flow or Alfven waves in MHD or RMHD flow are examples of such discontinuities. Because of their self-steepening character, shocks in a fluid flow simulation do not suffer from this problem. Various methods have been devised to reintroduce the steepness in linearly degenerate discontinuities. The *contact discontinuity steepener* in PPM (Colella and Woodward 1984) and the *artificial compression method* (Harten 1977, Harten 1989, Yang 1990) represent efforts in that direction. Such methods try to artificially steepen flow profiles when a discontinuity is detected in a linearly degenerate wave field. However, divining the existence of such a discontinuity proves to be a tricky task and the steepener can do more harm than good if it is improperly invoked. As a result, the modern trend consists of forgoing excessive reliance on such steepening techniques.

**Parallel Efficiency of Higher Order Schemes**

It should be noted that the third order WENO and the PPM reconstructions have the same stencil size. I.e., reconstructing the solution in a zone only requires the availability of a solution in the two zones adjoining the zone of interest. The fourth order WENO reconstruction requires one more zone on either side. In schemes that reconstruct the solution, the size of the stencil influences two important aspects of the solution strategy. First, when enforcing boundary conditions, it is best to have one layer of zones outside the physical domain where the reconstruction can be carried out in full. Thus for the third order WENO and PPM schemes, the solution needs to be specified in a layer of three zones outside the physical boundary of the problem; for fourth order WENO, the size of the layer increases to four zones. This ensures that reconstruction can be fully carried out in the one layer of exterior zones that abut the physical boundary. Second, when parallelizing a code, the solution is almost always partitioned into contiguous chunks of zones that are farmed out to individual processors. The solution on each processor will have to have the same sized halo of zones around it as was needed for the physical boundaries in the previous point. On modern parallel machines, this halo of zones have their data exchanged via a very efficient message passing process. Thus for most reasonably sized problems, the increase in cost associated with the messaging is almost negligible.

The figure below, from Garain, Balsara and Reid (2015), presents a weak scaling study using the RIEMANN framework for computational astrophysics. In recent years, one-sided messaging has become a reality with the advent of Coarray Fortran (CAF) and the third generation Message Passing Interface standard (MPI-3). Weak scalability studies that compare CAF and MPI-3 are presented on up to 65,536 processors. Both parallel programming paradigms scale well, showing that they are well-suited for Petascale-class applications. They both require rethinking the messaging strategies from the ground-up. However, once that investment is made, the resulting scalability is substantially better than that of MPI-2. The one-sided messaging in CAF is much more expressive and, therefore, substantially easier to implement than the one-sided messaging in MPI-3. Both those modern parallelization paradigms show very comparable scalability on a range of applications that were documented in Garain, Balsara and Reid (2015). Best-usage strategies for both those paradigms are also documented in that paper. The figure below shows the scalability from a 3D MHD-based ADER-WENO application that used four halo zones. We see that both CAF and MPI-3 operate at the highest levels of parallelism with comparable parallel efficiency on a large class of applications. Furthermore, that efficiency is close to optimal even for the largest numbers of processors. In that paper it is also shown that the one-sided messaging in CAF is slightly more efficient that the one-sided messaging from MPI-3 for all numbers of cores that we tested. The same simulations were also run with MPI-2 and the results that are documented in the above-mentioned paper show that CAF and MPI-3 show a significant improvement over MPI-2.

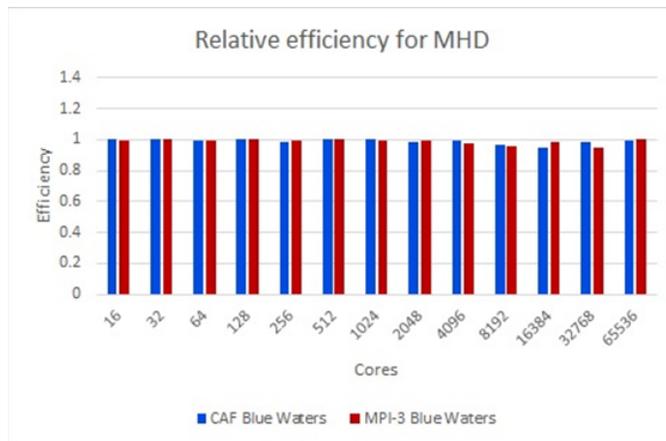

*This Figure, from Garain, Balsara and Reid (2015), shows the parallel efficiency relative to every doubling of processors for CAF and one-sided MPI-3. The results are based on MHD simulations with an ADER-WENO scheme. The blue bars show the relative scalability of CAF, the red bars show the relative scalability of MPI-3. Both paradigms show optimal scalability at PetaScale performance levels.*

**III.2) WENO Reconstruction in Multiple Dimensions**

Let us now consider the general, higher order, finite volume reconstruction in a zone ($i,j$) of the two-dimensional mesh shown in Fig. 1. Specifically, let us consider third order accurate reconstruction. The desired moments in the zone ($i,j$) are then given by

$$u_{i,j}(\tilde{x}, \tilde{y}) = \bar{u}_{i,j} + \hat{u}_x \tilde{x} + \hat{u}_{xx}\left(\tilde{x}^2 - \frac{1}{12}\right) + \hat{u}_y \tilde{y} + \hat{u}_{yy}\left(\tilde{y}^2 - \frac{1}{12}\right) + \hat{u}_{xy} \tilde{x} \tilde{y} \tag{30}$$

where $(\tilde{x}, \tilde{y})$ are the zone's local coordinates, as defined in eqn. (6) and the subscripts "*i,j*" in the moments are dropped just to keep the notation simple. We see, therefore, that a third order accurate finite volume reconstruction requires us to build all the moments that the WENO scheme from the previous section could provide us if it is applied in both the *x*- and *y*-directions. Thus the moments $\hat{u}_x$ and $\hat{u}_{xx}$ can be obtained by applying third order WENO reconstruction in the *x*-direction. Likewise, the moments $\hat{u}_y$ and $\hat{u}_{yy}$ can be obtained by applying the same WENO reconstruction in the *y*-direction. However, the moment $\hat{u}_{xy}$, which represents the cross-term in the reconstruction, is still left unspecified. This term is needed for true third order accurate finite volume reconstruction.

One possible way of obtaining all the moments in eqn. (30) might be to use several large multidimensional stencils and to try and obtain all the moments simultaneously from each of the stencils. This is, in fact, the favored method for WENO reconstruction on unstructured meshes, see Friedrichs (1998), Dumbser & Käser (2007). On structured meshes, a more economical method would be to obtain all the one-dimensional moments in the *x*- and *y*-directions using the dimension-by-dimension strategy outlined in the previous paragraph (Balsara *et al.* 2009). Assuming that this is done, we only need to specify the moment $\hat{u}_{xy}$ in a WENO sense in eqn. (30). For simplicity, assume that all the zones shown in Fig. 1 have unit extent in each direction. Furthermore, assume that the origin is located at the center of zone (*i,j*). Choosing the our first stencil with zones $\{(i, j),(i+1, j+1)\}$ and requiring the consistency condition

$$\int_{\tilde{y}=0.5}^{\tilde{y}=1.5} \int_{\tilde{x}=0.5}^{\tilde{x}=1.5} u_{i,j}(\tilde{x}, \tilde{y}) d\tilde{x}\, d\tilde{y} = \bar{u}_{i+1, j+1} \tag{31}$$

we get

$$\hat{u}_{xy} = \bar{u}_{i+1, j+1} - \bar{u}_{i,j} - \hat{u}_x - \hat{u}_y - \hat{u}_{xx} - \hat{u}_{yy} \tag{32}$$

Similarly, choosing our second, third and fourth stencils with zones $\{(i, j),(i+1, j-1)\}$, $\{(i, j),(i-1, j+1)\}$ and $\{(i, j),(i-1, j-1)\}$ we obtain three other alternative values for the cross term as

$$\begin{aligned}
\hat{u}_{xy} &= -\bar{u}_{i+1, j-1} + \bar{u}_{i,j} + \hat{u}_x - \hat{u}_y + \hat{u}_{xx} + \hat{u}_{yy} \\
\hat{u}_{xy} &= -\bar{u}_{i-1, j+1} + \bar{u}_{i,j} - \hat{u}_x + \hat{u}_y + \hat{u}_{xx} + \hat{u}_{yy} \\
\hat{u}_{xy} &= \bar{u}_{i-1, j-1} - \bar{u}_{i,j} + \hat{u}_x + \hat{u}_y - \hat{u}_{xx} - \hat{u}_{yy}
\end{aligned} \tag{33}$$

Since only the third order term is being reconstructed, we can exclusively focus on the second moments when constructing the smoothness measures. The four smoothness indicators for each of our four stencils are then given by the formula

$$IS = 4\,\hat{u}_{xx}^2 + 4\,\hat{u}_{yy}^2 + \hat{u}_{xy}^2 \tag{34}$$

where we use the four different choices for $\hat{u}_{xy}$ that are given in eqns. (32) and (33). The four smoothness measures can be used in the usual way, see eqn. (28), to obtain a non-linearly weighted value for $\hat{u}_{xy}$. Equal linear weights are ascribed to the four stencils considered in this section. This completes our description of third order accurate, finite-volume WENO reconstruction on structured meshes. Fourth order accurate, finite-volume, multi-dimensional WENO reconstruction strategies for structured meshes have also been catalogued in Balsara *et al.* (2009, 2013).

> **Implementing Multidimensional, Finite-volume, WENO Reconstruction:**
> The goal is to obtain the coefficients $\hat{u}_x$, $\hat{u}_{xx}$, $\hat{u}_y$, $\hat{u}_{yy}$, and $\hat{u}_{xy}$ in eqn. (30).
> <u>Step 1</u>: Use the results from the previous Sub-section to obtain $\hat{u}_x$, $\hat{u}_{xx}$, $\hat{u}_y$ and $\hat{u}_{yy}$ for each zone.
> <u>Step 2</u>: For each zone, use eqns. (32) and (33) to obtain four different choices of $\hat{u}_{xy}$, corresponding to the four different stencils.
> <u>Step 3</u>: Within each zone, use the four different choices of $\hat{u}_{xy}$ to build four different smoothness measures using eqn. (34). Use those smoothness measures to obtain nonlinear weights.
> <u>Step 4</u>: Use the nonlinear weights to obtain a non-linearly hybridized $\hat{u}_{xy}$ in eqn. (30).

**IV) Evolving Conservation Laws Accurately in Time – Part I, Runge-Kutta Methods**

The previous sections have shown us how to reconstruct the solution vector on a computational mesh. We saw that we could achieve second order accurate reconstruction in space with piecewise linear methods. We could also construct finite volume reconstructions that went beyond second order accuracy in space. Matching these spatial reconstruction techniques with methods that allow us to achieve a corresponding temporal accuracy is the goal of this section and the next. We tackle this section in three easy parts. First, we study the general philosophy and structure of Runge-Kutta time stepping; this is done in Sub-section IV.1. Second, we describe how a second order scheme is assembled with Runge-Kutta time stepping; done in Sub-section IV.2. Third, we understand the changes that have to be made in going beyond second order; we instantiate them with a third order scheme with Runge-Kutta time stepping. We do this in Sub-section IV.3. Runge-Kutta methods are perhaps the simplest way of achieving second and higher orders of temporal accuracy for hyperbolic problems.

**IV.1) Runge-Kutta Time Stepping**

Runge-Kutta time-discretizations have a logical simplicity which accounts for their great popularity. The Runge-Kutta methods are also referred to as method of lines approaches or semi-discrete approaches because they simplify the process of temporally updating the solution of a PDE to make it look very much like the time update of an ODE system. They are based on the viewpoint that we can write the PDE in eqn. (1) as

$$\frac{\partial U}{\partial t} = L(U) \equiv -F(U)_x - G(U)_y \tag{35}$$

Written this way, it has the semblance of an ordinary differential equation (ODE). The method of lines is not strictly speaking a "method" as much as it is a philosophy. It consists of using some semi-discrete approach for solving ODEs to achieve the temporal accuracy in eqn. (35). I.e., despite being inspired by ODEs, the method works for PDEs.

Not all second order Runge-Kutta methods have equally desirable attributes, especially as they apply to the TVD property. For example, if each of the stages for the improved Euler approximation is TVD then the final solution at the end of the two stages is also TVD. Unfortunately, the guarantees provided by the improved Euler approximation with respect to the TVD property only extend in their truest sense to scalar conservation laws, not to systems. However, a modified Euler approximation cannot even ensure such a TVD property for scalar conservation laws. (We have to balance this with the reality that the improved and modified Euler approximations produce results of comparable quality in practical problems.) Realize, therefore, that although several strong proofs are available for the stability properties of Runge-Kutta schemes, they are not as ironclad as one might like.

Several authors (Shu and Osher 1988, Shu 1988, Gottlieb and Shu 1998, Spiteri and Ruuth 2002, 2003, Gottlieb, Shu and Tadmor 2001, Gottlieb 2005, Gottlieb, Ketcheson and Shu 2011) have proved theorems showing that the time update in eqn. (35) can be carried out to higher orders of accuracy using a sequence of internal Runge-Kutta stages. Moreover, these time-update schemes have the same TVD property as the improved Euler approximation mentioned above. I.e., if each of the stages of the Runge-Kutta scheme is TVD then the final solution at the end of all the stages is also TVD. Runge-Kutta schemes having this special property are known as strong stability preserving (SSP) Runge-Kutta schemes. The SSP Runge-Kutta scheme at second order is indeed the improved Euler approximation given by

$$U^{(1)} = U^n + \Delta t \, L(U^n)$$
$$U^{n+1} = \frac{1}{2} U^n + \frac{1}{2} U^{(1)} + \frac{1}{2} \Delta t \, L(U^{(1)}) \tag{36}$$

The above Runge-Kutta scheme starts with a mesh function $\{U^n\}$ that is specified at a time $t^n$ and evolves it via the use of one internal stage $\{U^{(1)}\}$ to a mesh function $\{U^{n+1}\}$ that is specified at a time $t^{n+1} = t^n + \Delta t$. The third order accurate SSP Runge-Kutta scheme is given by

$$U^{(1)} = U^n + \Delta t\, L(U^n)$$
$$U^{(2)} = \frac{3}{4} U^n + \frac{1}{4} U^{(1)} + \frac{1}{4} \Delta t\, L(U^{(1)}) \tag{37}$$
$$U^{n+1} = \frac{1}{3} U^n + \frac{2}{3} U^{(2)} + \frac{2}{3} \Delta t\, L(U^{(2)})$$

The above second and third order SSP Runge-Kutta schemes are optimal in the sense that for one-dimensional flow they can support a CFL number of unity and, moreover, it is not possible to arrive at a time-explicit Runge-Kutta scheme of the same order that provides a larger CFL number per stage that is used in the scheme. For example, eqn. (37) is optimal because it is impossible to find another third order SSP Runge-Kutta scheme that increases its CFL number by more than one for every three stages used in the scheme. An almost optimal, fourth order accurate SSP Runge-Kutta scheme is given by

$$U^{(1)} = U^n + 0.391752226571890\, \Delta t\, L(U^n)$$
$$U^{(2)} = 0.444370493651235\, U^n + 0.555629506348765\, U^{(1)} + 0.368410593050371\, \Delta t\, L(U^{(1)})$$
$$U^{(3)} = 0.620101851488403\, U^n + 0.379898148511597\, U^{(2)} + 0.251891774271694\, \Delta t\, L(U^{(2)})$$
$$U^{(4)} = 0.178079954393132\, U^n + 0.821920045606868\, U^{(3)} + 0.544974750228521\, \Delta t\, L(U^{(3)})$$
$$U^{n+1} = 0.517231671970585\, U^{(2)} + 0.096059710526147\, U^{(3)} + 0.386708617503269\, U^{(4)}$$
$$\qquad + 0.063692468666290\, \Delta t\, L(U^{(3)}) + 0.226007483236906\, \Delta t\, L(U^{(4)})$$

$$\tag{38}$$

For one-dimensional flow, eqn. (38) can support a CFL number of 1.5. Notice that this is a five stage scheme. In contrast, the classical fourth order Runge-Kutta scheme is only a four stage scheme, thus saving the evaluation of one entire stage; but it is not SSP. The Butcher barriers that plague ordinary Runge-Kutta schemes at fifth and higher orders also plague SSP Runge-Kutta schemes at fourth and higher orders. The increasing number of extra stages in Runge-Kutta schemes make them progressively inefficient with increasing order. ADER schemes, which we will study in the next section, do not suffer from this deficiency.

Please note that for multidimensional problems, the permitted CFL number is divided by the dimensionality of the problem. Thus the second and third order schemes in eqns. (36) and (37) only support CFL numbers of 0.5 and 0.333 in two and three dimensions respectively. Please also recall that boundary conditions have to be applied consistently to each of the stages in eqns. (123) to (125). SSP Runge-Kutta schemes that go beyond fourth order have also been formulated by Spiteri and Ruuth (2002, 2003), but the ones presented here are the workhorses for most practical work.

**IV.2) Second Order Accurate Runge-Kutta Scheme**

Further specification of Runge-Kutta schemes requires us to provide a recipe for obtaining the fluxes at the zone boundaries at any stage of the multi-stage scheme. We start with the mesh function and use the methods from Section II to obtain the reconstructed profile $U_{i,j}(\tilde{x}, \tilde{y})$ within a zone. Eqns. (6) and (30) give us examples of such reconstructions at second and third order. It is traditional in this work to assume that a zone has been mapped to the unit square; eqn. (6) provides an example of how such a linear mapping is carried out. Consequently, $(\tilde{x}, \tilde{y}) \in [-1/2, 1/2] \times [-1/2, 1/2]$ form the local coordinates of each zone $(i,j)$. Each of the stages of a Runge-Kutta scheme is defined at only one time level. Consequently, observe from eqn. (4) that the time-averaging of the fluxes is not needed. In eqn. (35) we can discretize the spatial parts as

$$\frac{\partial \overline{U}_{i,j}}{\partial t} = L(\overline{U})_{i,j} = -\frac{1}{\Delta x}\left(\overline{F}_{i+1/2,j} - \overline{F}_{i-1/2,j}\right) - \frac{1}{\Delta y}\left(\overline{G}_{i,j+1/2} - \overline{G}_{i,j-1/2}\right), \qquad (39)$$

with the facially-averaged fluxes at the upper $x$- and $y$-faces of the zone $(i,j)$ defined by

$$\overline{F}_{i+1/2,j} \equiv \int_{\tilde{y}=-1/2}^{\tilde{y}=1/2} F(\tilde{x}=1/2, \tilde{y})\, d\tilde{y} \quad ; \quad \overline{G}_{i,j+1/2} \equiv \int_{\tilde{x}=-1/2}^{\tilde{x}=1/2} G(\tilde{x}, \tilde{y}=1/2)\, d\tilde{x}\ . \qquad (40)$$

Specification of the Runge-Kutta scheme requires specifying the above two integrals at each of the faces of the mesh. We detail the evaluation of the numerical fluxes in the next paragraph.

At second order, we assume that the vector of conserved variables $\overline{U}_{i,j}$ as well as its undivided differences, $\Delta_x \overline{U}_{i,j}$ and $\Delta_y \overline{U}_{i,j}$, are available for each zone $(i,j)$ of the mesh shown in Fig. 7. The left and right states needed for evaluating the Riemann problem at the top $x$-boundary of the zone being considered, i.e. at the $(i+1/2, j)$ location in Fig. 7, are given by

$$U_{L;i+1/2,j} = \overline{U}_{i,j} + \frac{1}{2}\Delta_x \overline{U}_{i,j} \quad ; \quad U_{R;i+1/2,j} = \overline{U}_{i+1,j} - \frac{1}{2}\Delta_x \overline{U}_{i+1,j} \qquad (41)$$

Likewise, the bottom and top states needed for evaluating the Riemann problem at the $(i, j+1/2)$ zone-boundary are given by

$$U_{B;i,j+1/2} = \overline{U}_{i,j} + \frac{1}{2}\Delta_y \overline{U}_{i,j} \quad ; \quad U_{T;i,j+1/2} = \overline{U}_{i,j+1} - \frac{1}{2}\Delta_y \overline{U}_{i,j+1} \qquad (42)$$

Fig. 7 shows a schematic representation of the four abutting zones $(i,j)$, $(i+1,j)$, $(i,j+1)$ and $(i+1,j+1)$ and illustrates various aspects of the construction that is catalogued in eqns (41) and (42). At second order, the integrals in eqn. (40) are just the values of the upwinded fluxes provided by any Riemann solver that is evaluated at the face centers. This dramatic simplification of the integrals does not carry over to higher orders. Thus at second order we get

$$\bar{F}_{i+1/2,j} = F_{RS}\left(U_{L;i+1/2,j},\, U_{R;i+1/2,j}\right) \quad;\quad \bar{G}_{i,j+1/2} = G_{RS}\left(U_{B;i,j+1/2},\, U_{T;i,j+1/2}\right) \tag{43}$$

Here $F_{RS}$ and $G_{RS}$ denote the Riemann solver, which is being used as a machine that accepts two states as inputs and provides the upwinded flux as an output. This completes our description of the Runge-Kutta method at second order.

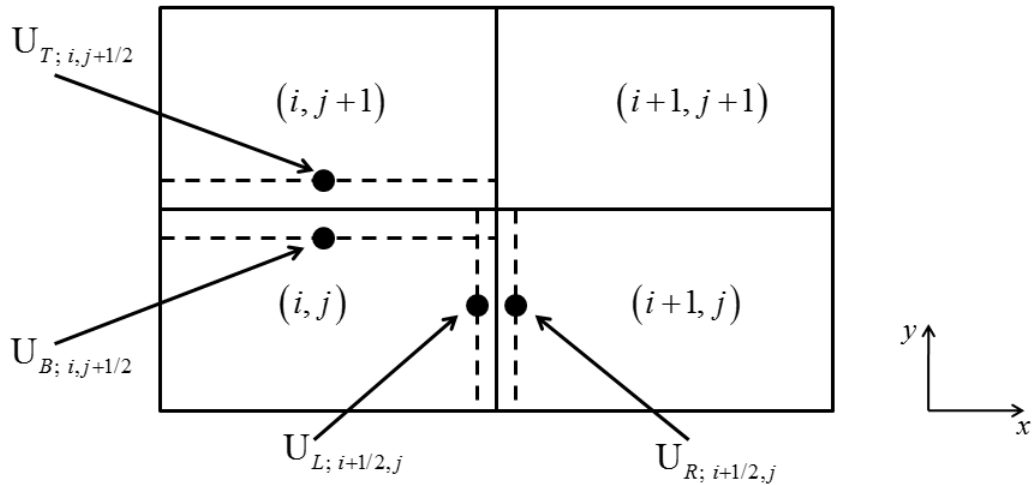

*Fig. 7 shows the construction for obtaining spatially second order accurate fluxes for the second order Runge-Kutta scheme. The quadrature points are shown with dots. The surfaces to the immediate left and immediate right of the zone boundaries are shown with dashes.*

---

**Stepwise Description of the Second Order Accurate Runge-Kutta Scheme**

We describe a single stage of the second order scheme described in eqn. (36).
<u>Step 1</u>: Impose boundary conditions and use the results from Section II to obtain the spatial modes in eqn. (6).
<u>Step 2</u>: Use eqn. (41) to obtain the left and right states at each *x*-boundary. Using an analogous construction, obtain the top and bottom states at each *y*-boundary using eqn. (42).
<u>Step 3</u>: Use the states from the previous step in the Riemann solvers shown in eqn. (43) to get the fluxes at each zone boundary.

---

**IV.3) Runge-Kutta Schemes at Higher Orders; Using Third Order as an Example**

When one tries to go beyond second order, eqns. (39) and (40) continue to be valid. As mentioned in the previous Sub-section, the difficulties at higher orders all arise from the integrals that have to be evaluated in eqn. (40). In order for the overall scheme to have third or fourth order of accuracy, the integrals in eqn. (40) have to be evaluated with the same order of accuracy. Let us consider the first integral in eqn. (40) which gives us the *x*-flux at the $(i+1/2, j)$ zone-boundary. To evaluate the integral with third order of accuracy using numerical quadrature, we would need to obtain the numerical flux at three suitably chosen quadrature points on that boundary. (Recall the third order accurate Simpson rule for numerical quadrature.) Each such flux would require an invocation of a Riemann solver, thus requiring three rather expensive solutions of the Riemann problem. At fourth order, one would have four quadrature points, thus requiring us to solve the Riemann problem four times. Clearly, if we continued this line of development, the higher order spatially-averaged fluxes in eqn. (40) would be very costly to evaluate because each call to the Riemann solver is itself quite expensive. Besides, three dimensional problems would be costlier yet, since they would have even more quadrature points in each face. Clearly, a more efficient approach would be very desirable.

A more efficient method for evaluating the spatially-averaged fluxes in eqn. (40) was presented in Atkins & Shu (1998), van der Ven & van der Vegt (2002a,b) and Dumbser *et al.* (2007). The method is called *quadrature free* because it avoids the use of a large number of quadrature points in the flux evaluation. It works for certain very useful classes of Riemann solvers, including the HLL, HLLC, HLLI and linearized Riemann solvers. Efficient third and fourth order approaches, with copious implementation-related details for three dimensional structured meshes, are documented in Balsara *et al.* (2012). Here we present the method for the HLL Riemann solver and focus on third order of accuracy in two dimensions. Say that the conservation law has "*M*" components. Let us start with an extension of eqn. (30), which we write for an "*M*" component vector of conserved variables in the zone $(i, j)$ as

$$U_{i,j}(\tilde{x}, \tilde{y}) = \overline{U}_{i,j} + \hat{U}_{i,j;x}\tilde{x} + \hat{U}_{i,j;y}\tilde{y} + \hat{U}_{i,j;xx}\left(\tilde{x}^2 - \frac{1}{12}\right) + \hat{U}_{i,j;yy}\left(\tilde{y}^2 - \frac{1}{12}\right) + \hat{U}_{i,j;xy}\tilde{x}\tilde{y}$$
(44)

Eqn. (44) is referred to as a *modal representation* in space and the vectors $\overline{U}_{i,j}$, $\hat{U}_{i,j;x}$, $\hat{U}_{i,j;y}$, $\hat{U}_{i,j;xx}$, $\hat{U}_{i,j;yy}$ and $\hat{U}_{i,j;xy}$ are called the *modes* of the reconstruction. In eqn. (44) we follow the convention that $(\tilde{x}, \tilde{y})$ are the local coordinates within the zone $(i, j)$ and that they are mapped to the unit square $[-.5,.5] \times [-.5,.5]$. We can use eqn. (44) to obtain the entire vector of conserved variables at any location within the square. The value of the conserved variables at any specific location within the zone of interest can be evaluated using eqn. (44). The locations within the zone where the values are evaluated are called *nodes*, and the values themselves are called *nodal* values.

To define an HLL Riemann solver at the *x*-face at $(i+1/2, j)$, we need to find the extremal wave speeds, $S_L$ and $S_R$, flowing in the x-direction at that zone boundary. We can obtain these speeds by evaluating eqn. (44) and its analogue from the zone (*i*+1,*j*) on either side of the center of the *x*-face being considered. To make this concrete, we build the two vectors of conserved variables given by

$$U^{(c)}_{L; i+1/2, j} = U_{i,j}(\tilde{x} = 1/2, \tilde{y} = 0) \quad ; \quad U^{(c)}_{R; i+1/2, j} = U_{i+1,j}(\tilde{x} = -1/2, \tilde{y} = 0) \tag{45}$$

We then use the left and right boundary values, $U^{(c)}_{L; i+1/2, j}$ and $U^{(c)}_{R; i+1/2, j}$ to obtain $S_L$ and $S_R$, i.e. the extremal wave speeds in the HLL Riemann solver. Fig. 8 shows a schematic representation of the two abutting zones (*i*,*j*) and (*i*+1,*j*) and illustrates various aspects of the construction that is catalogued here. The HLL Riemann solver at any location on the *x*-face at $(i+1/2, j)$ is written as

$$F_{i+1/2, j}(\tilde{y}) = \left[\frac{S_R}{S_R - S_L}\right] F_{L; i+1/2, j}(\tilde{y}) - \left[\frac{S_L}{S_R - S_L}\right] F_{R; i+1/2, j}(\tilde{y}) \\ + \left[\frac{S_R S_L}{S_R - S_L}\right] \left(U_{R; i+1/2, j}(\tilde{y}) - U_{L; i+1/2, j}(\tilde{y})\right) \tag{46}$$

Compare the above eqn. to the equation for the HLL flux and please review the HLL flux from Chapter 4 of the author's website. In principle, $S_L$ and $S_R$ can have different values at different points on the face being considered. The important insight from Dumbser *et al.* (2007) consists of freezing the wave speeds $S_L$ and $S_R$. I.e., we are freezing our wave model so that the extremal wave speeds all along the face being considered equal those evaluated at the center of the face. Freezing the wave model does not diminish the order of accuracy of the overall scheme. It does make the flux in eqn. (46) linear in terms of the left and right conserved variables, i.e. $U_{L; i+1/2, j}(\tilde{y})$ and $U_{R; i+1/2, j}(\tilde{y})$, as well as linear in terms of the left and right fluxes, i.e. $F_{L; i+1/2, j}(\tilde{y})$ and $F_{R; i+1/2, j}(\tilde{y})$. We show in the next paragraph that this small simplification makes it possible to spatially-average eqn. (46) over the *x*-face of interest.

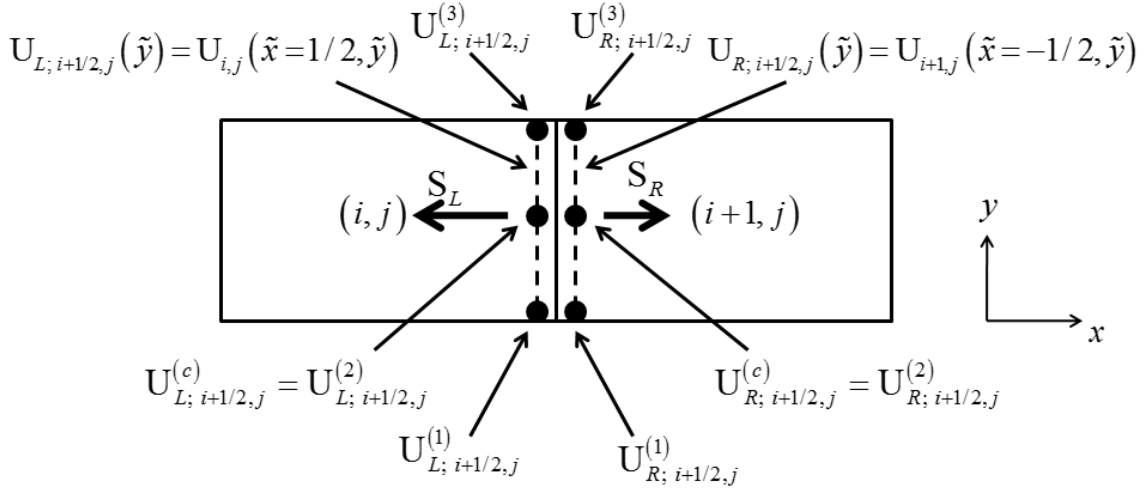

*Fig. 8 shows the construction for obtaining spatially third order accurate fluxes for the third order Runge-Kutta scheme. The wave model, along with the variables used for obtaining it, are shown. The quadrature points are shown with dots. The surfaces to the immediate left and immediate right of the zone boundary at (i+1/2,j) are shown with dashes.*

Notice now that $U_{L;\,i+1/2,j}(\tilde{y})$ and $U_{R;\,i+1/2,j}(\tilde{y})$ available as analytic expressions from eqn. (44) and its counterpart in zone $(i+1, j)$ as

$$U_{L;\,i+1/2,j}(\tilde{y}) = U_{i,j}(\tilde{x} = 1/2, \tilde{y})$$
$$= \left(\overline{U}_{i,j} + \frac{1}{2}\hat{U}_{i,j;x} + \frac{1}{6}\hat{U}_{i,j;xx}\right) + \left(\hat{U}_{i,j;y} + \frac{1}{2}\hat{U}_{i,j;xy}\right)\tilde{y} + \hat{U}_{i,j;yy}\left(\tilde{y}^2 - \frac{1}{12}\right)$$
$$U_{R;\,i+1/2,j}(\tilde{y}) = U_{i+1,j}(\tilde{x} = -1/2, \tilde{y})$$
$$= \left(\overline{U}_{i+1,j} - \frac{1}{2}\hat{U}_{i+1,j;x} + \frac{1}{6}\hat{U}_{i+1,j;xx}\right) + \left(\hat{U}_{i+1,j;y} - \frac{1}{2}\hat{U}_{i+1,j;xy}\right)\tilde{y} + \hat{U}_{i+1,j;yy}\left(\tilde{y}^2 - \frac{1}{12}\right)$$
(47)

$U_{L;\,i+1/2,j}(\tilde{y})$ is evaluated on the left dashed surface shown in Fig. 8. Similarly, using the analogue of eqn. (44) from the zone $(i+1, j)$, the second expression in eqn. (47) gives us $U_{R;\,i+1/2,j}(\tilde{y})$. $U_{R;\,i+1/2,j}(\tilde{y})$ is evaluated on the right dashed surface shown in Fig. 8. Thus the third term on the right hand side of eqn. (46) is analytically integrable over the *x*-face at $(i+1/2, j)$. To illustrate this explicitly, we have

$$\int_{\tilde{y}=-1/2}^{\tilde{y}=1/2} \left(U_{R;\,i+1/2,j}(\tilde{y}) - U_{L;\,i+1/2,j}(\tilde{y})\right) d\tilde{y}$$
$$= \left(\overline{U}_{i+1,j} - \frac{1}{2}\hat{U}_{i+1,j;x} + \frac{1}{6}\hat{U}_{i+1,j;xx}\right) - \left(\overline{U}_{i,j} + \frac{1}{2}\hat{U}_{i,j;x} + \frac{1}{6}\hat{U}_{i,j;xx}\right)$$
(48)

It is easily seen that the above integral is third order accurate. We now wish to obtain facially integrated versions of the x-flux on either side of the zone boundary being considered. In other words, we wish to obtain third order accurate integrals of the first two terms in eqn. (46). This can be accomplished if we have the x-fluxes at three quadrature points that lie immediately to the left of the *x*-boundary at $(i+1/2, j)$ in Fig. 8. To be specific, we use Simpson's rule as our third order accurate quadrature formula. Notice that the flux can only be evaluated at a quadrature point if we have the conserved variables at the same quadrature point. Eqn. (44) can now be evaluated at three quadrature points that lie immediately to the left of the *x*-face at $(i+1/2, j)$, as shown in Fig. 8. We then have

$$U^{(1)}_{L; i+1/2, j} = U_{i,j}(\tilde{x}=1/2, \tilde{y}=-1/2) \quad ; \quad U^{(2)}_{L; i+1/2, j} = U_{i,j}(\tilde{x}=1/2, \tilde{y}=0) \quad ; \quad U^{(3)}_{L; i+1/2, j} = U_{i,j}(\tilde{x}=1/2, \tilde{y}=1/2) \tag{49}$$

The analogue of eqn. (44) in zone (*i*+1,*j*) can also be evaluated at three quadrature points that lie immediately to the right of the *x*-face at $(i+1/2, j)$, as shown in Fig. 8. We then have

$$U^{(1)}_{R; i+1/2, j} = U_{i+1,j}(\tilde{x}=-1/2, \tilde{y}=-1/2) \quad ; \quad U^{(2)}_{R; i+1/2, j} = U_{i+1,j}(\tilde{x}=-1/2, \tilde{y}=0) \quad ; \quad U^{(3)}_{R; i+1/2, j} = U_{i+1,j}(\tilde{x}=-1/2, \tilde{y}=1/2) \tag{50}$$

Evaluating the x-fluxes from the three conserved variables in eqn. Eqn. (49) enables us to explicit the integral of the first term on the right hand side of eqn. (46) as

$$\int_{\tilde{y}=-1/2}^{\tilde{y}=1/2} F_{L; i+1/2, j}(\tilde{y}) d\tilde{y} = \frac{1}{6} F(U^{(1)}_{L; i+1/2, j}) + \frac{2}{3} F(U^{(2)}_{L; i+1/2, j}) + \frac{1}{6} F(U^{(3)}_{L; i+1/2, j}) \tag{51}$$

where the Simpson rule has been used to obtain third order accuracy. Eqn. (50) also enables us to explicit the integral of the second term on the right hand side of eqn. (46) as

$$\int_{\tilde{y}=-1/2}^{\tilde{y}=1/2} F_{R; i+1/2, j}(\tilde{y}) d\tilde{y} = \frac{1}{6} F(U^{(1)}_{R; i+1/2, j}) + \frac{2}{3} F(U^{(2)}_{R; i+1/2, j}) + \frac{1}{6} F(U^{(3)}_{R; i+1/2, j}) \tag{52}$$

and the above eqn. is again third order accurate. Eqns. (48), (51) and (52) can be used to obtain the third order accurate, facially-integrated HLL flux in the *x*-direction, i.e. the very same entity that we are evaluating via the first integral in eqn. (40). A similar construction can be used to make a rapid evaluation of the second integral in eqn. (40), giving us the facially-integrated HLL flux in the *y*-direction. When this is done at all faces, we can obtain a third order accurate representation of the right hand side of eqn.

(39). Using this term for each of the three stages in eqn. (37) completes our specification of a third order accurate Runge-Kutta scheme.

---

**Stepwise Description of the Third Order Accurate Runge-Kutta Scheme**

We describe a single stage of the third order scheme described in eqn. (37).

Step 1: Impose boundary conditions and use the results from Section III to obtain the spatial modes in eqn. (44).

Step 2: Use eqn. (44), along its analogue in zone $(i+1, j)$, to obtain $U^{(c)}_{L; i+1/2, j}$ and $U^{(c)}_{R; i+1/2, j}$ from eqn. (45). Use them in eqn. (6.56) to obtain the extremal wave speeds, $S_L$ and $S_R$, for use in eqn. (46).

Step 3: Use eqns. (47) to (52) to obtain the third order accurate spatially averaged integrals of the right hand side of eqn. (46). This yields a numerical flux at the *x*-faces. Construct similar numerical fluxes at the *y*-faces. Use the fluxes in eqn. (37) to complete the present stage of the Runge-Kutta scheme.

---

## V) Evolving Conservation Laws Accurately in Time – Part II, Predictor-Corrector Schemes

Despite their desirable simplicity, several of the tasks in a Runge-Kutta scheme have to be repeated at each internal stage. This increases the computational cost. Predictor-corrector schemes avoid some of this duplication of effort. Sub-section V.1 introduces predictor-corrector methods at second order. They yield the fastest schemes at second order and they also lay the foundation for ADER schemes. The formulation of higher order ADER schemes is difficult to understand. For this reason, we do it in two easier stages in the next two sub-sections. Sub-section V.2 introduces a very simple formulation of ADER schemes in one dimension at third order. Such a formulation is analytically tractable with a little bit of basic calculus. Sub-section V.3 introduces ADER methods in multidimensions, casting them in the role of higher order extensions of predictor-corrector type methods.

## V.1) Second Order Accurate Predictor-Corrector Schemes

A predictor-corrector scheme is made of two steps – the predictor step and the corrector step. In the predictor step, we construct the spatial variation, i.e. the undivided differences, within a zone and use it to obtain a measure of the time rate of change of the conserved variables within that zone. Thus starting with the conserved variables $\overline{U}^n_{i,j}$ in each of the zones $(i, j)$ at time $t^n$, we use a slope limiter to obtain $\Delta_x \overline{U}_{i,j}$ and $\Delta_y \overline{U}_{i,j}$ in eqn. (6). The predictor step then consists of using the variation within the zone to obtain a measure of $(\partial U/\partial t)^n_{i,j}$. There are various ways of doing this. Here we present a strategy that will prepare us for our study of ADER schemes in Sub-section IV.4. Thus consider the *nodal points* $(i+1/2, j)$, $(i-1/2, j)$, $(i, j+1/2)$ and $(i, j-1/2)$ *within* the zone $(i, j)$ that is under consideration. For example, the nodal points $(i+1/2, j)$ and $(i, j+1/2)$ associated with zone $(i, j)$ are shown in Fig. 7. We obtain the conserved variables at those points using just the values and slopes that are internal to the zone $(i, j)$. The variables $\overline{U}_{i,j}^n$, $\Delta_x \overline{U}_{i,j}$ and $\Delta_y \overline{U}_{i,j}$ can be thought of as the *modes* or the *modal values* of the solution within the zone $(i, j)$. We therefore use the modal values to obtain the *nodal values* as follows:

$$U_{i+1/2,j}^n = \overline{U}_{i,j}^n + \frac{1}{2}\Delta_x \overline{U}_{i,j} \quad ; \quad U_{i-1/2,j}^n = \overline{U}_{i,j}^n - \frac{1}{2}\Delta_x \overline{U}_{i,j} \quad ;$$
$$U_{i,j+1/2}^n = \overline{U}_{i,j}^n + \frac{1}{2}\Delta_y \overline{U}_{i,j} \quad ; \quad U_{i,j-1/2}^n = \overline{U}_{i,j}^n - \frac{1}{2}\Delta_y \overline{U}_{i,j} \tag{53}$$

The values $U_{i+1/2,j}^n$, $U_{i-1/2,j}^n$, $U_{i,j+1/2}^n$ and $U_{i,j-1/2}^n$ can be used to make an "in the small" approximation of $(\partial U / \partial t)_{i,j}^n$ as follows

$$\left(\frac{\partial U}{\partial t}\right)_{i,j}^n = -\frac{1}{\Delta x}\left(F\left(U_{i+1/2,j}^n\right) - F\left(U_{i-1/2,j}^n\right)\right) - \frac{1}{\Delta y}\left(G\left(U_{i,j+1/2}^n\right) - G\left(U_{i,j-1/2}^n\right)\right) \tag{54}$$

The above step can be carried out locally within all the zones, permitting us to predict the value of the conserved variable, not just in space, but also in time. As long at the time over which we seek to predict the values is less than the CFL limit on the time step, our predicted values will be accurate and stable.

The corrector part then consists of using $(\partial U / \partial t)_{i,j}^n$ and the undivided differences to obtain space- and time-centered values for the conserved variables on either side of each zone boundary. Thus at the zone boundary $(i+1/2, j)$ we obtain the left and right states given by:

$$U_{L;i+1/2,j}^{n+1/2} \equiv \overline{U}_{i,j}^n + \frac{1}{2}\Delta_x \overline{U}_{i,j}^n + \frac{1}{2}\Delta t \left(\frac{\partial U}{\partial t}\right)_{i,j}^n \quad ; \quad U_{R;i+1/2,j}^{n+1/2} \equiv \overline{U}_{i+1,j}^n - \frac{1}{2}\Delta_x \overline{U}_{i+1,j}^n + \frac{1}{2}\Delta t \left(\frac{\partial U}{\partial t}\right)_{i+1,j}^n$$

(55)

Likewise, at the zone boundary $(i, j+1/2)$ we obtain the top and bottom states given by

$$U_{B;i,j+1/2}^{n+1/2} \equiv \overline{U}_{i,j}^n + \frac{1}{2}\Delta_y \overline{U}_{i,j}^n + \frac{1}{2}\Delta t \left(\frac{\partial U}{\partial t}\right)_{i,j}^n \quad ; \quad U_{T;i,j+1/2}^{n+1/2} \equiv \overline{U}_{i,j+1}^n - \frac{1}{2}\Delta_y \overline{U}_{i,j+1}^n + \frac{1}{2}\Delta t \left(\frac{\partial U}{\partial t}\right)_{i,j+1}^n$$

(56)

We can think of eqns. (55) and (56) as endowing time evolution to the nodal values shown in Fig. 7.7. The final update step can now be written as

$$\bar{U}_{i,j}^{n+1} = \bar{U}_{i,j}^{n} - \frac{\Delta t}{\Delta x}\left(F_{RS}\left(U_{L;i+1/2,j}^{n+1/2}, U_{R;i+1/2,j}^{n+1/2}\right) - F_{RS}\left(U_{L;i-1/2,j}^{n+1/2}, U_{R;i-1/2,j}^{n+1/2}\right)\right)$$
$$- \frac{\Delta t}{\Delta y}\left(G_{RS}\left(U_{B;i,j+1/2}^{n+1/2}, U_{T;i,j+1/2}^{n+1/2}\right) - G_{RS}\left(U_{B;i,j-1/2}^{n+1/2}, U_{T;i,j-1/2}^{n+1/2}\right)\right) \quad (57)$$

The present scheme is stable up to a CFL number of 0.5 in two dimensions. (In three dimensions, the limiting CFL number becomes 1/3.) While this CFL number is less than the CFL number of some of the schemes which include a multidimensional wave model (Colella 1990, Saltzman 1994, LeVeque 1997, Abgrall 1994a,b, Balsara 2010, 2012a), its chief advantage is its simplicity, ease of implementation and its great speed. Notice that, unlike the second order Runge-Kutta schemes, the present scheme only invokes the limiters and the Riemann solver once per time step. As a result, it is also faster and slightly less diffusive than the second order Runge-Kutta scheme in certain instances.

---

**Stepwise Description of the Second Order Accurate Predictor-Corrector Scheme**

Step 1: Impose boundary conditions and use the results from Section II to obtain the spatial modes in eqn. (6).

Step 2: Construct the states in eqn. (53). Use these states to evaluate the appropriate fluxes in eqn. (54). This gives us an "in the small" time rate of update within each zone. It also completes the predictor step.

Step 3: Use eqns. (55) and (56) to obtain the left and right states at each *x*-boundary and the top and bottom states at each *y*-boundary.

Step 4: Feed the above states into the Riemann solver to obtain the desired fluxes in eqn. (57). This completes the corrector step as well as the time step.

---

**V.2) A Very Simple One-Dimensional ADER Scheme at Third Order**

ADER stands for Arbitrary DERivative in space and time. Only the ADER predictor step is pedagogically tricky, so in this Sub-section we restrict attention only to the ADER predictor step. The goal of this Sub-section to make the ADER predictor step accessible to the reader in its simplest setting. Let us consider a very simple ADER scheme for the one-dimensional conservation law $\partial_t U + \partial_x F = 0$. We can even take "U" to be a scalar for the sake of simplicity, though the logic in this Sub-section works even if "U" is a solution vector. Let the one dimensional mesh have zones of size $\Delta x$ and a timestep of size $\Delta t$. We wish to evolve the solution from time $t^n$ to a time $t^{n+1} = t^n + \Delta t$. For each zone with zone-center $x_i$ we can define a local and normalized spatial coordinate given by $\tilde{x} = (x - x_i)/\Delta x$ with a local time coordinate given by $\tilde{t} = (t - t^n)/\Delta t$. Consequently, in terms of the normalized coordinates, the PDE can be written as

$$\frac{\partial u}{\partial \tilde{t}} + \frac{\partial f}{\partial \tilde{x}} = 0 \quad \text{with} \quad u(\tilde{x},\tilde{t}) \equiv U(x,t) \quad \text{and} \quad f(\tilde{x},\tilde{t}) \equiv \frac{\Delta t}{\Delta x} F(x,t) \tag{58}$$

For the ADER predictor step we focus exclusively on the solution within the $i^{\text{th}}$ zone. We assume that third accurate spatial reconstruction has been carried out so that we start our ADER scheme with

$$u_i(\tilde{x},\tilde{t}=0) = \bar{u}_i + \hat{u}_{i,x}\tilde{x} + \hat{u}_{i,xx}\left(\tilde{x}^2 - \frac{1}{12}\right) \tag{59}$$

In the above equation, the mean value $\bar{u}_i$ in the $i^{\text{th}}$ zone is given by the time-update from the previous timestep; however, the modes $\hat{u}_{i,x}$ and $\hat{u}_{i,xx}$ are obtained by the third order accurate spatial reconstruction. The goal of the ADER predictor step is to predict the solution *within* the $i^{\text{th}}$ zone for all space-time points given by $(\tilde{x},\tilde{t}) \in [-1/2, 1/2] \times [0,1]$ in a fashion that is consistent with the governing dynamical equation given by eqn. (58). We want this time evolution to be third order accurate in space-time so that we want

$$u_i(\tilde{x},\tilde{t}) = \bar{u}_i + \hat{u}_{i,x}\tilde{x} + \hat{u}_{i,xx}\left(\tilde{x}^2 - \frac{1}{12}\right) + \hat{u}_{i,t}\tilde{t} + \hat{u}_{i,tt}\tilde{t}^2 + \hat{u}_{i,xt}\tilde{x}\tilde{t} \tag{60}$$

Eqn. (60) identifies a set of basis functions given by

$$\left\{ \phi_1(\tilde{x},\tilde{t}) = 1,\ \phi_2(\tilde{x},\tilde{t}) = \tilde{x},\ \phi_3(\tilde{x},\tilde{t}) = (\tilde{x}^2 - 1/12),\ \phi_4(\tilde{x},\tilde{t}) = \tilde{t},\ \phi_5(\tilde{x},\tilde{t}) = \tilde{t}^2,\ \phi_6(\tilde{x},\tilde{t}) = \tilde{x}\tilde{t} \right\} \tag{61}$$

Associated with this basis set, we have a set of modes given by $\{\bar{u}_i,\ \hat{u}_{i,x},\ \hat{u}_{i,xx},\ \hat{u}_{i,t},\ \hat{u}_{i,tt},\ \hat{u}_{i,xt}\}$. The first three basis functions in eqn. (61) are purely spatial, while the next three carry the temporal evolution. Realize, therefore, that the ADER predictor step that we seek should be a method for starting with eqn. (59) and yielding the coefficients $\hat{u}_{i,t}$, $\hat{u}_{i,tt}$ and $\hat{u}_{i,xt}$ in eqn. (60) in a fashion that is optimally consistent with the governing eqn. (58). We will devise an iterative strategy to achieve this convergence; the iterations are known to converge very fast.

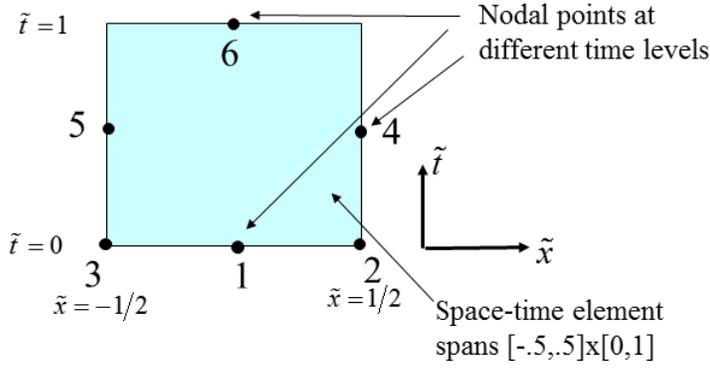

Fig. 9 shows the space-time element that is used in the ADER predictor step. The six nodal points in space-time are shown with the six dots.

Space-time element spans [-.5,.5]x[0,1]

We start the iterative solution process with $\hat{u}_{i,t} = 0$, $\hat{u}_{i,tt} = 0$ and $\hat{u}_{i,xt} = 0$. Let us first realize that the six coefficients in eqn. (60) constitute six modal coefficients. If we were to assign six meaningful sets of numbers to these coefficients, we would be specifying the entire space-time evolution of $u_i(\tilde{x}, \tilde{t})$ within the $i^{th}$ zone. Now please look at Fig. 9 and observe that it has six nodal points in space-time. Three of these six nodal points have been specified at $\tilde{t} = 0$. We assert that the specification of the six modal coefficients in eqn. (60), would be completely equivalent to specifying the six nodal values in space-time as shown in Fig. 9. Realize from eqn. (58) that it is not adequate to specify the modes in eqn. (60). Because the flux is also involved in eqn. (58), we should also specify the flux $f_i(\tilde{x}, \tilde{t})$ within the $i^{th}$ zone. The flux should be obtained with comparable space-time accuracy so that we write it as

$$f_i(\tilde{x}, \tilde{t}) = \hat{f}_i + \hat{f}_{i,x}\tilde{x} + \hat{f}_{i,xx}\left(\tilde{x}^2 - \frac{1}{12}\right) + \hat{f}_{i,t}\tilde{t} + \hat{f}_{i,tt}\tilde{t}^2 + \hat{f}_{i,xt}\tilde{x}\,\tilde{t} \qquad (62)$$

The flux coefficients in eqn. (62) should be consistent with the coefficients of the solution in eqn. (60) in a way that can only be arbitrated by the governing eqn. (58). This can be accomplished in a fashion that we will soon specify. In the rest of this paragraph, we show how the flux coefficients $\hat{f}_i$, $\hat{f}_{i,x}$ and $\hat{f}_{i,xx}$ can be obtained in a fashion that is consistent with the coefficients $\bar{u}_i$, $\hat{u}_{i,x}$ and $\hat{u}_{i,xx}$ at $\tilde{t} = 0$. To see how this is done, please focus on the first three nodal points in Fig. 9. These three nodal points are given by $(\tilde{x}_1, \tilde{t}_1) = (0,0)$, $(\tilde{x}_2, \tilde{t}_2) = (1/2, 0)$ and $(\tilde{x}_3, \tilde{t}_3) = (-1/2, 0)$. We evaluate the solution at these nodal points so that we can define nodal values of the solution within the zone being considered as $\tilde{u}^1 = u_i(\tilde{x}_1, \tilde{t}_1)$, $\tilde{u}^2 = u_i(\tilde{x}_2, \tilde{t}_2)$ and $\tilde{u}^3 = u_i(\tilde{x}_3, \tilde{t}_3)$. This is done by evaluating eqn. (59). Using these nodal values of the solution, we can evaluate nodal values of the fluxes within the zone being considered as $\tilde{f}^1 = f(\tilde{u}^1)$, $\tilde{f}^2 = f(\tilde{u}^2)$ and $\tilde{f}^3 = f(\tilde{u}^3)$. We wish to obtain the first three coefficients in eqn. (62). With these three nodal values of the fluxes in hand, we can specify the first three modal coefficients for the fluxes in the $i^{th}$ zone by asserting a system of three equations that is given by

$$f_i(\tilde{x}_1, \tilde{t}_1) = \tilde{f}^1 \quad ; \quad f_i(\tilde{x}_2, \tilde{t}_2) = \tilde{f}^2 \quad ; \quad f_i(\tilde{x}_3, \tilde{t}_3) = \tilde{f}^3 \tag{64}$$

On inverting the system, the result is

$$\hat{f}_i = \left(4\tilde{f}^1 + \tilde{f}^2 + \tilde{f}^3\right)/6 \quad ; \quad \hat{f}_{i,x} = \tilde{f}^2 - \tilde{f}^3 \quad ; \quad \hat{f}_{i,xx} = 2\left(\tilde{f}^2 - 2\tilde{f}^1 + \tilde{f}^3\right) \tag{64}$$

This completes the process of initializing the flux coefficients in eqn. (62) at $\tilde{t} = 0$. Fig. 10 shows us this initialization step in the form of a flowchart. This paragraph has also given us our first exposure to the process of transcribing from modal values to nodal values for the solution; using nodal solution values to evaluate nodal values for the fluxes; then using those nodal values for the fluxes to obtain the corresponding modal coefficients for the fluxes.

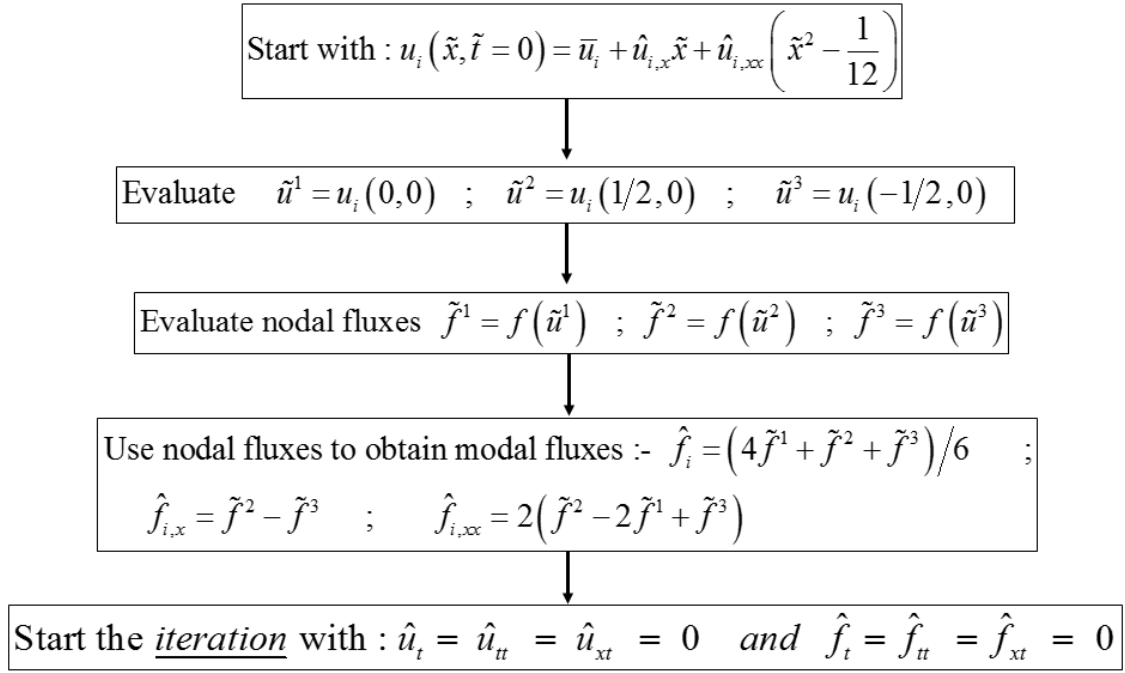

Fig. 10 shows a Flowchart for Initializing the ADER Iteration

We start the iterative process by setting $\hat{u}_{i,t} = 0$, $\hat{u}_{i,tt} = 0$ and $\hat{u}_{i,xt} = 0$. Correspondingly, we set $\hat{f}_{i,t} = 0$, $\hat{f}_{i,tt} = 0$ and $\hat{f}_{i,xt} = 0$. These time-dependent coefficients can only be set by appeal to the dynamics, i.e., by appealing to the governing eqn. (58). We enforce satisfaction of the governing equation via a Galerkin projection over the space-time element shown in Fig. 9. The test functions that we use are identical to the trial functions in eqn. (61). The projection can be explicitly written as

$$\int_{\tilde{t}=0}^{\tilde{t}=1} \int_{\tilde{x}=-1/2}^{\tilde{x}=1/2} \phi_i(\tilde{x},\tilde{t}) \left( \frac{\partial u_i(\tilde{x},\tilde{t})}{\partial \tilde{t}} + \frac{\partial f_i(\tilde{x},\tilde{t})}{\partial \tilde{x}} \right) d\tilde{x}\, d\tilde{t} = 0 \quad \text{for } i = 3,4,5 \tag{65}$$

Notice that only the three time-dependent test functions participate in the Galerkin projection in eqn. (65) because only those three test functions give us the time-dependent dynamics. Operationally, one substitutes $u_i(\tilde{x},\tilde{t})$ from eqn. (60) in eqn. (65) to obtain $\partial u_i(\tilde{x},\tilde{t})/\partial \tilde{t}$. Likewise, again operationally, one substitutes $f_i(\tilde{x},\tilde{t})$ from eqn. (62) in eqn. (65) to obtain $\partial f_i(\tilde{x},\tilde{t})/\partial \tilde{x}$. Then one takes one of the three time-evolutionary bases from eqn. (61) and one carries out the integration in eqn. (65) using a computer algebra system. It is not difficult to verify that the resulting equations give

$$\hat{u}_{i,t} = -\hat{f}_{i,x} \quad ; \quad \hat{u}_{i,tt} = -\hat{f}_{i,xt}/2 \quad ; \quad \hat{u}_{i,xt} = -2\hat{f}_{i,xx} \tag{66}$$

We see, therefore, that if we have a set of coefficients for $\hat{f}_{i,x}$, $\hat{f}_{i,xt}$ and $\hat{f}_{i,xx}$ during any stage in the iterative process then eqn. (66) gives an improved set of time-dependent coefficients $\hat{u}_{i,t}$, $\hat{u}_{i,tt}$ and $\hat{u}_{i,xt}$. To complete this iterative strategy, we have only to find a way to take these improved time-dependent coefficients and use them to build an improved set of flux coefficients in eqn. (62). We do that next.

Notice from Fig. 9 that we have three more quadrature points in space-time given by $(\tilde{x}_4,\tilde{t}_4) = (1/2, 1/2)$, $(\tilde{x}_5,\tilde{t}_5) = (-1/2, 1/2)$ and $(\tilde{x}_6,\tilde{t}_6) = (0,1)$. Immediately after the first iteration, all the coefficients in eqn. (60) will indeed be non-zero for most typical variations in the initial conditions. We evaluate the solution at the fourth, fifth and sixth nodal points in Fig. 9 so that we can define nodal values of the solution within the space-time element being considered as $\tilde{u}^4 = u_i(\tilde{x}_4,\tilde{t}_4)$, $\tilde{u}^5 = u_i(\tilde{x}_5,\tilde{t}_5)$ and $\tilde{u}^6 = u_i(\tilde{x}_6,\tilde{t}_6)$. Using these nodal values of the solution, we can evaluate nodal values of the fluxes within the zone being considered as $\tilde{f}^4 = f(\tilde{u}^4)$, $\tilde{f}^5 = f(\tilde{u}^5)$ and $\tilde{f}^6 = f(\tilde{u}^6)$. To close the loop, we should relate these nodal values of the fluxes to the modal coefficients for the fluxes in eqn. (62). Realize too that the first three nodal values for the fluxes, which were evaluated at $\tilde{t} = 0$, have not changed. We can write a system of three equations that is analogous to eqn. (64) for the fourth, fifth and sixth nodes. On inverting the system, the result is

$$\hat{f}_t = \tilde{f}^1 - 2\tilde{f}^2 - 2\tilde{f}^3 + 2\tilde{f}^4 + 2\tilde{f}^5 - \tilde{f}^6 \quad ; \quad \hat{f}_{tt} = -2\left(\tilde{f}^1 - \tilde{f}^2 - \tilde{f}^3 + \tilde{f}^4 + \tilde{f}^5 - \tilde{f}^6\right) \quad ;$$
$$\hat{f}_{xt} = -2\left(\tilde{f}^2 - \tilde{f}^3 - \tilde{f}^4 + \tilde{f}^5\right)$$
$$\tag{67}$$

This shows us how to start with an improved solution from eqn. (66); obtain from it an improved set of fluxes at the nodal points in Fig. 9 and to then use those improved nodal fluxes to obtain improved modal fluxes from eqn. (67). Once that is done, we can return to eqn. (66) and the iteration resumes. Fig. 11 shows us this iteration in the ADER predictor step in the form of a flowchart.

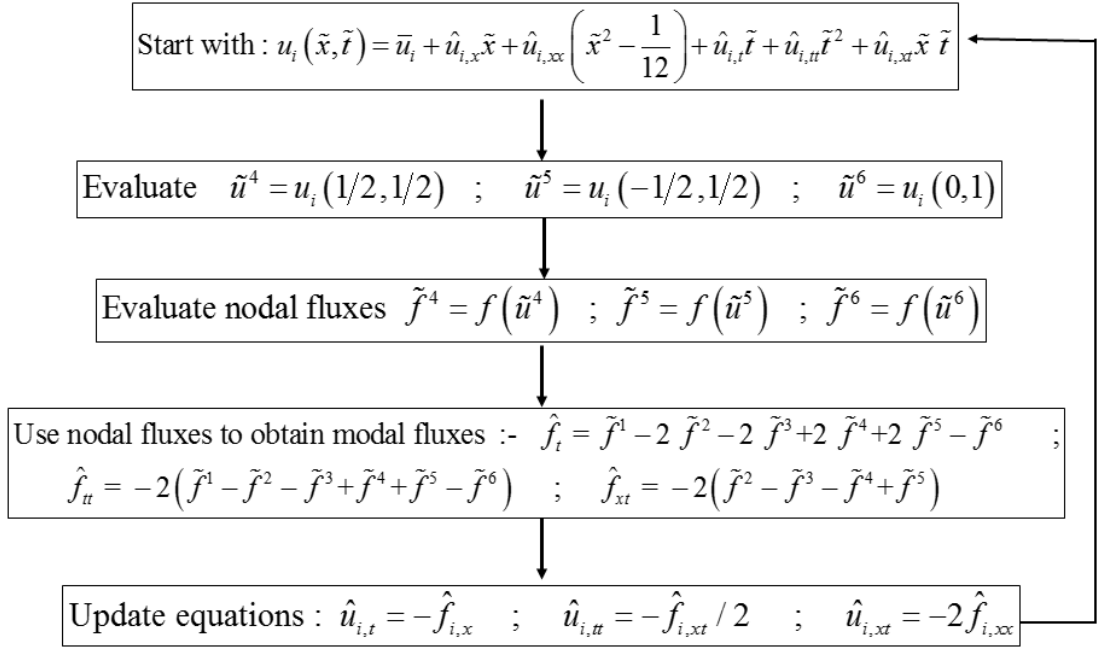

Fig. 11 shows a Flowchart for the ADER Iteration

The simple demonstration in this Sub-section is just meant to make the iterative ADER predictor step very accessible and easy to understand. Further information for multidimensional structured meshes is available in Balsara *et al.* (2009, 2013) and Dumbser *et al.* (2013). The iteration described in the previous two paragraphs converges very fast. For a temporally $N^{th}$ order scheme, one only needs (N-1) steps to converge to the level of discretization error. There is even a theoretical proof (Jackson 2017) that the method converges very fast. The next Sub-section provides all the details for higher order ADER schemes on multidimensional structured meshes.

**V.3) ADER Time Stepping for Second, and Higher Order, Time Accuracy**

In the previous Sub-sections we saw that predictor-corrector schemes at second order can be faster than their Runge-Kutta counterparts at the same order. This efficiency is due to the fact that each predictor-corrector time step only needs one reconstruction step and one solution of the Riemann problem. The ADER schemes presented here are the more efficient counterparts of the Runge-Kutta schemes at second and higher order. The ADER methodology is a time update strategy in the same way that Runge-Kutta schemes give us a method for evolving the PDE in time. Just like Runge-Kutta schemes, ADER schemes can be used for reconstruction-based schemes as well as discontinuous Galerkin schemes. Thus one can have ADER-WENO or ADER-DG schemes, analogous to RK-WENO schemes with Runge-Kutta time stepping or RKDG schemes respectively. ADER schemes represent a very economical method for arranging the time update and recent head-to-head comparisons have shown ADER time stepping to be faster than Runge-Kutta time stepping by a factor of up to two (Balsara *et al.* 2013) for the same order of accuracy.

Just like WENO schemes, ADER schemes have seen a few generations of development in the literature. Methods leading up to ADER schemes have been presented by several authors (van Leer 1979, Ben-Artzi and Falcovitz 1984). The above authors focused on the *generalized Riemann problem*. It consists of realizing that any second order scheme will have piecewise linear variations in the zones to the left and right of a zone boundary. As a result, we not just have a jump at the zone boundary but also have linear variations in the fluid quantities on either side of the jump. Consequently, the Riemann problem will no longer be a similarity solution in space-time. Instead, the wave structures in the Riemann problem will curve in response to the spatially varying states that they propagate into. Titarev & Toro (2002, 2005) and Toro & Titarev (2002) found a method for extending the generalized Riemann problem to higher orders and coined the ADER acronym. As a result, the left and right states could have any sort of polynomial variation at the zone boundary. Modern ADER schemes have been formulated more in the style of predictor-corrector schemes (Dumbser *et al.* 2008, Balsara *et al.* 2009, 2013). The predictor and corrector steps are indeed higher order extensions of the second order predictor-corrector scheme described in Sub-section V.1. The next two sub-sections describe the ADER predictor and corrector steps independently. In this review we describe a variant of ADER schemes, called ADER-CG, that is suited for problems with non-stiff source terms. The "CG" stands for continuous Galerkin and refers to the fact that the solution cannot take on abrupt temporal changes during a time step in response to stiff source terms.

**V.3.i) Multidimensional ADER-CG Predictor Step**

The ADER predictor step consists of developing a space-time representation of the vector of conserved variable in each zone. At second order, the ADER predictor step becomes identical to the predictor step from the predictor-corrector scheme in Sub-section V.1. To retain second order accuracy in time, one only needs to obtain the piecewise linear variation in time, which is explicited in eqn. (54). The extension of the second order predictor-corrector scheme to higher orders is entirely non-trivial. For that reason, we illustrate the ADER construction in two dimensions at third order. Extensions of the present section to even higher orders on three-dimensional structured and unstructured meshes have been presented in (Dumbser *et al.* 2008, Balsara *et al.* 2009, 2013 and Dumbser *et al.* 2013).

In order to write a third order accurate space-time dependence within a zone $(i, j)$ we first need to identify a set of local space-time basis functions that are defined in a local space-time coordinate system within each zone. Just as we developed local spatial coordinates $(\tilde{x}, \tilde{y})$ in eqn. (6), we now develop a local time coordinate system given by $\tilde{t} \equiv (t - t^n)/\Delta t$. Here we assume that we are evolving the solution from time $t^n$ to time $t^{n+1} = t^n + \Delta t$ in a zone with size $\Delta x$ and $\Delta y$ in the x- and y-directions. In terms of our local space-time coordinate system we have $(\tilde{x}, \tilde{y}, \tilde{t}) \in [-1/2, 1/2] \times [-1/2, 1/2] \times [0,1]$; we refer to this as a *reference space-time element*. Examining the mappings $\tilde{x} \equiv (x - x_i)/\Delta x$, $\tilde{y} \equiv (y - y_j)/\Delta y$ and $\tilde{t} \equiv (t - t^n)/\Delta t$, we see that the reference space-time element is an element obtained by linearly mapping the zone under consideration to a zone with unit zone size and time step. The spatial dependence can be analogous to that in eqn. (44) so that we can use the same spatial bases that were developed there. To develop space-time basis that retain third order accuracy within the zone being considered, we upgrade eqn. (44) to obtain

$$U_{i,j}(\tilde{x}, \tilde{y}, \tilde{t}) = \bar{U}_{i,j} + \hat{U}_{i,j;x}\tilde{x} + \hat{U}_{i,j;y}\tilde{y} + \hat{U}_{i,j;xx}\left(\tilde{x}^2 - \frac{1}{12}\right) + \hat{U}_{i,j;yy}\left(\tilde{y}^2 - \frac{1}{12}\right) + \hat{U}_{i,j;xy}\tilde{x}\tilde{y}$$
$$+ \hat{U}_{i,j;t}\tilde{t} + \hat{U}_{i,j;tt}\tilde{t}^2 + \hat{U}_{i,j;xt}\tilde{x}\tilde{t} + \hat{U}_{i,j;yt}\tilde{y}\tilde{t}$$
(68)

Notice that the spatial modes in eqn. (68), i.e. $\bar{U}_{i,j}$, $\hat{U}_{i,j;x}$, $\hat{U}_{i,j;y}$, $\hat{U}_{i,j;xx}$, $\hat{U}_{i,j;yy}$ and $\hat{U}_{i,j;xy}$, are available via some form of non-oscillatory reconstruction. (In the next section we will see that the spatial modes can even be evolved via some form of DG scheme.) The time-dependent modes, i.e. $\hat{U}_{i,j;t}$, $\hat{U}_{i,j;tt}$, $\hat{U}_{i,j;xt}$ and $\hat{U}_{i,j;yt}$ are chosen so that eqn. (68) retains all the terms in a Taylor expansion that are needed to provide a third order accurate reconstruction in space and time. Our task, in describing the ADER-CG predictor step, is to obtain the time-dependent modes within a zone when the spatial modes are available in that zone.

Reasoning by analogy with the second order case, i.e. eqn. (54), we realize that we will have to implicate the x- and y-fluxes in order to obtain an "in the small" time update within the zone of interest. Also notice that the terms $\left(F(U_{i+1/2,j}^n) - F(U_{i-1/2,j}^n)\right)/\Delta x$ and $\left(G(U_{i,j+1/2}^n) - G(U_{i,j-1/2}^n)\right)/\Delta y$ in eqn. (54) are just the x- and y-slopes of the x- and y-fluxes respectively. We, therefore, realize that in order to obtain the time-dependent modes in eqn. (68) we will need to study the moments of the fluxes. Reasoning in analogy with eqn. (54) we realize that only certain modes of the fluxes might eventually be needed. However, it is best to explicitly write out the entire modal representation of the x-flux in a reference space-time element as

$$\tilde{F}_{i,j}(\tilde{x}, \tilde{y}, \tilde{t}) = \hat{F}_{i,j} + \hat{F}_{i,j;x}\tilde{x} + \hat{F}_{i,j;y}\tilde{y} + \hat{F}_{i,j;xx}\left(\tilde{x}^2 - \frac{1}{12}\right) + \hat{F}_{i,j;yy}\left(\tilde{y}^2 - \frac{1}{12}\right) + \hat{F}_{i,j;xy}\tilde{x}\tilde{y}$$
$$+ \hat{F}_{i,j;t}\tilde{t} + \hat{F}_{i,j;tt}\tilde{t}^2 + \hat{F}_{i,j;xt}\tilde{x}\tilde{t} + \hat{F}_{i,j;yt}\tilde{y}\tilde{t}$$
(69)

and the entire modal representation of the y-flux in a reference space-time element as

$$\tilde{G}_{i,j}(\tilde{x},\tilde{y},\tilde{t}) = \hat{G}_{i,j} + \hat{G}_{i,j;x}\tilde{x} + \hat{G}_{i,j;y}\tilde{y} + \hat{G}_{i,j;xx}\left(\tilde{x}^2 - \frac{1}{12}\right) + \hat{G}_{i,j;yy}\left(\tilde{y}^2 - \frac{1}{12}\right) + \hat{G}_{i,j;xy}\tilde{x}\tilde{y}$$
$$+ \hat{G}_{i,j;t}\tilde{t} + \hat{G}_{i,j;tt}\tilde{t}^2 + \hat{G}_{i,j;xt}\tilde{x}\tilde{t} + \hat{G}_{i,j;yt}\tilde{y}\tilde{t}$$
(70)

In the next paragraph we will demonstrate how the governing equation is linearly mapped to the reference space-time element, which also shows the usefulness of working with the reference element. We will see in the next sub-section that having all the modes of the two fluxes above can be used to advantage in the corrector step of the ADER scheme. We also observe that since the entire spatial variation in eqn. (68) is assumed to be known at the beginning of the ADER step, we can obtain the spatial modes $\hat{F}_{i,j}$, $\hat{F}_{i,j;x}$, $\hat{F}_{i,j;y}$, $\hat{F}_{i,j;xx}$, $\hat{F}_{i,j;yy}$ and $\hat{F}_{i,j;xy}$ in eqn. (69) at the beginning of the ADER step. Likewise, we can obtain the spatial modes $\hat{G}_{i,j}$, $\hat{G}_{i,j;x}$, $\hat{G}_{i,j;y}$, $\hat{G}_{i,j;xx}$, $\hat{G}_{i,j;yy}$ and $\hat{G}_{i,j;xy}$ in eqn. (70) at the beginning of the ADER step. We will soon see how this evaluation can be carried out in a computationally efficient manner.

We will soon see that the variation of the conserved variables within a zone, i.e. eqn. (68), can be used to obtain the fluxes in the same zone, i.e. eqns. (69) and (70). The gradients of the fluxes, in turn, govern the time evolution of the conserved variables. To relate the modes of the fluxes to the time-dependent modes in eqn. (68), one has to utilize the governing equation, i.e. eqn. (1). The governing equation can be transformed to the local space-time coordinates of the zone $(i,j)$ as

$$\frac{\partial U_{i,j}(\tilde{x},\tilde{y},\tilde{t})}{\partial \tilde{t}} + \frac{\Delta t}{\Delta x}\frac{\partial F_{i,j}(\tilde{x},\tilde{y},\tilde{t})}{\partial \tilde{x}} + \frac{\Delta t}{\Delta y}\frac{\partial G_{i,j}(\tilde{x},\tilde{y},\tilde{t})}{\partial \tilde{y}} = 0 \quad \leftrightarrow$$
$$\frac{\partial U_{i,j}(\tilde{x},\tilde{y},\tilde{t})}{\partial \tilde{t}} + \frac{\partial \tilde{F}_{i,j}(\tilde{x},\tilde{y},\tilde{t})}{\partial \tilde{x}} + \frac{\partial \tilde{G}_{i,j}(\tilde{x},\tilde{y},\tilde{t})}{\partial \tilde{y}} = 0$$
(71)

This is tantamount to scaling the x-fluxes by $(\Delta t/\Delta x)$ and the y-fluxes by $(\Delta t/\Delta y)$ during the calculation so that we do not need to multiply too many factors of $(\Delta t/\Delta x)$ or $(\Delta t/\Delta y)$ all over. This scaling takes us from the physical zone to the reference element. When we reach the end of our calculation, i.e. when we have obtained converged modes in eqns. (68) to (70), we can always return to the physical zone by rescaling the fluxes as $F_{i,j}(\tilde{x},\tilde{y},\tilde{t}) \to \tilde{F}_{i,j}(\tilde{x},\tilde{y},\tilde{t})(\Delta x/\Delta t)$ and $G_{i,j}(\tilde{x},\tilde{y},\tilde{t}) \to \tilde{G}_{i,j}(\tilde{x},\tilde{y},\tilde{t})(\Delta y/\Delta t)$. In principle, we could substitute eqns. (68) to (70) into eqn. (71) and try to find a match to the polynomial terms, but this would become increasingly intractable as the order of accuracy increases. A simpler approach would be to project eqn. (71) into a basis space and require the *projection* to hold in a weak form. Thus let $\phi(\tilde{x},\tilde{y},\tilde{t})$ be a test function in space and time. We obtain a weak formulation of eqn. (71) by asserting that

$$\int_{\tilde{t}=0}^{\tilde{t}=1} \int_{\tilde{y}=-1/2}^{\tilde{y}=1/2} \int_{\tilde{x}=-1/2}^{\tilde{x}=1/2} \phi(\tilde{x},\tilde{y},\tilde{t}) \left[ \frac{\partial\, U_{i,j}(\tilde{x},\tilde{y},\tilde{t})}{\partial \tilde{t}} + \frac{\partial\, \tilde{F}_{i,j}(\tilde{x},\tilde{y},\tilde{t})}{\partial \tilde{x}} + \frac{\partial\, \tilde{G}_{i,j}(\tilde{x},\tilde{y},\tilde{t})}{\partial \tilde{y}} \right] d\tilde{x}\, d\tilde{y}\, d\tilde{t} = 0$$

(72)

While this can be asserted for any space-time test function $\phi(\tilde{x},\tilde{y},\tilde{t})$, it is best to use the test functions that are associated with the time-dependent modes in eqn. (68). Since we are interested in the time evolution of eqn. (68), the time-dependent basis functions are, in some sense, the best basis functions to use. The theoretical underpinnings of the finite element method also support this choice. Thus with $\phi(\tilde{x},\tilde{y},\tilde{t}) = \tilde{t}$ we have:

$$\hat{U}_{i,j;t} + \frac{4}{3} \hat{U}_{i,j;tt} = -\hat{F}_{i,j;x} - \hat{G}_{i,j;y} - \frac{2}{3} \hat{F}_{i,j;xt} - \frac{2}{3} \hat{G}_{i,j;yt} \tag{73}$$

Similarly, with $\phi(\tilde{x},\tilde{y},\tilde{t}) = \tilde{t}^2$ we have:

$$\hat{U}_{i,j;t} + \frac{3}{2} \hat{U}_{i,j;tt} = -\hat{F}_{i,j;x} - \hat{G}_{i,j;y} - \frac{3}{4} \hat{F}_{i,j;xt} - \frac{3}{4} \hat{G}_{i,j;yt} \tag{74}$$

Furthermore, with $\phi(\tilde{x},\tilde{y},\tilde{t}) = \tilde{x}\,\tilde{t}$ we get:

$$\hat{U}_{i,j;xt} = -2\hat{F}_{i,j;xx} - \hat{G}_{i,j;xy} \tag{75}$$

Likewise, with $\phi(\tilde{x},\tilde{y},\tilde{t}) = \tilde{y}\,\tilde{t}$ we get:

$$\hat{U}_{i,j;yt} = -\hat{F}_{i,j;xy} - 2\hat{G}_{i,j;yy} \tag{76}$$

Please note that the above four equations hold in the reference space-time element. The above four equations can then be rewritten as a more meaningful set as follows:

$$\begin{aligned}
\hat{U}_{i,j;t} &= -\hat{F}_{i,j;x} - \hat{G}_{i,j;y} \\
\hat{U}_{i,j;tt} &= -\left(\hat{F}_{i,j;xt} + \hat{G}_{i,j;yt}\right)/2 \\
\hat{U}_{i,j;xt} &= -2\hat{F}_{i,j;xx} - \hat{G}_{i,j;xy} \\
\hat{U}_{i,j;yt} &= -\hat{F}_{i,j;xy} - 2\hat{G}_{i,j;yy}
\end{aligned} \tag{77}$$

These are the equations that relate the modes of the x- and y-fluxes to the time-dependent modes in eqn. (68). Although they have been derived by a finite element-like procedure, it is possible to discern the finite-difference like structure for the time evolution in these equations. It turns out that they can be solved via iteration. The iterative procedure can be started by zeroing out all of the time-dependent terms in eqns. (68), (69) and (70). Each iteration yields an improved set of terms $\hat{U}_{i,j;t}$, $\hat{U}_{i,j;tt}$, $\hat{U}_{i,j;xt}$ and $\hat{U}_{i,j;yt}$. These terms can then be used to improve our approximations for $\hat{F}_{i,j;t}$, $\hat{F}_{i,j;tt}$, $\hat{F}_{i,j;xt}$ and $\hat{F}_{i,j;yt}$ in eqn. (69). Similarly, we can improve our approximations for $\hat{G}_{i,j;t}$, $\hat{G}_{i,j;tt}$, $\hat{G}_{i,j;xt}$ and $\hat{G}_{i,j;yt}$ in eqn. (70). Notice that each iteration is designed to sharpen our fidelity to the weak form of the governing equation because each iteration is an application of the projection in eqn. (72). Furthermore, it is an amazing result owing to the contractive nature of the *Picard iteration* that, at third order, only two iterations of eqn. (77) are needed to achieve third order accuracy. At second order, the Picard iteration theory requires only one iteration, which is why we did not iterate on eqn. (54). At fourth order, one would require three iterations, and so on.

We still have to specify how the spatial modes are to be obtained in eqns. (69) and (70). The idea is to identify a set of nodal points in the local space-time coordinate system at $\tilde{t} = 0$. The nine black circles in Fig. 12 show one possible set of such spatial nodes and are given by the ordered set of nodal points in the reference space-time element:

$$\{(0,0,0),(1/2,0,0),(-1/2,0,0),(0,1/2,0),(0,-1/2,0),$$
$$(1/2,1/2,0),(-1/2,1/2,0),(1/2,-1/2,0),(-1/2,-1/2,0)\} \tag{78}$$

The above set of nodes is labeled from 1 to 9 in Fig. 12. Once the conserved variables are obtained at these nodes using eqn. (68), they can be used to construct the nodal values of the x- and y-fluxes. Denoting the nodal location with a superscript, we now list the transcription from the nodal values to the spatial modes of the x-flux in eqn. (69) as follows

$$\hat{F}_{i,j;xx} = 2\left(\tilde{F}^2_{i,j} - 2\,\tilde{F}^1_{i,j} + \tilde{F}^3_{i,j}\right) \quad ; \quad \hat{F}_{i,j;yy} = 2\left(\tilde{F}^4_{i,j} - 2\,\tilde{F}^1_{i,j} + \tilde{F}^5_{i,j}\right) \quad ;$$
$$\hat{F}_{i,j;xy} = \tilde{F}^6_{i,j} - \tilde{F}^7_{i,j} - \tilde{F}^8_{i,j} + \tilde{F}^9_{i,j} \quad ; \tag{79}$$
$$\hat{F}_{i,j;x} = \tilde{F}^2_{i,j} - \tilde{F}^3_{i,j} \quad ; \quad \hat{F}_{i,j;y} = \tilde{F}^4_{i,j} - \tilde{F}^5_{i,j} \quad ; \quad \hat{F}_{i,j} = \tilde{F}^1_{i,j} + \left(\hat{F}_{i,j;xx} + \hat{F}_{i,j;yy}\right)/12$$

Notice how reminiscent the above expressions are to finite difference approximations for the moments. A similar transcription can be used for obtaining the spatial modes of the y-flux in eqn. (70). The spatial modes in eqns. (69) and (70) should be computed only once before the iteration described in the above paragraph is started. Notice that our choice of time-dependent basis functions in eqns. (68) to (70) is such that the time-dependent modes in eqn. (68) do not change the spatial modes in eqns. (69) and (70). We have arrived at a better appreciation of the nomenclature *ADER-CG*, where the "CG" refers to the fact that the scheme is continuous Galerkin in time.

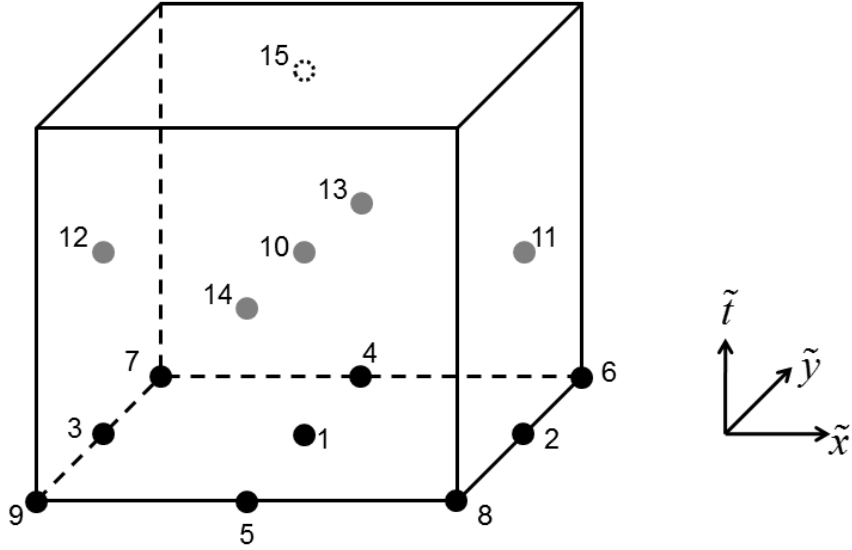

*Fig. 12 shows the placement of nodal points in the reference space-time element. The origin of our local coordinate system is actually centered at the first nodal point. The nine black circles, labeled 1 to 9, correspond to the nodes at time = 0. The five grey circles, labeled 10 to 14, correspond to the nodes at time = 1/2. Node 10 is collocated at the center of the space-time cube. Nodes 11 and 12 are centered in the x-faces; nodes 13 and 14 are centered in the y-faces. The dashed circle corresponds to the node 15 at time = 1. It is collocated at the center of the top face of the space-time cube.*

During each iteration, we start with the existing modes $\hat{F}_{i,j;x}$, $\hat{G}_{i,j;y}$, $\hat{F}_{i,j;xt}$, $\hat{G}_{i,j;yt}$, $\hat{F}_{i,j;xx}$, $\hat{G}_{i,j;xy}$, $\hat{F}_{i,j;xy}$ and $\hat{G}_{i,j;yy}$; please see the right hand sides of eqn. (77). Evaluating the right hand sides of eqn. (77) will then give us an improved set of time-dependent modes $\hat{U}_{i,j;t}$, $\hat{U}_{i,j;tt}$, $\hat{U}_{i,j;xt}$ and $\hat{U}_{i,j;yt}$. These can be used to build an improved set of time-dependent modes in eqns. (69) and (70) for use in the next iteration. We now pick a set of nodal points in the local space-time coordinate system with $\tilde{t} > 0$. The grey and dashed circles in Fig. 12 show one possible set of such nodes in space and time. They are given by the ordered set of nodal points in the reference space-time element:

$$\{(0,0,1/2),(1/2,0,1/2),(-1/2,0,1/2),(0,1/2,1/2),(0,-1/2,1/2),(0,0,1)\} \qquad (80)$$

The above set is labeled from 11 to 15 in Fig. 12. As before, conserved variables can be obtained at those nodes by using our best available approximation of eqn. (68). The conserved variables at these nodes can, in turn, be used to obtain a better approximation for the fluxes at the same nodes. Denoting the nodal location with a superscript, we now list the transcription from the nodal values to the time-dependent modes of the x-flux in eqn. (69) as follows

$$\hat{F}_{i,j;xt} = 2\left(\tilde{F}_{i,j}^{11} - \tilde{F}_{i,j}^{12} - \tilde{F}_{i,j}^{2} + \tilde{F}_{i,j}^{3}\right) \quad ; \quad \hat{F}_{i,j;yt} = 2\left(\tilde{F}_{i,j}^{13} - \tilde{F}_{i,j}^{14} - \tilde{F}_{i,j}^{4} + \tilde{F}_{i,j}^{5}\right) \quad ;$$

$$\hat{F}_{i,j;tt} = 2\left(\tilde{F}_{i,j}^{15} - 2\,\tilde{F}_{i,j}^{10} + \tilde{F}_{i,j}^{1}\right) \quad ; \quad \hat{F}_{i,j;t} = \tilde{F}_{i,j}^{15} - \tilde{F}_{i,j}^{1} - \hat{F}_{i,j;tt}$$

(81)

A similar transcription can be used for obtaining the time-dependent modes of the y-flux in eqn. (70). This completes our description of the predictor step of the ADER-CG scheme in two dimensions at third order.

---

**Stepwise Description of the ADER-CG Predictor Step**

The predictor step consists of an initialization which is described in Steps 1 to 3 below. The initialization is then followed by two iterations at third order. Each iteration consists of repeating Steps 4 to 6 below. Steps 7 and 8 are meant to polish the space-time representation of the fluxes after the iteration.

<u>Step 1</u>: Impose boundary conditions and use the results from Section III to obtain the spatial modes in eqn. (68).

<u>Step 2</u>: Using eqn. (68), evaluate the conserved variables at the spatial nodal points given in eqn. (78). Use those conserved variables to obtain the fluxes at the same nodal points.

<u>Step 3</u>: Use the nodal values of the fluxes in eqn. (79) to obtain the spatial modes in eqns. (69) and (70). The time-dependent modes in eqns. (68) to (70) are set to zero. This completes the initialization step that is done only once before the start of the iteration.

<u>Step 4</u>: This is the start of the iteration. Evaluate the right hand sides of eqn. (77). This gives an improved set of time-dependent modes for eqn. (68).

<u>Step 5</u>: Use the improved eqn. (68) to evaluate the conserved variables at the space-time nodal points given in eqn. (80). Use those conserved variables to obtain the fluxes at the same nodal points.

<u>Step 6</u>: Use the nodal values of the fluxes in eqn. (81) to obtain the time-dependent modes in eqns. (69) and (70). Go back to the start of the iteration. Iterate twice for third order.

<u>Step 7</u>: Repeat Step 5.

<u>Step 8</u>: Repeat Step 6.

---

**V.3.ii) Multidimensional ADER-CG Corrector Step**

Sub-section IV.3 demonstrated a very efficient quadrature-free strategy for starting with a higher order spatial variation, i.e. eqn. (44), and using it to obtain a spatially averaged numerical flux. In other words, we devised a computationally efficient strategy for integrating eqn. (46) over the face of interest. The Runge-Kutta schemes that were documented in Sub-section IV.3 use multiple stages to build a time-accurate update. The predictor step of the ADER-CG scheme documented above yields the space-time variation of the conserved variables and the fluxes in eqns. (68) to (70). Eqns. (3) and (4) show how the space-time integration of the fluxes at zone boundaries yields a one-step update. It is our goal to demonstrate how such an update can be carried out at high orders in a quadrature-free fashion by using an ADER formulation. For the sake of simplicity, we make our demonstration specific to third order on a two-dimensional structured mesh.

However, the ideas readily extend to three dimensions and unstructured meshes (Dumbser *et al.* 2008, Balsara *et al.* 2009, 2013).

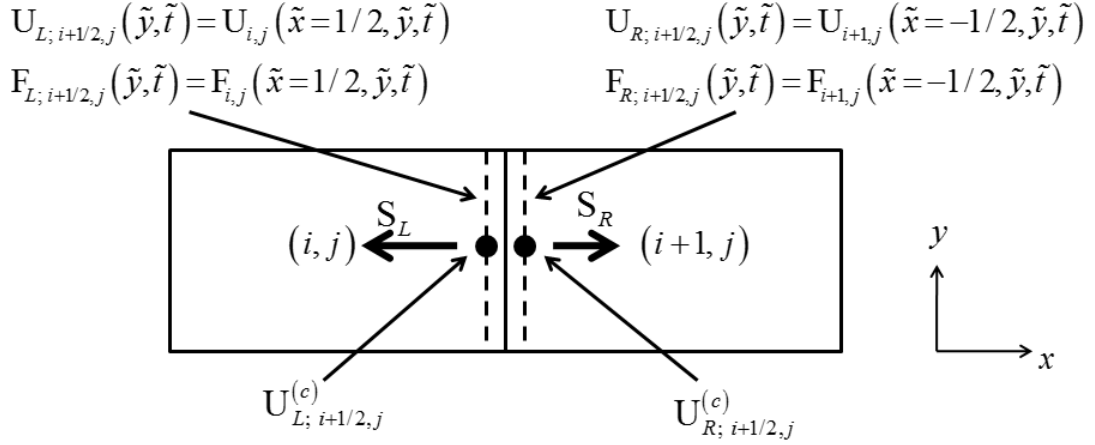

Fig. 13 shows the construction for obtaining spatially and temporally accurate fluxes for a higher order ADER scheme. The variables used for obtaining the wave model are shown with dots. The surfaces to the immediate left and immediate right of the zone boundary at (i+1/2,j) are shown with dashes.

Let us begin by extending eqn. (46) for the HLL flux in the $x$-direction to include space and time variations in the upper $x$-boundary of the zone $(i,j)$ as

$$F_{i+1/2,j}(\tilde{y},\tilde{t}) = \left[\frac{S_R}{S_R - S_L}\right] F_{L;\,i+1/2,j}(\tilde{y},\tilde{t}) - \left[\frac{S_L}{S_R - S_L}\right] F_{R;\,i+1/2,j}(\tilde{y},\tilde{t}) \\ + \left[\frac{S_R S_L}{S_R - S_L}\right] \left(U_{R;\,i+1/2,j}(\tilde{y},\tilde{t}) - U_{L;\,i+1/2,j}(\tilde{y},\tilde{t})\right) \tag{82}$$

To define an HLL Riemann solver at the $x$-face indexed by $(i+1/2,j)$, we need to find the extremal wave speeds, $S_L$ and $S_R$, flowing in the x-direction at that zone boundary. We can obtain these speeds by evaluating eqn. (68) and its analogue from the zone $(i+1,j)$ on either side of the spatial and temporal center of the $x$-face being considered. To make this concrete, we build the two vectors of conserved variables given by

$$U^{(c)}_{L;\,i+1/2,j} = U_{i,j}(\tilde{x}=1/2,\tilde{y}=0,\tilde{t}=1/2) \quad;\quad U^{(c)}_{R;\,i+1/2,j} = U_{i+1,j}(\tilde{x}=-1/2,\tilde{y}=0,\tilde{t}=1/2)$$
(83)

The above two states are analogous to the states in eqn. (45) with the exception that eqn. (83) is also centered in time. We then use the left and right boundary values, $U^{(c)}_{L;\,i+1/2,j}$ and $U^{(c)}_{R;\,i+1/2,j}$ to obtain $S_L$ and $S_R$. Observe that eqn. (83) is different from eqn. (45) in that it corresponds to extending Fig. 8 in the temporal direction. This is shown in Fig. 13. Notice that Fig. 8 only includes spatial variation whereas Fig. 13 includes the spatial and temporal variation of the conserved variables and fluxes on either side of the zone boundary. With $S_L$ and $S_R$ being frozen for this time step, eqn. (82) becomes a linear function of $U_{L;\,i+1/2,j}(\tilde{y},\tilde{t})$ and $F_{L;\,i+1/2,j}(\tilde{y},\tilde{t})$, which are evaluated by using the space-time variation in the zone that lies to the left of the zone boundary shown in Fig. 13, and $U_{R;\,i+1/2,j}(\tilde{y},\tilde{t})$ and $F_{R;\,i+1/2,j}(\tilde{y},\tilde{t})$, which are evaluated by using the space-time variation in the zone that lies to the right of the zone boundary shown in Fig. 13. The utility of having the space-time representation of the solution and fluxes in eqns. (68) to (70) now becomes readily apparent. Using eqn. (68), the spatially and temporally integrated value of $U_{L;\,i+1/2,j}(\tilde{y},\tilde{t})$ is then given by using the space-time variation in the zone $(i,j)$ as

$$\int_{\tilde{t}=0}^{\tilde{t}=1}\int_{\tilde{y}=-1/2}^{\tilde{y}=1/2} U_{L;\,i+1/2,j}(\tilde{y},\tilde{t})\,d\tilde{y}\,d\tilde{t} = \left(\bar{U}_{i,j} + \frac{1}{2}\hat{U}_{i,j;x} + \frac{1}{6}\hat{U}_{i,j;xx}\right) + \frac{1}{2}\left(\hat{U}_{i,j;t} + \frac{1}{2}\hat{U}_{i,j;xt}\right) + \frac{1}{3}\hat{U}_{i,j;tt} \quad (84)$$

An analogous expression can be written for the space-time integral of $F_{L;\,i+1/2,j}(\tilde{y},\tilde{t})$ by using eqn. (69). Similarly, the spatially and temporally integrated value of $U_{R;\,i+1/2,j}(\tilde{y},\tilde{t})$ is given by using the space-time representation of the conserved variables in the zone $(i+1,j)$ as

$$\int_{\tilde{t}=0}^{\tilde{t}=1}\int_{\tilde{y}=-1/2}^{\tilde{y}=1/2} U_{R;\,i+1/2,j}(\tilde{y},\tilde{t})\,d\tilde{y}\,d\tilde{t} = \left(\bar{U}_{i+1,j} - \frac{1}{2}\hat{U}_{i+1,j;x} + \frac{1}{6}\hat{U}_{i+1,j;xx}\right) + \frac{1}{2}\left(\hat{U}_{i+1,j;t} - \frac{1}{2}\hat{U}_{i+1,j;xt}\right) + \frac{1}{3}\hat{U}_{i+1,j;tt} \quad (85)$$

An analogous expression can be written for the space-time integral of $F_{R;\,i+1/2,j}(\tilde{y},\tilde{t})$. These integrals enable us to obtain a third order accurate space-time integration of the numerical flux in eqn. (82). Eqns. (84) and (85), along with their analogues for the x-flux, enable us to write down that space-time integration explicitly. By applying these ideas in both dimensions we get space-time averaged numerical fluxes that can be directly used in eqn. (3) to obtain a one-step update for our conservation law. This completes our description of the ADER-CG corrector step. The three dimensional extension has more terms but is easily accomplished with the help of a symbolic manipulation package.

**Stepwise Description of the ADER-CG Corrector Step**

Step1: Use eqn. (68), along its analogue in zone $(i+1,j)$, to obtain $U^{(c)}_{L;\,i+1/2,j}$ and $U^{(c)}_{R;\,i+1/2,j}$ from eqn. (83). Use them to obtain the extremal wave speeds, $S_L$ and $S_R$, for use in eqn. (82).

> Step 2: Use eqns. (84) and (85), as well as their analogues for the *x*-fluxes, to obtain the space-time integrals of the right hand side of eqn. (82). This yields a numerical flux at the *x*-faces. Construct similar numerical fluxes at the *y*-faces.
>
> Step 3: Use eqn. (3) to obtain a one-step, third order accurate update. This completes the ADER-CG corrector step as well as the entire time step.

## VI) Runge-Kutta Discontinuous Galerkin (RKDG) Schemes

*Galerkin schemes* refer to a class of schemes that posit a set of basis functions on the entire computational domain and then solve the problem in terms of the modes of the basis functions. Fourier techniques for solving PDEs can be thought of as an example of Galerkin methods. Sine and cosine functions form the basis functions in this example and the solution is expressed in terms of the Fourier modes, i.e. the coefficients of the sines and cosines. Because we are interested in hyperbolic conservation laws that can give rise to discontinuities, it is not advantageous to have a set of basis functions that span the entire computational domain. For example, if a discontinuous function is represented in terms of a discrete set of Fourier basis functions, we would encounter a Gibbs phenomenon at the location of the discontinuity. (In all fairness, spectral methods can handle problems with a few weak and isolated shocks, but it becomes increasingly difficult to handle the general case where strong shocks may form at several locations.)

The rest of this section is split into three sub-sections. The first sub-section provides a basic description of DG methods. The second sub-section describes recent WENO limiters; the sub-section after that describes MOOD limiters.

## VI.1) Basic Description of discontinuous Galerkin (DG) Methods

*Runge-Kutta Discontinuous Galerkin* (RKDG) methods are based on the idea that within each zone one can have a small set of basis functions that may indeed become discontinuous at zone boundaries (Cockburn & Shu 1989, 1998, Cockburn, Hou & Shu 1990, Cockburn, Karniadakis and Shu 2000). The discontinuities at zone boundaries can then be treated by solving a Riemann problem. The moments of the basis functions then become the independent variables that are to be evolved by the scheme. (The basis functions are also sometimes called trial functions.) Let us consider eqn. (44) to appreciate the difference between a scheme that is based on reconstruction and a discontinuous Galerkin scheme. A third order scheme that reconstructs the solution would reconstruct all the moments in eqn. (44), except of course the zone-averaged value. This would have to be done at each stage of the Runge-Kutta time evolution strategy. Only one evolutionary equation is solved for the vector of conserved variables $\overline{U}_{i,j}$ in zone $(i, j)$. I.e., the components of $\overline{U}_{i,j}$ in zone $(i, j)$ are the only *degrees of freedom* in that zone. In contrast, a third order discontinuous Galerkin (DG) method is based on the viewpoint that all the moments in eqn. (44) are degrees of freedom in zone $(i, j)$ and should be evolved in time. This is to be done in a fashion that is consistent with the governing equations, i.e. the hyperbolic conservation law. Six evolutionary equations are then developed for the six vectors $\bar{U}_{i,j}$, $\hat{U}_{i,j;x}$, $\hat{U}_{i,j;y}$, $\hat{U}_{i,j;xx}$, $\hat{U}_{i,j;yy}$ and $\hat{U}_{i,j;xy}$. I.e., we now have six times as many degrees of freedom as we would have in a reconstruction-based algorithm. Thus, in place of eqn. (44), we can extend our notation to show the time-dependence as

$$U_{i,j}(\tilde{x}, \tilde{y}, t) = \bar{U}_{i,j}(t) + \hat{U}_{i,j;x}(t)\tilde{x} + \hat{U}_{i,j;y}(t)\tilde{y} + \hat{U}_{i,j;xx}(t)\left(\tilde{x}^2 - \frac{1}{12}\right)$$
$$+ \hat{U}_{i,j;yy}(t)\left(\tilde{y}^2 - \frac{1}{12}\right) + \hat{U}_{i,j;xy}(t)\tilde{x}\,\tilde{y} \tag{86}$$

All the modes in eqn. (86) have, therefore, been endowed with time-evolution. A third order Runge-Kutta time stepping scheme can be used to discretize their evolution in time.

Recall that reconstruction-based schemes build all the moments of the zone-centered variable in eqn. (44). This reconstruction is carried out at the start of every time step, and yet, only the zone-averaged variable is updated at the end of a time step. In contrast, because all the moments are evolved in an RKDG scheme, and the evolution is consistent with the governing equation, the method can be very accurate. If the solution is smooth to begin with and remains so in most of the zones of the computational domain, then evolving all the moments can really help improve accuracy. Our guiding philosophy in a DG scheme would therefore be to do as little limiting as possible within a zone, because any such limiting would damage the information that is contained in some or all of the higher moments. In regions of smooth flow, no limiting is needed so that the method retains its theoretical accuracy. In practice, the presence of discontinuities forces us to restrict the higher moments in eqn. (86), with the result that RKDG schemes, quite like their finite volume brethren, have to be non-linearly stabilized. However, the philosophy is to apply non-linear stabilization to the moments as sparingly as possible. Practical experience has shown that RKDG schemes can be stabilized with a minimal amount of limiting.

Also recall that as the order of accuracy increases, reconstruction-based schemes use increasingly larger stencils that impede parallelism. If nonlinear stabilizaion is not needed in the physical problem, the RKDG method requires a very small stencil. The small stencil can become an advantage on parallel computers.

In the course of this Section we will see that DG methods are generalizations of finite volume methods in the sense that they use all the concepts of limiting and Riemann solvers that were initially developed within the context of finite volume schemes. However, DG methods recast these ideas within the context of a finite element framework. This makes the method very proficient at handling flow problems with complicated, body-fitted geometries (Bassi and Rebay 1997**,** Warburton *et al.* 1998). DG methods have also been used for solving problems on arbitrary Lagrangian Eulerian (ALE) meshes where the zone boundaries of the mesh can move in response to flow features or other dynamics (van der Vegt and van der Ven 2002a,b, Boscheri and Dumbser 2014, Boscheri *et al.* 2014, 2015, Boscheri and Dumbser 2017). When dealing with problems with geometric complexity, one has to go through the complication of working with a set of boundary-conforming elements though (Dubiner 1991, Warburton 1990, Karniadakis and Sherwin 1999). The Galerkin formulation also makes DG methods very useful for solution-dependent space and time adaptivity (Biswas, Devine and Flaherty 1994). DG methods enable one to simultaneously have *h-adaptivity*, where the size of the mesh (denoted by "*h*") is locally refined, and *p-adaptivity*, where the order of the method (denoted by "*p*") is increased on refined patches. Collectively, this is known as *hp-adaptivity*. The *hp*-adaptive methods can offer spectral-like convergence to the physical solution of a scientific or engineering problem. As a result, DG methods are very popular in engineering applications where one simultaneously has complicated boundaries and a need to refine with increasing accuracy around local surfaces of interest.

Consider the hyperbolic system in eqn. (1) which has to be solved on the mesh shown in Fig. 1. Say we have to take a time step on zones with sizes $\Delta x$ and $\Delta y$ in each direction. In terms of the local coordinates within a zone, eqn. (1) can be written as

$$\frac{\partial U}{\partial t} = -\frac{1}{\Delta x} F(U)_{\tilde{x}} - \frac{1}{\Delta y} G(U)_{\tilde{y}} \tag{87}$$

Notice that the derivatives on the right hand side have been written in the zone's local coordinates. Comparing eqn. (87) to eqn. (35), it is easy to see how the time discretization might be carried out with a Runge-Kutta method. However, we need to find evolutionary equations for all the moments of eqn. (86). To that end, realize that the basis functions in eqn. (86) are actually a set of orthogonal Legendre polynomials. As with any set of basis functions, we can obtain their coefficients, i.e. the modes or the degrees of freedom, by making an orthogonal projection. When the basis functions are not orthogonal, the derivation becomes only slightly more involved. Thus, in general, we multiply the above equation by an arbitrary test function $\varphi(\tilde{x}, \tilde{y})$ that is defined over the zone $(i, j)$ of interest and integrate over the zone of interest. Using integration by parts we get:

$$\frac{\partial}{\partial t}\left[\int_{\tilde{y}=-1/2}^{\tilde{y}=1/2}\int_{\tilde{x}=-1/2}^{\tilde{x}=1/2} \varphi(\tilde{x},\tilde{y})\, U(\tilde{x},\tilde{y},t)\, d\tilde{x}\, d\tilde{y}\right] = \frac{1}{\Delta x}\left[\int_{\tilde{y}=-1/2}^{\tilde{y}=1/2}\int_{\tilde{x}=-1/2}^{\tilde{x}=1/2} F(\tilde{x},\tilde{y},t)\, \partial_{\tilde{x}}\varphi(\tilde{x},\tilde{y})\, d\tilde{x}\, d\tilde{y}\right]$$

$$-\frac{1}{\Delta x}\left[\int_{\tilde{y}=-1/2}^{\tilde{y}=1/2} F(\tilde{x}=1/2,\tilde{y},t)\, \varphi(\tilde{x}=1/2,\tilde{y})\, d\tilde{y}\right] + \frac{1}{\Delta x}\left[\int_{\tilde{y}=-1/2}^{\tilde{y}=1/2} F(\tilde{x}=-1/2,\tilde{y},t)\, \varphi(\tilde{x}=-1/2,\tilde{y})\, d\tilde{y}\right]$$

$$+\frac{1}{\Delta y}\left[\int_{\tilde{y}=-1/2}^{\tilde{y}=1/2}\int_{\tilde{x}=-1/2}^{\tilde{x}=1/2} G(\tilde{x},\tilde{y},t)\, \partial_{\tilde{y}}\varphi(\tilde{x},\tilde{y})\, d\tilde{x}\, d\tilde{y}\right]$$

$$-\frac{1}{\Delta y}\left[\int_{\tilde{x}=-1/2}^{\tilde{x}=1/2} G(\tilde{x},\tilde{y}=1/2,t)\, \varphi(\tilde{x},\tilde{y}=1/2)\, d\tilde{x}\right] + \frac{1}{\Delta y}\left[\int_{\tilde{x}=-1/2}^{\tilde{x}=1/2} G(\tilde{x},\tilde{y}=-1/2,t)\, \varphi(\tilde{x},\tilde{y}=-1/2)\, d\tilde{x}\right]$$

$$\tag{88}$$

If the basis functions form an orthogonal set, as they do for eqn. (86), then it is always best to draw the test functions from that set.

It is worthwhile to make four observations about eqn. (88). First, notice that the integrals are applied componentwise for a hyperbolic conservation law. Second, notice that when $\varphi(\tilde{x}, \tilde{y})$ is set to unity, i.e. our test function is a constant, then the first and fourth terms on the right hand side of eqn. (88) become zero. Eqn. (88) just yields an evolutionary equation for the conserved quantity, i.e. the first term on the right hand side of eqn. (86). Thus we get

$$\frac{\partial \overline{U}_{i,j}(t)}{\partial t} = - \frac{1}{\Delta x} \left[ \int_{\tilde{y}=-1/2}^{\tilde{y}=1/2} F(\tilde{x}=1/2, \tilde{y}, t) \, d\tilde{y} \right] + \frac{1}{\Delta x} \left[ \int_{\tilde{y}=-1/2}^{\tilde{y}=1/2} F(\tilde{x}=-1/2, \tilde{y}, t) \, d\tilde{y} \right]$$
$$- \frac{1}{\Delta y} \left[ \int_{\tilde{x}=-1/2}^{\tilde{x}=1/2} G(\tilde{x}, \tilde{y}=1/2, t) \, d\tilde{x} \right] + \frac{1}{\Delta y} \left[ \int_{\tilde{x}=-1/2}^{\tilde{x}=1/2} G(\tilde{x}, \tilde{y}=-1/2, t) \, d\tilde{x} \right]$$
(89)

In that case, the boundary integrals in eqn. (89) should match up with the ones in eqns. (39) and (40). In other words, the boundary integrals on the right hand side of eqn. (88) should retrieve the upwinded fluxes evaluated at the appropriate order at the boundaries of the zone being considered. Since we are illustrating RKDG schemes at third order, we must retrieve eqns. (46) to (52) if we are using an HLL flux, and analogous expressions if we are using a different flux function. This is achieved if $F(\tilde{x}=1/2, \tilde{y}, t)$ and $F(\tilde{x}=-1/2, \tilde{y}, t)$ are actually the resolved x-fluxes coming from a properly upwinded Riemann solver applied to the upper and lower x-boundaries of the zone being considered. Similarly, $G(\tilde{x}, \tilde{y}=1/2, t)$ and $G(\tilde{x}, \tilde{y}=-1/2, t)$ are resolved y-fluxes provided by an upwinded Riemann solver applied to the upper and lower y-boundaries of the zone being considered. This is referred to as a *weak formulation* of the hyperbolic system. Our reinterpretation of the surface integrals in eqn. (88) provides a properly upwinded flux which, in turn, enables the variables in one zone to interact with their neighbors across the zone boundaries. These upwinded fluxes are used in the update of all the boundary integrals in eqn. (88). Third, notice that when $\varphi(\tilde{x}, \tilde{y})$ has spatial variation, the first and fourth terms on the right hand side of eqn. (88) pick up non-trivial contributions from the area integrals. Those terms are needed for accurate time-evolution of higher moments, as we will see in the next paragraph. Fourth, notice that when basis functions are non-orthogonal, one has to invert a small matrix, known as a *mass matrix*, in order to obtain the modal time evolution. Since our basis set is orthogonal, our mass matrix is a diagonal matrix and we do not face this problem here.

We now write out the time-evolution of the modes in eqn. (86) explicitly. This is most easily done by using the Legendre polynomials as our test functions. The zeroth moment is already catalogued in eqn. (89). Using $\varphi(\tilde{x}, \tilde{y}) = \tilde{x}$ in eqn. (88) we get

$$\frac{1}{12}\frac{\partial \hat{U}_{i,j;x}(t)}{\partial t} = \frac{1}{\Delta x}\left[\int_{\tilde{y}=-1/2}^{\tilde{y}=1/2}\int_{\tilde{x}=-1/2}^{\tilde{x}=1/2} F(\tilde{x},\tilde{y},t)\, d\tilde{x}\, d\tilde{y}\right]$$

$$-\frac{1}{2\Delta x}\left[\int_{\tilde{y}=-1/2}^{\tilde{y}=1/2} F(\tilde{x}=1/2,\tilde{y},t)\, d\tilde{y}\right] - \frac{1}{2\Delta x}\left[\int_{\tilde{y}=-1/2}^{\tilde{y}=1/2} F(\tilde{x}=-1/2,\tilde{y},t)\, d\tilde{y}\right] \quad (90)$$

$$-\frac{1}{\Delta y}\left[\int_{\tilde{x}=-1/2}^{\tilde{x}=1/2} G(\tilde{x},\tilde{y}=1/2,t)\, \tilde{x}\, d\tilde{x}\right] + \frac{1}{\Delta y}\left[\int_{\tilde{x}=-1/2}^{\tilde{x}=1/2} G(\tilde{x},\tilde{y}=-1/2,t)\, \tilde{x}\, d\tilde{x}\right]$$

Notice that the factor 1/12 on the left hand side of eqn. (90) comes from the mass matrix. Also notice the area integration, which is an extra term that one has to evaluate with an appropriate order of accuracy. In RKDG schemes, this is usually done by numerical quadrature. Furthermore, observe that both the x-flux terms contribute with the same signs at the boundaries. I.e. while there is a conservation principle for the conserved variables, see eqn. (89), there is no conservation principle for the higher moments. Using $\varphi(\tilde{x},\tilde{y})=\tilde{y}$ in eqn. (88) we get

$$\frac{1}{12}\frac{\partial \hat{U}_{i,j;y}(t)}{\partial t} = \frac{1}{\Delta y}\left[\int_{\tilde{y}=-1/2}^{\tilde{y}=1/2}\int_{\tilde{x}=-1/2}^{\tilde{x}=1/2} G(\tilde{x},\tilde{y},t)\, d\tilde{x}\, d\tilde{y}\right]$$

$$-\frac{1}{\Delta x}\left[\int_{\tilde{y}=-1/2}^{\tilde{y}=1/2} F(\tilde{x}=1/2,\tilde{y},t)\, \tilde{y}\, d\tilde{y}\right] + \frac{1}{\Delta x}\left[\int_{\tilde{y}=-1/2}^{\tilde{y}=1/2} F(\tilde{x}=-1/2,\tilde{y},t)\, \tilde{y}\, d\tilde{y}\right] \quad (91)$$

$$-\frac{1}{2\Delta y}\left[\int_{\tilde{x}=-1/2}^{\tilde{x}=1/2} G(\tilde{x},\tilde{y}=1/2,t)\, d\tilde{x}\right] - \frac{1}{2\Delta y}\left[\int_{\tilde{x}=-1/2}^{\tilde{x}=1/2} G(\tilde{x},\tilde{y}=-1/2,t)\, d\tilde{x}\right]$$

Using $\varphi(\tilde{x},\tilde{y})=\left(\tilde{x}^2-1/12\right)$ in eqn. (88) yields

$$\frac{1}{180}\frac{\partial \hat{U}_{i,j;xx}(t)}{\partial t} = \frac{2}{\Delta x}\left[\int_{\tilde{y}=-1/2}^{\tilde{y}=1/2}\int_{\tilde{x}=-1/2}^{\tilde{x}=1/2} F(\tilde{x},\tilde{y},t)\, \tilde{x}\, d\tilde{x}\, d\tilde{y}\right]$$

$$-\frac{1}{6\Delta x}\left[\int_{\tilde{y}=-1/2}^{\tilde{y}=1/2} F(\tilde{x}=1/2,\tilde{y},t)\, d\tilde{y}\right] + \frac{1}{6\Delta x}\left[\int_{\tilde{y}=-1/2}^{\tilde{y}=1/2} F(\tilde{x}=-1/2,\tilde{y},t)\, d\tilde{y}\right]$$

$$-\frac{1}{\Delta y}\left[\int_{\tilde{x}=-1/2}^{\tilde{x}=1/2} G(\tilde{x},\tilde{y}=1/2,t)\, \left(\tilde{x}^2-1/12\right) d\tilde{x}\right] + \frac{1}{\Delta y}\left[\int_{\tilde{x}=-1/2}^{\tilde{x}=1/2} G(\tilde{x},\tilde{y}=-1/2,t)\, \left(\tilde{x}^2-1/12\right) d\tilde{x}\right]$$
$$(92)$$

Using $\varphi(\tilde{x},\tilde{y})=\left(\tilde{y}^2-1/12\right)$ in eqn. (88) yields

$$\frac{1}{180} \frac{\partial \hat{U}_{i,j;yy}(t)}{\partial t} = +\frac{2}{\Delta y}\left[\int_{\tilde{y}=-1/2}^{\tilde{y}=1/2}\int_{\tilde{x}=-1/2}^{\tilde{x}=1/2} G(\tilde{x},\tilde{y},t)\,\tilde{y}\,d\tilde{x}\,d\tilde{y}\right]$$
$$-\frac{1}{\Delta x}\left[\int_{\tilde{y}=-1/2}^{\tilde{y}=1/2} F(\tilde{x}=1/2,\tilde{y},t)\,(\tilde{y}^2-1/12)\,d\tilde{y}\right] + \frac{1}{\Delta x}\left[\int_{\tilde{y}=-1/2}^{\tilde{y}=1/2} F(\tilde{x}=-1/2,\tilde{y},t)\,(\tilde{y}^2-1/12)\,d\tilde{y}\right]$$
$$-\frac{1}{6\Delta y}\left[\int_{\tilde{x}=-1/2}^{\tilde{x}=1/2} G(\tilde{x},\tilde{y}=1/2,t)\,d\tilde{x}\right] + \frac{1}{6\Delta y}\left[\int_{\tilde{x}=-1/2}^{\tilde{x}=1/2} G(\tilde{x},\tilde{y}=-1/2,t)\,d\tilde{x}\right]$$
(93)

Lastly, using $\varphi(\tilde{x},\tilde{y}) = \tilde{x}\,\tilde{y}$ in eqn. (88) yields

$$\frac{1}{144}\frac{\partial \hat{U}_{i,j;xy}(t)}{\partial t} = \frac{1}{\Delta x}\left[\int_{\tilde{y}=-1/2}^{\tilde{y}=1/2}\int_{\tilde{x}=-1/2}^{\tilde{x}=1/2} F(\tilde{x},\tilde{y},t)\,\tilde{y}\,d\tilde{x}\,d\tilde{y}\right] + \frac{1}{\Delta y}\left[\int_{\tilde{y}=-1/2}^{\tilde{y}=1/2}\int_{\tilde{x}=-1/2}^{\tilde{x}=1/2} G(\tilde{x},\tilde{y},t)\,\tilde{x}\,d\tilde{x}\,d\tilde{y}\right]$$
$$-\frac{1}{2\Delta x}\left[\int_{\tilde{y}=-1/2}^{\tilde{y}=1/2} F(\tilde{x}=1/2,\tilde{y},t)\,\tilde{y}\,d\tilde{y}\right] - \frac{1}{2\Delta x}\left[\int_{\tilde{y}=-1/2}^{\tilde{y}=1/2} F(\tilde{x}=-1/2,\tilde{y},t)\,\tilde{y}\,d\tilde{y}\right]$$
$$-\frac{1}{2\Delta y}\left[\int_{\tilde{x}=-1/2}^{\tilde{x}=1/2} G(\tilde{x},\tilde{y}=1/2,t)\,\tilde{x}\,d\tilde{x}\right] - \frac{1}{2\Delta y}\left[\int_{\tilde{x}=-1/2}^{\tilde{x}=1/2} G(\tilde{x},\tilde{y}=-1/2,t)\,\tilde{x}\,d\tilde{x}\right]$$
(94)

This completes our derivation of the time-dependence of the modes in eqn. (86). Third order Runge-Kutta time stepping from eqn. (37) can be used to evolve the eqns. from (89) to (94)).

    While the DG methods have several genuine advantages in certain circumstances, they also have their drawbacks. As the number of moments that one evolves increases, the permitted explicit time step decreases (Cockburn, Shu and Karniadakis 2000). One way to rectify this consists of evolving only a few of the lower moments while reconstructing the higher moments (Qiu and Shu 2004, 2005, Balsara *et al.* 2007, Dumbser *et al.* 2008). This does increase the permitted time step while relinquishing only a small amount of the accuracy. In the vicinity of discontinuities, a limiter does need to be applied to the higher moments in eqn. (86). The high resolution that comes from evolving the higher moments is only realized if most of the moments are not changed by the limiting process. Thus in problems with several strong, interacting shocks, these methods might lose their advantage. Several limiters have been presented over the years (Biswas, Devine and Flaherty 1994, Burbeau, Sagaut, Bruneau 2001, Qiu and Shu 2004, 2005, Balsara *et al.* 2007, Krivodonova 2007, Zhu *et al.* 2008, Xu, Liu & Shu 2009a,b,c, Xu & Lin 2009, Xu *et al.* 2011, Zhu and Qiu 2011, Zhong and Shu 2013, Zhu *et al.* 2013, Dumbser *et al.* 2014). The problem is that there has been no coalescence of consensus around any one particular limiter. In the next Sub-Section we present a WENO limiter by Zhong and Shu (2013). In the Sub-section after that, we present the MOOD limiter of Dumbser *et al.* (2014). Storing the large number of degrees of freedom can also be problematical if computer memory is limited.

**VI.2) WENO Limiter for DG Methods**

We now describe the simplest form of WENO limiting (Zhong and Shu 2013, Zhu *et al*. 2013) with several modifications made here to make it amenable to seamless implementation. This limiting strategy is to be used with some form of discontinuity detector so that one only invokes the limiter in zones that have a discontinuity, i.e. zones that are denoted "troubled" zones. (We discuss positivity preserving reconstruction in higher order schemes in a subsequent section. In that section, we will have occasion to discuss discontinuity detectors.) The underlying idea is that one should only invoke this limiter in as few zones as possible. The other design philosophy is that even if the limiter is invoked in a zone where it may not truly be needed, it should not damage the higher order accuracy of the DG algorithm. Let us denote the zone that has to be limited on a two-dimensional Cartesian mesh with a subscript "$i, j$". We illustrate the third order limiting procedure for this zone. We assume that from the previous timestep the DG scheme has left us with a polynomial given by

$$U_{i,j}(\tilde{x}, \tilde{y}) = \bar{U}_{i,j} + \hat{U}_{i,j;x}\tilde{x} + \hat{U}_{i,j;y}\tilde{y} + \hat{U}_{i,j;xx}\left(\tilde{x}^2 - \frac{1}{12}\right) + \hat{U}_{i,j;yy}\left(\tilde{y}^2 - \frac{1}{12}\right) + \hat{U}_{i,j;xy}\tilde{x}\tilde{y}$$
(95)

Notice that eqn. (95) resembles eqn. (86), the only difference being that the time dependence has been dropped so as to yield a more compact notation. Displaying the DG limiter at third order in 2D is general enough to enable the reader to extend these ideas to any order and also 3D with the help of a computer algebra system. Notice too that conservation requires that only the mode $\bar{U}_{i,j}$ in the above equation should be kept intact, the remaining modal coefficients, i.e. $\hat{U}_{i,j;x}$, $\hat{U}_{i,j;y}$, $\hat{U}_{i,j;xx}$, $\hat{U}_{i,j;yy}$ and $\hat{U}_{i,j;xy}$, can be modified via the limiting process.

Because our limiting is based on the WENO philosophy, we first define smoothness indicators. For the polynomial in eqn. (95) we can construct a smoothness indicator that is given by

$$IS_{i,j} = \int_{z=-1/2}^{1/2} \int_{y=-1/2}^{1/2} \int_{x=-1/2}^{1/2} \left[ \begin{array}{c} \left(\frac{\partial u_{i,j}(\tilde{x}, \tilde{y})}{\partial \tilde{x}}\right)^2 + \left(\frac{\partial u_{i,j}(\tilde{x}, \tilde{y})}{\partial \tilde{y}}\right)^2 + \left(\frac{\partial^2 u_{i,j}(\tilde{x}, \tilde{y})}{\partial \tilde{x}^2}\right)^2 + \\ \left(\frac{\partial^2 u_{i,j}(\tilde{x}, \tilde{y})}{\partial \tilde{y}^2}\right)^2 + \left(\frac{\partial^2 u_{i,j}(\tilde{x}, \tilde{y})}{\partial \tilde{x}\partial \tilde{y}}\right)^2 \end{array} \right] dx\, dy\, dz$$
(96)

The square bracket in the above equation is not a matrix, but just contains a summation of perfect squares. Here the $u_{i,j}(\tilde{x}, \tilde{y})$ denotes a component of $U_{i,j}(\tilde{x}, \tilde{y})$, which can be taken literally to be a component or it can even be taken to be an eigenweight that has been obtained via a characteristic projection. Because the polynomial in eqn. (95) is written in terms of an orthogonal basis set, the integration in eqn. (96) yields a nice closed form expression given by

$$IS_{i,j} = (\hat{u}_{i,j;x})^2 + (\hat{u}_{i,j;y})^2 + \frac{13}{3}(\hat{u}_{i,j;xx})^2 + \frac{13}{3}(\hat{u}_{i,j;yy})^2 + \frac{7}{6}(\hat{u}_{i,j;xy})^2$$
(97)

Here, $\hat{u}_{i,j;x}$ can be a component of $\hat{U}_{i,j;x}$ if we want to limit on the conserved variables. Alternatively, if we want to limit on the characteristic variables, $\hat{u}_{i,j;x}$ can be an eigenweight that is obtained by a characteristic projection of $\hat{U}_{i,j;x}$ in one of the two principal directions of the mesh.

The next step consists of realizing that the zone $(i, j)$ has four immediate von Neumann neighbors given by zones $(i+1, j)$, $(i-1, j)$, $(i, j+1)$ and $(i, j-1)$. Because solutions of hyperbolic PDEs propagate from one zone to the next, it is likely that even when the solution in zone $(i, j)$ is troubled, the solution in one of these neighboring zones is salient. As a result, the moments from that neighboring zone, suitably shifted to the current zone, could help to limit the zone $(i, j)$. We have now to study what a suitable shift is. Let us focus on the zone $(i+1, j)$ and examine how its reconstruction polynomial would appear if it were shifted one zone to the left so as to coincide with the zone $(i, j)$. We then have

$$U_{i+1,j}(\tilde{x}, \tilde{y}) = \overline{U}_{i+1,j} + \hat{U}_{i+1,j;x}(\tilde{x}-1) + \hat{U}_{i+1,j;y}\tilde{y} \\ + \hat{U}_{i+1,j;xx}\left((\tilde{x}-1)^2 - \frac{1}{12}\right) + \hat{U}_{i+1,j;yy}\left(\tilde{y}^2 - \frac{1}{12}\right) + \hat{U}_{i+1,j;xy}(\tilde{x}-1)\tilde{y} \tag{98}$$

Notice that the polynomial in zone $(i+1, j)$ is now written in the local coordinates of the zone $(i, j)$. A little simplification yields

$$U_{i+1,j}(\tilde{x}, \tilde{y}) = \left(\overline{U}_{i+1,j} - \hat{U}_{i+1,j;x} + \hat{U}_{i+1,j;xx}\right) + \left(\hat{U}_{i+1,j;x} - 2\hat{U}_{i+1,j;xx}\right)\tilde{x} + \left(\hat{U}_{i+1,j;y} - \hat{U}_{i+1,j;xy}\right)\tilde{y} \\ + \hat{U}_{i+1,j;xx}\left(\tilde{x}^2 - \frac{1}{12}\right) + \hat{U}_{i+1,j;yy}\left(\tilde{y}^2 - \frac{1}{12}\right) + \hat{U}_{i+1,j;xy}\tilde{x}\tilde{y} \tag{99}$$

Notice that the mean value of eqn (99), averaged over zone $(i, j)$, does not equal $\overline{U}_{i,j}$. However, if we were to replace $\left(\overline{U}_{i+1,j} - \hat{U}_{i+1,j;x} + \hat{U}_{i+1,j;xx}\right)$ by $\overline{U}_{i,j}$ in eqn. (99) then it could potentially be a polynomial that one could use to replace eqn. (95). Therefore, analogous to eqn. (99), we can define a shifted polynomial from zone $(i+1, j)$ which has the correct zone average for zone $(i, j)$. It is given by

$$\tilde{U}_{i+1,j}(\tilde{x}, \tilde{y}) = \overline{U}_{i,j} + \left(\hat{U}_{i+1,j;x} - 2\hat{U}_{i+1,j;xx}\right)\tilde{x} + \left(\hat{U}_{i+1,j;y} - \hat{U}_{i+1,j;xy}\right)\tilde{y} \\ + \hat{U}_{i+1,j;xx}\left(\tilde{x}^2 - \frac{1}{12}\right) + \hat{U}_{i+1,j;yy}\left(\tilde{y}^2 - \frac{1}{12}\right) + \hat{U}_{i+1,j;xy}\tilde{x}\tilde{y} \tag{100}$$

Eqn. (100) is the suitably shifted polynomial that is shifted from zone $(i+1, j)$ to zone $(i, j)$. In general, we do not want to replace eqn. (95) with eqn. (100). However, if zone $(i, j)$ is a troubled zone with a bad (i.e. a seriously TVD violating) solution, then this might be warranted. We now see that a WENO-style weighting between all the available von Neumann neighbors might help us decide whether to replace the troubled polynomial and by how much. We should do this weighting in a WENO style in order to avoid very rapid switching of the stencil. Relating eqn. (100) to eqns. (95) and (97) we can also write down a smoothness indicator for the shifted polynomial in eqn. (100). Note that even though $\tilde{U}_{i+1,j}(\tilde{x}, \tilde{y})$ relates to zone $(i+1, j)$, its smoothness indicator should be evaluated over zone $(i, j)$ when we seek a limiting procedure for zone $(i, j)$. Analogous to eqn. (97) we write the smoothness indicator for eqn. (100) explicitly as

$$IS_{i+1,j} = \left(\hat{u}_{i+1,j;x} - 2\hat{u}_{i+1,j;xx}\right)^2 + \left(\hat{u}_{i+1,j;y} - \hat{u}_{i+1,j;xy}\right)^2 + \frac{13}{3}\left(\hat{u}_{i+1,j;xx}\right)^2 + \frac{13}{3}\left(\hat{u}_{i+1,j;yy}\right)^2 + \frac{7}{6}\left(\hat{u}_{i+1,j;xy}\right)^2$$
(101)

This shows us how to shift a polynomial by one zone to a neighboring zone and how to evaluate its smoothness indicator.

We have three remaining immediate neighbors for the zone $(i, j)$. Now that we understand the concept, we quickly write down the analogues of eqn. (100) and the corresponding smoothness indicators. From the zone $(i-1, j)$ we obtain

$$\tilde{U}_{i-1,j}(\tilde{x}, \tilde{y}) = \overline{U}_{i,j} + \left(\hat{U}_{i-1,j;x} + 2\hat{U}_{i-1,j;xx}\right)\tilde{x} + \left(\hat{U}_{i-1,j;y} + \hat{U}_{i-1,j;xy}\right)\tilde{y}$$
$$+ \hat{U}_{i-1,j;xx}\left(\tilde{x}^2 - \frac{1}{12}\right) + \hat{U}_{i-1,j;yy}\left(\tilde{y}^2 - \frac{1}{12}\right) + \hat{U}_{i-1,j;xy}\tilde{x}\,\tilde{y}$$
(102)

and the corresponding smoothness indicator is

$$IS_{i-1,j} = \left(\hat{u}_{i-1,j;x} + 2\hat{u}_{i-1,j;xx}\right)^2 + \left(\hat{u}_{i-1,j;y} + \hat{u}_{i-1,j;xy}\right)^2 + \frac{13}{3}\left(\hat{u}_{i-1,j;xx}\right)^2 + \frac{13}{3}\left(\hat{u}_{i-1,j;yy}\right)^2 + \frac{7}{6}\left(\hat{u}_{i-1,j;xy}\right)^2$$
(103)

From the zone $(i, j+1)$ we obtain

$$\tilde{U}_{i,j+1}(\tilde{x}, \tilde{y}) = \overline{U}_{i,j} + \left(\hat{U}_{i,j+1;x} - \hat{U}_{i,j+1;xy}\right)\tilde{x} + \left(\hat{U}_{i,j+1;y} - 2\hat{U}_{i,j+1;yy}\right)\tilde{y}$$
$$+ \hat{U}_{i,j+1;xx}\left(\tilde{x}^2 - \frac{1}{12}\right) + \hat{U}_{i,j+1;yy}\left(\tilde{y}^2 - \frac{1}{12}\right) + \hat{U}_{i,j+1;xy}\tilde{x}\,\tilde{y}$$
(104)

and the corresponding smoothness indicator is

$$IS_{i,j+1} = \left(\hat{u}_{i,j+1;x} - \hat{u}_{i,j+1;xy}\right)^2 + \left(\hat{u}_{i,j+1;y} - 2\hat{u}_{i,j+1;yy}\right)^2 + \frac{13}{3}\left(\hat{u}_{i,j+1;xx}\right)^2 + \frac{13}{3}\left(\hat{u}_{i,j+1;yy}\right)^2 + \frac{7}{6}\left(\hat{u}_{i,j+1;xy}\right)^2$$
(105)

From the zone $(i, j-1)$ we obtain

$$\tilde{U}_{i,j-1}(\tilde{x}, \tilde{y}) = \bar{U}_{i,j} + \left(\hat{U}_{i,j-1;x} + \hat{U}_{i,j-1;xy}\right)\tilde{x} + \left(\hat{U}_{i,j-1;y} + 2\hat{U}_{i,j-1;yy}\right)\tilde{y} \\ + \hat{U}_{i,j-1;xx}\left(\tilde{x}^2 - \frac{1}{12}\right) + \hat{U}_{i,j-1;yy}\left(\tilde{y}^2 - \frac{1}{12}\right) + \hat{U}_{i,j-1;xy}\tilde{x}\tilde{y}$$
(106)

and the corresponding smoothness indicator is

$$IS_{i,j-1} = \left(\hat{u}_{i,j-1;x} + \hat{u}_{i,j-1;xy}\right)^2 + \left(\hat{u}_{i,j-1;y} + 2\hat{u}_{i,j-1;yy}\right)^2 + \frac{13}{3}\left(\hat{u}_{i,j-1;xx}\right)^2 + \frac{13}{3}\left(\hat{u}_{i,j-1;yy}\right)^2 + \frac{7}{6}\left(\hat{u}_{i,j-1;xy}\right)^2$$
(107)

This completes our description of the shifted reconstruction polynomials and their corresponding smoothness indicators.

Now that the smoothness indicators from the neighboring zones are in hand, we can develop the corresponding non-linear weights. To zone $(i, j)$ we ascribe a central linear weight given by $\gamma_C = 0.96$; and to the four immediate neighbors we ascribe linear weights given by $\gamma_N = 0.01$. The non-linear weights are then given by

$$w_{i,j} = \frac{\gamma_C}{\left(IS_{i,j} + \varepsilon\right)^p} \; ; \; w_{i+1,j} = \frac{\gamma_N}{\left(IS_{i+1,j} + \varepsilon\right)^p} \; ; \; w_{i-1,j} = \frac{\gamma_N}{\left(IS_{i-1,j} + \varepsilon\right)^p} \; ;$$

$$w_{i,j+1} = \frac{\gamma_N}{\left(IS_{i,j+1} + \varepsilon\right)^p} \; ; \; w_{i,j-1} = \frac{\gamma_N}{\left(IS_{i,j-1} + \varepsilon\right)^p} \; ;$$

$$\bar{w}_{i,j} = w_{tot}^{-1} w_{i,j} \; ; \; \bar{w}_{i+1,j} = w_{tot}^{-1} w_{i+1,j} \; ; \; \bar{w}_{i-1,j} = w_{tot}^{-1} w_{i-1,j} \; ; \; \bar{w}_{i,j+1} = w_{tot}^{-1} w_{i,j+1} \; ; \; \bar{w}_{i,j-1} = w_{tot}^{-1} w_{i,j-1} \; ;$$

$$w_{tot} = w_{i,j} + w_{i+1,j} + w_{i-1,j} + w_{i,j+1} + w_{i,j-1}$$

(108)

In the above equation, we can set $\varepsilon = 10^{-12}$ and p=2 as suggested by Zhong and Shu (2013). The reconstructed polynomial, with limiting, is then given by

$$u_{i,j}(\tilde{x}, \tilde{y}) \to \bar{w}_{i,j} u_{i,j}(\tilde{x}, \tilde{y}) + \bar{w}_{i+1,j} \tilde{u}_{i+1,j}(\tilde{x}, \tilde{y}) + \bar{w}_{i-1,j} \tilde{u}_{i-1,j}(\tilde{x}, \tilde{y}) \\ + \bar{w}_{i,j+1} \tilde{u}_{i,j+1}(\tilde{x}, \tilde{y}) + \bar{w}_{i,j-1} \tilde{u}_{i,j-1}(\tilde{x}, \tilde{y})$$
(109)

Here $u_{i,j}(\tilde{x}, \tilde{y})$ is a component of eqn. (95); $\tilde{u}_{i+1,j}(\tilde{x}, \tilde{y})$ is a component of eqn. (100); and $\tilde{u}_{i-1,j}(\tilde{x}, \tilde{y})$, $\tilde{u}_{i,j+1}(\tilde{x}, \tilde{y})$ and $\tilde{u}_{i,j-1}(\tilde{x}, \tilde{y})$ are components of eqns. (102), (104) and (106) respectively. If characteristic variables are being used, they could be the eigenweights that are obtained from characteristic projection. This completes our description of the WENO limiting for DG schemes. The limiting strategy described here is implemented by exactly following the sequence of equations that is described in this section.

**VI.3) MOOD Limiter for DG Methods**

MOOD stands for Multidimensional Optimal Order Detection. It is based on the realization that a higher order scheme may return an oscillatory result at the end of a timestep. Alternatively, a higher order scheme may return an unphysical result at the end of a timestep. In both those situations, we realize that a lower order scheme would have served us better in those zones that turned out to be pathological (or troubled). The catch is that if one is taking a timestep from a time of $t^n$ to a time of $t^{n+1}$ then one does not know which zone might produce a troubled result till the timestep has completed. The MOOD philosophy argues that *a priori limiting* of the solution at time $t^n$ may indeed result in excessive limiting in zones where this is not needed. For a WENO scheme, where the reconstruction step includes a non-linear hybridization (i.e. a limiting) procedure, this is not much of an issue. However, DG schemes may, in principle, not need any limiting at all. In such circumstances, falsely invoking the limiter at time $t^n$ can lead to excessive limiting. The MOOD philosophy, therefore, suggests that it is best to hold off on the limiting process till the timestep has completed, i.e. till a time of $t^{n+1}$. At that advanced time, the solution itself can be polled to see if it violates physical admissibility (i.e. a loss of pressure or density positivity) or numerical admissibility (i.e. production of an oscillatory profile on the mesh). In all such cases, the troubled zone can be tagged and its time integration can be redone in an a posteriori sense. This is called *a posteriori limiting*. Operationally, this limiting involves using a known and stable TVD or low order WENO scheme to evolve the offending zone again from a time of $t^n$ to a time of $t^{n+1}$. Realize, therefore, that the solution has to be available at both times. Furthermore, the data from the troubled zones in the DG scheme has to be extracted in a suitable fashion and handed over to a different solver. The data from that different solver has to be handed back in a suitable fashion to the troubled zones in the DG scheme.

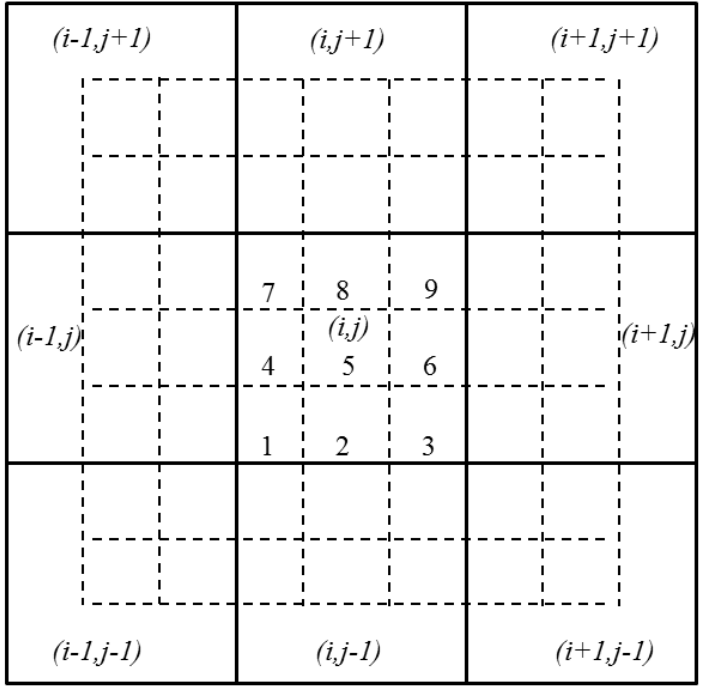

*Fig. 14 shows the situation where MOOD limiting is applied to the zone (i,j) on a 2D mesh. The DG scheme is third order accurate. The nine zones that participate in the limiting are shown with solid lines. Since the DG scheme is third order accurate, the zone (i,j) is split into nine sub-cells. These cells are labeled 1,…,9 and are shown with dashed lines. These nine sub-cells are updated with a TVD or WENO scheme that is lower order and very stable. At the end of the sub-cell update the updated moments of the DG scheme can be recovered.*

MOOD limiting for DG is based on viewing the DG polynomial over one DG zone as being equivalent to specifying just one volume-averaged solution vector on a set of sub-cells of that zone. An early sub-cell based approach to DG limiting was first developed in Balsara *et al.* (2007). MOOD-type limiting for DG schemes has been developed by Dumbser *et al.* (2014) and Sonntag and Munz (2014). We have provided several extra clarifications here to make it easy to understand and implement. It works well with both ADER and RK time update strategies. In our explanation of MOOD limiting, let us go from the specific to the general. In this paragraph let us specifically describe in words the process of applying MOOD limiting on a 2D mesh on which we are evolving a third order DG scheme. At a time of $t^n$ and a time of $t^{n+1}$ we have stored all the modes that are given in eqn. (86) for all the zones of the mesh. The DG scheme evolves these modes so that we have six modes within each zone for each of the conserved variables at each of the two time levels. The DG solution at time level $t^{n+1}$ may have some troubled zones. Please look at Fig. 14 and let us assume that zone $(i, j)$ will eventually be found to be a troubled zone, so that we wish to first detect the pathology and then redo the timestep for that zone with a more stable method. We choose a simple TVD or lower order WENO scheme as our stable method. In order for the lower order method to have the same amount of information as the six modes that we have evolved with the DG method, we split each zone of the entire mesh into at least nine sub-cells. Each of these sub-cells will receive a volume-averaged solution from its parent DG cell at time $t^{n+1}$. Because each DG zone has six spatially-varying modes, it can easily supply a unique volume-averaged solution vector to each of its nine sub-cells shown in Fig. 14. We now apply a *physical admissibility detector* (PAD) and a *numerical admissibility detector* (NAD) to each sub-cell. (The PAD and NAD are described in detail in a later paragraph.) If all the sub-cells associated with a parent DG zone are salient at time $t^{n+1}$, we say that the DG zone had a successful update and we don't consider that cell any further. However, if any of the sub-cells has an unphysical solution (i.e. it triggers the PAD) or an oscillatory solution (i.e. it triggers the NAD), we declare the zone to be a troubled zone. Let us say, for the sake of argument that the zone $(i, j)$ is found to be a troubled zone. We will have to, therefore, redo the time update from a time of $t^n$ to a time of $t^{n+1}$ for those nine sub-cells in Fig. 14 with a simple TVD or low order WENO scheme that is known to be very stable. The TVD or WENO reconstruction might require a halo of two or three zones, which is why we show a layer of two sub-cells around the nine sub-cells that we identified in Fig. 14. The DG solution in zone $(i, j)$ at time $t^n$ is then imparted (*scattered*) to the nine sub-cells in Fig. 14. The nine sub-cells shown in Fig. 14 are then evolved from a time of $t^n$ to a time of $t^{n+1}$ with a TVD or WENO scheme. The nine sub-cells now hold a salient solution at time $t^{n+1}$. From these nine sub-cell averages at time $t^{n+1}$ we can retrieve (*gather*) the DG polynomial in zone $(i, j)$ at time $t^{n+1}$. We can now say that the zone $(i, j)$ has undergone an a posteriori MOOD limiting for this timestep.

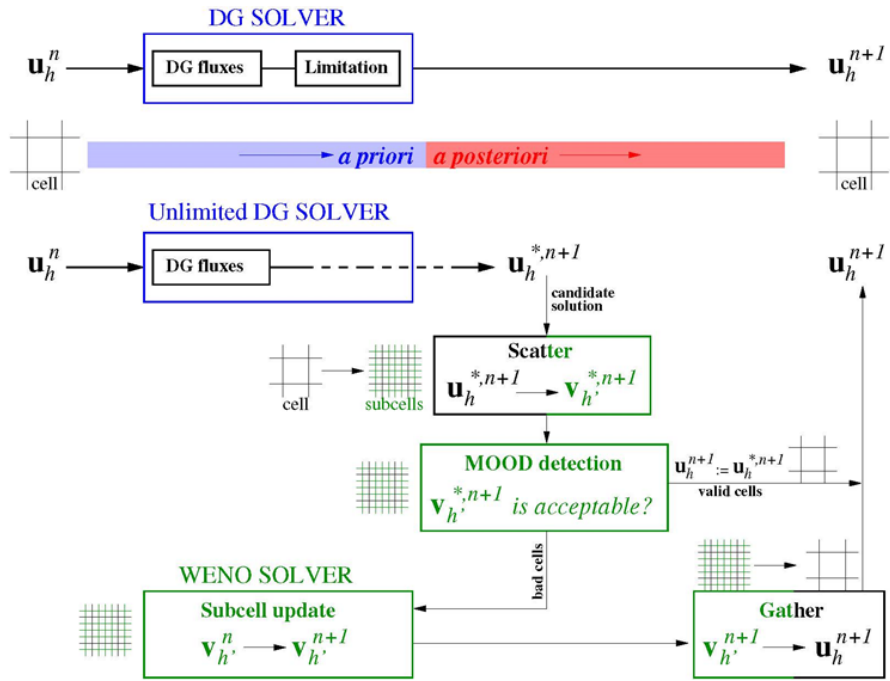

*Fig. 15, which is Fig. 2 of Dumbser et al. (2014), shows a flowchart of the MOOD limiting process. The a posteriori limiting consists of a scatter to a finer sub-cell mesh and a detection step for detecting admissible zones. Zones that are inadmissible have to redo their timestep on the sub-cells using a TVD or WENO scheme. The resulting sub-cell solution is then gathered back to the DG mesh.*

Now let us consider the general case. Say, for the sake of discussion that we want to represent the same amount of information that is contained in an $N^{th}$ order DG polynomial in one dimension. To represent the same amount of information in a finite volume sense we will need (N+1) sub-cells within each DG zone. These sub-cells will have featureless slabs of fluid. These (N+1) sub-cells will have only one solution vector each. We, therefore, say that the DG polynomial at $N^{th}$ order has as much information as the volume-averaged solution vectors in each of the (N+1) sub-cells. For an $N^{th}$ order DG scheme in "d" dimensions, we will have to split each DG zone into at least $(N+1)^d$ sub-cells. Fig. 15, which is Fig. 2 from Dumbser *et al*. (2014), shows a flowchart that describes the MOOD limiting of DG schemes. The solution at time $t^{n+1}$ within each DG zone is scattered to its sub-cells. On those sub-cells, we apply the PAD and the NAD. If none of the sub-cells associated with a DG zone shows any pathology, the DG solution in that zone is deemed acceptable. If not, we flag the zone and scatter the DG solution at the earlier time $t^n$ from the troubled zone as well as its halo of neighbors. This information is now available on the sub-cell mesh. Such a mesh will contain the sub-cells associated with a troubled zone and also all halo sub-cells that are needed for a time update. The sub-cells associated with the troubled zone then undergo a time update from time $t^n$ to time $t^{n+1}$ with the help of a TVD or low order WENO scheme. The sub-cell solution at the advanced time $t^{n+1}$ is then gathered back to the troubled zone on the DG mesh. These gather and scatter steps are arranged so that they can be done very efficiently. We see, therefore, that there are two crucial parts that we need to describe further. First, we need to describe the scatter and gather steps and their efficient implementation. Second, we need to give some useful information about the PAD and NAD. We do that in the ensuing paragraphs.

First, let us describe the scatter and gather steps. We will do this for a third order DG scheme in 2D. The reader will see that with the help of a computer algebra system the procedure can be extended to any order and to 3D. Let us start with eqn. (86) and describe the scatter step. From Fig. 14 we realize that specifying just the volume-averaged solution vector in each of the nine sub-cells in Fig. 14 would give us more than sufficient information to retrieve all the moments in eqn. (86). The sub-cells have uniform size so that the first sub-cell is given by $[-1/2,-1/6] \times [-1/2,-1/6]$, the second sub-cell is given by $[-1/6,1/6] \times [-1/2,-1/6]$, the third sub-cell is given by $[1/6,1/2] \times [-1/2,-1/6]$ and so on. We now show how the volume-averaged solution vectors in the nine sub-cells in Fig. 14 can be obtained from the DG polynomial. We will denote those volume-averaged solution vectors by $U_{(k)}$ with $k=1,..,9$. Suppressing the time-dependence in eqn. (86), or just using eqn. (95) for the sake of convenience, we have

$$\overline{U}_{(1)} = 9 \int_{\tilde{y}=-1/2}^{-1/6} \int_{\tilde{x}=-1/2}^{-1/6} U_{i,j}(\tilde{x},\tilde{y}) d\tilde{x}\, d\tilde{y} = \overline{U}_{i,j} - \frac{1}{3}\hat{U}_{i,j;x} - \frac{1}{3}\hat{U}_{i,j;y} + \frac{1}{27}\hat{U}_{i,j;xx} + \frac{1}{27}\hat{U}_{i,j;yy} + \frac{1}{9}\hat{U}_{i,j;xy}$$

$$\overline{U}_{(2)} = 9 \int_{\tilde{y}=-1/2}^{-1/6} \int_{\tilde{x}=-1/6}^{1/6} U_{i,j}(\tilde{x},\tilde{y}) d\tilde{x}\, d\tilde{y} = \overline{U}_{i,j} - \frac{1}{3}\hat{U}_{i,j;y} - \frac{2}{27}\hat{U}_{i,j;xx} + \frac{1}{27}\hat{U}_{i,j;yy}$$

$$\overline{U}_{(3)} = 9 \int_{\tilde{y}=-1/2}^{-1/6} \int_{\tilde{x}=1/6}^{1/2} U_{i,j}(\tilde{x},\tilde{y}) d\tilde{x}\, d\tilde{y} = \overline{U}_{i,j} + \frac{1}{3}\hat{U}_{i,j;x} - \frac{1}{3}\hat{U}_{i,j;y} + \frac{1}{27}\hat{U}_{i,j;xx} + \frac{1}{27}\hat{U}_{i,j;yy} - \frac{1}{9}\hat{U}_{i,j;xy}$$

$$\overline{U}_{(4)} = 9 \int_{\tilde{y}=-1/6}^{1/6} \int_{\tilde{x}=-1/2}^{-1/6} U_{i,j}(\tilde{x},\tilde{y}) d\tilde{x}\, d\tilde{y} = \overline{U}_{i,j} - \frac{1}{3}\hat{U}_{i,j;x} + \frac{1}{27}\hat{U}_{i,j;xx} - \frac{2}{27}\hat{U}_{i,j;yy}$$

$$\overline{U}_{(5)} = 9 \int_{\tilde{y}=-1/6}^{1/6} \int_{\tilde{x}=-1/6}^{1/6} U_{i,j}(\tilde{x},\tilde{y}) d\tilde{x}\, d\tilde{y} = \overline{U}_{i,j} - \frac{2}{27}\hat{U}_{i,j;xx} - \frac{2}{27}\hat{U}_{i,j;yy}$$

$$\overline{U}_{(6)} = 9 \int_{\tilde{y}=-1/6}^{1/6} \int_{\tilde{x}=1/6}^{1/2} U_{i,j}(\tilde{x},\tilde{y}) d\tilde{x}\, d\tilde{y} = \overline{U}_{i,j} + \frac{1}{3}\hat{U}_{i,j;x} + \frac{1}{27}\hat{U}_{i,j;xx} - \frac{2}{27}\hat{U}_{i,j;yy}$$

$$\overline{U}_{(7)} = 9 \int_{\tilde{y}=1/6}^{1/2} \int_{\tilde{x}=-1/2}^{-1/6} U_{i,j}(\tilde{x},\tilde{y}) d\tilde{x}\, d\tilde{y} = \overline{U}_{i,j} - \frac{1}{3}\hat{U}_{i,j;x} + \frac{1}{3}\hat{U}_{i,j;y} + \frac{1}{27}\hat{U}_{i,j;xx} + \frac{1}{27}\hat{U}_{i,j;yy} - \frac{1}{9}\hat{U}_{i,j;xy}$$

$$\overline{U}_{(8)} = 9 \int_{\tilde{y}=1/6}^{1/2} \int_{\tilde{x}=-1/6}^{1/6} U_{i,j}(\tilde{x},\tilde{y}) d\tilde{x}\, d\tilde{y} = \overline{U}_{i,j} + \frac{1}{3}\hat{U}_{i,j;y} - \frac{2}{27}\hat{U}_{i,j;xx} + \frac{1}{27}\hat{U}_{i,j;yy}$$

$$\overline{U}_{(9)} = 9 \int_{\tilde{y}=1/6}^{1/2} \int_{\tilde{x}=1/6}^{1/2} U_{i,j}(\tilde{x},\tilde{y}) d\tilde{x}\, d\tilde{y} = \overline{U}_{i,j} + \frac{1}{3}\hat{U}_{i,j;x} + \frac{1}{3}\hat{U}_{i,j;y} + \frac{1}{27}\hat{U}_{i,j;xx} + \frac{1}{27}\hat{U}_{i,j;yy} + \frac{1}{9}\hat{U}_{i,j;xy}$$

(110)

This completes our description of the scatter step.

Let us now focus on the gather step. The gather step reverses the scatter step. In other words, given nine sub-cells with solution vectors that have been evolved up to a time $t^{n+1}$, we wish to find the best set of coefficients that we use in eqn. (95). Realize that the mean value has to be preserved between the nine sub-cells and the one parent DG zone. As a result, for the sake of conservation, we have

$$\overline{U}_{i,j} = \frac{1}{9}\left(\sum_{k=1}^{9}\overline{U}_{(k)}\right) \tag{111}$$

The other moments should be set so as to have the best values that they can have. Eqn. (110) then gives us a system of nine equations in five unknowns. The optimal solution can be obtained via a least squares minimization of the following $9\times 5$ overdetermined system

$$\begin{pmatrix} -1/3 & -1/3 & 1/27 & 1/27 & 1/9 \\ 0 & -1/3 & -2/27 & 1/27 & 0 \\ 1/3 & -1/3 & 1/27 & 1/27 & -1/9 \\ -1/3 & 0 & 1/27 & -2/27 & 0 \\ 0 & 0 & -2/27 & -2/27 & 0 \\ 1/3 & 0 & 1/27 & -2/27 & 0 \\ -1/3 & 1/3 & 1/27 & 1/27 & -1/9 \\ 0 & 1/3 & -2/27 & 1/27 & 0 \\ 1/3 & 1/3 & 1/27 & 1/27 & 1/9 \end{pmatrix} \begin{pmatrix} \hat{U}_{i,j;x} \\ \hat{U}_{i,j;y} \\ \hat{U}_{i,j;xx} \\ \hat{U}_{i,j;yy} \\ \hat{U}_{i,j;xy} \end{pmatrix} = \begin{pmatrix} \overline{U}_{(1)} - \overline{U}_{i,j} \\ \overline{U}_{(2)} - \overline{U}_{i,j} \\ \overline{U}_{(3)} - \overline{U}_{i,j} \\ \overline{U}_{(4)} - \overline{U}_{i,j} \\ \overline{U}_{(5)} - \overline{U}_{i,j} \\ \overline{U}_{(6)} - \overline{U}_{i,j} \\ \overline{U}_{(7)} - \overline{U}_{i,j} \\ \overline{U}_{(8)} - \overline{U}_{i,j} \\ \overline{U}_{(9)} - \overline{U}_{i,j} \end{pmatrix} \quad (112)$$

Since it is very easy to solve this least squares system by inverting a small $5 \times 5$ matrix, the solution can be efficiently obtained. In fact, since the matrix only has constant coefficients, the inversion has only to be done once. From Fig. 14 it is also important to realize that the fluxes across the boundaries of zone $(i, j)$ change when the sub-cells are updated. As a result, the values of the solutions in zones $(i+1, j)$, $(i-1, j)$, $(i, j+1)$ and $(i, j-1)$ will also change.

The physical admissibility detector (PAD) consists simply of realizing that the sub-cells associated with each DG zone should each have positive density and pressure. If the flow is relativistic, the zone should also have sub-luminal velocities. In other words, the PAD is just dependent on the physics of the problem. The numerical admissibility detector (NAD) consists of just requiring no new extrema to develop in the solution and it is applied component-wise to the sub-cells. To apply the NAD to the zone $(i, j)$ in Fig. 14, we go to all the sub-cells associated with all the nine zones shown in Fig. 14. Let us say that $u_{(k)}^m$ is the m$^{th}$ component of the vector of conserved variables in the k$^{th}$ sub-cell of zone $(i, j)$. For the m$^{th}$ component in the solution vector, we find the minimum and maximum value that this component assumes in all of the sub-cells in all the nine DG zones shown in Fig. 14. Let $u_{min}^m$ be that minimum value and let $u_{max}^m$ be that maximum value. In order to avoid a clipping of physical extrema, we want to allow the solution to slightly exceed the minimum and maximum ranges if needed. So we require the m$^{th}$ component of the solution vector in all the sub-cells of zone $(i, j)$ to lie in the following range in order to be numerically admissible. The range is given by

$$u_{min}^m - \delta^m \leq u_{(k)}^m \leq u_{max}^m + \delta^m \quad \forall \quad k = 1,..,9 \quad (113)$$

The extent by which the minimum or maximum can be exceeded is given by "$\delta^m$". Unfortunately, the value of "$\delta^m$" is set by heuristic considerations. However, a reasonable suggestion from Dumbser *et al.* (2014) is to use

$$\delta^m = \max\left(\delta_0, \varepsilon\left(u_{\max}^m - u_{\min}^m\right)\right) \tag{114}$$

with $\delta_0 = 10^{-4}$ and $\varepsilon = 10^{-3}$ being used in the above equation. With the arrangement of terms in eqns. (113) and (114), the solution is allowed to develop some new extrema as long as the extrema are bounded. If the conditions in eqn. (113) are passed by all the components "m" of all the sub-cells, we say that the DG zone $(i, j)$ has passed the NAD. If the DG zone $(i, j)$ also passes through the PAD, we say that the zone $(i, j)$ is acceptable and does not need any further MOOD limiting. If a zone does not pass the PAD and NAD conditions, we use the scatter and gather process from eqns. (110) to (112) to redo the time-evolution in the troubled zone with a lower order TVD or WENO scheme.

**Stepwise Description of the Third Order Accurate RKDG Scheme**

Step 1: Apply the boundary conditions and, if needed, limit the modes in eqn. (86).
Step 2: Limit the solution within each zone. For example, the WENO limiter for DG schemes can be applied by exactly following the steps in Sub-section VI.2.
Step 3: Use eqn. (44), along its analogue in zone $(i+1, j)$, to obtain $U_{L; i+1/2, j}^{(c)}$ and $U_{R; i+1/2, j}^{(c)}$ from eqn. (45). Use them to obtain the extremal wave speeds, $S_L$ and $S_R$, for use in eqn. (46).
Step 4: Use ideas similar to those in eqns. (47) to (52) to obtain the third order accurate evaluation of the boundary integrals in eqns. (89) to (94)). Add their contributions into the time rate of update.
Step 5: Evaluate the areal integrals in eqn. (90) to (94)) and add their contributions into the time rate of update. This completes one stage in a third order Runge-Kutta update.

**PNPM Schemes; Putting DG Schemes in Perspective**

Sections III presented schemes that were based on WENO reconstruction. In such schemes, we start with the conserved variable, $\bar{U}_{i,j}$, in each zone and reconstruct all the moments shown in eqn. (44) at each time step. Only $\bar{U}_{i,j}$ is evolved in each zone using the fluxes. In Section VI for DG schemes, we are doing something very different. We are endowing time-evolution to all the moments of eqn. (86). As a result, we have many evolutionary equations, eqns. (89) to (94). This increases the use of computer memory and adds to the computational complexity of the scheme. The obvious questions are: Does this yield a tangible advantage in accuracy? Is this advantage obtained in all circumstances?

Schemes that reconstruct all moments using WENO reconstruction and schemes that evolve all the moments, like the DG schemes, sit on opposite ends of the spectrum. The majority of the variation in eqn. (86) is contained in the first few moments. As a result, one can conceive of a scheme that is based on $M^{th}$ order polynomials that evolves the polynomials up to $N^{th}$ order. Such a scheme would be $(M+1)^{th}$ order accurate and we call it a *PNPM scheme*. Here we have $M \leq N$ so that we have the option of evolving fewer moments than the full complement of moments that are required for $(M+1)^{th}$ order accuracy. The moments that are not evolved would have to be reconstructed at each time step. For example, in eqn. (86), we can endow time evolution exclusively to $\bar{U}_{i,j}(t)$, $\bar{U}_{i,j;x}(t)$ and $\bar{U}_{i,j;y}(t)$ while reconstructing the remaining moments using a WENO-like strategy. The evolutionary equations would then just be eqns. (89), (90) and (91). This would yield a third order accurate P1P2 scheme. Such a scheme would sit between the third order RK-WENO scheme described in Sections III and IV and the third order RKDG scheme described in Section VI. It would evolve the conserved variable and its first moments, but it would also reconstruct the second moments at each time step. A P1P2 scheme would cost more than a third order RK-WENO scheme but less than an RKDG scheme. Thus at third order, we can have a P0P2 scheme, which is just the RK-WENO scheme, or we can have the P2P2 scheme, which is just the RKDG scheme, or we can have the P1P2 scheme, which sits between the previous two. A P1P2 scheme would also use an intermediate amount of memory. For lower orders, the savings in memory and computational complexity are not dramatic. For higher orders, they can be substantial.

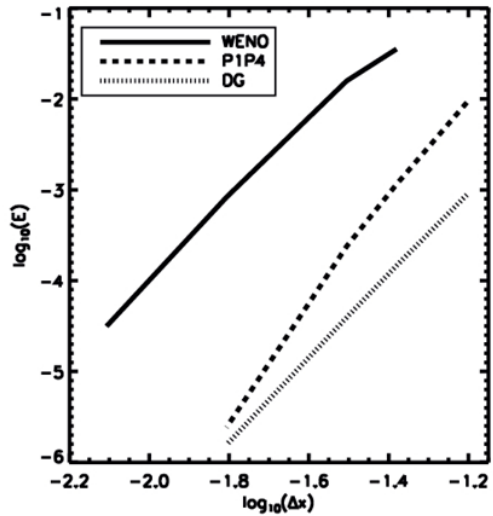

*This figure shows the accuracies of various fifth order schemes as the mesh is refined. The same smooth two-dimensional vortex flow problem was used for all the schemes and all the schemes were formally fifth order accurate. The solid curve shows the error in the WENO scheme with increasing resolution. The dashed curve shows the error in the P1P4 scheme and the dotted curve shows the error in the DG scheme.*

The real question is: How does dropping the higher moments impact accuracy? The figure shown above gives a partial answer for fifth order schemes. See Dumbser *et al.* (2008) for a more complete answer. Dumbser *et al.* (2008) built on prior work by Qiu & Shu (2004, 2005), Schwartzkopff, Dumbser & Munz (2004) and Balsara et al. (2007). A fifth order WENO scheme (solid curve), fifth order P1P4 scheme (dashed curve) and fifth order DG scheme (dotted curve) were run on the same smooth, two-dimensional hydrodynamical vortex problem. While all schemes meet their design accuracies, we see that the WENO scheme has substantially larger error at the same resolution. The error in the P1P4 scheme is intermediate between that in the WENO and DG schemes. Thus, for a fixed resolution, the DG and PNPM schemes give smaller errors than the WENO schemes. They do carry a higher cost though. The DG results shown are without the use of a limiter because the problem is smooth. If the problem is such as to need a minimum amount of limiting, then the higher cost of the PNPM and DG schemes is well justified. If a large fraction of the computational domain is dominated by shocks, many of the zones will have to take on some amount of limiting and then the advantages of the PNPM and DG schemes will be diminished. Thus the choice of method may often depend on the application area.

| $CFL_{max}$ | N=0 | N=1 | N=2 | N=3 | N=4 |
|---|---|---|---|---|---|
| M=1 | 1.00 | 0.33 | | | |
| M=2 | 1.00 | 0.32 | 0.17 | | |
| M=3 | 1.00 | 0.32 | 0.17 | 0.10 | |
| M=4 | 1.00 | 0.32 | 0.17 | 0.10 | 0.069 |

As explained in the introduction, PNPM methods also permit larger timesteps than DG schemes of comparable accuracy. The above table, from Dumbser *et al.* (2008), shows the limiting CFL number of various one-dimensional PNPM schemes from second to fifth order. Please note that for multidimensional problems, the permitted CFL number is divided by the dimensionality of the problem. The CFL numbers in the above table are based on using a time-update strategy with a temporal order of accuracy that matches the spatial order of accuracy. P0PM is the same thing as a WENO scheme of $(M+1)^{th}$ order in space and time; PMPM is the same thing as a DG scheme of $(M+1)^{th}$ order in space and time. It is easy to see that P1PM and P2PM methods provide robustly large CFL numbers while offering accuracies that are comparable to DG schemes of the same order.

**VII) Positivity Preserving Reconstruction**

Obtaining numerical solutions for the Euler equation that retain positive densities and pressures is incredibly important. Some Riemann solvers can guarantee a positive resolved state while others cannot make such a guarantee. A Riemann solver that guarantees positivity can be very useful in obtaining a physical solution. When either the density or pressure become negative, the Euler system loses its convexity property, handicapping our ability to obtain physical solutions. However, a loss of *positivity* does not arise exclusively from the Riemann solver. It can even arise due to the kind of reconstruction that is used. The TVD property only guarantees positivity of the reconstructed profile in one dimension. In multiple dimensions, certain parts of a reconstructed profile within a zone can lose positivity even when TVD reconstruction is used. This loss of positivity usually occurs near the vertices of a zone, where the piecewise linear profile reaches its extremal values. For higher order reconstruction, the problem becomes a little worse because the reconstructed profile can also attain extremal values inside the zone. For that reason, we focus attention on obtaining a reconstructed profile that retains positive density and pressure. There are several papers where the topic of positivity has been discussed, both for Euler and MHD flow (Barth and Frederickson 1990, Barth 1995, Liu and Lax 1996, Lax and Liu 1998, Balsara and Spicer 1999b, Zhang and Shu 2010, Balsara 2012b). The positivity preserving method we present here derives from the latter two references. A video introduction to this work is included in Balsara (2012b). Recently, Balsara and Kim (2016) have presented a scheme for RMHD that tries to preserve the sub-luminal velocity of the flow.

We describe the method on a two dimensional structured mesh, though it extends naturally to three dimensions and it could also extend naturally to unstructured meshes. Let $\rho$ and P be the density and pressure and let **v** be the velocity vector. Let $\gamma$ be the ratio of specific heats. Let **m** denote the momentum density and $\mathcal{E}$ the energy density. For Euler flow we can write $\mathcal{E} = P/(\gamma-1) + \rho\, \mathbf{v}^2/2$.

We first need to define a *flattener function* that can identify regions of strong shocks within our computational domain. The method, therefore, begins by obtaining the divergence of the velocity, $(\nabla \bullet \mathbf{v})_{i,j}$, and the sound speed, $c_{s;i,j} \equiv \sqrt{\gamma P_{i,j}/\overline{\rho}_{i,j}}$, within a zone $(i,j)$ as shown in Fig. 1. To identify a shock, the undivided divergence of the velocity within a zone has to be compared with the minimum of the sound speed in the zone $(i,j)$ and all its immediate neighbors. Thus we need the minimum sound speed from all the neighbors, see Fig. 1. It is defined by

$$c_{s;i,j}^{\min-nbr} = \min\left(c_{s;i-1,j-1}, c_{s;i-1,j}, c_{s;i-1,j+1}, c_{s;i,j-1}, c_{s;i,j}, c_{s;i,j+1}, c_{s;i+1,j-1}, c_{s;i+1,j}, c_{s;i+1,j+1},\right) \quad (115)$$

In each zone, which is assumed to have an extent $\Delta x$, we define the flattener as

$$\eta_{i,j} = \min\left[1,\ \max\left[0,\ -\left(\Delta x\, (\nabla \bullet \mathbf{v})_{i,j} + \kappa_1\, c_{s;i,j}^{\min-nbr}\right)\Big/\left(\kappa_1\, c_{s;i,j}^{\min-nbr}\right)\right]\right] \quad (116)$$

While there is some flexibility in the value of $\kappa_1$, here we take $\kappa_1 = 0.4$. Numerical experimentation has shown this value to work well at several orders and for a large range of problems. Notice from the structure of the above equation that when the flow develops rarefactions, i.e. $(\nabla \bullet \mathbf{v})_{i,j} \geq 0$, the reconstruction is left completely untouched by the flattener. For compressive motions of modest strength, i.e when $-\kappa_1\, c_{s;i,j}^{\min-nbr} < \Delta x\, (\nabla \bullet \mathbf{v})_{i,j} < 0$, the flattener also leaves the reconstruction untouched. We, therefore, see that $\eta_{i,j} = 0$ when the flow is smooth and it goes to $\eta_{i,j} = 1$ in a continuous fashion when strong shocks are present. It is possible to improve on the previous flattener. Zones that are about to be run over by a shock but have not yet entered the shock would also be stabilized if they were to experience some flattening. We identify such situations by looking at the pressure variation. We describe the method for the x-direction as

$$\begin{aligned}&if\left((\eta_{i,j} > 0)\,and\,(\eta_{i+1,j} = 0)\,and\,(P_{i,j} > P_{i+1,j})\right) then\quad \eta_{i+1,j} = \eta_{i,j}\\ &if\left((\eta_{i,j} > 0)\,and\,(\eta_{i-1,j} = 0)\,and\,(P_{i,j} > P_{i-1,j})\right) then\quad \eta_{i-1,j} = \eta_{i,j}\end{aligned} \quad (117)$$

Please note that eqn. (116) is applied first to the entire mesh in order to identify zones that are already inside a shock. Eqn. (117) is applied subsequently in order to identify zones that are about to be run over by a shock. It is trivial to extend the above equation to the y-direction. For multidimensional problems, the above strategy can be applied to each of the principal directions of the mesh.

We now wish to obtain the minimum and maximum values of the density and pressure variables from the neighboring zones. For Fig. 1, we can do this for the density variable by setting

$$\rho_{i,j}^{\min-nbr} = \min\left(\overline{\rho}_{i-1,j-1}, \overline{\rho}_{i-1,j}, \overline{\rho}_{i-1,j+1}, \overline{\rho}_{i,j-1}, \overline{\rho}_{i,j}, \overline{\rho}_{i,j+1}, \overline{\rho}_{i+1,j-1}, \overline{\rho}_{i+1,j}, \overline{\rho}_{i+1,j+1},\right)$$
$$\rho_{i,j}^{\max-nbr} = \max\left(\overline{\rho}_{i-1,j-1}, \overline{\rho}_{i-1,j}, \overline{\rho}_{i-1,j+1}, \overline{\rho}_{i,j-1}, \overline{\rho}_{i,j}, \overline{\rho}_{i,j+1}, \overline{\rho}_{i+1,j-1}, \overline{\rho}_{i+1,j}, \overline{\rho}_{i+1,j+1},\right)$$
(118)

where the overbars indicate zone-averaged values. A multidimensional TVD limiting strategy would have demanded that $\rho_{i,j}^{\min-nbr} \leq \rho_{i,j}(x,y) \leq \rho_{i,j}^{\max-nbr}$ where $\rho_{i,j}(x,y)$ is the reconstructed density in the zone of interest. Similar expressions should be obtained for the pressure.

To accommodate non-oscillatory reconstruction schemes, we need to extend the range $\left[\rho_{i,j}^{\min-nbr}, \rho_{i,j}^{\max-nbr}\right]$ in a solution-dependent way. Using the flattener variable, this is easily done as:

$$\rho_{i,j}^{\min-extended} = \rho_{i,j}^{\min-nbr}\left(1 - \kappa_2 + \kappa_2 \eta_{i,j}\right) \quad ; \quad \rho_{i,j}^{\max-extended} = \rho_{i,j}^{\max-nbr}\left(1 + \kappa_2 - \kappa_2 \eta_{i,j}\right) \quad (119)$$

For this work we took $\kappa_2 = 0.4$ based on extensive numerical experimentation. Observe that it does extend the range of permitted densities to $\left[(1-\kappa_2)\rho_{i,j}^{\min-nbr}, (1+\kappa_2)\rho_{i,j}^{\max-nbr}\right]$ in regions of smooth flow. If strong shocks are present in the vicinity of the zone of interest, the range of permitted densities is smoothly reduced to $\left[\rho_{i,j}^{\min-nbr}, \rho_{i,j}^{\max-nbr}\right]$ as the strength of the shocks become progressively larger. We can do similarly for the pressures. As a result, within each zone (i,j) we obtain a range of densities $\left[\rho_{i,j}^{\min-extended}, \rho_{i,j}^{\max-extended}\right]$ and demand that the reconstructed density profile satisfy $\rho_{i,j}^{\min-extended} \leq \rho_{i,j}(x,y) \leq \rho_{i,j}^{\max-extended}$. Similarly, we obtain a range of pressures $\left[P_{i,j}^{\min-extended}, P_{i,j}^{\max-extended}\right]$ and demand that the pressure variable that can be derived at any point within the zone of interest be bounded by $P_{i,j}^{\min-extended} \leq P_{i,j}(x,y) \leq P_{i,j}^{\max-extended}$. In practice, it might be valuable to also provide absolute floor values for $\rho_{i,j}^{\min-extended}$ and $P_{i,j}^{\min-extended}$.

Notice from the previous two paragraphs that the density is a conserved variable and the zone-averaged density is already contained within the range $\left[\rho_{i,j}^{\min-extended}, \rho_{i,j}^{\max-extended}\right]$ by construction. Thus bringing the reconstructed density within the range simply requires us to reduce the spatially varying part of the density. The pressure, on the other hand, is a derived variable. While the zone-averaged pressure still lies within the range $\left[P_{i,j}^{\min-extended}, P_{i,j}^{\max-extended}\right]$, bringing the reconstructed pressure within this range is harder, especially since the reconstruction is almost always expressed in terms of the conserved variables. The next insight comes from Zhang & Shu (2010) who presented an implementable strategy for doing this. For any conserved variable, say for instance the density in the zone $(i,j)$, we can write

$$\tilde{\rho}_{i,j}(x,y) = (1-\tau)\, \bar{\rho}_{i,j} + \tau\, \rho_{i,j}(x,y) \tag{120}$$

Here $\rho_{i,j}(x,y)$ is the original reconstructed profile in the zone of interest, $\bar{\rho}_{i,j}$ is the zone-averaged density and $\tau \in [0,1]$. Please do not confuse "$\tau$" with the time variable. In this section it will refer exclusively to a parameter we use to restore positivity. When $\tau = 1$, the corrected profile $\tilde{\rho}_{i,j}(x,y)$ is exactly equal to $\rho_{i,j}(x,y)$. Thus if the entire reconstructed density lies within the desired range then such a situation is equivalent to setting $\tau = 1$ within that zone. If the reconstructed profile lies outside the range, one can always bring the corrected profile $\tilde{\rho}_{i,j}(x,y)$ within the range by finding some $\tau < 1$ which accomplishes this. For $\tau = 0$, this is always satisfied, ensuring that any conserved variable can be brought within the desired range by progressively reducing the value of "$\tau$" from unity till the variable is within the range.

The implementable strategy, which draws on Sanders (1988), Barth (1995) and Zhang & Shu (2010), consists of having a set of "$Q$" nodal points $\{(x^q, y^q); q=1,..,Q\}$ within each zone and evaluating the entire vector of conserved variables at those points. The index "$q$" tags the nodal points within each zone. It is worth pointing out that the present strategy requires a judicious choice of nodal points in order to work well. We will give some further details about the choice of nodal points for a structured mesh at the end of this section. Thus we have $\rho_{i,j}^q \equiv \rho_{i,j}(x^q, y^q)$ and we can also use them to find $\rho_{i,j}^{\min-zone} = \min\left(\rho_{i,j}^1, \rho_{i,j}^2, .., \rho_{i,j}^Q\right)$ and $\rho_{i,j}^{\max-zone} = \max\left(\rho_{i,j}^1, \rho_{i,j}^2, .., \rho_{i,j}^Q\right)$. As shown by Barth (1995), within each zone $(i,j)$ we can obtain a variable

$$\tau_{i,j} = \min\left(1,\ \min\left(\frac{\rho_{i,j}^{\max-extended} - \bar{\rho}_{i,j}}{\rho_{i,j}^{\max-zone} - \bar{\rho}_{i,j}},\, \frac{\bar{\rho}_{i,j} - \rho_{i,j}^{\min-extended}}{\bar{\rho}_{i,j} - \rho_{i,j}^{\min-zone}}\right)\right) \tag{121}$$

Then the corrected profile for the density, which lies within the desired solution-dependent range and has sufficient leeway to be a non-oscillatory reconstruction, is given by

$$\tilde{\rho}_{i,j}(x,y) = (1-\tau_{i,j})\, \bar{\rho}_{i,j} + \tau_{i,j}\, \rho_{i,j}(x,y) \tag{122}$$

Notice that eqns. (120) and (122) differ in their import, because $\tau_{i,j}$ from eqn. (121) is used in eqn. (122). For most practical calculations, this correction will only be invoked in an extremely small fraction of zones and, that too, for a very small fraction of the total number of time steps. In practice, the physical velocity should not change when the density profile is corrected. Since the momentum density scales as the density, when the variation in the density is reduced, it also helps to reduce the variation in the momentum density by the same amount. Similarly, the total energy density should also be reduced by the same amount.

The previous paragraph has shown how the density is brought within the desired range. We now describe the process of bringing the pressure within the desired range for Euler flow. The analogous demonstration for MHD flow is presented in Balsara (2012b). The positivity for the pressure variable is enforced after the positivity fixes for the density variable have been incorporated, as described in the previous paragraph. The philosophy applied here is quite similar to the one used for the density. The only difference is that the pressure is a derived variable. Thus we write

$$\tilde{\rho}_{i,j}(x,y) = (1-\tau)\,\bar{\rho}_{i,j} + \tau\,\rho_{i,j}(x,y)\ ;\ \tilde{\mathbf{m}}_{i,j}(x,y) = (1-\tau)\,\bar{\mathbf{m}}_{i,j} + \tau\,\mathbf{m}_{i,j}(x,y)\ ;$$
$$\tilde{\mathcal{E}}_{i,j}(x,y) = (1-\tau)\,\bar{\mathcal{E}}_{i,j} + \tau\,\mathcal{E}_{i,j}(x,y) \tag{123}$$

As before, we have $\tau \in [0,1]$, and we observe that with $\tau = 0$ the pressure is guaranteed to be within the desired range. Our positivity enforcing method relies on the fact that the zone-averaged value is always assumed to retain positive density and pressure, which can indeed be guaranteed by using a positivity preserving Riemann solver. Working with the previously defined nodal points, we can define $\rho_{i,j}^q \equiv \rho_{i,j}(x^q, y^q)$, $\mathbf{m}_{i,j}^q \equiv \mathbf{m}_{i,j}(x^q, y^q)$ and $\mathcal{E}_{i,j}^q \equiv \mathcal{E}_{i,j}(x^q, y^q)$. We can then define the pressure at each nodal point by

$$\mathrm{P}_{i,j}^q = (\gamma-1)\left(\mathcal{E}_{i,j}^q - \frac{\left(\mathbf{m}_{i,j}^q\right)^2}{2\rho_{i,j}^q}\right) \tag{124}$$

If $\mathrm{P}_{i,j}^q$ lies within the desired range of pressures, we set a nodal variable $\tau_{i,j}^q = 1$. If $\mathrm{P}_{i,j}^q$ is not within the desired range, we wish to find a nodal variable $\tau_{i,j}^q < 1$ which brings it within the desired range. We illustrate the case where the $\mathrm{P}_{i,j}^{\min-extended}$ bound is violated by the $q^{\text{th}}$ nodal point. The variable $\tau_{i,j}^q < 1$ which brings that nodal pressure back within the desired range is given by solving

$$(\gamma-1)\left\{\left[(1-\tau_{i,j}^q)\bar{\mathcal{E}}_{i,j} + \tau_{i,j}^q \mathcal{E}_{i,j}^q\right] - \frac{1}{2}\frac{\left[(1-\tau_{i,j}^q)\bar{\mathbf{m}}_{i,j} + \tau_{i,j}^q \mathbf{m}_{i,j}^q\right]^2}{\left[(1-\tau_{i,j}^q)\bar{\rho}_{i,j} + \tau_{i,j}^q \rho_{i,j}^q\right]}\right\} = \mathrm{P}_{i,j}^{\min-extended} \tag{125}$$

The above equation is easy to solve for $\tau_{i,j}^q$ because it is actually a quadratic, a fact made apparent by writing it explicitly as

$$\left(\tau_{i,j}^{q}\right)^{2}\left[2\left(\rho_{i,j}^{q}-\bar{\rho}_{i,j}\right)\left(\mathcal{E}_{i,j}^{q}-\bar{\mathcal{E}}_{i,j}\right)-\left(\mathbf{m}_{i,j}^{q}-\bar{\mathbf{m}}_{i,j}\right)^{2}\right]$$
$$+\left(\tau_{i,j}^{q}\right)\left[2\bar{\rho}_{i,j}\left(\mathcal{E}_{i,j}^{q}-\bar{\mathcal{E}}_{i,j}\right)+2\bar{\mathcal{E}}_{i,j}\left(\rho_{i,j}^{q}-\bar{\rho}_{i,j}\right)-2\left(\mathbf{m}_{i,j}^{q}-\bar{\mathbf{m}}_{i,j}\right)\bullet\bar{\mathbf{m}}_{i,j}-2\mathrm{e}_{i,j}^{\min-extended}\left(\rho_{i,j}^{q}-\bar{\rho}_{i,j}\right)\right]$$
$$+\left[2\bar{\rho}_{i,j}\bar{\mathcal{E}}_{i,j}-\left(\bar{\mathbf{m}}_{i,j}\right)^{2}-2\mathrm{e}_{i,j}^{\min-extended}\bar{\rho}_{i,j}\right]=0$$

$$\text{with } \mathrm{e}_{i,j}^{\min-extended} \equiv \mathrm{P}_{i,j}^{\min-extended}/(\gamma-1)$$

(126)

The above step should be done for all the defective nodes within a zone. As before, we expect that only a very small fraction of zones in a practical computation will need this pressure positivity fix. We can then find $\tau_{i,j} = \min\left(\tau_{i,j}^{1}, \tau_{i,j}^{2}, ..., \tau_{i,j}^{Q},\right)$. As before, $\tau_{i,j}$ can now be used to shrink the spatially varying part of all the conserved variables in zone (*i,j*); i.e. as shown in eqn. (9). Indeed note from the above two equations that one has to shrink the spatial variation of *all* the conserved variables in order to bring *all* the nodal pressures within the desired range. This completes our description of the positivity preserving scheme for Euler flow.

The method described above needs to be implemented on a set of nodal points within a zone. The nodes should be picked in such a way that they bring out the extremal variation within a zone. For piecewise linear reconstruction, the extrema in the reconstructed function are always obtained at the vertices of the zone. Because piecewise linear reconstruction is a special sub-case of any higher order reconstruction, the vertices should always be included in the set of nodes within a zone, even at higher orders. For higher order reconstruction, Balsara (2012b) provides a detailed description of how the nodes are to be picked. For a two-dimensional mesh at third order, Fig. 12 provides a good example of nodal points that might be used. We only use the black circles in Fig. 12 for enforcing positivity.

## VIII) Accuracy Analysis on Multidimensional Test Problems

Since we have catalogued several high accuracy schemes, it becomes interesting to demonstrate the difference that order of accuracy makes in the solution of problems with smooth flow. When demonstrating the order of accuracy of a method, it is very helpful (though not essential) to pick test problems whose initial conditions and time-evolution can be specified analytically. We demonstrate the order of accuracy of the higher order schemes that were catalogued in the previous sections.

## VIII.1) Hydrodynamical Vortex with ADER-WENO Schemes

In the hydrodynamic vortex problem, presented by Jiang & Shu (1996), an isentropic vortex propagates at 45° to the grid lines in a two-dimensional domain with periodic boundaries given by [-5, 5] x [-5, 5]. The unperturbed flow at the initial time can be written as $(\rho, P, v_x, v_y) = (1, 1, 1, 1)$. The ratio of the specific heats is given by $\gamma = 1.4$. The entropy and the temperature are defined as $S = P/\rho^\gamma$ and $T = P/\rho$. The vortex is set up as a fluctuation of the unperturbed flow with the fluctuations given by

$$(\delta v_x, \delta v_y) = \frac{\varepsilon}{2\pi} e^{0.5(1-r^2)}(-y, x)$$

$$\delta T = -\frac{(\gamma-1)\varepsilon^2}{8\gamma\pi^2} e^{(1-r^2)}$$

$$\delta S = 0$$

Its strength is controlled by the parameter $\varepsilon$, and we set $\varepsilon = 5$. The radius "$r$" from the origin of the domain and can be written as $r^2 = x^2 + y^2$. Because the vortex represents a self-similar flow profile, it undergoes a form-preserving translation along the diagonal of the computational domain. As a result, the above initial conditions can be used to specify the fluid variables at any later time.

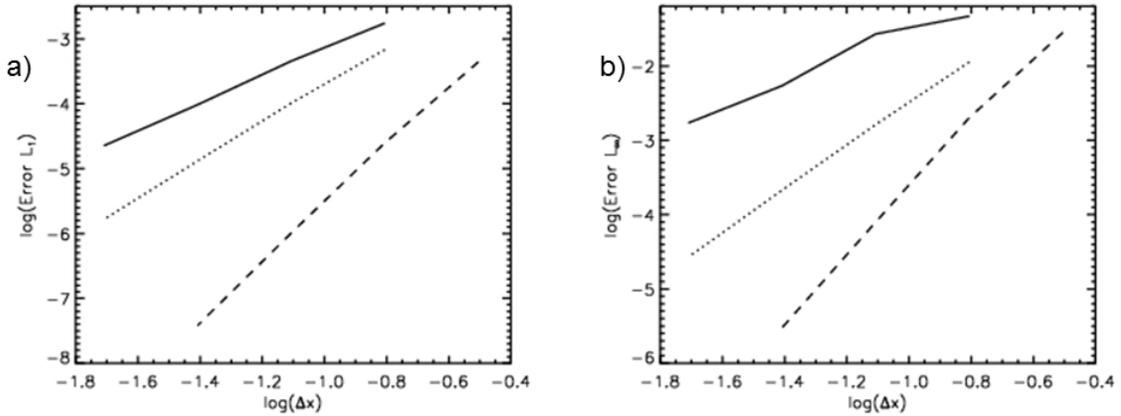

Fig. 16 shows the variation of the $L_1$ (left panel) and $L_\infty$ (right panel) errors as a function of zone size for the 2d vortex problem. The solid lines show the second order predictor-corrector scheme with minmod limiter; the dotted and dashed lines show third and fourth order ADER-WENO schemes respectively.

The analytically predicted conserved variables can be compared to the numerically computed conserved variables in order to demonstrate accuracy. Note though that once one goes past second order, initializing the zone-averaged conserved variables on a mesh is a non-trivial exercise. The reason is easy to illustrate at third order. The above equations can be used to predict the conserved variables associated with the vortex for all times and at any point within a zone. Thus one can predict them at the nodes defined by eqn. (78). Notice though that at third order, the zone-averaged conserved variables are not well-approximated by the conserved variables that are evaluated at the central node within a zone. It actually requires a numerical quadrature to evaluate the zone-averaged conserved variables. Thus if we take $U^1_{i,j},...,U^9_{i,j}$ to be the values of the conserved variables that are evaluated at the nine nodes from eqn. (78) within zone $(i, j)$, we have

$$\overline{U}_{i,j} = U^1_{i,j} + \left( U^2_{i,j} - 2\, U^1_{i,j} + U^3_{i,j} + U^4_{i,j} - 2\, U^1_{i,j} + U^5_{i,j} \right) \big/ 6$$

Figs. 16a and 16b show the logarithms of the errors measured in the $L_1$ and $L_\infty$ norms for the vortex test problem as a function of the logarithm of the zone size $\Delta x$. This is done for the second, third and fourth order schemes. We see that the higher order schemes produce a smaller error on the coarsest meshes. Moreover, as the mesh is refined, the error in the higher order schemes decreases much faster with mesh refinement. Schemes with WENO reconstruction and ADER time stepping were used to generate the third and fourth order results. The second order scheme used an MC limiter with a predictor-corrector formulation.

**VIII.2) MHD Vortex with DG and PNPM Schemes**

In the previous sub-section we presented a genuinely two-dimensional Euler problem associated with a fluid vortex that was made to propagate at $45^0$ to the computational mesh. The problem was extended to MHD in Balsara (2004). It is especially good for accuracy testing because it consists of a smoothly-varying and dynamically stable configuration that carries out non-trivial motion in the computational domain. The problem is set up on a two-dimensional domain given by [-5,5]X[-5,5]. The domain is periodic in both directions. An unperturbed magnetohydrodynamic flow with ( $\rho$, P, $v_x$, $v_y$, $B_x$, $B_y$ ) = (1, 1, 1, 1, 0, 0) is initialized on the computational domain. The ratio of specific heats is given by $\gamma = 5/3$. The vortex is initialized at the center of the computational domain by way of fluctuations in the velocity and magnetic fields given by

$$\left( \delta v_x, \delta v_y \right) = \frac{\kappa}{2\pi} e^{0.5(1-r^2)} \left( -y, x \right)$$

$$\left( \delta B_x, \delta B_y \right) = \frac{\mu}{2\pi} e^{0.5(1-r^2)} \left( -y, x \right)$$

We used $\mu = 2\pi$ in the above equation for the results shown here. The magnetic vector potential in the z-direction associated with the magnetic field in the previous equation is given by

$$\delta A_z = \frac{\mu}{2\pi} e^{0.5(1-r^2)}$$

The magnetic vector potential plays an important role in the divergence-free initialization of the magnetic field on the computational domain. The circular motion of the vortex produces a centrifugal force. The tension in the magnetic field lines provides a centripetal force. The magnetic pressure also contributes to the dynamical balance in addition to the gas pressure. The condition for dynamical balance is given by

$$\frac{\partial P}{\partial r} = \left[ \rho \left(\frac{\kappa}{2\pi}\right)^2 - \frac{1}{2\pi}\left(\frac{\mu}{2\pi}\right)^2 \right] r\, e^{(1-r^2)} + \frac{1}{4\pi}\left(\frac{\mu}{2\pi}\right)^2 r^3\, e^{(1-r^2)}$$

For the fluid case, Jiang and Shu (1996) provide an isentropic solution for the above equation. For the MHD case it is simplest to set the density to unity and solve the above equation for the pressure. The fluctuation in the pressure is then given by

$$\delta P = \frac{1}{8\pi}\left(\frac{\mu}{2\pi}\right)^2 (1-r^2)\, e^{(1-r^2)} - \frac{1}{2}\left(\frac{\kappa}{2\pi}\right)^2 e^{(1-r^2)}$$

As a result all aspects of the flow field are available in analytical form for all time which makes this problem very useful for accuracy analysis. The vortex can be set up with any strength because it is an exact solution of the MHD equations. It is worth pointing out that this test problem is easily extended to three dimensions by having a non-zero value for the z-component of the magnetic field. The simplest extension consists of giving the magnetic field a constant pitch angle with respect to the z-axis.

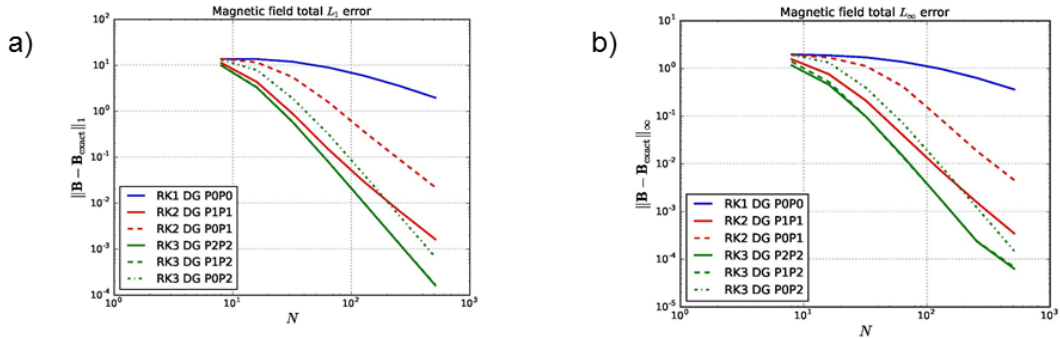

*Fig. 17a and 17b from Balsara and Käppeli (2017) shows the $L_1$ and $L_\infty$ errors for the propagation of a magnetized vortex as a function of mesh size measured along one of the edges of the mesh. P0P1 and P1P1 schemes are shown at second order. At third order we show P0P2, P1P2 and P2P2 schemes. The results for P1P2 and P2P2 schemes coincide with one another.*

Accuracy analysis of this test problem using RK-WENO and ADER-WENO schemes has been presented in Balsara (2009) and Balsara *et al*. (2009). Here we present results from Balsara and Käppeli (2017) involving just the magnetic field but with the variation in the velocity and pressure suppressed. As a bonus though, we show the error as measured in the $L_1$ and $L_\infty$ norms for several PNPM schemes in Fig. 17. We see that there is a quality gap between the P0P1 scheme and the P1P1 scheme (which is indeed the P=1 DG scheme). Likewise, we see a quality gap between the P0P2 scheme (WENO scheme) and the P2P2 scheme (which is indeed the P=2 DG scheme). However, the P1P2 and P2P2 schemes produce results in Fig. 17 that are virtually indistinguishable! Despite having comparable accuracy, the third order P1P2 scheme was able to take substantially larger timesteps than the third order P2P2 scheme, showing that it offers some advantages.

**VIII.3) RHD and RMHD Vortices with ADER-WENO Schemes**

For classical hydrodynamics and MHD, there are several very nice, non-trivial multidimensional test problems for demonstrating that a numerical method meets its design accuracy. The present RHD and RMHD test problems, first described in Balsara and Kim (2017), are the relativistic analogues of the classical hydrodynamical and MHD vortices. They should prove very useful for accuracy testing of RHD and RMHD codes.

The problem is set up on a periodic domain that spans $[-5,5] \times [-5,5]$. We first describe the velocity and magnetic field in the rest frame of the vortex. For nonrelativistic hydrodynamics or MHD, making the vortex move on the mesh is just a matter of adding a net velocity. For relativistic hydrodynamics and MHD, one has to include the additional complications of relativistic velocity addition and Lorentz transformation. These additional tasks are entirely non-trivial for relativistic flow. For that reason, we initially focus on the description of the vortex in its own rest frame. In a subsequent paragraph we will describe the velocity addition and Lorentz transformation. The velocity of the vortex (before it is made to move relative to the mesh) is given by

$$(v_x, v_y) = v_{max}^\phi e^{0.5(1-r^2)}(-y, x)$$

For both the hydrodynamical and RMHD test problems we have used $v_{max}^\phi = 0.7$. Notice that the velocity diminishes rapidly far away from the center of the vortex. This rapid drop in the velocity ensures that the boundaries of the domain have a negligible effect on the dynamics of the vortex. The magnetic field of the vortex (before it is made to move relative to the mesh) is given by

$$(B_x, B_y) = B_{max}^\phi e^{0.5(1-r^2)}(-y, x)$$

For the RMHD test problem we set $B_{max}^\phi = 0.7$. Notice that the magnetic field diminishes rapidly far away from the center of the vortex. This rapid drop in magnetic pressure and magnetic tension ensures that the boundaries of the domain have a negligible effect on the dynamics of the vortex. The corresponding magnetic vector potential, which is very useful for setting up a divergence-free vector field, is given by

$$A_z = B_{max}^\phi e^{0.5(1-r^2)}$$

The pseudo-entropy is defined by $S = P/\rho^\Gamma$ with polytropic index $\Gamma = 5/3$. The pressure and density of the vortex are also set to unity at the center of the vortex. The vortex is initialized to be isentropic so that $\delta S = 0$; i.e., the entropy is a constant throughout the vortex. Consistent with this velocity field and magnetic field, the steady state equation for the radial momentum of the vortex yields a pressure balance condition. This pressure balance condition for an RMHD vortex is given by

$$r \frac{dP_{Tot}}{dr} = (\rho h + b^2)\gamma^2 (v^\phi)^2 - (b^\phi)^2$$

For the hydrodynamical case, the above equation simplifies to become.

$$r \frac{dP_g}{dr} = \rho h \gamma^2 (v^\phi)^2$$

Depending on the circumstance, one of the above two equations is numerically integrated radially outwards from the center of the vortex. Along with the isentropic condition, this equation fully specifies the run of the density and pressure in the vortex as a function of radius. Fig. 18a shows the run of thermal pressure as a function of radius for the vortices used here in the relativistic hydrodynamics and RMHD cases. Notice that a specification of the pressure at all radial points in the vortex also yields the density because of the isentropic condition. Observe that the thermal pressure profile for the magnetized vortex is less steep in Fig. 18a because the magnetic pressure supplements the gas pressure. This completes the description of the vortex in its own rest frame. Because the next steps associated with relativistic velocity addition and Lorentz transformation are non-trivial, we recommend that the run of density and pressure for the vortices should be tabulated on a very fine one-dimensional radial mesh. Typically, this radial mesh should have resolution that is much finer than the two-dimensional mesh on which the problem is computed.

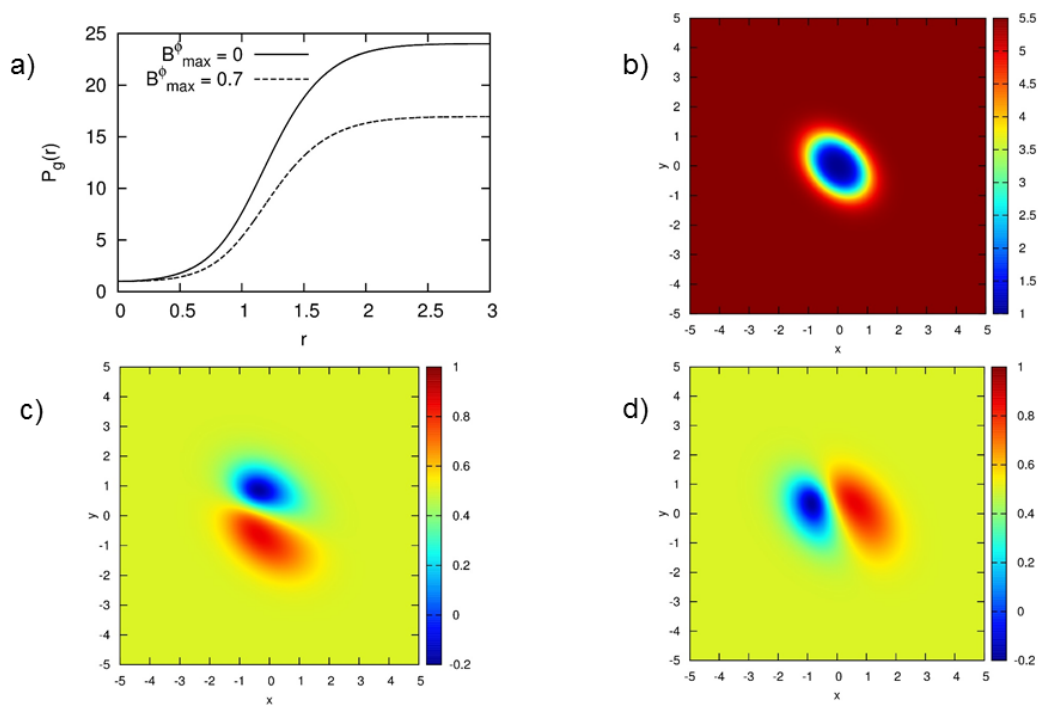

*Fig. 18 is from Kim and Balsara (2016). Fig. 18a shows the run of thermal pressure as a function of radius for the vortices used here in the relativistic hydrodynamics and RMHD cases. Fig. 18b we show the density profile of the RMHD vortex on the computational mesh at the initial time. Figs. 18c and 18d show the x-velocity and y-velocity respectively.*

We now describe the process of mapping the vortex to a computational mesh on which it moves with a speed $\beta_x \hat{x} + \beta_y \hat{y}$. We use $\beta_x = \beta_y = 0.5$ for our vortex; i.e. our vortex moves on the mesh with a speed that is $1/\sqrt{2}$ times the speed of light. Let us define $\gamma_\beta \equiv 1/\sqrt{1 - \beta_x^2 - \beta_y^2}$ to be the Lorentz factor associated with this velocity. In reality, this mapping of the vortex to a computational mesh is achieved by making the computational mesh move with a speed $-\beta_x \hat{x} - \beta_y \hat{y}$ relative to the rest frame of the vortex. Let the rest frame of the vortex be described by the unprimed spacetime coordinates $(t, x, y, z)^T$. The coordinates of the computational mesh, therefore, correspond to primed spacetime coordinates given by $(t', x', y', z')^T$. In practice, $t' = 0$ when initializing the computational mesh and realize too that the equations that describe the vortex in its own rest frame are also time-independent, i.e. they do not depend on "$t$". Thus for any chosen coordinate $(t' = 0, x', y', z' = 0)^T$ on the computational mesh we can find the corresponding unprimed coordinates via the following Lorentz transformation

$$\begin{pmatrix} t' \\ x' \\ y' \\ z' \end{pmatrix} = \begin{pmatrix} \gamma_\beta & \gamma_\beta \beta_x & \gamma_\beta \beta_y & 0 \\ \gamma_\beta \beta_x & 1+(\gamma_\beta-1)\beta_x^2/\beta^2 & (\gamma_\beta-1)\beta_x\beta_y/\beta^2 & 0 \\ \gamma_\beta \beta_y & (\gamma_\beta-1)\beta_x\beta_y/\beta^2 & 1+(\gamma_\beta-1)\beta_y^2/\beta^2 & 0 \\ 0 & 0 & 0 & 1 \end{pmatrix} \begin{pmatrix} t \\ x \\ y \\ z \end{pmatrix}$$

The unprimed coordinates refer to the rest frame of the vortex. In the unprimed frame, all the flow variables associated with the vortex have already been specified via the discussion in the previous paragraph. Scalar variables, like density and thermal pressure, are referred to the rest frame of the fluid parcel, i.e. they are proper variables that transform as scalars. Consequently, they transform unchanged as long as the Lorentz transform in the previous equation is properly applied. Three-velocities have to be suitably transformed using the relativistic addition of velocities. The appropriate formulae that give us the velocities in the primed frame from the original velocities in the unprimed frame are given below as:

$$v'_x = \frac{\gamma_\beta \beta_x + \left[1+\frac{(\gamma_\beta-1)\beta_x^2}{\beta^2}\right]v_x + \frac{(\gamma_\beta-1)\beta_x\beta_y}{\beta^2}v_y}{\gamma_\beta(1+\beta_x v_x + \beta_y v_x)}$$

and

$$v'_y = \frac{\gamma_\beta \beta_y + \frac{(\gamma_\beta-1)\beta_x\beta_y}{\beta^2}v_x + \left[1+\frac{(\gamma_\beta-1)\beta_y^2}{\beta^2}\right]v_y}{\gamma_\beta(1+\beta_x v_x + \beta_y v_x)}$$

With the relativistic velocity addition formulae described above, we can obtain the velocities at any point on our computational mesh. Since we use a magnetic vector potential to initialize our magnetic field, we point out that the electric field potential, $\Phi$, and the magnetic vector potential ($\vec{A}$) together form a four-vector $(\Phi, A^i)^T$. Being a four-vector, it transforms just like a four-coordinate. We can, therefore, obtain the magnetic vector potential in the primed frame. In the specific instance of the vortex that we describe here, the z-component of the magnetic vector potential is unchanged as we transform from the unprimed frame back to the primed frame. Even in the primed frame, the previously described Lorentz transformation is such that only the z-component of the magnetic vector potential will be non-zero. Likewise, the value of $\Phi$ is immaterial and set to zero. We see therefore that it is easy to initialize the divergence-free magnetic field for the vortex on the computational mesh. This completes our discussion of the set-up for relativistically boosted hydrodynamical and RMHD vortices on a computational mesh. Because these relativistic vortices are new in the literature, in Fig. 18b we show the density profile of the RMHD vortex on the computational mesh at the initial time. Notice that the boosted vortex shows substantial Lorentz contraction in its density variable. Figs. 18c and 18d show the x-velocity and y-velocity respectively. Notice that the velocity profiles are not symmetrical about the northeast-pointing diagonal of the mesh owing to the relativistic velocity addition formulae.

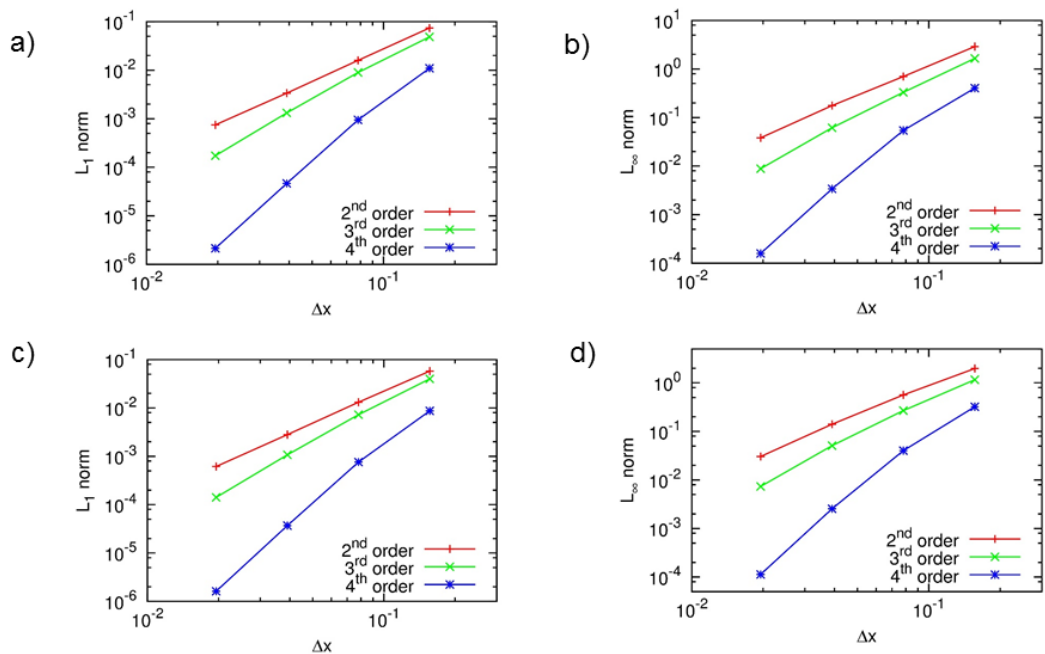

*Fig. 19 is from Balsara and Kim (2016). Figs. 19a and 19b show the $L_1$ and $L_\infty$ errors from ADER-WENO schemes for the RHD vortex problem; the density variable is shown. Figs. 19c and 19d show the $L_1$ and $L_\infty$ errors from ADER-WENO schemes for the RMHD vortex problem; the x-component of the magnetic field is shown.*

Figs. 19a and 19b show the errors measured in the $L_1$ and $L_\infty$ norms for the RHD vortex. The error is measured in the density variable, i.e. the proper density times the Lorentz factor. Figs. 19c and and 19d show the errors measured in the $L_1$ and $L_\infty$ norms for the RMHD vortex. In this instance, we show the error measured in the x-component of the magnetic field. ADER-WENO schemes at second, third and fourth order were used. We see that the schemes meet their design accuracies.

**IX) Test Problems**

In this section we do not focus on one-dimensional test problems. Good libraries of one-dimensional test problems for Euler flow have been provided in Woodward and Colella (1984). For analogous catalogues of one-dimensional Riemann problems for MHD flow, please see Ryu and Jones (1995), Dai and Woodward (1994) and Falle (2001). For a list of one-dimensional Riemann problems for RHD flow, please see Martí and Müller (2003) and also Rezzolla and Zanotti (2001). For an analogous catalogue of RMHD problems, please see Balsara (2001) and Giacomazzo and Rezzolla (2006).

In the rest of this section, we present several stringent multidimensional test problems for Euler, MHD, RHD and RMHD flow that were all done with higher order schemes.

**IX.1) Euler Flow: The Forward-Facing Step Test Problem with ADER-DG Schemes**

This problem was first presented by Woodward and Colella (1984). The problem consists of a two-dimensional wind tunnel that spans a domain of [0, 3] x [0, 1]. A forward-facing step is set up at a location given by the coordinates (0.6,0.2). Inflow boundary conditions are applied at the left boundary, where the gas enters the wind tunnel at Mach 3.0 with a density of 1.4 and a pressure of unity. The right boundary is an outflow boundary. The walls of the wind tunnel and the step are set to be reflective boundaries. The singularity at the corner was treated with the same technique that was suggested by Woodward and Colella (1984). The simulation was run until a time of 4.0 time units and the ratio of specific heats is given by 1.4.

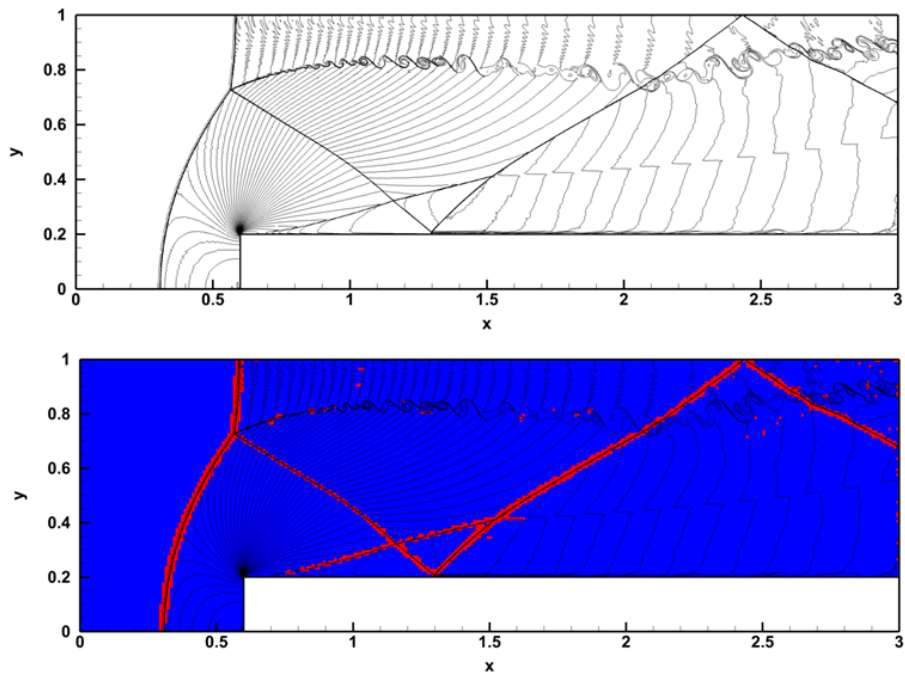

Fig. 20 from Dumbser et al. (2014) shows the density variable from the forward facing step problem using an ADER-DG scheme at sixth order. The result in the upper panel was computed on a 300×100 zone mesh and is shown at a time of 4 units. We see that the simulation captures the roll-up of the vortex very clearly. The lower panel shows the zones that were flagged for MOOD limiting in red. We see that only a very small fraction of zones were limited by the MOOD limiting procedure.

Fig. 20 from Dumbser *et al*. (2014) shows the density variable from the forward facing step problem using an ADER-DG scheme at sixth order. The result in the upper panel was computed on a 300×100 zone mesh and is shown at a time of 4 units. Even though the mesh seems to have only 30,000 zones, a high order DG scheme can capture substantial amounts of sub-structure within each zone. We see that the simulation captures the roll-up of the vortex very clearly. The lower panel shows the zones that were flagged for MOOD limiting in red. We see that only a very small fraction of zones were limited by the MOOD limiting procedure. The CFL number was set to 0.4.

The step induces a forward-facing bow shock, which interacts with the upper wall. The interaction of the bow shock with the upper wall initiates a Mach stem. All the shocks are properly captured on the computing grid and have sharp profiles. The vortex sheet that emanates from the Mach stem is correctly resolved with only a few zones across the sheet. We notice that the vortex sheet shows little or no spreading over the length of the computational domain. This demonstrates the ability of the high order schemes to provide a better resolution for a smaller number of zones.

**IX.2) Euler Flow: Double Mach Reflection Problem with ADER-WENO Scheme**

This problem was presented by Woodward and Colella (1984). We use the same setup for this test problem as the above authors. A Mach 10 shock hits a reflecting wall which spreads from $x = 1/6$ to $x = 4$ at the bottom of the domain. The two-dimensional computational mesh spans $[0, 4] \times [0, 1]$. The angle between the shock and the wall is 60°. At the start of the computation, the position of the shock is given by $(x, y) = (1/6, 0)$. The undisturbed fluid in front of the shock is initialized with a density of 1.4 and a pressure of 1. The exact post-shock condition is used for the bottom boundary from $x = 0$ to $x = 1/6$ to mimic an angled wedge. For the remaining boundary at the bottom of the domain we used a reflective boundary condition. The top boundary condition imposes the exact motion of a Mach 10 shock in the flow variables. The left and right boundaries are set to be inflow and outflow boundaries.

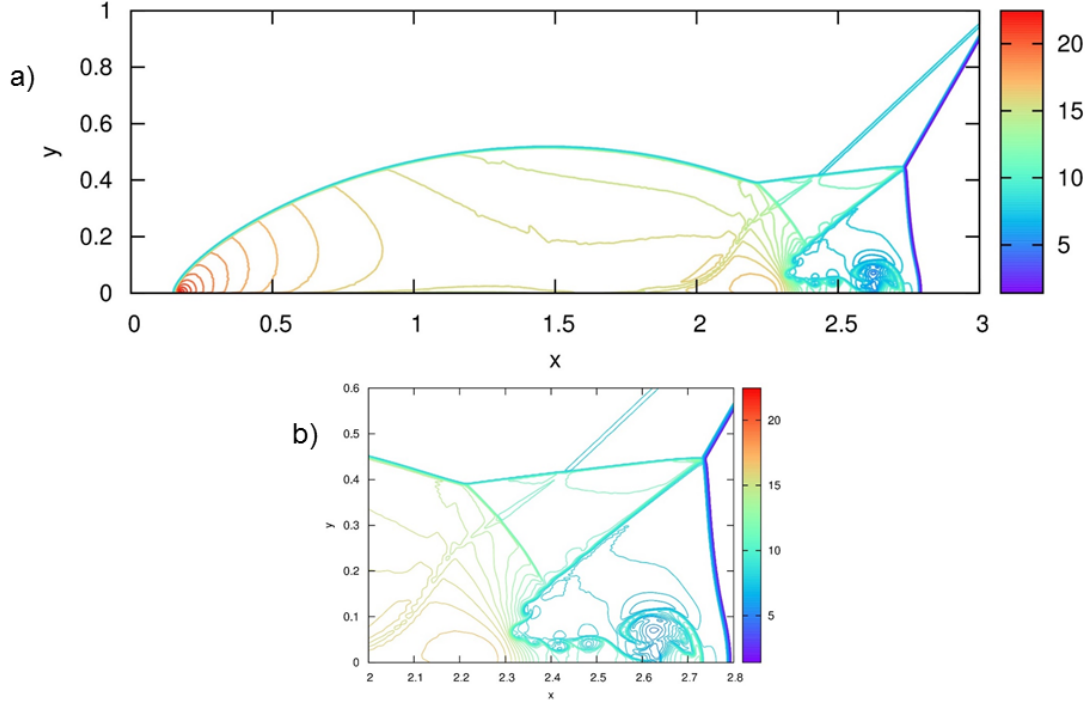

*Fig. 21 from Balsara and Nkonga (2017) shows the density variable from a 4$^{rd}$ order ADER-WENO scheme for the double Mach reflection problem. We clearly see the roll up of the Mach stem due to Kelvin-Helmholtz instability in the zoom-in figure shown in Fig. 21b.*

Fig. 21 shows the density variable at $t = 0.2$ in the sub-domain given by [0, 3] × [0, 1]. The upper panel shows a simulation with a resolution of 1920 × 480 zones. At the high resolution, the Mach stem displays a roll-up due to the operation of the Kelvin-Helmholtz instability. We used the fourth order ADER-WENO scheme for both simulations. Notice that the fourth order ADER-WENO scheme resolves all the structures that form in this problem. According to Cockburn & Shu (1998), a second order scheme would need at least four times as many zones in each direction to resolve the instability and for such a simulation we would need much more CPU time than the fourth order scheme shown in Fig. 21. That demonstrates the efficiency of the higher order schemes presented here.

**IX.3) MHD Flow: 2D Rotor Test Problem with ADER-WENO Scheme**

This problem was suggested in Balsara and Spicer (1999) and Balsara (2004). The problem is set up on a two dimensional unit square. It consists of having a dense, rapidly spinning cylinder, in the center of an initially stationary, light ambient fluid. The two fluids are threaded by a magnetic field that is uniform to begin with and has a value of 2.5 units. The pressure is set to 0.5 in both fluids; though it can also be set to unity. The ambient fluid has unit density. The rotor has a constant density of 10 units out to a radius of 0.1. Between a radius of 0.1 and $0.1 + 6\,\Delta x$ a linear taper is applied to the density so that the density in the cylinder linearly joins the density in the ambient. The taper is, therefore, spread out over six computational zones and it is a good idea to keep that number fixed as the resolution is increased or decreased. The ambient fluid is initially static. The rotor rotates with a uniform angular velocity that extends out to a radius of 0.1.

At a radius of 0.1 it has a toroidal velocity of one unit. Between a radius of 0.1 and $0.1+6\,\Delta x$ the rotor's toroidal velocity drops linearly in the radial velocity from one unit to zero so that at a radius of $0.1+6\,\Delta x$ the velocity blends in with that of the ambient fluid. The ratio of specific heats is taken to be 5/3. The problem is stopped at a time of 0.29.

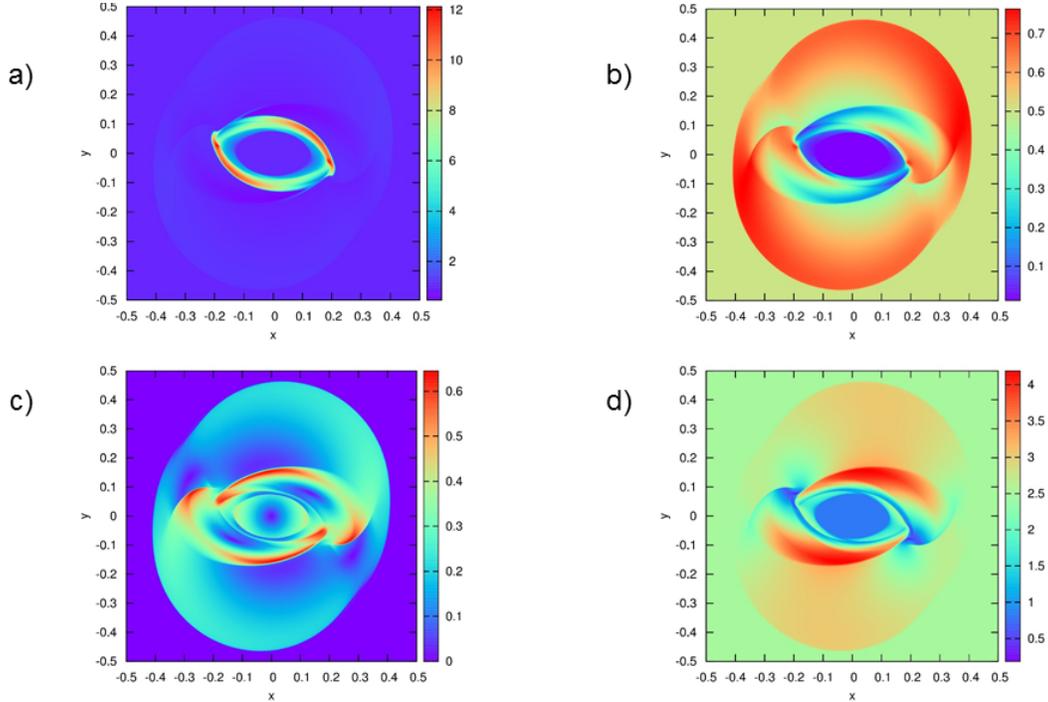

*Fig. 22 from Balsara and Nkonga (2017) shows the results from the MHD Rotor test problem. Figs. 22a, 22b, 22c and 22d show the density, pressure, magnitude of the fluid velocity and magnitude of the magnetic field at the final time. A fourth order ADER-WENO scheme with 1000×1000 zone resolution was used.*

The RIEMANN framework for computational astrophysics was applied to this problem. Fig. 22, which is drawn from Balsara and Nkonga (2017), shows the results from the MHD Rotor test problem. Figs. 22a, 22b, 22c and 22d show the density, pressure, magnitude of the fluid velocity and magnitude of the magnetic field at the final time. A fourth order ADER-WENO scheme with 1000×1000 zone resolution was used.

**IX.4) MHD Flow: 3D Extreme Blast Test Problem with ADER-WENO Scheme**

This test problem is a more extreme extension of a 2D blast test problem from Balsara and Spicer (1999). The present test problem was described in Balsara and Nkonga (2017) and uses a multidimensional Riemann solver described in that same paper. The plasma $\beta$ measures the ratio of the thermal pressure to the magnetic pressure. As the plasma's $\beta$ becomes smaller, this problem becomes increasingly stringent. The problem consists of a $\gamma=1.4$ gas with unit density and a pressure of 0.1 initialized on a $257^3$ zone mesh spanning the unit cube. Initially we have $B_x = B_y = B_z = 150/\sqrt{3}$. The pressure is initially reset to a value of 1000 inside a central region with a radius of 0.1. The plasma's $\beta$ is initially given by $1.117\times10^{-4}$. A CFL number of 0.4 was used. The problem is run up to a time of 0.0075, by which time a strong magnetosonic blast wave propagates through the domain. The problem was run with a third order ADER-WENO scheme with the MuSIC Riemann solver applied at the edges of the mesh. (The term MuSIC in the Riemann solver stands for a Riemann solver that is "Multidimensional, Self-similar, strongly-Interacting, Consistent".) Methods to ensure pressure positivity from Balsara (2012b) were used.

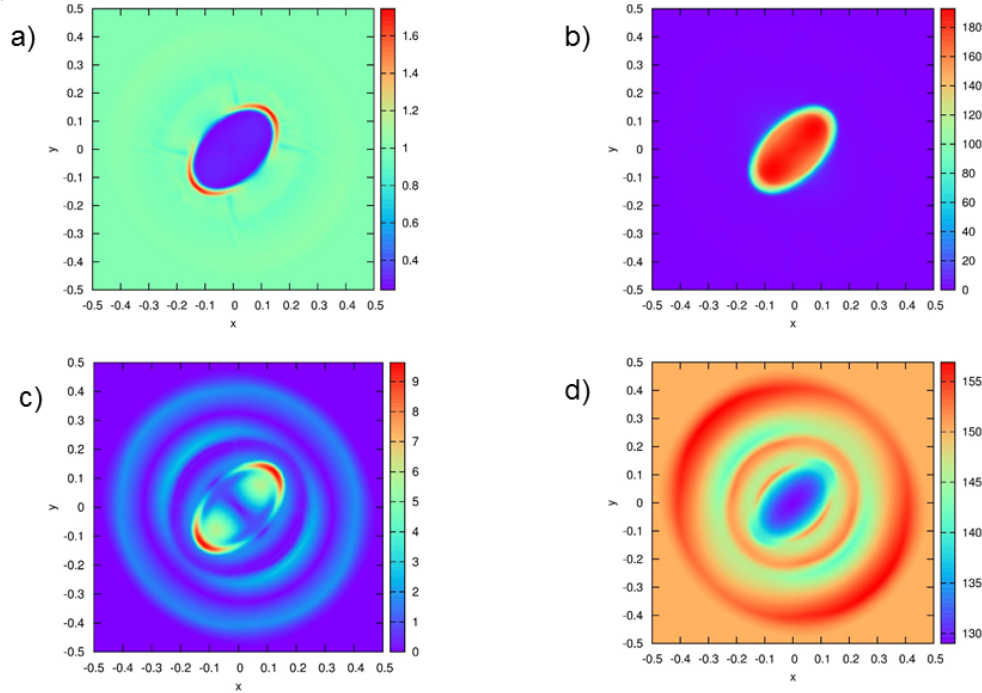

*Fig. 23 from Balsara and Nkonga (2017) shows the variables from the 3D blast problem in the z=0 mid-plane of the computational domain. Fig. 23a shows the plot of the density for the mid-plane in the z-direction. Fig. 23b shows the plot for the pressure in the same plane. Figs. 23c and 23d show the magnitude of the velocity and the magnitude of the magnetic field, in the z=0 plane.*

Fig. 23 shows the variables from the 3D blast problem in the $z = 0$ mid-plane of the computational domain. Fig. 23a shows the plot of the density for the mid-plane in the z-direction. Fig. 23b shows the same for the pressure in the same plane. Figs. 23c and 23d show the magnitude of the velocity and the magnitude of the magnetic field, again in the same plane. We see that despite this being a very stringent problem, the densities and pressures are positive, as expected.

**IX.5) MHD Flow: Decay of Finite Amplitude Torsional Alfven Waves with ADER-WENO Scheme**

Turbulence studies play an increasingly important role in several fields, like astrophysics or space physics. The ability to propagate finite amplitude Alfven waves over large distances and long times on a computational mesh is crucial for carrying out simulations of MHD turbulence. If the Alfven waves are damped strongly because of inherent numerical dissipation in a code, the code will fail to capture the resulting turbulence. This is because MHD turbulence is mainly sustained by Alfven waves. The Alfven wave decay test problem, first presented by Balsara (2004), examines the numerical dissipation of torsional Alfven waves in two dimensions. In this test problem torsional Alfven waves propagate at an angle of 9.462º to the y-axis through a domain given by [-3, 3] x [-3, 3] . The domain was set up with 120×120 zones and has periodic boundary conditions. We do not present further details of the set-up, because the problem is already well-described in the above-mentioned paper. The simulation was stopped at 129 time units by which time the Alfven waves had crossed the domain several times. Depending on the dissipation properties of the scheme, the amplitude of the torsional Alfven wave will, of course, decay. A more dissipative method will cause greater dissipation of the Alfven wave; a less dissipative method will reduce that dissipation.

It is often said that the quality of the Riemann solver is not very important, especially when high order schemes are used. But practitioners have not quantified the precise order of accuracy of the scheme at which the quality of the Riemann solver becomes immaterial. We set out to quantify this order of accuracy for MHD simulations. To that end, we simulated the torsional Alfven wave decay problem with second, third and fourth order schemes with the 1D HLLI Riemann solver along with the 2D MuSIC Riemann solver with sub-structure. Used in this fashion, both the 1D and 2D Riemann solvers are complete; i.e. they fully represent all the waves that arise in the MHD system. We then simulated the same problem again with the same second, third and fourth order schemes. However, this time we used a 1D HLL Riemann solver along with the 2D MuSIC Riemann solver without any sub-structure. In other words, in our second set of simulations both Riemann solvers did not resolve any intermediate waves.

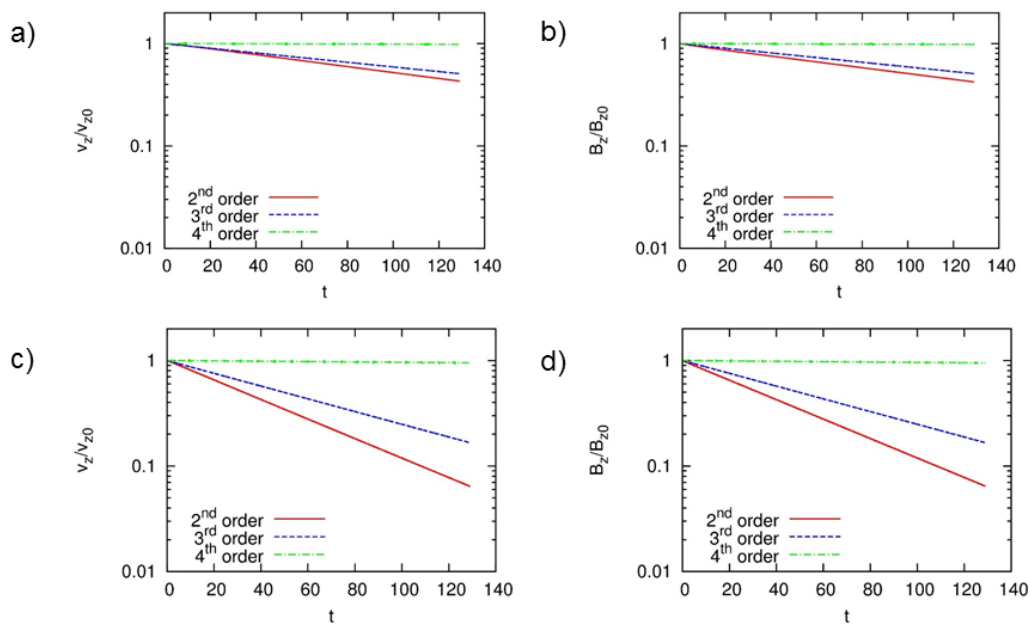

*Figs. 24a and 24b show the evolution of the maximum z-velocity and maximum z-component of the magnetic field in the torsional Alfven wave as a function of time. For the simulations shown in Figs. 24a and 24b we used the 1D HLLI Riemann solver along with the 2D MuSIC Riemann solver with sub-structure. Figs. 24c and 24d show the same information as Figs. 24a and 24b, the only difference being that we used the 1D HLL Riemann solver along with the 2D MuSIC Riemann solver without sub-structure.*

Figs. 24a and 24b show the evolution of the maximum z-velocity and maximum z-component of the magnetic field in the torsional Alfven wave as a function of time. For the simulations shown in Figs. 24a and 24b we used the 1D HLLI Riemann solver along with the 2D MuSIC Riemann solver with sub-structure. Figs. 24c and 24d show the same information as Figs. 24a and 24b, the only difference being that we used the 1D HLL Riemann solver along with the 2D MuSIC Riemann solver without sub-structure. Comparing the two sets of figures, we see that the inferior Riemann solvers produce a six-times larger decay in the amplitude of the Alfven wave at second order. At third order, the inferior Riemann solvers produce a three-times larger decay in the amplitude of the Alfven wave. Notice that the second order scheme with superior Riemann solvers is less dissipative than the third order scheme with inferior Riemann solvers! At fourth order, the difference between the inferior Riemann solvers and the exact Riemann solvers is almost negligible. We, therefore, conclude that second and third order schemes are greatly benefited by the quality of the Riemann solver. It is only at fourth and higher orders of accuracy that the difference between a superior and an inferior Riemann solver begins to become quite small! However, please note that a fourth order scheme has computational complexity that is substantially higher than a second or third order scheme. The Riemann solver with substructure has a computational complexity that is only marginally higher than a Riemann solver without substructure. As a result, it is very advantageous to improve the quality of all schemes at all orders.

**IX.6) RMHD Flow: 2D Relativistic Rotor Test Problem with ADER-WENO Scheme**

The rotor test problem was initially presented for classical MHD by Balsara & Spicer (1999) and it has been adapted to RMHD by Del Zanna *et al*. (2003) in two-dimensions and Mignone *et. al.* (2009) in three-dimensions. Balsara and Kim (2016) pointed out that there are nuances in setting up this problem on a mesh. In order for a mesh to actually represent the high Lorentz factor flows in this problem, they showed that the mesh resolution had to be comparably high. The problem is set up on a unit domain in two dimensions which spans $[-0.5, 0.5] \times [-0.5, 0.5]$. A unit x-magnetic field is set up all over the domain with a unit thermal pressure. There is a unit density in the problem everywhere except within a radius of 0.1, where the density becomes ten times larger. The high density region is set into rapid rotation with a velocity given by $\vec{v}(x, y) = -w\, y\, \hat{x} + w\, x\, \hat{y}$, thus forming a rotor. The parameter "$w$" controls the rotation speed. Because very small changes in "$w$" can result in very large changes in the Lorentz factor, the problems arise when one tries to set up this problem on a computational mesh. The high Lorentz factor flows are confined to a very thin ring at the outer boundary of the rotor. We used $w = 9.9944$ which corresponds to a maximal Lorentz factor of 30, which requires the use of a mesh with at least $3500 \times 3500$ zones.

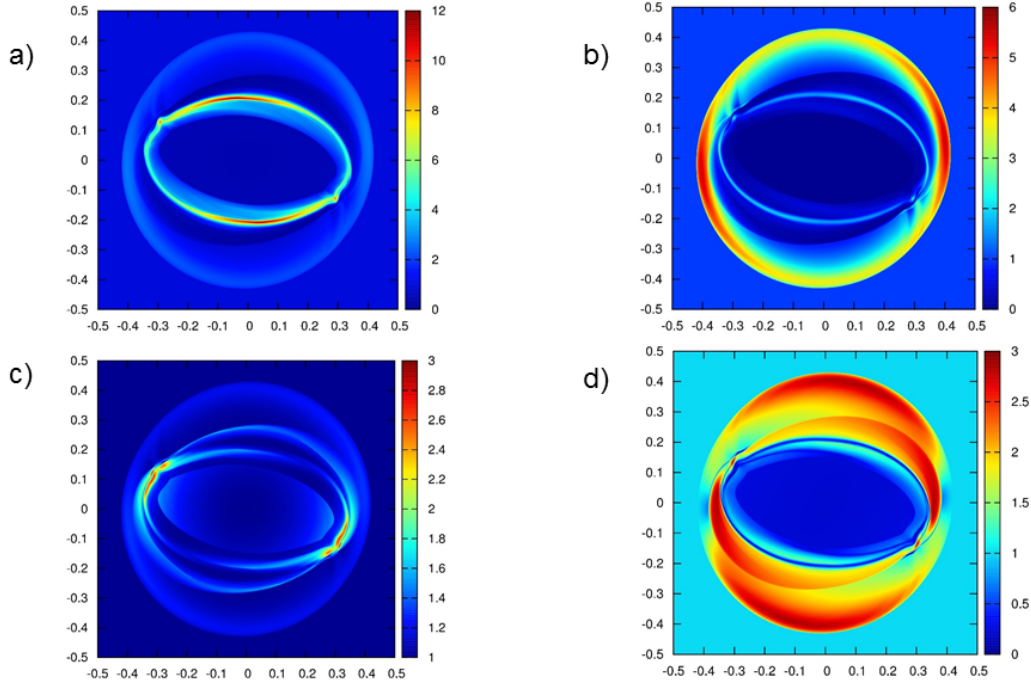

*Figs. 25a, 25b, 25c and 25d from Balsara and Kim (2016) show the density, gas pressure, Lorentz factor and magnetic field strength for the RMHD rotor problem with a starting Lorentz factor of 30. The simulation was run on a 4700×4700 zone mesh with a third order ADER-WENO scheme and stopped at a time of 0.4.*

We used a mesh with 4700×4700 zones for this simulation. Figs. 25a through 25d show the density, gas pressure, Lorentz factor and magnetic field magnitude at a final time of 0.4. Despite the very large initial Lorentz factor, we see that all the flow variables are well-represented. The large Lorentz factor produces a substantial outward expansion in the density owing to the large centrifugal effect in the fast-rotating flow. The magnetic field in Fig. 25d is strongly compressed due to the high Lorentz factor. The simulation in Fig. 5 was run with a CFL of 0.4 using a third order accurate ADER-WENO scheme along with the MuSIC Riemann solver.

**IX.7) RMHD Flow: 2D Relativistic Orzag-Tang Test Problem with ADER-WENO Scheme**

The Orzag Tang test problem (Orzag and Tang 1979) is designed to illustrate the transition to turbulence for MHD flows. The RMHD variant of that test problem has been proposed by Beckwith and Stone (2011). We do not repeat the set-up here. The problem was set up on a unit square with 1000×1000 zones and run to a final time of 0.8. The problem was run with a fourth order ADER-WENO scheme with the MuSIC Riemann solver applied at the edges of the mesh. Figs. 26a, 26b, 26c and 26d show the density, pressure, magnitude of the velocity and magnitude of the magnetic field at the final time for the relativistic Orzag Tang problem. All the requisite RMHD flow features are captured nicely in our simulations.

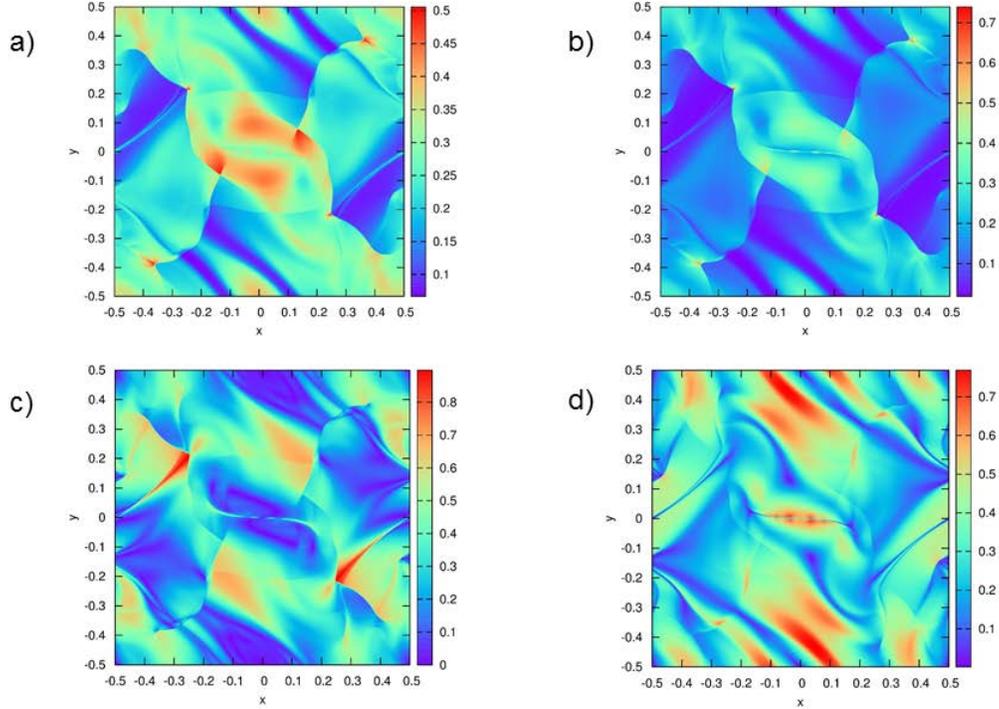

*Figs. 26a, 26b, 26c and 26d from Balsara and Nkonga (2017) show the density, pressure, magnitude of the velocity and magnitude of the magnetic field at the final time for the relativistic Orzag Tang problem. The problem was run with a fourth order ADER-WENO scheme with the MuSIC Riemann solver on a 1000×1000 zone mesh to a final time of 0.8.*

## IX.8) RMHD Flow: Long-Term Decay of Relativistic Alfven Waves with ADER-WENO Scheme

Turbulence in non-relativistic and relativistic plasmas is currently one of the hot topics in astrophysics. We know that the turbulence in magnetized plasmas is Alfvenic; i.e., the propagation and interaction of Alfven waves gives rise to turbulence. In order for RMHD turbulence to be correctly represented, we need to ensure that isolated, torsional Alfven waves can propagate with minimal numerical dissipation on a computational mesh. The RMHD wave families can propagate at 45° to the mesh lines with minimum dissipation. It is much more difficult to achieve good propagation of waves that are required to propagate at a small angle to one of the mesh lines (Balsara 2004).

We construct an RMHD version of a test problem that examines the dissipation of torsional Alfven waves when they propagate at a small angle to the mesh. See Balsara (2004) for non-relativistic test. We use a uniform $120 \times 120$ zone mesh that spans $[-3,3] \times [-3,3]$ in the xy-plane. An uniform density, $\rho_0 = 1$, and pressure, $P_0 = 1$, are initialized on the mesh. The unperturbed velocity is $v_0 = 0$, and the unperturbed magnetic field is $B_0 = 0.5$. A polytropic index of $\Gamma = 4/3$ is used in this simulation. The amplitude of the Alfven wave fluctuation ($B_1$) can be parameterized in terms of the velocity fluctuation, which has a value of 0.1 in this problem. The Alfven wave is designed to propagate along the wave vector, $\mathbf{k} = k_x \hat{x} + k_y \hat{y}$, where $k_x = 1/6$, $k_y = 1$. The velocity and magnetic field are given as follows:

$$\mathbf{v} = v_1 n_y \cos\phi \hat{x} - v_1 n_x \cos\phi \hat{y} + v_1 \sin\phi \hat{z},$$

$$\mathbf{B} = [B_0 n_x + B_1 n_y \cos\phi]\hat{x} + [B_0 n_y - B_1 n_x \cos\phi]\hat{y} + B_1 \sin\phi \hat{z}.$$

Here, the unit vector, $\mathbf{n} = n_x \hat{x} + n_y \hat{y} = (k_x \hat{x} + k_y \hat{y})/\sqrt{k_x^2 + k_y^2}$, the phase of the wave at initial time, $\phi = 2\pi(k_x x + k_y y)$, and the perturbation amplitude of the magnetic field is given by $B_1 = v_1 \sqrt{\rho_0 + \frac{\Gamma}{\Gamma-1}P_0 + B_0^2}$. The corresponding vector potential for the magnetic field is given by

$$\mathbf{A} = \frac{B_1}{2\pi k_x} \cos\phi \hat{y} + \left[ B_0(n_x y - n_y x) - \frac{B_1}{2\pi\sqrt{k_x^2 + k_y^2}} \sin\phi \right] \hat{z}.$$

The entire simulation is run to a time of $t = 130$ by which time the Alfven waves have crossed the computational domain five times. A CFL of 0.4 is used.

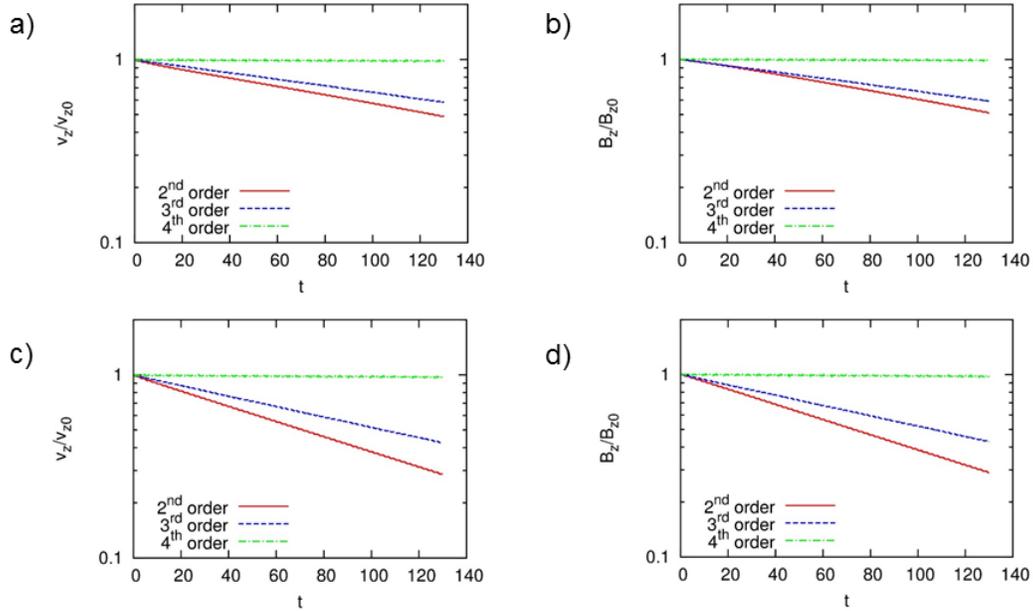

*Fig. 27 is from Balsara and Nkonga (2017). Figs. 27a and 27b show the evolution of the maximum z-velocity and maximum z-component of the magnetic field in the relativistic torsional Alfven wave as a function of time. For the simulations shown in Figs. 27a and 27b we used the 1D HLLI Riemann solver along with the 2D MuSIC Riemann solver with sub-structure. Figs. 27c and 27d show the same information as Figs. 27a and 27b, the only difference being that we used the 1D HLL Riemann solver along with the 2D MuSIC Riemann solver without sub-structure.*

Figs. 27a and 27b show the evolution of the maximum z-velocity and maximum z-component of the magnetic field in the relativistic torsional Alfven wave as a function of time. For the simulations shown in Figs. 27a and 27b we used the 1D HLLI Riemann solver along with the 2D MuSIC Riemann solver with sub-structure. Figs. 27c and 27d show the same information as Figs. 27a and 27b, the only difference being that we used the 1D HLL Riemann solver along with the 2D MuSIC Riemann solver without sub-structure. Comparing the two sets of figures, we see that the inferior Riemann solvers again show substantially larger dissipation at second and third orders. It is only at fourth order that we find a much-reduced difference between a Riemann solver with sub-structure and a Riemann solver that does not resolve any intermediate waves. As before, notice that the second order scheme with superior Riemann solvers is less dissipative than the third order scheme with inferior Riemann solvers! We, therefore, conclude that a Riemann solver that resolves intermediate waves is very important for reducing dissipation in second and third order schemes. At fourth and higher orders, that importance is diminished. The incremental costs of including sub-structure in a Riemann solver are only slight, making it advantageous to improve the quality of all schemes at all orders.

**X) Conclusion**

There is a great need for precision in computational astrophysics. The greater computational astrophysics community has roused itself into producing some very good methods for the PDE systems that are of interest in astrophysics, cosmology and numerical relativity. This review seeks to bring together the astrophysics community and the computational physics community, showing that great strides of progress can be made by the inter-diffusion of knowledge.

At second order, we have seen the value of TVD reconstruction. PPM schemes incorporate many aspects of TVD reconstruction while aiming for higher orders in the reconstructing polynomials. WENO schemes give us a method for carrying out reconstruction at successively higher orders. It is important to realize though that order of accuracy is not all-important. The ability to maintain other physical principles, such as positivity of density and pressure, also play an important role in the design of numerical schemes. It is also valuable to realize that reconstructing all the moments is not the only pathway to higher order. RKDG, HWENO and PNPM schemes offer us methods for retaining higher moments and evolving them in time. For problems with relatively smooth flows over the entire computational domain, such methods can provide a significant advantage over schemes that resort to a complete reconstruction of all the moments at each and every timestep.

Higher order spatial reconstruction should indeed be matched with higher order time evolution. Such a balanced accuracy in spatial and temporal accuracy is most desirable since a diminished time accuracy certainly results in a decreased overall accuracy of the numerical scheme. We have displayed two competing methodologies in time accurate simulation – Runge-Kutta timestepping and ADER timestepping. The former has the advantage of simplicity in programming, even if it requires extra stages at orders beyond third order. Reasonably simple formulations of the ADER timestepping have also become commonplace and they do offer the advantage of increased code speed.

The methods presented here are all based on finite volume formulations. If the computational emphasis is on uniform, structured mesh simulations, finite difference formulations may well offer a speed advantage. However, the finite volume formulations presented in this review are more versatile. They take well to complex geometries and extend seamlessly to unstructured meshes. ALE meshes, where the boundaries of the mesh can move, are also treated successfully by these methods. They can be used as base-level algorithms for adaptive mesh refinement calculations. They are quite fast and parallelize well. There is a rich literature and a wealth of practical experience associated with these methods. Their pitfalls, when they exist, are well-documented in the literature along with possible remedies. This makes them reliable workhorses for practical computation. The examples provided in this review have illustrated their excellent performance on a range interesting problems.

**Acknowledgements**

DSB acknowledges support via NSF grants NSF-DMS-1361197, NSF-ACI-1533850, NSF-DMS-1622457. Several simulations were performed on a cluster at UND that is run by the Center for Research Computing. Computer support on NSF's XSEDE and Blue Waters computing resources is also acknowledged. It is a pleasure to acknowledge Michael Dumbser, Vladimir Florinski, Sudip Garain, Katie Gurski, Roger Käppeli, Jinho Kim, Jiequan Li, Gino Montecinos, Boniface Nkonga, Chi-Wang Shu and Zhiliang Xu for help with the figures and also for numerous helpful discussions. The continued encouragement by Luciano Rezzolla was very helpful in completing this review, and it is also gratefully acknowledged.

S. Gottlieb, *On higher order strong stability preserving Runge-Kutta and multistep time discretizations*, Journal of Scientific Computing, 25 (1/2) (2005) 105

S. Gottlieb, C.-W. Shu, E. Tadmor, *Strong stability-preserving higher order time discretization methods*, SIAM Review, 43(1) (2001) 89-112

S. Gottlieb, D. Ketcheson and C.-W. Shu, *Strong Stability Preserving Runge-Kutta and Multistep Time Discretizations*, World Scientific (2011) (ISBN-10: 9814289264)

K.F. Gurski, *An HLLC-type approximate Riemann solver for ideal magnetohydrodynamics*, SIAM J. Sci. Comput. 25 (2004) 2165

Harten, A., *The artificial compression method for computation of shocks and contact discontinuities. I. Single conservation laws*, Communications in Pure and Applied Mathematics 30 (1977) 611

Harten, A., *High resolution schemes for conservation laws*, Journal of Computational Physics, 49 (1983) 357-393

Harten, A. Lax, P.D. and van Leer, B, *On upstream differencing and Godunov-type schemes for hyperbolic conservation laws*, SIAM Review, 25 (1983) 289-315

Harten, A., Engquist, B., Osher, S. and Chakravarthy, S., *Uniformly high order essentially non-oscillatory schemes III*, Journal of Computational Physics, 71 (1987) 231-303

Harten, A., *ENO Schemes with subcell resolution*, Journal of Computational Physics, 83 (1989) 148

Hawley, J. F.; Smarr, L. L.; Wilson, J. R., *A numerical study of nonspherical black hole accretion. I Equations and test problems*, Astrophysical Journal, 277 (1984) 296

Henrick, A.K., Aslam, T.D. and Powers, J.M., *Mapped weighted essentially non-oscillatoriy schemes: Achieving optimal order near critical points*, Journal of Computational Physics, 207 (2006) 542-567

Hesthaven J. and Warburton, T., *Nodal Discontinuous Galerkin Methods: Algorithms, Analysis and Applications*, Springer, Berlin, New York (2008)

Hirsch, C., *Numerical computation of internal and external flows, vol. I: Fundamentals of Numerical Discretization*, Wiley (1988)

Honkkila, V. and Janhunen, P., *HLLC solver for relativistic MHD*, J. Comput. Phys., 223 (2007) 643-656

## Appendix A) The Eigenstructure of the Euler Equations

The three dimensional Euler equations in Cartesian geometry can be written in conservation form

$$\partial_t U + \partial_x F(U) + \partial_y G(U) + \partial_z H(U) = 0 \tag{A.1}$$

as

$$\frac{\partial}{\partial t}\begin{pmatrix} \rho \\ \rho v_x \\ \rho v_y \\ \rho v_z \\ \mathcal{E} \end{pmatrix} + \frac{\partial}{\partial x}\begin{pmatrix} \rho v_x \\ \rho v_x^2 + P \\ \rho v_x v_y \\ \rho v_x v_z \\ (\mathcal{E}+P)v_x \end{pmatrix} + \frac{\partial}{\partial y}\begin{pmatrix} \rho v_y \\ \rho v_x v_y \\ \rho v_y^2 + P \\ \rho v_y v_z \\ (\mathcal{E}+P)v_y \end{pmatrix} + \frac{\partial}{\partial z}\begin{pmatrix} \rho v_z \\ \rho v_x v_z \\ \rho v_y v_z \\ \rho v_z^2 + P \\ (\mathcal{E}+P)v_z \end{pmatrix} = 0 \tag{A.2}$$

Here $\rho$ is the density, $v_x$, $v_y$ and $v_z$ are the three velocity components, $\mathcal{E}$ is the total energy density and "P" is the pressure. To solve the equations we have to assume an equation of state and we use the simplest polytropic equation of state here which allows us to write

$$\mathcal{E} = e + \frac{1}{2}\rho \mathbf{v}^2 \quad \text{with} \quad e \equiv \frac{P}{\Gamma - 1} \tag{A.3}$$

where "e" is the thermal energy density and $\Gamma$ is the polytropic index. To study the eigenstructure, we can consider one-dimensional variations. As a result, we suppress the $y$ and $z$-variations in eqn. (A.2). The equation with $x$-directional variation can be written in characteristic form

$$\partial_t U + A\, \partial_x U = 0 \tag{A.4}$$

To write the above equation in characteristic form, we need the Jacobian matrix for the flux, in other words we need $A \equiv \partial F(U)/\partial U$. The Jacobian matrix allows us to rewrite the above equation as

$$\frac{\partial}{\partial t}\begin{pmatrix} \rho \\ \rho v_x \\ \rho v_y \\ \rho v_z \\ \mathcal{E} \end{pmatrix} + \begin{pmatrix} 0 & 1 & 0 & 0 & 0 \\ -v_x^2 + \frac{(\Gamma-1)}{2}\mathbf{v}^2 & 2v_x - (\Gamma-1)v_x & -(\Gamma-1)v_y & -(\Gamma-1)v_z & (\Gamma-1) \\ -v_x v_y & v_y & v_x & 0 & 0 \\ -v_x v_z & v_z & 0 & v_x & 0 \\ -v_x H + \frac{(\Gamma-1)}{2}v_x \mathbf{v}^2 & H - (\Gamma-1)v_x^2 & -(\Gamma-1)v_x v_y & -(\Gamma-1)v_x v_z & \Gamma v_x \end{pmatrix} \frac{\partial}{\partial x}\begin{pmatrix} \rho \\ \rho v_x \\ \rho v_y \\ \rho v_z \\ \mathcal{E} \end{pmatrix} = 0$$

$$\tag{A.5}$$

where $\mathbf{v}^2 = v_x^2 + v_y^2 + v_z^2$. In eqn. (A.5) we define the total enthalpy "H" by

$$\rho H \equiv e + P + \frac{1}{2}\rho \mathbf{v}^2 \quad \Leftrightarrow \quad P = \frac{(\Gamma-1)}{\Gamma}\rho\left[H - \frac{1}{2}\mathbf{v}^2\right] \tag{A.6}$$

where the second equation in eqn. (A.6) assumes a polytropic gas. Eqn. (A.5) is still rather complicated and the best way to simplify it is to write it in terms of primitive variables, i.e. the density, velocity components and the pressure. The update equations for the primitive variables are usually not in conservation form, but they do make the system easier to analyze. Thus we define our vector of primitive variables as

$$V \equiv \begin{pmatrix} \rho & v_x & v_y & v_z & P \end{pmatrix}^T \tag{A.7}$$

Eqn. (A.4) can then be written as

$$\partial_t V + A_p\, \partial_x V = 0 \quad \text{where} \quad A_p \equiv \frac{\partial V}{\partial U} A \frac{\partial U}{\partial V} \tag{A.8}$$

Notice that $\partial U/\partial V$ and $\partial V/\partial U$ in eqn. (A.8) are Jacobian matrices that permit us to transform from the vector of primitive variables to the vector of conserved variables and vice versa. For the purposes of this section, a subscript of "p" applied to any matrix or eigenvector will denote that it pertains to a primitive variable. For the Euler equations we get

$$\frac{\partial}{\partial t}\begin{pmatrix} \rho \\ v_x \\ v_y \\ v_z \\ P \end{pmatrix} + \begin{pmatrix} v_x & \rho & 0 & 0 & 0 \\ 0 & v_x & 0 & 0 & 1/\rho \\ 0 & 0 & v_x & 0 & 0 \\ 0 & 0 & 0 & v_x & 0 \\ 0 & \Gamma P & 0 & 0 & v_x \end{pmatrix} \frac{\partial}{\partial x}\begin{pmatrix} \rho \\ v_x \\ v_y \\ v_z \\ P \end{pmatrix} = 0 \tag{A.9}$$

with the Jacobian matrices given by

$$\frac{\partial U}{\partial V} = \begin{pmatrix} 1 & 0 & 0 & 0 & 0 \\ v_x & \rho & 0 & 0 & 0 \\ v_y & 0 & \rho & 0 & 0 \\ v_z & 0 & 0 & \rho & 0 \\ \mathbf{v}^2/2 & \rho v_x & \rho v_y & \rho v_z & 1/(\Gamma-1) \end{pmatrix} \text{ and}$$

$$\frac{\partial V}{\partial U} = \begin{pmatrix} 1 & 0 & 0 & 0 & 0 \\ -v_x/\rho & 1/\rho & 0 & 0 & 0 \\ -v_y/\rho & 0 & 1/\rho & 0 & 0 \\ -v_z/\rho & 0 & 0 & 1/\rho & 0 \\ (\Gamma-1)\mathbf{v}^2/2 & -(\Gamma-1)v_x & -(\Gamma-1)v_y & -(\Gamma-1)v_z & (\Gamma-1) \end{pmatrix}$$

(A.10)

Comparing eqns. (A.9) to (A.5) clearly shows that it is much easier to obtain the eigenvalues and eigenvectors using the primitive variables. The eigenvectors for the primitive variables can subsequently be transformed back to their conserved counterparts using the transformation matrices in eqn. (A.10). I.e. if $r_p$ is a right eigenvector in primitive variables then its counterpart in terms of the conserved variables is easily obtained by $(\partial U/\partial V)r_p$. Similarly, if $l_p$ is a left eigenvector in the primitive variables then its counterpart in terms of conserved variables is given by $l_p(\partial V/\partial U)$.

The eigenvalues are easily found and are given by the ordered set

$$\{v_x - c_s, \ v_x, \ v_x, \ v_x, \ v_x + c_s\} \quad \text{where } c_s \equiv \sqrt{\frac{\Gamma P}{\rho}} \tag{A.11}$$

The matrix of right eigenvectors in the primitive variables is then given by

$$R_p = \begin{pmatrix} 1 & 1 & 0 & 0 & 1 \\ -c_s/\rho & 0 & 0 & 0 & c_s/\rho \\ 0 & 0 & 1 & 0 & 0 \\ 0 & 0 & 0 & 1 & 0 \\ c_s^2 & 0 & 0 & 0 & c_s^2 \end{pmatrix} \tag{A.12}$$

The first and fifth columns of $R_p$ in eqn. (A.12) give us eigenvectors for left-going and right-going sound waves. The sound waves are genuinely non-linear and can self-steepen as they propagate. The eigenvectors tell us that if a wave is to be a sound wave then the fluctuation in its density, x-velocity and pressure should be proportional to the components of the corresponding eigenvector. The remaining eigenvectors are linearly degenerate. The second column of $R_p$ corresponds to an entropy wave and tells us that an entropy pulse consists of a change in density while the x-velocity and pressure remain unchanged. The third and fourth columns of $R_p$ correspond to shear waves with a shear in the y and z-components of velocity. The matrix of left eigenvectors is now given by

$$L_p = \begin{pmatrix} 0 & -\rho/(2c_s) & 0 & 0 & 1/(2c_s^2) \\ 1 & 0 & 0 & 0 & -1/c_s^2 \\ 0 & 0 & 1 & 0 & 0 \\ 0 & 0 & 0 & 1 & 0 \\ 0 & \rho/(2c_s) & 0 & 0 & 1/(2c_s^2) \end{pmatrix} \tag{A.13}$$

The rows of $L_p$ give us the left eigenvectors and we see that they are arranged in the same sequence as the columns of $R_p$ in eqn. (A.12). Thus the first and fifth rows of eqn. (A.13) correspond to left and right-going sound waves respectively. The second row of eqn. (A.13) corresponds to the entropy wave and the third and fourth rows correspond to shear waves in the y and z-velocities. It is also easy to verify that the left and right eigenvectors are orthonormal, i.e. $L_p R_p = I$ where "$I$" is the identity matrix. This property is very useful when projecting a solution into its characteristic variables as was already seen in Section 3.4. While eqns. (A.12) and (A.13) give us the eigenvectors in the space of primitive variables, the transformation matrices in eqn. (A.10) can be used to obtain the eigenvectors in the space of conserved variables.

**Appendix B) The Eigenstructure of the Non-Relativistic MHD**

The three dimensional MHD equations in Cartesian geometry can be written in conservation form as

$$\frac{\partial}{\partial t}\begin{pmatrix}\rho \\ \rho v_x \\ \rho v_y \\ \rho v_z \\ \varepsilon \\ B_x \\ B_y \\ B_z\end{pmatrix} + \frac{\partial}{\partial x}\begin{pmatrix}\rho v_x \\ \rho v_x^2 + P + \mathbf{B}^2/8\pi - B_x^2/4\pi \\ \rho v_x v_y - B_x B_y/4\pi \\ \rho v_x v_z - B_x B_z/4\pi \\ (\varepsilon + P + \mathbf{B}^2/8\pi)v_x - B_x(\mathbf{v}\cdot\mathbf{B})/4\pi \\ 0 \\ (v_x B_y - v_y B_x) \\ -(v_z B_x - v_x B_z)\end{pmatrix}$$

$$+ \frac{\partial}{\partial y}\begin{pmatrix}\rho v_y \\ \rho v_x v_y - B_x B_y/4\pi \\ \rho v_y^2 + P + \mathbf{B}^2/8\pi - B_y^2/4\pi \\ \rho v_y v_z - B_y B_z/4\pi \\ (\varepsilon + P + \mathbf{B}^2/8\pi)v_y - B_y(\mathbf{v}\cdot\mathbf{B})/4\pi \\ -(v_x B_y - v_y B_x) \\ 0 \\ (v_y B_z - v_z B_y)\end{pmatrix} + \frac{\partial}{\partial z}\begin{pmatrix}\rho v_z \\ \rho v_x v_z - B_x B_z/4\pi \\ \rho v_y v_z - B_y B_z/4\pi \\ \rho v_z^2 + P + \mathbf{B}^2/8\pi - B_z^2/4\pi \\ (\varepsilon + P + \mathbf{B}^2/8\pi)v_z - B_z(\mathbf{v}\cdot\mathbf{B})/4\pi \\ (v_z B_x - v_x B_z) \\ -(v_y B_z - v_z B_y) \\ 0\end{pmatrix} = 0$$

(B.1)

Here $\rho$ is the density, $v_x$, $v_y$ and $v_z$ are the three velocity components, $B_x$, $B_y$ and $B_z$ are the three magnetic field components, $\varepsilon$ is the total energy density and "P" is the pressure. The equations are written in CGS units. We also assume a polytropic equation of state for the thermal energy "e" with polytropic index $\Gamma$ so that we get

$$\varepsilon = e + \frac{1}{2}\rho \mathbf{v}^2 + \frac{\mathbf{B}^2}{8\pi} \quad \text{with} \quad e \equiv \frac{P}{(\Gamma - 1)} \tag{B.2}$$

The presence of a magnetic field makes the total pressure anisotropic. The magnetic fields can also exert tensional forces parallel to the field lines in the dynamical equations. Just as in eqn. (A.2), the first five rows of eqn. (B.1) express mass, momentum and energy conservation with the Lorenz force terms contributing to the momentum and energy fluxes. The induction equation is given by

$$\frac{\partial \mathbf{B}}{\partial t} + c\, \nabla \times \mathbf{E} = 0 \tag{B.3}$$

The electric field vector, $\mathbf{E}$, is defined in the ideal MHD limit by

$$\mathbf{E} = -\frac{1}{c}\mathbf{v} \times \mathbf{B} \tag{B.4}$$

Notice that the last three rows of eqn. (B.1) actually recast the induction equation in conservation form. This enables us to take all of the higher order Godunov scheme machinery that we have developed for hydrodynamics and reuse it for the solution of the MHD equations. With the constraint $\nabla \cdot \mathbf{B} = 0$ enforced at the start of a calculation, eqn. (B.3) shows that it should remain so throughout the calculation. Several early authors, Brackbill and Barnes (1980) and Brackbill (1985) have shown that violating the $\nabla \cdot \mathbf{B} = 0$ constraint leads to unphysical plasma transport orthogonal to the magnetic field. Yee (1966), Brecht *et al.* (1981) Evans and Hawley (1989) and DeVore (1991) showed the utility of satisfying this constraint at a discrete level in a numerical code. We will explore this issue further in the context of higher order Godunov schemes in a later review. For now, it is important to point out that for one-dimensional variations the divergence-free constraint also implies that the magnetic field in that direction is a constant. In other words, for situations where the entire variation in the flow variables is along the *x*-axis, the *x*-component of the magnetic field must remain a constant. For that reason, when we consider *x*-directional variations of the MHD equation we will assume that $B_x$ is a constant.

As with the Euler equations, we restrict the variations in eqn. (B.1) to the *x*-direction. With that restriction, $B_x$ ceases to have a variation along the *x*-direction. Consequently, for this section and the next, we can drop it from the vector of conserved variables and the flux vector in the *x*-direction. We then arrive at a seven component vector of conserved variables. The 7×7 characteristic matrix A, see eqns. (A.4) and (A.5), can be written as

$$A = \begin{pmatrix} 0 & 1 & 0 & 0 & 0 & 0 & 0 \\ -v_x^2 + \frac{(\Gamma-1)}{2}\mathbf{v}^2 & 2v_x - (\Gamma-1)v_x & -(\Gamma-1)v_y & -(\Gamma-1)v_z & (\Gamma-1) & (2-\Gamma)\frac{B_y}{4\pi} & (2-\Gamma)\frac{B_z}{4\pi} \\ -v_x v_y & v_y & v_x & 0 & 0 & -\frac{B_x}{4\pi} & 0 \\ -v_x v_z & v_z & 0 & v_x & 0 & 0 & -\frac{B_x}{4\pi} \\ \delta_{51} & \delta_{52} & \delta_{53} & \delta_{54} & \delta_{55} & \delta_{56} & \delta_{57} \\ -\frac{1}{\rho}(B_y v_x - B_x v_y) & \frac{B_y}{\rho} & -\frac{B_x}{\rho} & 0 & 0 & v_x & 0 \\ -\frac{1}{\rho}(B_z v_x - B_x v_z) & \frac{B_z}{\rho} & 0 & -\frac{B_x}{\rho} & 0 & 0 & v_x \end{pmatrix}$$

(B.5)

with

$$\delta_{51} = v_x \left( -H + \frac{(\Gamma-1)}{2} \mathbf{v}^2 \right) + \frac{B_x}{4\pi\rho}(\mathbf{v}\cdot\mathbf{B}) \ ; \ \ \delta_{52} = H - (\Gamma-1)v_x^2 - \frac{B_x^2}{4\pi\rho} \ ;$$

$$\delta_{53} = -(\Gamma-1)v_x v_y - \frac{B_x B_y}{4\pi\rho} \ ; \ \ \delta_{54} = -(\Gamma-1)v_x v_z - \frac{B_x B_z}{4\pi\rho} \ ; \ \ \delta_{55} = \Gamma v_x \ ; \quad (B.6)$$

$$\delta_{56} = (2-\Gamma)v_x \frac{B_y}{4\pi} - v_y \frac{B_x}{4\pi} \ ; \ \ \delta_{57} = (2-\Gamma)v_x \frac{B_z}{4\pi} - v_z \frac{B_x}{4\pi}$$

The total enthalpy for MHD flows can be written as

$$\rho H \equiv \mathcal{E} + P + \frac{\mathbf{B}^2}{8\pi} \ \Leftrightarrow \ P = \frac{(\Gamma-1)}{\Gamma}\left[\rho H - \frac{1}{2}\rho \mathbf{v}^2 - \frac{\mathbf{B}^2}{4\pi}\right] \quad (B.7)$$

The reader should compare the above equation to eqn. (A.6) for the hydrodynamical case.

As with the Euler equations, eqn. (B.5) for the conserved variables is quite complicated and the easiest simplifications occur when it is written in terms of the primitive variables. The vector of primitive variables is given by

$$V \equiv \begin{pmatrix} \rho & v_x & v_y & v_z & P & B_y & B_z \end{pmatrix}^T \quad (B.8)$$

The characteristic matrix for the hyperbolic system $\partial_t V + A_p \partial_x V = 0$ can now be written in primitive variables, see eqn. (A.8), as

$$A_p = \begin{pmatrix} v_x & \rho & 0 & 0 & 0 & 0 & 0 \\ 0 & v_x & 0 & 0 & \frac{1}{\rho} & \frac{B_y}{4\pi\rho} & \frac{B_z}{4\pi\rho} \\ 0 & 0 & v_x & 0 & 0 & -\frac{B_x}{4\pi\rho} & 0 \\ 0 & 0 & 0 & v_x & 0 & 0 & -\frac{B_x}{4\pi\rho} \\ 0 & \rho c_s^2 & 0 & 0 & v_x & 0 & 0 \\ 0 & B_y & -B_x & 0 & 0 & v_x & 0 \\ 0 & B_z & 0 & -B_x & 0 & 0 & v_x \end{pmatrix} \quad (B.9)$$

The large number of zeros in eqn. (B.9) clearly shows that it is easier to work with when finding eigenvectors. The Jacobian matrices $\partial U/\partial V$ and $\partial V/\partial U$ that allow us to transform from the vector of primitive variables to the vector of conserved variables are now given by

$$\frac{\partial U}{\partial V} = \begin{pmatrix} 1 & 0 & 0 & 0 & 0 & 0 & 0 \\ v_x & \rho & 0 & 0 & 0 & 0 & 0 \\ v_y & 0 & \rho & 0 & 0 & 0 & 0 \\ v_z & 0 & 0 & \rho & 0 & 0 & 0 \\ \frac{\mathbf{v}^2}{2} & \rho v_x & \rho v_y & \rho v_z & \frac{1}{(\Gamma-1)} & \frac{B_y}{4\pi} & \frac{B_z}{4\pi} \\ 0 & 0 & 0 & 0 & 0 & 1 & 0 \\ 0 & 0 & 0 & 0 & 0 & 0 & 1 \end{pmatrix}$$

$$\frac{\partial V}{\partial U} = \begin{pmatrix} 1 & 0 & 0 & 0 & 0 & 0 & 0 \\ -\frac{v_x}{\rho} & \frac{1}{\rho} & 0 & 0 & 0 & 0 & 0 \\ -\frac{v_y}{\rho} & 0 & \frac{1}{\rho} & 0 & 0 & 0 & 0 \\ -\frac{v_z}{\rho} & 0 & 0 & \frac{1}{\rho} & 0 & 0 & 0 \\ (\Gamma-1)\frac{\mathbf{v}^2}{2} & -(\Gamma-1)v_x & -(\Gamma-1)v_y & -(\Gamma-1)v_z & (\Gamma-1) & -(\Gamma-1)\frac{B_y}{4\pi} & -(\Gamma-1)\frac{B_z}{4\pi} \\ 0 & 0 & 0 & 0 & 0 & 1 & 0 \\ 0 & 0 & 0 & 0 & 0 & 0 & 1 \end{pmatrix}$$

(B.10)

We have now built up all the requisite matrices for evaluating the eigenvalues and eigenvectors of the MHD system and we take that task up next.

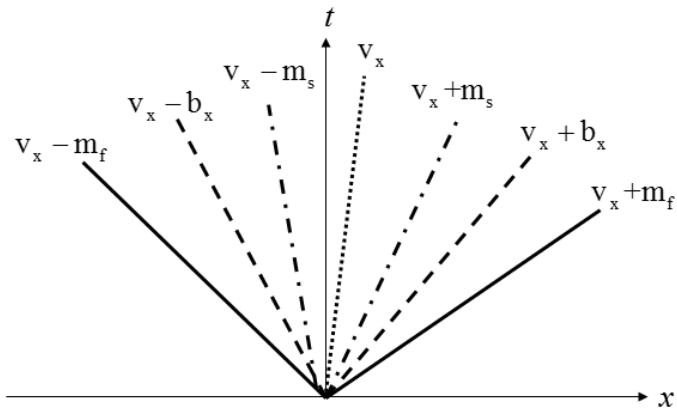

*Fig. B.1 shows the seven waves for the MHD system. They foliate symmetrically about the fluid's x-velocity. The fast waves are shown by solid lines. The Alfven waves are shown by dashed lines. The slow waves are shown by the dot-dashed lines. The entropy wave is shown by the dotted line.*

We now define the Alfvenic speeds in each of the coordinate directions as

$$b_x \equiv \frac{B_x}{\sqrt{4\pi\rho}} \quad ; \quad b_y \equiv \frac{B_y}{\sqrt{4\pi\rho}} \quad ; \quad b_z \equiv \frac{B_z}{\sqrt{4\pi\rho}} \quad ; \quad b \equiv \sqrt{b_x^2 + b_y^2 + b_z^2} \quad ; \quad b_\perp \equiv \sqrt{b_y^2 + b_z^2}$$
(B.11)

The ordered set of eigenvalues is given by

$$\{v_x - m_f, \quad v_x - b_x, \quad v_x - m_s, \quad v_x, \quad v_x + m_s, \quad v_x + b_x, \quad v_x + m_f\}$$
(B.12)

Here $m_f$ and $m_s$ are the speeds of the *fast and slow magnetosonic waves* relative to the fluid's advection speed, $v_x$. They are obtained by solving the quartic

$$m^4 - \left(c_s^2 + b^2\right)m^2 + c_s^2 b_x^2 = 0$$
(B.13)

where the sound speed $c_s$ is defined as in eqn. (A.11). To maintain the ordering in eqn. (B.12) we pick the roots of the quartic with the additional requirement that $m_f \geq m_s \geq 0$. Eqn. (B.12) shows us that the MHD system has seven different waves that are placed symmetrically about the flow speed $v_x$, as shown in Fig. B.1 for the canonical case where $B_x \neq 0$. The MHD waves can all be defined by their speed relative to $v_x$, as can be seen from eqns. (B.12) and (B.13). Thus we have a pair of left and right-going fast waves that propagate with a speed $m_f$ relative to the x-velocity. Consistent with their name, the fast waves are the fastest leftward and rightward propagating waves. Next we have a pair of left and right-going Alfven waves that propagate with a speed $b_x$ relative to the x-velocity. We then have a pair of left and right-going slow waves that propagate with a speed $m_s$ relative to the x-velocity. Lastly, we have an entropy wave that propagates with the x-velocity. Notice that $m_s \leq b_x \leq m_f$ so that the eigenvalues do form an ordered set most of the times. However, it cannot be guaranteed that $m_s < b_x < m_f$ for all values of the primitive variables, with the result that the eigenvalues can become degenerate and the system is, therefore, not strictly hyperbolic. Based on Fig. B.1 we see that the seven waves divide space-time into eight regions. We, therefore, anticipate that the Riemann problem for MHD will do the same.

    The eigensystem for MHD has a very intricate wave structure which has been explored in depth by Jefferey and Taniuti (1964). The eigenvectors catalogued in Jefferey and Taniuti (1964) are prone to singularities, which makes it impossible to implement them as-is in a numerical code. Consequently, Brio and Wu (1988) and Roe and Balsara (1996) carried out a study of the MHD eigenvectors and formulated them in a manner that makes them computationally useful. Because we have already studied the eigensystem for the Euler equations in detail, we provide a qualitative introduction to the wave families in MHD by comparing and contrasting them with the wave families in the Euler system. The entropy wave is linearly degenerate as in the Euler case. However, when $B_x \neq 0$, a contact discontinuity cannot simultaneously have a jump in the transverse velocities. The shear in the transverse velocities that we expect from the Euler

equations is now carried by the two Alfven waves which are also linearly degenerate. However, in the canonical case with $B_x \neq 0$, each Alfven wave requires that a specific relationship hold between the variation in the transverse velocities and the variation in the transverse magnetic fields. This makes it possible for the MHD system to sustain finite amplitude torsional Alfven waves. (Though see Goldstein (1978), Jayanti and Hollweg (1993) and Del Zanna, Velli and Londrillo (2001) for a study of the stability properties of these torsional Alfven waves.) The fast and slow waves are genuinely non-linear and compressive, i.e. an increase in density in either of those wave families results in a corresponding increase in the pressure. The sound waves in the Euler system are similarly compressive. As a result, we expect fast and slow magnetosonic shocks to produce a simultaneous increase in density and pressure. The propagation of sound waves in the Euler equations is isotropic relative to the fluid velocity. The presence of a magnetic field breaks this isotropy. As a result, the propagation speeds for fast and slow magnetosonic waves do depend on the direction of the magnetic field. In the limit where the magnetic field smoothly goes to zero in the MHD system, the two fast magnetosonic waves go over to the two sound waves in the Euler system. In that same limit, the slow magnetosonic waves combine with the Alfven waves to produce the shear waves of the Euler system.

To study the anisotropic propagation of MHD waves further, let the magnetic field be aligned with the x-axis and let us examine the wave propagation in the rest frame of the fluid. Fig. B.2 shows the propagation speeds of the different families of waves relative to the magnetic field direction. The wave speeds shown are relative to the fluid velocity. The magnetic field is shown as the vector **B** and is aligned along the *x*-axis in this figure. The propagation direction for the waves is shown by the arrow that makes an angle θ with respect to the magnetic field direction. The distance of the curve associated with a given wave family from the origin in any direction θ gives the speed of that wave family. As in Fig. B.1, the solid, dashed and dot-dash curves in Fig. B.2 pertain to fast, Alfven and slow waves. Note that for a fixed value of the sound speed $c_s$ Figs. B.2a, B.2b and B.2c correspond to a sequence with increasing magnetic field, with the result that the fast magnetosonic wave speed also increases correspondingly. In the interest of showing Figs. B.2a, B.2b and B.2c clearly, they have been rescaled to have roughly the same size. We see that the fast wave always propagates so that it provides the outer bound on the wave speeds in all directions. Similarly, the slow wave propagates so that it provides the inner bound on the wave speeds in all directions. However, there are several situations when two, and even three, wave families propagate with the same speed. The case where $b/c_s < 1$ is shown in Fig. B.2a. We see that when $b_\perp = 0$ for this case, the Alfven waves and the fast waves become degenerate, i.e. they have the same wave speed. As a result, we should expect a degeneracy of the eigenvectors in that limit. Fig. B.2b shows the case with $b/c_s = 1$. When $b_\perp = 0$ for this case, we see that all the fast, slow and Alfven waves all become degenerate. As a result, this is also known as the triple umbilic case. This is also the limit in which the eigenvectors develop their worst singularities unless something special is done to cure the singularities. Fig. B.2c displays the case where $b/c_s > 1$, showing us again that when $b_\perp = 0$ the Alfven waves and the slow waves become degenerate. Figs. B.2a, B.2b and B.2c also show us that degeneracies arise between the Alfven waves and slow waves when the wave propagation is orthogonal to the direction of the magnetic field. Lastly, as **B** → 0 degeneracies can also be shown to arise between the Alfven waves and slow waves. The wave diagrams in Fig. B.2, therefore, give us a good understanding of the degeneracies in the eigenvalues that can prevail in the MHD system.

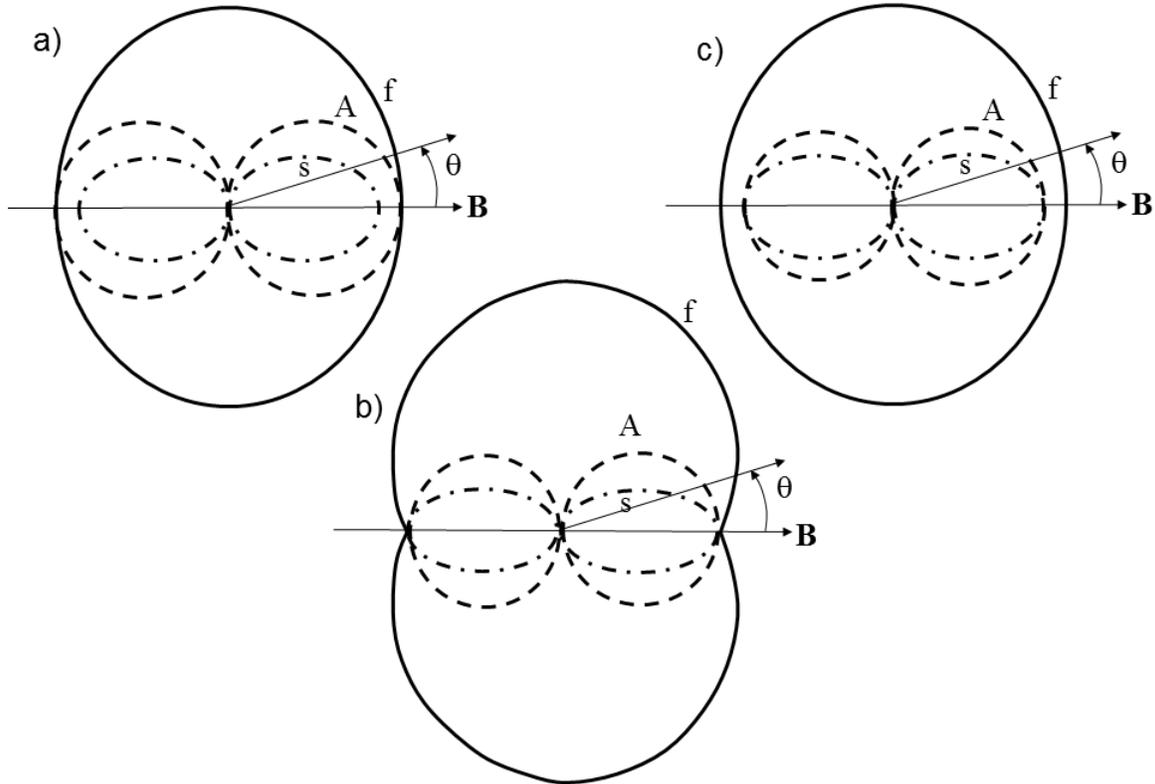

*Fig. B.2 showing surfaces of normal speeds for a) $b/c_s < 1$, b) $b/c_s = 1$, c) $b/c_s > 1$. The solid, dashed and dot-dashed curves show the fast, Alfven and slow wave speeds.*

      We have seen that the eigenvalues can become degenerate in certain limits. Consequently, the eigenvectors can become indeterminate in some of those limits. Early progress in the design of higher order Godunov schemes for numerical MHD (Brio and Wu 1988, Zachary, Malagoli and Colella 1992) had been impeded by the fact that the eigenvectors could indeed become indeterminate in some of the limits. In those limits, it therefore becomes impossible to project the solution into its constituent waves. The eigenvector degeneracy it difficult to carry out a characteristic reconstruction or to formulate a linearized Riemann solver. The eigenvector indeterminacy turned out to be a multiplicative one, i.e. by multiplying the left and right eigenvectors by suitable combinations of factors it is possible to arrive at eigenvectors that are orthonormal and retain saliency in all the limits where the original eigenvectors become indeterminate. A complete, orthonormal set of left and right eigenvectors that always retain saliency was given in Roe and Balsara (1996). The regularized, orthonormal eigenvectors for the 7×7 MHD system are reproduced below

$$R_p = \begin{pmatrix} \alpha_f\,\rho & 0 & \alpha_s\,\rho & 1 & \alpha_s\,\rho & 0 & \alpha_f\,\rho \\ -\alpha_f\,m_f & 0 & -\alpha_s\,m_s & 0 & \alpha_s\,m_s & 0 & \alpha_f\,m_f \\ \alpha_s\,m_s\,\beta_{y,s} & -\beta_z & -\alpha_f\,m_f\,\beta_{y,s} & 0 & \alpha_f\,m_f\,\beta_{y,s} & \beta_z & -\alpha_s\,m_s\,\beta_{y,s} \\ \alpha_s\,m_s\,\beta_{z,s} & \beta_y & -\alpha_f\,m_f\,\beta_{z,s} & 0 & \alpha_f\,m_f\,\beta_{z,s} & -\beta_y & -\alpha_s\,m_s\,\beta_{z,s} \\ \alpha_f\,\rho\,c_s^2 & 0 & \alpha_s\,\rho\,c_s^2 & 0 & \alpha_s\,\rho\,c_s^2 & 0 & \alpha_f\,\rho\,c_s^2 \\ \alpha_s\,\sqrt{4\pi\rho}\,c_s\,\beta_y & -\sqrt{4\pi\rho}\,\beta_{z,s} & -\alpha_f\,\sqrt{4\pi\rho}\,c_s\,\beta_y & 0 & -\alpha_f\,\sqrt{4\pi\rho}\,c_s\,\beta_y & -\sqrt{4\pi\rho}\,\beta_{z,s} & \alpha_s\,\sqrt{4\pi\rho}\,c_s\,\beta_y \\ \alpha_s\,\sqrt{4\pi\rho}\,c_s\,\beta_z & \sqrt{4\pi\rho}\,\beta_{y,s} & -\alpha_f\,\sqrt{4\pi\rho}\,c_s\,\beta_z & 0 & -\alpha_f\,\sqrt{4\pi\rho}\,c_s\,\beta_z & \sqrt{4\pi\rho}\,\beta_{y,s} & \alpha_s\,\sqrt{4\pi\rho}\,c_s\,\beta_z \end{pmatrix}$$
(B.14)

$$L_p = \begin{pmatrix} 0 & -\dfrac{\alpha_f\,m_f}{2\,c_s^2} & \dfrac{\alpha_s\,m_s\,\beta_{y,s}}{2\,c_s^2} & \dfrac{\alpha_s\,m_s\,\beta_{z,s}}{2\,c_s^2} & \dfrac{\alpha_f}{2\,c_s^2\,\rho} & \dfrac{\alpha_s\,\beta_y}{2\,c_s\,\sqrt{4\pi\rho}} & \dfrac{\alpha_s\,\beta_z}{2\,c_s\,\sqrt{4\pi\rho}} \\ 0 & 0 & -\dfrac{\beta_z}{2} & \dfrac{\beta_y}{2} & 0 & -\dfrac{\beta_{z,s}}{2\sqrt{4\pi\rho}} & \dfrac{\beta_{y,s}}{2\sqrt{4\pi\rho}} \\ 0 & -\dfrac{\alpha_s\,m_s}{2\,c_s^2} & -\dfrac{\alpha_f\,m_f\,\beta_{y,s}}{2\,c_s^2} & -\dfrac{\alpha_f\,m_f\,\beta_{z,s}}{2\,c_s^2} & \dfrac{\alpha_s}{2\,c_s^2\,\rho} & -\dfrac{\alpha_f\,\beta_y}{2\,c_s\,\sqrt{4\pi\rho}} & -\dfrac{\alpha_f\,\beta_z}{2\,c_s\,\sqrt{4\pi\rho}} \\ 1 & 0 & 0 & 0 & -\dfrac{1}{c_s^2} & 0 & 0 \\ 0 & \dfrac{\alpha_s\,m_s}{2\,c_s^2} & \dfrac{\alpha_f\,m_f\,\beta_{y,s}}{2\,c_s^2} & \dfrac{\alpha_f\,m_f\,\beta_{z,s}}{2\,c_s^2} & \dfrac{\alpha_s}{2\,c_s^2\,\rho} & -\dfrac{\alpha_f\,\beta_y}{2\,c_s\,\sqrt{4\pi\rho}} & -\dfrac{\alpha_f\,\beta_z}{2\,c_s\,\sqrt{4\pi\rho}} \\ 0 & 0 & \dfrac{\beta_z}{2} & -\dfrac{\beta_y}{2} & 0 & -\dfrac{\beta_{z,s}}{2\sqrt{4\pi\rho}} & \dfrac{\beta_{y,s}}{2\sqrt{4\pi\rho}} \\ 0 & \dfrac{\alpha_f\,m_f}{2\,c_s^2} & -\dfrac{\alpha_s\,m_s\,\beta_{y,s}}{2\,c_s^2} & -\dfrac{\alpha_s\,m_s\,\beta_{z,s}}{2\,c_s^2} & \dfrac{\alpha_f}{2\,c_s^2\,\rho} & \dfrac{\alpha_s\,\beta_y}{2\,c_s\,\sqrt{4\pi\rho}} & \dfrac{\alpha_s\,\beta_z}{2\,c_s\,\sqrt{4\pi\rho}} \end{pmatrix}$$
(B.15)

All that remains is to catalogue some of the coefficients in eqns. (B.14) and (B.15). This is done as

$$\beta_y = \frac{b_y}{b_\perp} \;;\; \beta_z = \frac{b_z}{b_\perp} \;;\; \beta_{y,s} = \beta_y\,\mathrm{sgn}(b_x) \;;\; \beta_{z,s} = \beta_z\,\mathrm{sgn}(b_x) \;;$$

$$\alpha_f = \sqrt{\frac{c_s^2 - m_s^2}{m_f^2 - m_s^2}} \;;\; \alpha_s = \sqrt{\frac{m_f^2 - c_s^2}{m_f^2 - m_s^2}}$$
(B.16)

The terms $\alpha_f$ and $\alpha_s$ in eqn. (B.16) are a measure of how closely the fast and slow waves approximate the behavior of sound waves. For example, if $\alpha_f \approx 1$ then the eigenvector for the fast waves behaves almost like the eigenvector for the sound waves.

There are a couple of limiting cases where the expressions in eqn. (B.16) need to be modified. In the first limiting case we have $b_\perp \to 0$ so that $\beta_y$ and $\beta_z$ need to be redefined. In the second case we simultaneously have $b \to c_s$ and $b_\perp \to 0$, which requires a modification of $\alpha_f$ and $\alpha_s$. In either of those two limits it helps to realize that the eigenvalues become degenerate so that these terms may take on different values depending on how the limits are approached. The way in which these limits are approached is not known a priori in a numerical code. Consequently, one must provide a numerical code with any one reasonable choice. All reasonable choices are acceptable as long as they yield a complete and non-singular eigenspace into which the solution and fluxes can be projected. Since we know that this is the case for eqns. (B.14) and (B.15), we provide the following choices. In the limit $b_\perp \to 0$ we use

$$\beta_y = \beta_z = \frac{1}{\sqrt{2}} \tag{B.17}$$

In the limit where we simultaneously have $b \to c_s$ and $b_\perp \to 0$, i.e. when we are very close to the triple umbilic point shown in Fig. B.2b, we set

$$\alpha_f = \sin\left(\frac{\phi}{2}\right) \quad ; \quad \alpha_s = \cos\left(\frac{\phi}{2}\right) \quad \text{with} \quad \tan(\phi) \equiv \frac{b_\perp}{|b_x| - c_s} \tag{B.18}$$

This completes our description of the MHD eigensystem.

**Appendix C) Relativistic Hydrodynamics and Magnetohydrodynamics**

The equations of relativistic hydrodynamics and MHD are used to model high speed flows. While some nuclear collisions have been modeled by the equations of relativistic hydrodynamics, most of the applications derive from high energy astrophysics. These equations are primarily used to model phenomena that take place at speeds approaching the speed of light. Such speeds are reached in astrophysical settings, especially when considering flows around neutron stars and black holes. As a result, special and general relativistic effects have to be considered. For all other situations, the regular Euler and MHD equations prove to be very serviceable. In studying this topic it is quite advantageous to arrive at it in gradual stages. For that reason, in this section we introduce the special relativistic form of the hydrodynamic and MHD equations. General relativistic effects, which incorporate the effects of a curved space-time, have been considered in some of the cited references.

The special relativistic hydrodynamic equations have been very nicely discussed in the text by Synge (1957) and the physics of relativistic shock waves arising from those equations have been nicely presented in Taub (1948). The first thing to realize about a parcel of fluid that is moving with a velocity **v** that is close to the speed of light "*c*" is that the parcel will experience length contraction when viewed in the frame of reference of a stationary observer, i.e. the *lab frame*. Thus in the lab frame, one considers the Lorentz contraction which is given by the *Lorentz factor* $\gamma \equiv 1/\sqrt{1 - \mathbf{v}^2/c^2}$. If the fluid has a density $\rho$ in its own rest frame, the rest frame density increases to a value of $\rho\gamma$ in the lab frame. The continuity equation is, therefore, an expression of the conservation of the total number of atoms and is given by

$$\frac{\partial}{\partial t}(\rho \gamma) + \frac{\partial}{\partial x_i}(\rho \gamma v_i) = 0 \qquad (C.1)$$

Fluids that are flowing at relativistic speeds can only be accelerated to these speeds by very energetic processes. As a result, they often have unusually large amounts of internal energy and pressure. That internal energy and pressure can also contribute to the fluid's inertia. As a result, we define the specific enthalpy as $h \equiv 1 + \Gamma P/c^2(\Gamma - 1)$ which provides a further multiplicative contribution from the fluid's internal energy to the rest mass. Here $\Gamma$ is the polytropic index of the gas, which is assumed to be ideal for the sake of simplicity. As a result, the fluid has $\rho h \gamma$ amount of mass density when viewed from the lab frame. The specific momentum of the fluid is given by $\gamma \mathbf{v}$. The momentum density of the fluid is then given by $\rho h \gamma^2 \mathbf{v}$ and the equation that describes its evolution can be written as

$$\frac{\partial}{\partial t}(\rho h \gamma^2 v_i) + \frac{\partial}{\partial x_j}(\rho h \gamma^2 v_i v_j + P \gamma \delta_{ij}) = 0 \qquad (C.2)$$

The energy density of the fluid includes just the contribution of the internal energy to the fluid's inertia and is therefore given by $\rho h \gamma^2 - P/c^2$. The equation for the energy density is then given by

$$\frac{\partial}{\partial t}(\rho h \gamma^2 - P/c^2) + \frac{\partial}{\partial x_i}(\rho h \gamma^2 v_i) = 0 \qquad (C.3)$$

Synge (1957) provides an extensive derivation of the relativistic continuity equation as well as the relativistic momentum and energy equations. Pons *et al.* (1998) have shown a very interesting connection between general and special relativistic hydrodynamics based on analyzing locally flat space-times. Aloy *et al.* (1999) have provided an extensive review of numerical methods for special and general relativistic hydrodynamics.

The parallels between eqns. (C.1) to (C.3) and the Euler equations are easy to spot. Setting $\gamma = h = 1$ for the non-relativistic limit in eqns. (C.1) and (C.2) then gives back the continuity and momentum equations for Euler flow. Reducing eqn. (C.3) to yield the energy equation for Euler flow is a little more subtle, because the rest mass of a particle contributes to the energy density when considering relativistic flows whereas that energy can be cleanly subtracted away for non-relativistic flows. The relativistic flow equations also form a hyperbolic set of equations and have the same foliation of waves as the Euler equations. While there are many parallels between the Euler equations and their relativistic extensions, there are two prominent points of difference. First, while it is quite easy to obtain the primitive variables from the conserved variables for Euler flow, doing so for relativistic flow involves solving a transcendental equation. Second, carrying out the eigenmodal analysis for relativistic flow is a lot harder. These two attributes, which make the relativistic flow equations harder to work with, also carry over to RMHD.

The text by Anile (1989) provides an excellent introduction to RMHD. Several excellent formulations for general relativistic MHD have recently been presented in the literature, see Komissarov (2004), McKinney (2006) and DelZanna et al. (2007). General relativists usually use a set of geometrized units where $G=c=1$ and we use those units here in describing the equations of RMHD. Here G is Newton's constant and $c$ is the speed of light. The factor of $4\pi$ that we met in classical MHD is also absorbed via a redefinition of the magnetic field. All the same considerations that we made for relativistic hydrodynamics also have to be made here, with the result that the continuity equation is identical to eqn. (C.1). The introduction of a magnetic field **B** also introduces a motional emf, thus resulting in an electric field in the plasma which is given by $\mathbf{E} = -\mathbf{v} \times \mathbf{B}$ even in the relativistic limit. The *Poynting flux* $\mathbf{E} \times \mathbf{B}$ is a measure of the momentum flux density of the electromagnetic field and so it's time evolution has also to be factored in when accounting for the total momentum density. The energy density of the electric and magnetic fields can also make a significant contribution to the magnetofluid's pressure. Thus the momentum equation becomes

$$\frac{\partial}{\partial t}\left(\rho h \gamma^2 \, \mathrm{v}_i + (\mathbf{E} \times \mathbf{B})_i\right)$$
$$+ \frac{\partial}{\partial x_j}\left(\rho h \gamma^2 \, \mathrm{v}_i \, \mathrm{v}_j - \mathrm{E}_i \, \mathrm{E}_j - \mathrm{B}_i \, \mathrm{B}_j + \left(\mathrm{P} + \frac{1}{2}\left(\mathbf{E}^2 + \mathbf{B}^2\right)\right) \gamma \, \delta_{ij}\right) = 0 \quad \text{(C.4)}$$

Just as the magnetic energy contributed to the energy density for classical MHD, the electric and magnetic energy densities now contribute to the energy density of a magnetofluid. In electromagnetism, the Poynting flux also represents the flux of energy. Consequently, it makes a further contribution to the energy flux. The energy equation is therefore given by

$$\frac{\partial}{\partial t}\left(\rho h \gamma^2 - \mathrm{P} + \frac{1}{2}\left(\mathbf{E}^2 + \mathbf{B}^2\right)\right) + \frac{\partial}{\partial x_i}\left(\rho h \gamma^2 \, \mathrm{v}_i + (\mathbf{E} \times \mathbf{B})_i\right) = 0 \quad \text{(C.5)}$$

Faraday's law is already relativistically invariant. As a result, the evolution equation for the relativistic magnetic field is still given by

$$\frac{\partial \mathbf{B}}{\partial t} = \nabla \times (\mathbf{v} \times \mathbf{B}) \quad \text{(C.6)}$$

The magnetic field is still divergence-free, i.e. $\nabla \cdot \mathbf{B} = 0$. This completes our description of the special relativistic MHD equations.

The above equations for RMHD can be compared to the equations of classical MHD. The parallels are easy to spot. The relativistic flow equations also form a hyperbolic set of equations and have the same foliation of waves as the classical MHD equations. The same eigenvector degeneracies that plague classical MHD also plague relativistic MHD. The degeneracies have been catalogued in Anile (1989) and a set of eigenvectors that are suitable for computational work has been catalogued in Balsara (2001) and Anton et al. (2010).

**Appendix D) Brief Introduction to the HLL Riemann Solver**

The easiest way to describe an HLL Riemann solver is to resort to a wave model where all the flow structures between the two extremal states are replaced by a single constant state $U^*$ which corresponds to a single flux $F^*$. This extreme simplification of the one-dimensional Riemann problem is illustrated in Fig. D.1a. The three constant states of the HLL Riemann solver are given by

$$U_{HLL}^{(RS)} = \begin{cases} U_L & \text{if } S_L > 0 \\ U^* & \text{if } S_L \leq 0 \leq S_R \\ U_R & \text{if } S_R < 0 \end{cases} \qquad (D.1)$$

Integrating the one-dimensional conservation law in its weak form over the rectangle ABCD in space and time, and using Gauss' Law, we get an expression for the constant resolved state $U^*$ as

$$U^* = \frac{S_R U_R - S_L U_L - (F_R - F_L)}{S_R - S_L} \qquad (D.2)$$

The flux from the HLL Riemann solver can now be written as

$$F_{HLL}^{(RS)} = \begin{cases} F_L & \text{if } S_L > 0 \\ F^* & \text{if } S_L \leq 0 \leq S_R \\ F_R & \text{if } S_R < 0 \end{cases} \qquad (D.3)$$

Integrating the one-dimensional conservation law in its weak form over the rectangle ABFE in space and time, and using Gauss' Law, we get an expression for the constant resolved flux $F^*$ as

$$F^* = \left[\frac{S_R}{S_R - S_L}\right] F_L - \left[\frac{S_L}{S_R - S_L}\right] F_R + \left[\frac{S_R S_L}{S_R - S_L}\right] (U_R - U_L) \qquad (D.4)$$

Notice that the derivations of eqns. (D.2) and (D.4) are based on strictly formal considerations of conservation and so it would be wrong to assert that $F^* = F(U^*)$. Observe too that when $S_L \leq 0 \leq S_R$ the first two terms on the right hand side of eqn. (D.4) constitute a convex combination of left and right fluxes while the third term carries the dissipation.

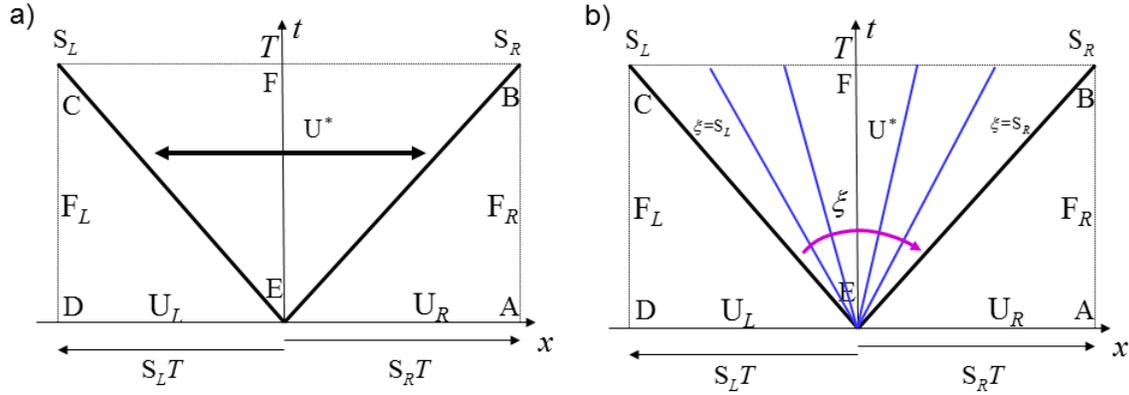

Fig. D.1a shows the wave model that is used for the HLL Riemann solver. Fig. D.1b shows an analogous wave model that is used for the HLLI Riemann solver. The straight lines in Fig. D.1b indicate similarity variable $\xi=x/t$. The solution inside the Riemann fan can take on different values along each different value of the similarity variable. Judicious introduction of sub-structure in the Riemann fan can reduce the numerical diffusion from the Riemann solver.

The description of the HLL Riemann solver is still incomplete. If the solution of an exact Riemann problem were available then it would be possible to specify $S_L$ and $S_R$ in eqns. (D.2) and (D.4). However, it would be self-defeating to solve the exact Riemann problem in order to specify the extremal wave speeds in an approximate Riemann solver. Thus it is advantageous to arrive at those two wave speeds via some other strategy. A suggestion by Einfeldt (1988) consists of using the extremal speeds of the linearized Riemann solver. These speeds are easily obtained without making a computationally costly evaluation of the eigenvectors. Let us intantiate for the Euler system. Let $\lambda^1(U_L)$ be the left-going sound speed in the state $U_L$; and let $\lambda^M(U_R)$ be the right-going sound speed in the state $U_R$. Let $\bar{v}_x$ be the Roe-averaged x-velocity and let $\bar{c}_s$ be the Roe-averaged sound speed. Thus we have two choices:-

$$S_L \equiv \min\left(\lambda^1(U_L), \bar{v}_x - \bar{c}_s, -\varepsilon\right) \quad ; \quad S_R \equiv \max\left(\lambda^M(U_R), \bar{v}_x + \bar{c}_s, \varepsilon\right) \quad \text{(D.5a)}$$

or

$$S_L \equiv \min\left(\lambda^1(U_L), \bar{v}_x - \bar{c}_s\right) \quad ; \quad S_R \equiv \max\left(\lambda^M(U_R), \bar{v}_x + \bar{c}_s\right) \quad \text{(D.5b)}$$

Here "$\varepsilon$" is some very tiny positive number. If eqn. (D.5a) is used, one does not need to explore the three cases in eqns. (D.1) and (D.3); this results in a very simple computer implementation. In most reasonable situations, the above equations provide a good estimate of the extremal signal speeds. As a matter of practical usage, eqn. (D.5) works well.

The HLL Riemann solver can represent extremal shock waves exactly. I.e. right and left going fast magnetosonic shocks in MHD flows can be represented exactly. Because the Riemann fan is opened, it also enforces entropy properly in situations where rarefaction fans might be present. It also has good positivity properties, so that the resolved state U* will have positive density and pressure if the left and right states are physical. Its one failing is that it washes out all the intermediate waves in the Riemann fan. As a result, the intermediate waves are treated diffusively. This means that an entropy wave will be diffused on the mesh. Likewise, an Alfven wave in MHD or RMHD flow will be treated diffusively by the HLL Riemann solver. The HLLI Riemann solver, which we describe in the next appendix, overcomes this failing.

**Appendix E) Brief Introduction to the HLLI Riemann Solver**

The HLLI Riemann solver was developed by Dumbser and Balsara (2016) based on insights derived from Balsara (2014) and Einfeldt et al. (1991). We do not discuss all the details here. But it is worth pointing out that this is a very general purpose Riemann solver that can apply to all manner of hyperbolic systems. Here we present details associated with the HLLI Riemann solver for hyperbolic conservation laws since many of the PDEs of use in astrophysics have such a conservation law structure. The HLLI Riemann solver can be retrofitted to any HLL Riemann solver can will improve its solution quality. Furthermore, if we only wish to improve a subset of the full set of waves in the hyperbolic system then we only need to evaluate the eigenvalues and eigenvectors for that subset of waves. This is an especially desirable feature for computational astrophysics because we usually only want to improve the quality of the entropy wave and the Alfven waves in MHD and RMHD simulations and the eigenstructure for that subset of waves is quite easy to evaluate.

We begin by evaluating all the terms in the HLL Riemann solver. The HLLI Riemann solver is based on the realization that the similarity variable $\xi = x/t$ demarcates the sub-structure in the Riemann fan. Instead of having discrete jumps associated with each intermediate wave family in the Riemann fan, one can give each intermediate family of waves a linear profile, i.e. a profile that varies linearly with the similarity variable $\xi = x/t$ within the Riemann fan. Of course, the profile has to follow a very specific form if it is to fulfil our goals of reducing the dissipation associated with the intermediate wave family that is being considered. Say we want to improve the representation of the $p^{\text{th}}$ wave family. We can then modify eqn. (D.1) so that we have

$$U_{\text{HLLI}}^{(RS)}(\xi) = \begin{cases} U_L & \text{if } \xi < S_L \\ U^* + \underbrace{2\,\delta^p}_{\text{Weight}}\; \underbrace{r^p \left[ l^p \cdot (U_R - U_L) \right]}_{\text{Contribution from } p^{th} \text{ wave}} \underbrace{\frac{(\xi - (S_R + S_L)/2)}{(S_R - S_L)}}_{\text{Linear profile}} & \text{if } S_L \leq \xi \leq S_R \\ U_R & \text{if } S_R < \xi \end{cases}$$

(E.1)

The state $U^*$ is still the HLL state and is given by eqn. (D.2). Here $l^p$ and $r^p$ are orthonormalized left and right eigenvectors respectively for that wave family that propagates with a speed $\lambda^p$. The eigenvectors $l^p$ and $r^p$ and the eigenvalue $\lambda^p$ can be evaluated by using the state $U^*$. The term $\delta^p$ is a special weight that we will soon specify. Notice that $\left[l^p \cdot (U_R - U_L)\right]$ is just an eigenweight so that $r^p \left[l^p \cdot (U_R - U_L)\right]$ is just the contribution from the $p^{\text{th}}$ wave family. The linear profile is given by $\left(\xi - (S_R + S_L)/2\right)/(S_R - S_L)$ in the above formula. The weight $\delta^p$ is then given by

$$\delta^p = 1 - \frac{\min(\lambda^p, 0)}{S_L} - \frac{\max(\lambda^p, 0)}{S_R} \qquad (E.2)$$

This specific form of the weight is designed to produce the least amount of dissipation. The theory supporting this claim is provided in Appendix B of Dumbser and Balsara (2016). The flux corresponding to eqn. (E.1) is evaluated at the zone boundary $(\xi = 0)$ and it is given by

$$F_{\text{HLLI}}^{(RS)} = \begin{cases} F_L & \text{if } 0 < S_L \\ F^* - \frac{S_R S_L}{(S_R - S_L)} \delta^p \underbrace{r^p \left[l^p \cdot (U_R - U_L)\right]}_{\text{Contribution from } p^{\text{th}} \text{ wave}} & \text{if } S_L \leq 0 \leq S_R \\ F_R & \text{if } S_R < 0 \end{cases} \qquad (E.3)$$

Here $F^*$ is the numerical HLL flux obtained from eqn. (D.4). We can now clearly see how the HLLI Riemann solver is built on top of the HLL Riemann solver. It consists of making additional contributions to the numerical flux that we obtain from the HLL Riemann solver. The "I" in HLLI refers to the intermediate family of waves that are represented in the Riemann solver.

The good news is that the contribution of each individual wave family can be treated additively. Furthermore, we may only be interested in a subset of the intermediate waves associated with a hyperbolic system. Let us say that we are interested in "$M$" waves. In that case, we only need to evaluate the eigenvalues and eigenvectors for the subset of waves that are of interest to us. We can then write the numerical flux from the HLLI Riemann solver as

$$F_{\text{HLLI}}^{(RS)} = \begin{cases} F_L & \text{if } 0 < S_L \\ F^* - \frac{S_R S_L}{(S_R - S_L)} \sum_{m=1}^{M} \delta^m \, r^m \left[l^m \bullet (U_R - U_L)\right] & \text{if } S_L \leq 0 \leq S_R \\ F_R & \text{if } S_R < 0 \end{cases} \qquad (E.4)$$

Appendix C of Dumbser and Balsara (2016) gives pseudocode that is suitable for computer implementation.